\documentclass[twoside,12pt]{article}
\usepackage{epsfig}
\usepackage{rotating}

\def\Journal#1#2#3#4{{#1} {#2} (#4) #3 }
\def\NPBPS{{\em Nucl. Phys.} B (Proc. Suppl.)}
\def\PPNP{\em Prog. Part. Nucl. Phys.}
\def\PLB{{\em Phys. Lett.} B}
\def\RMP{\em Rev. Mod. Phys.}
\def\IJMPA{{\em Int. J. Mod. Phys.} A}
\def\PRC{{\em Phys. Rev.} C}
\def\PRD{{\em Phys. Rev.} D}
\def\PRL{\em Phys. Rev. Lett.}

\def\AP{\em Ann. Phys.}
\def\PREP{\em Phys. Rep.}
\def\PR{\em Phys. Rev.}
\def\PL{\em Phys. Lett.}
\def\EPJA{{\em Eur. Phys. J.} A}
\def\EPJC{{\em Eur. Phys. J.} C}
\def\ZPA{{\em Z. Phys.} A}
\def\ZPC{{\em Z. Phys.} C}
\def\ZP{\em Z. Phys.}
\def\NPA{{\em Nucl. Phys.} A}
\def\NPB{{\em Nucl. Phys.} B}
\def\JPG{{\em J. Phys.} G}
\def\FBSS{\em Few-Body Systems Suppl.}
\def\ZNF{\em Z. Naturforsch.}
\def\SJNP{\em Sov. J. Nucl. Phys.}
\def\NIMA{{\em Nucl. Instr. and Meth.} A}
\def\NCA{{\em Nuovo Cimento} A}

\def\LNC{\em Lett. Nuovo Cimento}
\def\IEEE{\em IEEE Trans. Nucl. Sci.}
\def\PINL{\em $\pi$N Newslett.}
\def\JPSJ{\em J. Phys. Soc. Japan}
\def\APPB{{\em Acta Phys. Pol.} B}
\def\BAEX{\em Baryon Excitations, Lectures of the COSY Workshop held at the
Forschungszentrum J{\"u}lich, 2 to 3 May 2000, Edts. T. Barnes and H.P. Morsch,
ISBN 3-89336-273-8}

\def\d{\mbox{D$_{13}$(1520)}}
\def\s{\mbox{S$_{11}$(1535)}}
\def\ss{\mbox{S$_{11}$(1650)}}
\def\f{\mbox{F$_{15}$(1680)}}
\def\ff{\mbox{F$_{37}$(1950)}}
\def\p{\mbox{P$_{33}$(1232)}}
\def\pp{\mbox{P$_{11}$(1440)}}

\begin{document}

\title{\vspace{1cm} Study of Non-Strange Baryon Resonances with Meson 
Photoproduction}
\author{B.\ Krusche$^{1}$ and S.\ Schadmand$^2$ 
\\
$^1$Institut f\"ur Physik, Universit\"at Basel,
CH-4056 Basel, Switzerland\\
$^2$II Physikalisches Institut, Universit\"at Giessen, D-35392 Giessen, 
Germany 
}
\maketitle

\vspace*{-1.1cm}
\begin{abstract}
\noindent{Photoproduction} of mesons is an excellent tool for the study of 
nucleon resonances. Complementary to pion induced reactions, photoproduction 
on the free proton contributes to the determination of the basic properties 
of nucleon resonances like excitation energy, decay widths, spin, and the 
coupling to the photon. Photoproduction from light nuclei, in particular from 
the deuteron, reveals the isospin structure of the electromagnetic excitation
of the nucleon. During the last few years, progress in this field has been 
substantial. New accelerator facilities combined with state-of-the-art detector 
technologies have pushed the experiments to unprecedented sensitivity and 
precision. The experimental progress has been accompanied by new developments 
for the reaction models, necessary to extract the properties of the 
nucleon states, and for modern hadron models which try to connect these 
properties to QCD. The emphasis of this review lies on the experimental side 
and focuses on experiments aiming at precise studies of the low-lying nucleon 
resonances. 
\end{abstract}
\vspace*{-0.7cm}
\tableofcontents
 
\section{Introduction}
During the last 30 years Quantum Chromo Dynamics (QCD), the formal theory 
of the color interactions between quarks, emerged as the theory of the 
strong interaction. The perturbative approach to this theory has been 
extremely successful in the high energy regime where it has been tested by 
numerous experiments. However, in the low energy regime, the perturbative 
approach is meaningless, and a solution of QCD is not known on a scale 
typical for the mass of the nucleon and its excited states, where the 
strong coupling constant becomes large. Lattice gauge calculations have 
provided results for the ground state properties and very recently also 
for excited states of the nucleon 
(see e.g. 
\cite{Lee_02}-\cite{Richards_01}). 
However, the prediction of the excitation spectrum of nucleons is still out 
of reach even for the most powerful computer systems. This situation offers 
both a challenge and a chance: we do want to understand the physics laws 
governing the building blocks of matter at low energies, in the regime 
where we encounter them in nature. On the other hand, it is obvious that 
the complex many-body system `nucleon' offers the ideal testing ground for 
concepts of the strong interaction in the non-perturbative regime. 

Perturbative QCD at high energies deals with the interactions of quarks and 
gluons. However, our picture of the nucleon is more related to effective 
constituent quarks and mesons that somehow subsume the complicated low 
energy aspects of the interaction generating the nucleon many-body structure 
of valence quarks, sea quarks, and gluons. Therefore, the most important 
step towards an understanding of nucleon structure is the identification 
of the relevant effective degrees-of-freedom, which naturally must reflect 
the internal symmetries of the underlying fundamental interaction. This 
step is attempted in the framework of constituent quark models of baryons, 
which have contributed substantially to our understanding of the strong 
interaction. In a sense, these models were the starting point for the 
development of QCD. 

The most basic version of the constituent quark model, using a harmonic 
oscillator potential, had its origin in the work of Gell-Mann 
\cite{Gell-Mann_64}, Greenberg \cite{Greenberg_64}, 
Dalitz \cite{Dalitz_66}, and collaborators. Copley, Karl, and Obryk 
\cite{Copley_69} and Feynman, Kislinger, and Ravndal \cite{Feynman_71} 
gave the first clear evidence for the underlying SU(6)$\otimes$O(3) 
symmetry of the hadron spectrum. Later, Koniuk and Isgur \cite{Koniuk_80a} 
laid the basis for the description of the electromagnetic and strong decays 
in the framework of the quark model. The classification of the mesons and 
baryons into the well-known multiplet structures as derived from the 
symmetry, and the description of the hadronic excitation spectrum with only 
few fitting parameters were a striking success of the model. An excellent 
overview over modern quark models is given by Capstick and Roberts 
\cite{Capstick_00}. Most of the models start from three equivalent 
constituent quarks in a collective potential. The masses of the up and down 
constituent quarks range from 220 MeV for relativistic models to 330 MeV 
for non-relativistic models. Here, the quarks are not point-like but have 
electric and strong form factors. The potential is generated by a confining 
interaction, for example in the flux tube picture, and the quarks interact via 
a short range residual interaction. This fine-structure interaction, usually 
taken as color magnetic dipole-dipole interaction mediated via 
one-gluon-exchange (OGE), is responsible for the spin-spin and spin-orbit 
dependent terms. 
%
%
%
\begin{figure}[hbt]
\begin{minipage}{10.0cm}
{\mbox{\epsfysize=2.8cm \epsffile{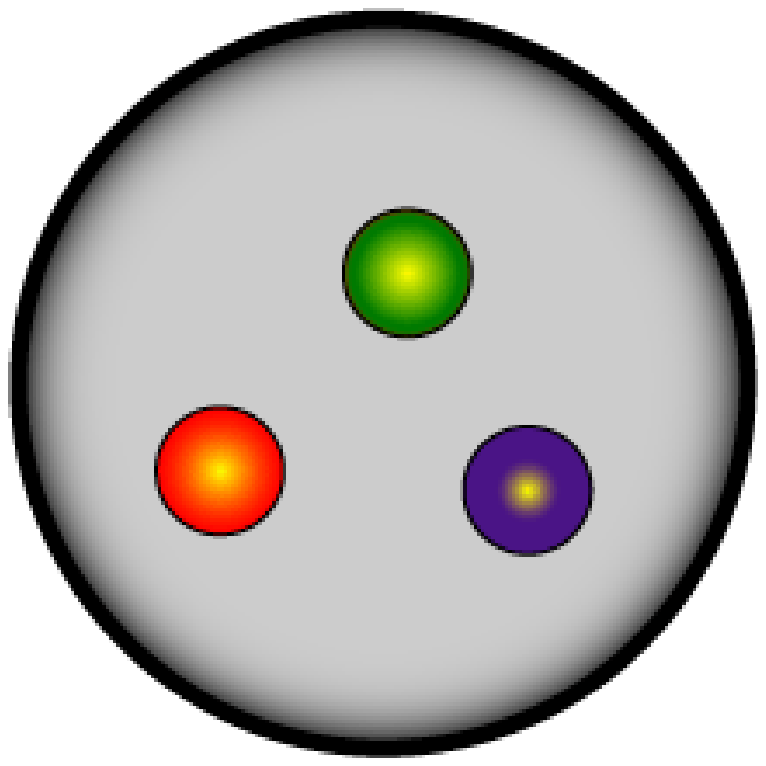}
\hspace*{0.5cm}
\epsfysize=2.8cm \epsffile{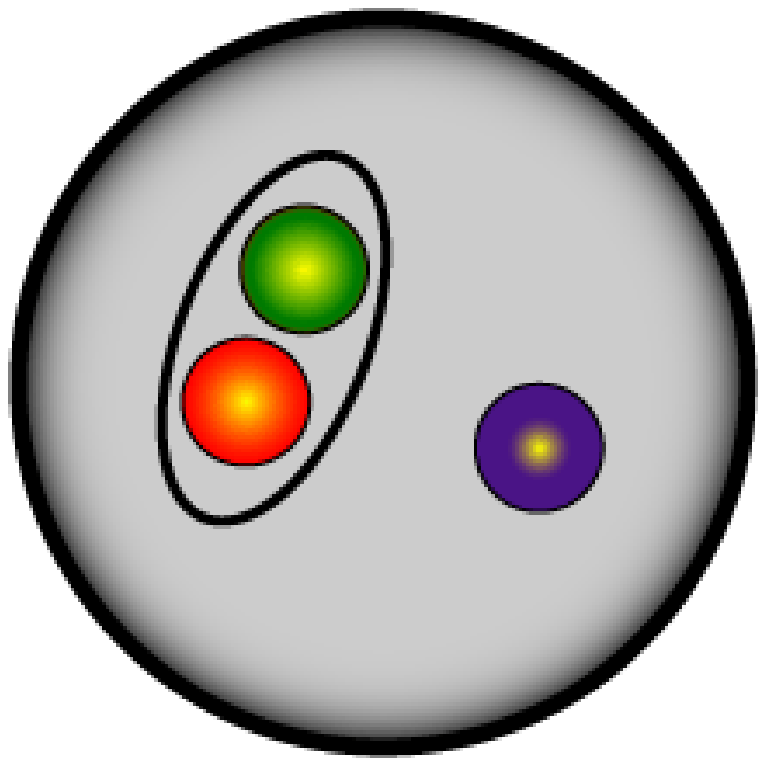}
\hspace*{0.5cm}
\epsfysize=2.8cm \epsffile{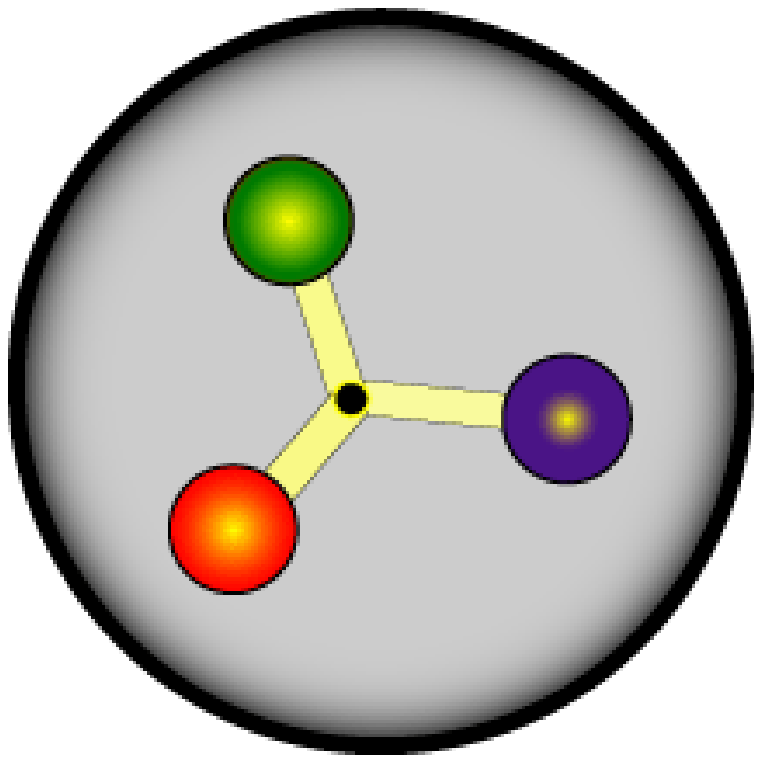}}}
\end{minipage}
\hspace*{0.5cm}
\begin{minipage}{7.5cm}
\caption{Effective degrees-of-freedom in quark models: three equivalent
constituent quarks, quark - diquark structure, quarks and flux tubes?
}       
\label{fig_01}
\end{minipage}
\end{figure}

%
However, alternative concepts are not a priori ruled out. In fact, models 
have been proposed which are based on other degrees-of-freedom 
(see fig.\ref{fig_01}).
One group of models describes the nucleon structure in terms of a 
quark - diquark ($q-q^2$) cluster (see Anselmino et al. 
for a review \cite{Anselmino_93}).
If the diquark is sufficiently strongly bound, low lying excitations of the 
nucleon will not include excitations of the diquark. Therefore, these models  
predict {\em fewer} low-lying excited states of the nucleon than the 
conventional quark models. On the other hand, the number of states would be 
{\em increased} in an algebraic model proposed by Bijker et al. 
\cite{Bijker_94,Bijker_97}. The model is based on collective excitations of 
string-like objects carrying the quantum numbers of the quarks. Radial 
excitations arise from rotations and vibrations of these strings. Alternative 
models are available not only in view of the `constituents' but also in 
view of the residual interaction. In conventional models, this interaction 
is due to OGE. Meanwhile, Glozman and Riska \cite{Glozman_96} have developed 
a model where the residual interaction is due to the exchange of Goldstone 
bosons, taken to be the pseudo-scalar octet mesons. This is a radically 
different picture since in this case gluons do not contribute at all to the 
nucleon structure. In addition, this model leads to a quark - diquark 
clustering effect giving rise to specific selection rules for the decay 
properties of excited baryons \cite{Glozman_96a}.

%
%
%
\begin{figure}[hbt]
\begin{minipage}{10.cm}
\hspace*{1.3cm}{\mbox{\epsfysize=6.3cm \epsffile{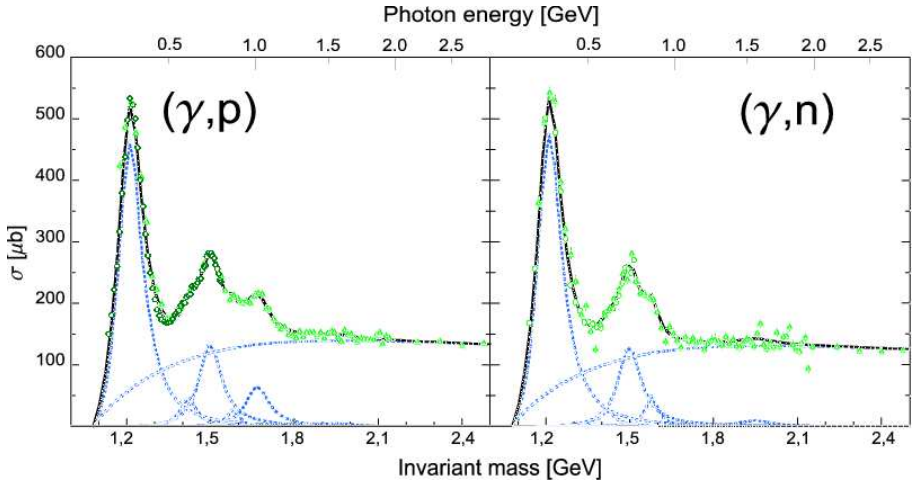}}}
\end{minipage}
\hspace*{2.cm}
\begin{minipage}{6.0cm}
\caption{Cross section for total photoabsorption on the proton (left hand side)
and the neutron (right hand side) \cite{Bianchi_96}. Points: 
measured data, curves: fit of Breit-Wigner shapes of nucleon 
resonances (\p, \pp, \d, \s, \f (only for proton), and \ff) and a 
smoothly varying background.}
\label{fig_02}       
\end{minipage}
\end{figure}
%

The number of excited states with definite quantum numbers follows directly
from the number of effective degrees-of-freedom and their quantum numbers 
in the models. Consequently, a comparison of the experimentally determined 
excitation spectrum to the model predictions should allow the determination 
of the number of degrees-of-freedom. However, from the experimental point of 
view the situation is quite different from atomic or nuclear physics.
The dominant decay channel of nucleon resonances is the hadronic decay via the 
emission of mesons. Thus, the lifetimes of the excited states are typical for 
the strong interaction ($\tau\approx10^{-24}$s) with corresponding widths of 
a few 100 MeV. The spacing of the resonances is often no more than a few 10 
MeV such that the overlap is large. This makes it difficult to identify and 
investigate individual states, as demonstrated in fig.~\ref{fig_02}. 
The figure shows the cross section for total photoabsorption 
\cite{Bianchi_96}, i.e. for the reaction $\gamma N\rightarrow NX$ on the
proton and the neutron. The latter was measured in quasifree kinematics from 
the deuteron. In this simple picture the cross section was fitted with a 
smooth background and Breit-Wigner curves for the excited states which are 
labeled in the usual notation as $L_{2I2J}(W)$. Here, $W$ is the mass, 
$L$=0,1,2,... the angular momentum for the decay into the $N\pi$-channel given 
in the spectroscopic notation as S,P,D,... and $I$, $J$ are isospin and spin of 
the resonances, respectively. Only the lowest lying excited state of the 
nucleon, the $\Delta$-resonance (\p), corresponds to an isolated peak in the 
spectrum. At masses around 1500 MeV, several resonances (\pp, \d, \s) contribute
to the broad resonance structure observed in the spectrum. 
This energy regime is called the second resonance region.
Inclusive measurements like photoabsorption do not allow a detailed 
investigation of such closely spaced resonances.

Which experimental tools are available for the study of nucleon resonances?
The dominant decay of any excited nucleon state is the emission of mesons
via the strong interaction. Electromagnetic decays via photon emission have
typical branching ratios below the 1\% level and are extremely difficult to
identify in the presence of large background levels. Meson production in 
hadron induced reactions profits from large cross sections and has been 
intensely used for the study of nucleon resonances. 

Beams of stable baryons like protons, deuterons, and $\alpha$-particles have 
been used at many accelerators. Recently, the cooler rings at CELSIUS and COSY 
have been active in this field. Purely baryon induced reactions favor the 
exploration of the isospin degree of freedom. However, the interpretation is 
quite involved since initial and final state are governed by the strong 
interaction. Here, the presence of at least two baryons in the final state 
gives rise to complex final state interaction effects. Furthermore, due to 
the large mass of the beam particles, high beam energies must be employed in 
order to access the resonance regions. Much of the more recent work with baryon 
beams concentrates on high sensitivity studies of meson production thresholds. 
An excellent overview over this topic is given by Moskal, Wolke, Khoukaz, and 
Oelert \cite{Moskal_02}.

The most widely used reactions for the study of nucleon resonances use 
beams of long-lived mesons. In particular, the elastic scattering of charged 
pions off the nucleon, and inelastic pion induced reactions, contributed to 
the experimental data base. Again, the hadronic cross sections are large, and 
the isospin degree of freedom is accessible. Meanwhile, the final state with 
only one baryon is less complicated. Sophisticated multipole analyses of pion 
induced reactions, followed by a parameterization of the partial waves in terms of 
resonances and background contributions, have been performed by different 
groups (see e.g.
\cite{Cutkosky_79}-\cite{Vrana_00}).
These results still form the backbone of nucleon resonance properties. 
However, the exclusive use of pion induced reactions would bias the data base 
for resonances coupling weakly to the $\pi N$ channel. Indeed, a comparison of 
the excitation spectrum predicted 
by modern quark models to the experimentally established set of nucleon 
resonances results in the problem of `missing resonances': many more states 
are predicted than have been observed. But is this evidence for inept effective
degrees-of-freedom in the models or a simple experimental bias? Already more 
than 20 years ago, Koniuk and Isgur have argued in a paper \cite{Koniuk_80} 
entitled `Where have all the resonances gone?' that the reason for the mismatch 
is the decoupling of many resonances from the partial wave analysis of pion 
scattering. These resonances can only be found when other initial and/or final 
states are investigated. In fact, recent quark models \cite{Capstick_94}, 
predict a number of the unobserved resonances to have large decay branching 
ratios for the emission of mesons other than pions.
In this case, the nucleon should ideally be excited by scattering of the 
respective mesons. However, most of them are short lived making the preparation 
of secondary beams impossible. The use of reactions induced by the 
electromagnetic interaction offers an alternative. 

Detailed tests of quark models cannot be achieved with 
excitation spectra alone. In this sense, the situation for nucleon physics is 
similar to nuclear physics. In both cases, the excitation energies and quantum 
numbers of the states do not provide the most sensitive observables. More 
crucial tests come from the transitions between the states which reflect their 
internal structure and are more sensitive to the model wave functions. 
Photo- and electroproduction of mesons is particularly interesting for this 
purpose since the rich information connected to the electromagnetic transition 
amplitudes can be accessed in addition to the dominant hadronic decay modes. 
The photon couples only to the spin-flavor degrees-of-freedom of the quarks 
revealing their spin-flavor correlations which are related to the configuration
mixing predicted by the models. There is a price to pay for the advantage. The
electromagnetic cross sections are naturally much smaller than the hadronic 
ones. More importantly, photon induced reactions can have significant 
non-resonant contributions, called `background'. For example, nucleon Born 
terms or vector meson exchange complicate the extraction of the resonance 
properties. 
%
%
%
\begin{figure}[thb]
\begin{minipage}{0.cm}
{\mbox{\epsfysize=6.2cm \epsffile{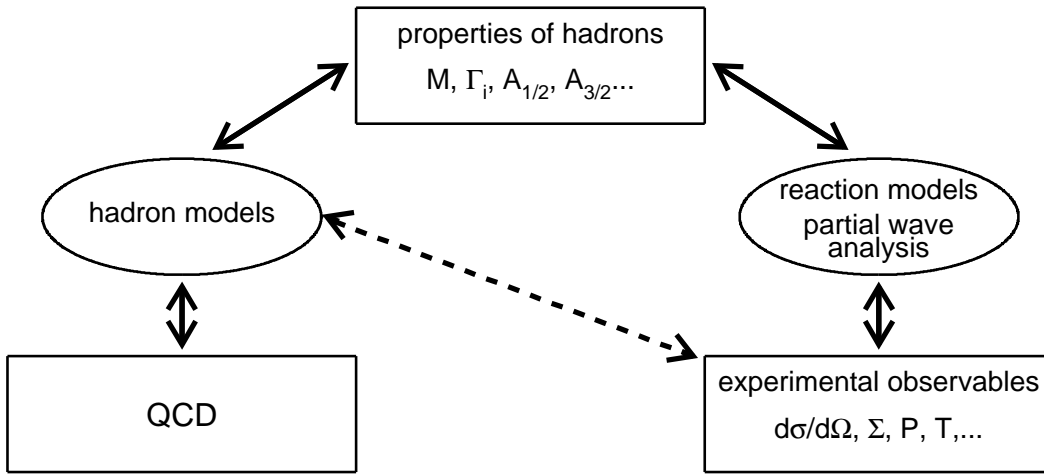}}}
\end{minipage}
\hspace*{13.8cm}
\begin{minipage}{4.2cm}
\vspace*{-1.3cm}
\caption{Schematic representation of the relation between experimental 
observables, baryon properties, and QCD via reaction and hadron models.}
\label{fig_03}       
\end{minipage}
\end{figure}
%
Therefore, it is mandatory to use reliable reaction models for the 
analysis of the photoproduction data. This situation is schematically 
illustrated in fig.~\ref{fig_03}. 
Reaction models are used to 
extract hadron properties like excitation energies, widths, and branching 
ratios to different decay channels from the  physical observables. QCD inspired
models of the hadron are used to make predictions for these properties. 
More recently, quark models of the nucleon have been developed which directly 
make predictions for the physical observables (see sec. \ref{sec:nucleon}).

The detailed understanding of the elementary process of resonance excitation 
on the free nucleon is the basis for the investigation of baryon resonances 
in the nuclear medium. In the case of bound nucleons, even properties like 
mass and width, which may be influenced by the nuclear medium, are mostly 
unknown. Such effects arise, for example, from the additional decay channel 
$R N\rightleftharpoons NN$ and from Pauli blocking of the $R\rightarrow N\pi$ 
decay where $R$ stands for a $N^{\star}$ or $\Delta$ resonance. However, it 
came as a surprise when measurements of the total photoabsorption cross 
section on nuclei indicated a depletion of the resonance structure in the 
second resonance region \cite{Frommhold_92,Bianchi_93}. Bianchi et al.
\cite{Bianchi_94} reported that, while in the $\Delta$-resonance region 
strength is only redistributed by broadening effects, strength is {\em missing}
in the \d ~region. On the other hand, measurements of exclusive reaction 
channels like $\eta$ photoproduction \cite{Roebig_96} or single $\pi^o$ 
photoproduction \cite{Krusche_01} did not find any effects beyond trivial 
final state interactions. An overview over the in-medium properties of nucleon 
resonances will be given elsewhere.
 
\begin{figure}[hbt]
\centerline{\epsfysize=7.6cm \epsffile{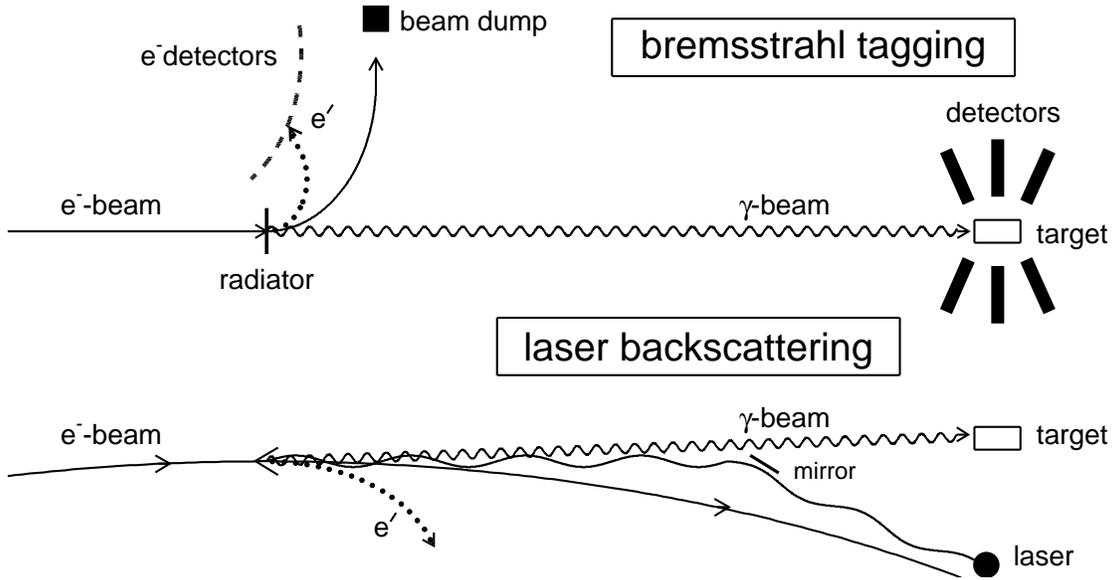}}
\caption{Schematic illustration of tagged photon facilities from 
bremsstrahlung (upper part) and Compton scattered laser beams (lower part).
The solid lines indicate the electron beams, the dotted lines scattered
electrons, and the curled lines the photon beams. The scattered electrons are
deflected by magnetic fields (not indicated). Position of detectors is only
indicated in the upper part. In both cases electron and production detectors 
are operated in coincidence.}
\label{fig_00}       
\end{figure}

In the present paper, we will review recent progress in the investigation
of nucleon resonances with meson photoproduction. The restriction to reactions
induced by real photons - as opposed to electron scattering experiments where 
the nucleon is excited by virtual photons - is not kept strictly. 
Results from electroproduction experiments are included when the dependence of 
the observables on the four-momentum transfer $Q^2$ is directly relevant for 
the discussion. On the experimental side, the progress made in accelerator and 
detector technology during the last fifteen years has considerably enhanced our 
possibilities to investigate the nucleon with different probes. In particular, 
the new generation of electron accelerators, CEBAF at JLab in Newport News, 
ELSA in Bonn, ESRF in Grenoble, MAMI in Mainz, and SPring8 in Osaka are 
equipped with tagged photon facilities and state-of-the-art detector systems.
Quasi monochromatic photon beams provided by photon tagging are the 
working horse of the real photon programs. Two different techniques
are used to produce photon beams: bremsstrahlung and Compton backscattering.
The principles are sketched in fig. \ref{fig_00}. In the first case, the 
electron beam from the accelerator impinges on a radiator (usually a thin metal
foil). Scattered electrons produce bremsstrahlung with the typical 
$1/E_{\gamma}$ spectral distribution. In the second case, photons from a 
laser are scattered from electrons circulating in a storage ring.
A certain advantage of this technique is that polarization degrees of freedom 
are transfered from the laser photons to the Compton back-scattered high energy 
photons. On the other hand, beam intensities are limited since high intensity
laser beams reduce the lifetime of the stored electron beams. In both
cases, the energies of the photon and the scattered electron are correlated
via the known incident electron (and photon) energy. The scattered electrons
are momentum selected with magnetic fields and detected in the focal 
plane of the magnetic spectrometers. The production detectors are operated in 
coincidence with the electron detectors so that the incident photon energies 
are known event-by-event within the resolution of the tagging detector. 
The bremsstrahlung technique is used at ELSA, JLab (CLAS), and MAMI.
Laser backscattering is employed at BNL (LEGS), at ESRF (GRAAL), and at SPring8
(LEPS).

%
%
%
%
    
These facilities have opened the way to meson photoproduction experiments of 
unprecedented sensitivity and precision. In some areas, the cross section of 
photon induced reactions is currently known more precisely than the cross 
section of the corresponding pion induced reactions. This is the case for 
$\eta$ meson production although the pion induced yields are larger by about 
two orders of magnitude. The experimental test of nucleon models can proceed 
along two different roads. The problem of missing resonances can be attacked 
by a large scale survey investigating many different final states 
($N\pi$, $N\pi\pi$, $N\eta$, $N\eta'$, $N\omega$, $N\rho$ etc.) 
over a large energy range. An instructive discussion of `Guidelines for 
identifying new Baryon 
Resonances' \cite{Manley_03} and `How many $N^{\star}$ do we need?' 
\cite{Bennhold_03} is given by Bennhold and Manley. Secondly, the 
low lying resonances can be studied in great detail providing data for 
precision tests of the models. Here, the availability of linearly and 
circularly polarized photon beams and polarized targets has provided access 
to observables, which are sensitive to specific resonances. The 
present review concerns itself with the second approach where the results
are in a more mature state. This situation reflects the 
fact that the Mainz MAMI accelerator, which is limited in energy to 880 MeV, 
has been operational for more than ten years. The investigation 
of high lying resonances, in particular at JLab and ELSA, started more recently. 

\begin{figure}[hbt]
\centerline{\epsfysize=11.5cm \epsffile{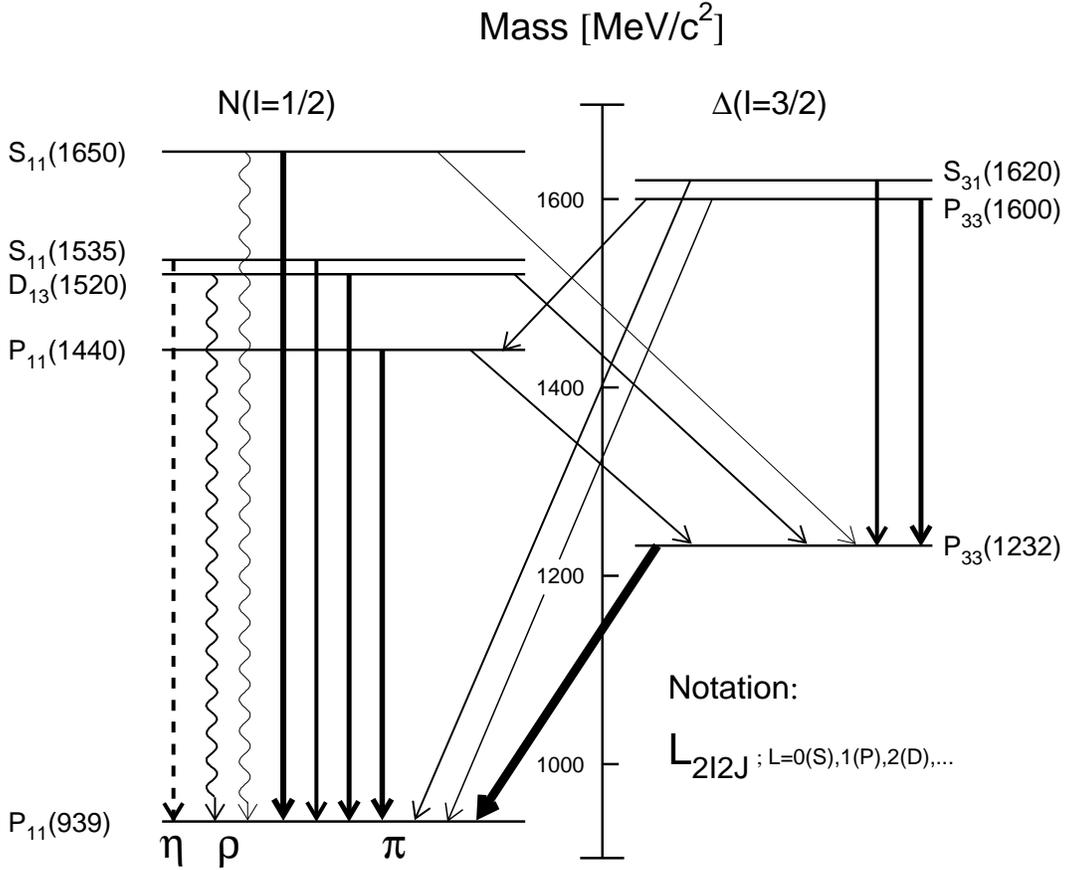}}
\caption{Low lying excited states of the nucleon \cite{PDG} with isospin 
I=1/2 (left hand side) and isospin I=3/2 (right hand side). The arrows indicate
the decays via pion emission (solid), $\eta$-emission (dashed) and 
$\rho$-emission (curled). The thickness of the arrows scales with the
branching ratios of the respective decays. 
Weak decay branchings have been omitted.}
\label{fig_04}       
\end{figure}

The relevant low energy excitation scheme of the nucleon with prominent 
transitions via meson emission is summarized in fig.~\ref{fig_04}. 
%
%
%
%
Most resonances have comparable branching ratios into the $N\pi$ final state, 
so that their contribution to pion photoproduction is mainly determined by 
their photon couplings. Single pion production will be discussed in detail 
in sec. \ref{sec:Delta} where new precise studies of the properties of the 
\p -resonance are presented. In the second resonance region, single pion 
production is dominated by the decay of the \d -resonance, which is discussed 
in sec. \ref{ssec:D13}.   

Even in this low energy region, the emission of heavier mesons is important 
for some states. The most selective channel is the photoproduction of 
$\eta$-mesons which is dominated by the excitation of the 
\s-resonance. This selectivity comes partly from the fact that 
$\Delta$-resonances cannot decay to the nucleon ground state via emission of 
the isoscalar $\eta$. More importantly, the \pp - and \d-resonances have 
very small decay branching ratios into $N\eta$ since they need to decay with 
relative orbital angular momentum $l=1,2$. Close to threshold, these $l$-values
are strongly suppressed as compared to the s-wave decay of the \s. The 
\pp-resonance even lies below the $\eta$-production threshold so that only 
its high energy tail could contribute.  On the other hand, the large difference
in the $N\eta$ decay branching ratios of the first and second 
S$_{11}$-resonance  must reflect a different structure of the states. 
Recent results for the photoproduction of $\eta$ mesons are discussed in 
\ref{ssec:S11}. 

In addition to single pion and $\eta$ emission, double pion production
contributes to resonance decays in this energy region. The pion emission of 
some resonances does not only lead to the nucleon ground state, but also to 
higher lying states, in particular to the \p ~which subsequently decays via 
emission of a second pion. The other contribution comes from decays into the 
$N\rho$ channel with a final state pion pair from the $\rho$ meson decay.
The detailed experimental study of the double pion production reaction, 
allowing to a large extend the extraction of resonance contributions, is 
discussed in \ref{ssec:twopi}.

At excitation energies above the second resonance region further meson
production thresholds open. $\omega$ mesons are produced off the free proton 
above $E_{\gamma}$=1108 MeV, $\eta'$ mesons above 1447 MeV, and $\Phi$ mesons 
above 1573 MeV. In addition, the thresholds for open strangeness production 
open around 1 GeV (915 MeV for $K\Lambda$ and 1052 MeV for $K\Sigma$). 
Here, final states with open strangeness can couple to intermediate non-strange 
nucleon resonances. These couplings carry important information about the 
internal structure of the states. Finally, sequential decay chains of 
resonances, such as $\Delta^*\rightarrow\Delta\eta\rightarrow N\eta\pi$ will
become important. Progress in this field is rapid, and we will give examples
of recent results and ongoing efforts in the Conclusions and Outlook section.
The main part of this review concentrates on the low lying states in view of 
the fact that the available data base for pion, double pion, and 
$\eta$-photoproduction is by far superior.  

\newpage   
\section{Photoexcitation of Free and Quasi-free Nucleons}
\label{sec:nucleon}

The most prominent decay channel of excited states of the nucleon is the
emission of light mesons belonging to the ground state nonets of pseudo-scalar 
and vector mesons. Photoproduction of pions is by far the best explored channel.
However, as already mentioned, it is difficult to isolate the contributions 
from individual resonances if only pion production is studied, and bias arises 
for resonances which couple only weakly to $N\pi$. For this reason, the study
of photoproduction reactions involving heavier mesons or multiple meson 
production has attracted a lot of attention. The 
properties of the relevant mesons are summarized in app. 6.1.

The formalism of the photoproduction of pseudo-scalar mesons from nucleons is 
well known. Detailed reviews can be found in \cite{Drechsel_92,Benmerrouche_95}.
Here, only a brief description is given in order to establish the notation of 
the observables which will be used for the discussion of the results. The most 
general Lorentz and gauge invariant amplitude for the photoproduction of a 
pseudo-scalar particle from a nucleon can be written in the 
Chew-Goldberger-Low-Nambu (CGLN) parameterization \cite{Chew_57}:
\begin{equation}
\label{eq421}
{\cal{F}}=
iF_1\cdot\vec{\sigma}\cdot\vec{\epsilon}+
F_2(\vec{\sigma}\cdot\vec{q})(\vec{\sigma}\cdot(\vec{k}\times\vec{\epsilon}))+
iF_3(\vec{\sigma}\cdot\vec{k})(\vec{q}\cdot\vec{\epsilon})
+iF_4(\vec{\sigma}\cdot\vec{q})(\vec{q}\cdot\vec{\epsilon})
\end{equation}
where $\vec{k}$, $\vec{q}$ are momentum unit vectors of the photon and meson,
$\vec{\epsilon}$ is the polarization vector for a real photon of helicity 
$\lambda_{\gamma}=\pm 1$, and $\vec{\sigma}$ are the nucleon's spin matrices.
A different parameterization of the amplitude exists in terms of the helicities
of the initial and final state particles (see app. 6.2).

The differential cross section in the center of momentum (cm) frame for an 
unpolarized target and an unpolarized photon beam is given in terms of the
CGLN-amplitudes by:
\begin{eqnarray}
\label{eq41}
\frac{k^{\star}}{q^{\star}}\frac{d\sigma}{d\Omega} & = &
[|F_1|^2+|F_2|^2+\frac{1}{2}|F_3|^2+\frac{1}{2}|F_4|^2+Re(F_1F_3^{\star})] \\
& & +[Re(F_3F_4^{\star})-2Re(F_1F_2^{\star})]cos(\Theta^{\star})\nonumber \\
& & -[\frac{1}{2}|F_3|^2+\frac{1}{2}|F_4|^2+Re(F_1F_4^{\star})
      +Re(F_2F_3^{\star})]cos^2(\Theta^{\star})\nonumber \\
& & -[Re(F_3F_4^{\star})]cos^3(\Theta^{\star})\nonumber
\end{eqnarray}
where $q^{\star}$, $k^{\star}$ are meson and photon cm momenta, respectively,
and $\Theta^{\star}$ is the cm polar angle of the meson 
(throughout the paper a `$\star$' at kinematical variables indicates the cm
system, but note that for the amplitudes it means `complex conjugated').
The expressions for 
all polarization observables in terms of the $F_i$ or the helicity
amplitudes $H_i$ can be found in \cite{Barker_75,Drechsel_92,Knochlein_95}. 

The amplitude involves four complex functions (the $F_i$), and consequently 
complete information about the reaction requires the determination of seven 
independent real quantities (the overall phase is arbitrary) at each incident 
photon energy and each meson emission angle. A `complete' experiment 
does not only require the measurement of the differential cross section
$d\sigma/d\Omega$, the photon beam asymmetry $\Sigma$, the target asymmetry 
$T$, and the recoil nucleon polarization $R$ (referred to as $\cal{S}$-type
experiments). In addition, several double polarization observables which are
characterized as $\cal{BT}$- (beam-target), $\cal{BR}$- (beam-recoil), and 
$\cal{TR}$- (target-recoil) type have to be determined. The question which set 
of observables allows 
a unique determination of the amplitudes is not trivial and has been intensely 
discussed in the literature. Barker, Donnachie, and Storrow \cite{Barker_75} 
showed that in the transversity representation (see app. 6.2)
the four $\cal{S}$-type experiments determine the magnitude of the amplitudes. 
The additional measurement of at least three double polarization observables, 
not all from the same group, determines the phases up to discrete ambiguities. 
Furthermore, they argued that the measurement of five double polarization 
observables with no more than three from one of the groups $\cal{BT}$, 
$\cal{BR}$, $\cal{TR}$ is sufficient for a unique determination of all 
amplitudes (BDS-rule). Thus, nine measurements are required in total. 
However, Keaton and Workman \cite{Keaton_96} have shown that certain 
combinations of double polarization observables, which satisfy the BDS-rule, 
do not resolve all discrete ambiguities. On the other hand, Chiang and Tabakin 
\cite{Chiang_97} have proven that in addition to the $\cal{S}$-type experiments,
already four appropriately chosen double spin observables are sufficient for 
a unique determination of the amplitudes. This means that the `complete' 
experiment requires at least eight measurements. However, in general it is 
impractical to do this ideal experiment. Therefore, the analysis of meson 
photoproduction data often relies on reaction models. In particular, close to 
production thresholds, where only few partial waves contribute, differential 
cross sections alone often provide valuable information.
%
%
%
\begin{figure}[htb]
\begin{minipage}{6.cm}
{\mbox{\epsfysize=2.5cm \epsffile{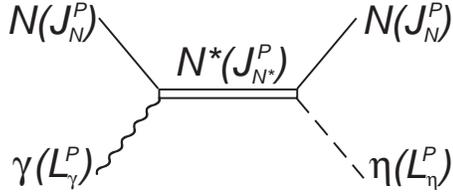}}}
\caption{Photoproduction of mesons via excitation of nucleon
resonances.} 
\label{fig_05}       
\end{minipage}
\end{figure}
%
\vspace*{-5.2cm}
\hspace*{6.4cm}
\begin{minipage}{12.cm}
We are primarily interested in the photoproduction of mesons via the 
intermediate excitation of resonances. This process is sketched in fig. 
\ref{fig_05} for $\eta$-production (the discussion is valid for 
any pseudo-scalar meson). It is advantageous to decompose initial and final 
state into multipole components since the intermediate resonance has definite 
parity and angular momentum. In the initial state, the photon with spin 
$\vec{s}_{\gamma}$ ($s=1$), orbital angular momentum $\vec{\tilde{l}}$ 
relative to the target nucleon, and total angular momentum 
$\vec{L}_{\gamma}=\vec{\tilde{l}}+\vec{s}_{\gamma}$ couples electromagnetically
to the nucleon with spin $\vec{J}_{N}$ ($J=1/2$) and parity ${P_N}=1$ to 
produce a resonance with spin $\vec{J}_{N^{\star}}$ and parity $P_{N^{\star}}$.
\end{minipage}

\vspace*{0.2cm}
\noindent{The} usual multipole expansion of the photon field gives rise to 
electric $EL$- and magnetic $ML$-multipoles with angular momentum $L$ and 
parity $P_{\gamma}=(-1)^L$ for the electric and $P_{\gamma}=(-1)^{L+1}$ for the 
magnetic case. This process obeys the following selection rules:
\begin{eqnarray}
|L_{\gamma}-J_N|=|L_{\gamma}-1/2|\leq J_{N^{\star}}\leq 
|L_{\gamma}+1/2|=|L_{\gamma}+J_N|\\
P_{N^{\star}}=P_N\cdot P_{\gamma}=P_{\gamma}\;. \nonumber
\end{eqnarray}
The resonance subsequently decays by strong interaction to the nucleon ground 
state via emission of the meson with spin 0, parity $P_{\eta}=-1$ and relative 
orbital angular momentum $L_{\eta}$. The respective selection rules 
must be fulfilled:
\begin{eqnarray}
|L_{\eta}-J_N| = |L_{\eta}-1/2| \leq J_{N^{\star}} \leq 
|L_{\eta}+1/2| = |L_{\eta}+J_N|\\
P_{N^{\star}} = P_N\cdot P_{\eta}\cdot(-1)^{L_{\eta}} = (-1)^{L_{\eta}+1}\;. 
\nonumber
\end{eqnarray}
Consequently, we have 
\begin{eqnarray}
P_{\gamma} & = P_{N^{\star}} & = (-1)^{L_{\eta}+1}\\
L_{\gamma}\pm 1/2 & = J_{N^{\star}} & = L_{\eta}\pm 1/2\;,\nonumber
\end{eqnarray}
where the two `$\pm$' are independent. Parity and angular momentum
conservation allow two possibilities:
\begin{eqnarray}
EL: & L & = L_{\eta}\pm 1 \\
ML: & L & = L_{\eta}\;.
\end{eqnarray} 
The corresponding photoproduction multipoles for pseudo-scalar mesons are 
denoted as $E_{l\pm}$ and $M_{l\pm}$, where $E$, $M$ stands for electric or 
magnetic photon multipoles, $l$ ($l=L_{\eta}$ in the above example) denotes 
the relative orbital angular momentum of the final meson - nucleon system, 
and `+' or `$-$' indicates whether the spin 1/2 of the nucleon must be added to 
or subtracted from $l$ to give the total angular momentum $J_{N^{\star}}$ of
the intermediate state. 

With the exception of $J_{N^{\star}}$=1/2 resonances, which can only be excited 
by one multipole ($E_{0+}$ for negative parity states and $M_{1-}$ for
positive parity states), each resonance can be excited by one electric and 
one magnetic multipole. Examples for the lowest order multipoles are given 
in tab.~\ref{tab_01}, app. 6.3.

The partial wave decomposition of the CGLN-amplitudes into the multipole 
amplitudes corresponding to definite parity and angular momentum states is 
given by:
\begin{eqnarray}
\label{eq42}
F_1(\Theta^{\star}) & = &
\sum_{l=0}^{\infty}[lM_{l+}+E_{l+}]P_{l+1}^{\prime}(cos(\Theta^{\star}))
      +[(l+1)M_{l-}+E_{l-}]P_{l-1}^{\prime}(cos(\Theta^{\star}))\nonumber\\
F_2(\Theta^{\star}) & = &
\sum_{l=0}^{\infty}[(l+1)M_{l+}+lM_{l-}]P_{l}^{\prime}(cos(\Theta^{\star}))\\
F_3(\Theta^{\star}) & = &
\sum_{l=0}^{\infty}[E_{l+}-M_{l+}]P_{l+1}^{\prime\prime}(cos(\Theta^{\star}))
                  +[E_{l-}-M_{l-}]P_{l-1}^{\prime\prime}(cos(\Theta^{\star}))
                                                                \nonumber\\
F_4(\Theta^{\star}) & = &
\sum_{l=0}^{\infty}[M_{l+}-E_{l+}-M_{l-}-E_{l-}]
                    P_{l-1}^{\prime\prime}(cos(\Theta^{\star}))\nonumber
\end{eqnarray}
where the $P_{l}^{\prime}$, $P_{l}^{\prime\prime}$ are derivatives of 
Legendre polynomials. The angular distributions reflect the quantum numbers 
of the excited state when the cross section is dominated by a resonance. The 
most familiar example from pion photoproduction is the excitation of the 
P$_{33}$(1232)-resonance ($\Delta$-resonance) via the $M_{1+}$-multipole 
which exhibits the characteristic $(5-3cos^2(\Theta^{\star}))$ angular 
distribution (see fig.~\ref{fig_09}). Therefore, the analysis of resonance 
contributions to pion photoproduction uses a parameterization of the cross 
section in terms of the multipole amplitudes. However, the differential cross 
sections by themselves do not allow a unique extraction of the multipoles. This 
is obvious from table \ref{tab_01} because the entries for the angular 
distributions exhibit a certain symmetry. They depend on the combination 
of the spin of the resonance and the order of the photon multipole but not on 
the combination of the parities of resonance and multipole. For example,
the excitation of a 5/2$^+$ resonance by an electric quadrupole has the same 
angular dependence as the excitation of a 5/2$^-$ resonance by a magnetic 
quadrupole. This ambiguity can be resolved with polarization observables.

The case of vector meson production is much more complicated. The general 
formalism of polarization observables in vector meson photoproduction and their
multipole analysis was discussed by Tabakin and coworkers 
(see e.g. 
\cite{Pichowsky_96}-\cite{Kloet_00}). 
The spin of the mesons contributes three additional degrees-of-freedom, and 12 
independent amplitudes have to be considered. This corresponds to the 
determination of 23 independent real quantities for each incident photon 
energy and each meson emission angle since the overall phase is again not 
needed. At the same time the number of observables increases. In the case of 
pseudo-scalar mesons, 12 double polarization observables are defined 
(see e.g. \cite{Barker_75}) apart from the four $\cal{S}$-type observables.
For vector mesons, the number of observables increases to 8 (independent) 
single, 51 (non-zero) double, 123 triple, and 108 quadruple polarization 
observables \cite{Zhao_98,Bennhold_01a,Pichowsky_96}. The extraction of 
resonance parameters will have to rely on model dependent analyses even more 
than in the case of the pseudo-scalar mesons. A `complete' measurement 
is certainly out of reach. In contrast to the pseudo-scalar case, the 
measurement of all $\cal{S}$-type observables does not even allow to fix the 
magnitude of the amplitudes. 

So far, we have ignored one complication of meson photoproduction, namely
the treatment of isospin. For a resonance excitation process, as
depicted in figure \ref{fig_05}, isospin must be conserved at the hadronic 
vertex. As a consequence only N$^{\star}$ resonances are 
allowed as intermediate states in $\eta$-photoproduction. For pion 
photoproduction, $\Delta$-resonances may also contribute. On the other hand, 
the electromagnetic interaction violates isospin conservation so that the 
production vertex is complicated by the presence of isoscalar ($\Delta I=$0) 
and isovector ($\Delta I=$0,$\pm$1) components of the electromagnetic current. 
Each multipole amplitude has to be reconstructed from the various isospin 
contributions. If the transition operator is split into an isoscalar part 
$\hat{S}$ and an isovector part $\hat{V}$, three independent matrix elements 
are obtained for the photoproduction of isovector mesons from the nucleon 
\cite{Watson_52} in the notation 
$\langle I_f,I_{f3}|\hat{A}|I_i,I_{i3}\rangle$:
\begin{equation}
\label{eq:iso_1}
A^{IS} = \langle \frac{1}{2},\pm \frac{1}{2}|\hat{S}|\frac{1}{2},\pm 
      \frac{1}{2}\rangle
\;\;\;\;\;\;\;\;\;\;\;\;\;      
\mp A^{IV} = \langle \frac{1}{2},\pm \frac{1}{2}|\hat{V}|\frac{1}{2},\pm
      \frac{1}{2}\rangle
\;\;\;\;\;\;\;\;\;\;\;\;\;      
A^{V3} = \langle \frac{3}{2},\pm \frac{1}{2}|\hat{V}|\frac{1}{2},
        \pm\frac{1}{2}\rangle\;. 
\end{equation}
The multipole amplitudes of the four possible photoproduction reactions 
can be expressed in terms of this isospin amplitudes as \cite{Walker_69}
(see also app. 6.4):
\begin{eqnarray}
\label{eq:iso_2}
A(\gamma p\rightarrow\pi^+ n) & = &
-\sqrt{\frac{1}{3}}\;A^{V3}+\sqrt{\frac{2}{3}}(A^{IV}-A^{IS}) \\
A(\gamma p\rightarrow\pi^o p) & = &
+\sqrt{\frac{2}{3}}\;A^{V3}+\sqrt{\frac{1}{3}}(A^{IV}-A^{IS})\nonumber\\
A(\gamma n\rightarrow\pi^- p) & = &
+\sqrt{\frac{1}{3}}\;A^{V3}-\sqrt{\frac{2}{3}}(A^{IV}+A^{IS})\nonumber\\
A(\gamma n\rightarrow\pi^o n) & = &
+\sqrt{\frac{2}{3}}\;A^{V3}+\sqrt{\frac{1}{3}}(A^{IV}+A^{IS})\;.\nonumber
\end{eqnarray}
The situation is simpler for isoscalar mesons like the $\eta$ meson. 
Here, the isospin changing part $V3$ cannot contribute, and all amplitudes
$X$ can be written as: 
\begin{equation}
\label{iso_3}
A(\gamma p\rightarrow\eta p) = (A^{IS}+A^{IV})\;\;\;\;\;\;\;\;\;\;\;\; 
A(\gamma n\rightarrow\eta n) = (A^{IS}-A^{IV})\;.
\end{equation}
The complication of isospin means, that a complete characterization of the 
photoproduction amplitudes for isovector as well as for isoscalar mesons
requires measurements off the neutron which must rely on meson photoproduction 
from light nuclei. 
This introduces additional uncertainties due to nuclear effects. In principle, 
two possibilities exist to learn about the isospin composition via 
photoproduction from nuclei. Photoproduction from bound nucleons in quasifree 
kinematics
can be used to extract the production cross section on the neutron. Here, the 
meson is produced on one nucleon which is subsequently knocked out of the 
nucleus. The other nucleons act only as spectators. The small binding 
energy and the comparatively well understood nuclear structure single out 
the deuteron as an exceptionally important target nucleus. However, for  
better control of systematic effects it is desirable to study the reaction 
for more strongly bound nuclei as well. Here, the extreme case is 
$^4$He. The different methods for the extraction of the neutron cross section 
will be discussed in detail for the example of $\eta$-photoproduction in 
\ref{ssec:S11}. Additional information may be obtained from coherent meson 
photoproduction, where the reaction amplitudes from all nucleons add coherently
and the nucleus remains in its ground state. Nuclei with 
different ground state quantum numbers may be used as spin/isospin filters of 
the production amplitude. The light nuclei $^2$H ($J$=1, $I_z$=0), $^3$H 
($J$=1/2, $I_z$=$-$1/2), $^3$He ($J$=1/2, $I_z$=+1/2), and $^4$He ($J$=0, $I_z$=0)
provide a selection of the relevant quantum numbers, although the $^3$H case 
is basically unexplored due to complications in the usage of tritium targets. 
Furthermore, with the exception of pion photoproduction, coherent cross 
sections are small due to the nuclear form factors. Incoherent 
excitations of the nuclei, where the final state nucleus is in an excited 
state, could provide even more flexible spin/isospin filters. 
These reactions have been even less exploited  due to the small size of the 
cross sections.  

\begin{figure}[h]
\centerline{\epsfysize=2.2cm \epsffile{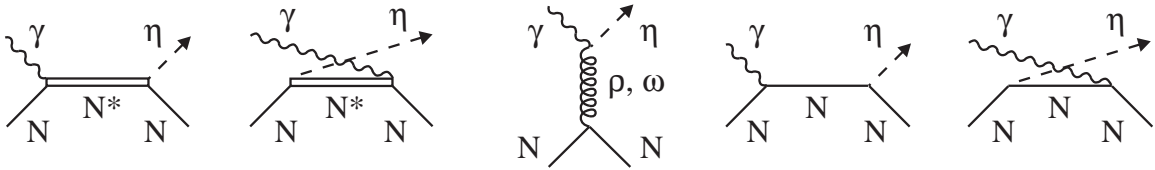}}
\caption{Tree level diagrams contributing to 
$\eta$-photoproduction from the nucleon.}
\label{fig_06}       
\end{figure}

Two obvious difficulties arise when extracting the properties of nucleon 
resonances from meson photoproduction data: in general, it is impractical
to carry out a `complete' experiment which allows the unique determination
of the photoproduction amplitudes and in most cases, important non-resonant 
background contributions must be separated from resonance excitations. 
As an example, fig.~\ref{fig_06} shows the most important contributions
to the low energy photoproduction of $\eta$-mesons.
%
%
%
%
Born terms and vector meson exchange terms are expected to contribute
in addition to nucleon resonance excitations. Therefore, reaction models
are needed which allow the extraction of the resonant multipoles in the 
presence of the background from incomplete data sets. 
More model assumptions are necessary for the analysis when fewer observables 
have been measured. Therefore, such models have been developed from different 
concepts and for different levels of sophistication. 

In the following, we will give an overview of the different methods 
which is certainly not complete. More detailed discussions for specific 
reactions will be given in the following sections. Partial wave analyses, 
trying to extract the photoproduction amplitudes without making assumptions 
about production processes, have mostly been limited to pion photoproduction. 
In this case, resonant contributions can be extracted in a second step by 
fitting the multipole amplitudes with resonance parameterizations and 
backgrounds. Fixed-t dispersion relations have frequently been used for such 
multipole analyses (see e.g. \cite{Arai_82,Crawford_83} and ref. therein).
Recently Hanstein et al. \cite{Hanstein_98} have updated this type of analysis 
for the now available extended and more precise data base at low incident 
photon energies ($E_{\gamma} < 450$ MeV). An extensive analysis of the world 
data base for pion photoproduction up to 2 GeV is given in the T-matrix 
formalism by Arndt and coworkers 
\cite{Arndt_90}-\cite{Arndt_96}.
The most recent update of this work \cite{Arndt_02}, includes the new 
precise data, in particular the polarization observables in the data base. 
This is the SAID partial wave
analysis \cite{Arndt_03}. The results, also for other reaction channels, 
are available online \cite{SAID}.

All other models start from a separation of the amplitudes into resonance
and background contributions. The first group of models, called Isobar 
Analyses, parameterizes the electric and magnetic multipole amplitudes in 
terms of Breit-Wigner curves for the resonances and smoothly varying 
phenomenological forms of the background amplitudes. The simplest of these 
analyses was used for the total photoabsorption data in ref. 
\cite{Bianchi_96} (see fig.~\ref{fig_02}). Breit-Wigner curves 
for the resonances and a phenomenological background were fitted to the total 
cross section data. Different reaction channels contribute to total 
photoabsorption where the total cross section is the only recorded observable. 
A more sophisticated analysis does not seem to be practical.  
Reliable, precise resonance properties cannot be extracted in this way. The 
interference between the contributions from the different resonances and the 
background, which in most cases is important, is not 
accounted for. These terms were included in early isobar analyses of 
pion- and $\eta$-photoproduction data \cite{Walker_69,Metcalf_74,Hicks_73}, 
where observables like angular distributions and polarization 
degrees-of-freedom were reconstructed from the multipole amplitudes. The 
advantage of the isobar models is that they have a simple, physically 
transparent form which is well suited to analyze the data. On the other hand, 
the models usually involve fits of many parameters, and the background terms 
are not treated in a very sophisticated manner.

Both problems were attacked in the framework of the Effective Lagrangian
Approach (ELA). In these models, all contributions to the photoproduction 
reaction are derived on an equal footing from the effective Lagrangian 
densities corresponding to the interaction vertices. In this way, the number 
of fit parameters is reduced to a smaller set of coupling constants which can 
partly be compared to or taken from other reactions. The analysis of pion- 
and $\eta$-photoproduction in this framework was mostly developed by 
N.C. Mukhopadhyay and coworkers \cite{Davidson_91,Benmerrouche_95} building 
on the work of Olsson and Osypowski \cite{Olsson_7578}. An ELA-model for pion 
photoproduction was also proposed by Garcilazo and Moya de Guerra 
\cite{Garcilazo_93}. A certain drawback of most of these models is their 
complexity and difficulty in handling. This holds in particular for nuclear 
applications like photoproduction from few nucleon systems which aim at the 
isospin structure of resonance excitations. In practice, the model of Blomqvist 
and Laget \cite{Blomqvist_77}, which is a non-relativistic reduction of 
the Olsson - Osypowski \cite{Olsson_7578} model, was used much more extensively
for such applications. Modern versions of isobar models constitute a compromise.
For pion and $\eta$-photoproduction \cite{Knochlein_95,Drechsel_99,Chiang_02} 
the effective Lagrangian parameterizations for the background processes have 
been adopted while keeping the Breit-Wigner forms for the resonances. Online
versions of these models are available at the MAID homepage \cite{MAID}.

In general, inclusion of the background terms at the tree level violates 
unitarity in the ELA-models as well as in the isobar models. It was argued 
(see e.g. \cite{Benmerrouche_95}) that this does not pose a severe problem for 
$\eta$-photoproduction, since the background terms are small, and the coupling 
of the $\eta$-meson to the nucleon is weak. On the other hand, Sauermann et al.
\cite{Sauermann_95} found important effects in $\eta$-photoproduction due to 
unitarity corrections. However, unitarity effects are much more important for 
pion photoproduction. The pion-nucleon coupling is large, and it is well known 
that background multipoles can have large imaginary parts, building up from 
pion-nucleon re-scattering contributions. Models for pion photoproduction like 
the Unitary Isobar Model of Drechsel et al. \cite{Drechsel_99} need a careful 
adjustment of the phases of the background and resonance contributions to the 
corresponding pion-nucleon scattering phase shifts.    

\begin{figure}[t]
\begin{minipage}{0.0cm}
{\mbox{\epsfysize=5.cm \epsffile{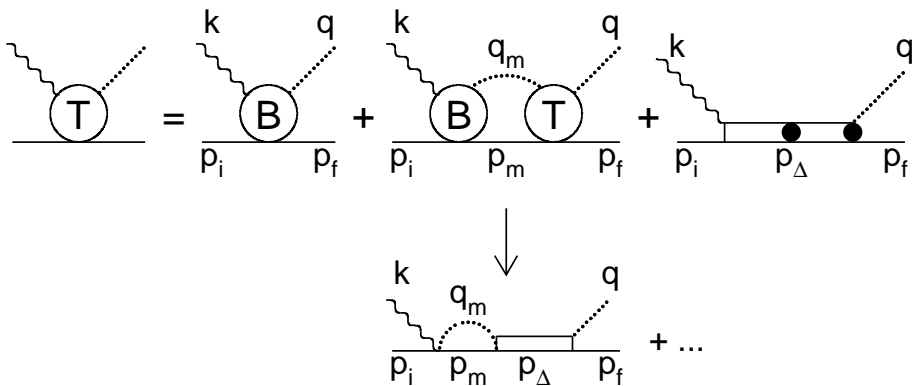}}}
\end{minipage}
\hspace*{13.cm}
\begin{minipage}{5.cm}
\caption{Schematic representation of the scattering equation used in
\cite{Nozawa_90} for the calculation of pion photoproduction from the nucleon
(upper part). The lower part shows one of the diagrams from the rescattering
contributions.}
\label{fig_07}       
\end{minipage}
\end{figure}

For a precise comparison of resonance properties to quark model predictions,
a further, subtle difficulty arises from the separation of resonance and 
background contributions. Let's consider, as an example, the properties of the 
$\gamma N\Delta$-vertex studied via pion photoproduction, to be discussed in 
detail in \ref{ssec:quadru}. Once sufficient observables have been measured, 
the reaction models as discussed above can be used to separate resonance from 
background contributions at  the tree level. However, off-shell re-scattering 
effects are neglected in the models. Usually, quark models predict the 
properties of the `bare' $\gamma N\Delta$-vertex in the absence of re-scattering
contributions. However, the data analysis in the framework of the reaction 
models cannot separate the bare interaction from contributions where an 
off-shell pion is produced in a non-resonant reaction (e.g. nucleon Born term) 
and then is re-scattered off the nucleon to produce a $\Delta$-resonance
(see lower part fig.~\ref{fig_07}).
%
%
%
%
The reaction models are thus sensitive to the properties of the `dressed'
vertex which includes the re-scattering terms. The solution to this problem
requires that either the hadron models make predictions for the dressed
vertex or the reaction models manage to extract the properties of the bare
vertex from the data. The reaction models attacked this problem with dynamical 
models of pion photoproduction. These models build on the experience from 
purely hadronic systems. On account of the strong interaction, a reasonable 
description with lowest order contributions is not possible and multiple 
scattering processes where carefully treated. The models developed by Tanabe 
and Otha \cite{Tanabe_85}, Yang \cite{Yang_85}, with later extension by Nozawa, 
Blankleider, and Lee \cite{Nozawa_90} include the final state $\pi N$ 
interaction explicitly so that unitarity is ensured. They treat the $\pi N$
interaction via the well-known meson exchange contributions and 
phenomenological two-body separable potentials for multiple scattering.
A schematic description of the scattering equation used in \cite{Nozawa_90}
is shown in fig.~\ref{fig_07}. Sato and Lee \cite{Sato_96} have further 
developed a consistent meson-exchange description of the $\pi N$ scattering 
and pion photoproduction which allows to extract the bare vertex couplings. 
However, all of these models are quite involved and have so far been almost 
exclusively applied to pion photoproduction in the $\Delta$-resonance region.  

The approaches discussed above analyze individual meson production reactions
and use input from other channels only in indirect ways, e.g. as constraints 
for resonance parameters or coupling constants. The resonance properties like 
excitation energy, total width and the couplings to the photon - nucleon or 
meson - nucleon states do not depend on the individual reactions. More 
efficient use of the available data is made in the framework of coupled channel
analyses. These models simultaneously fit photon- and pion-induced meson 
production reactions for many different final states 
($N\pi$, $N\pi\pi$, $N\eta$, $N\eta '$, $N\rho$, $K\Lambda$, $K\Sigma$,...).
The models have been developed based on different concepts for the treatment
of the individual reaction channels. Recent results from three approaches 
incorporate unitarity but differ in the background treatment, the 
parameterization of the resonance contributions, and the use of theoretical 
constraints like gauge invariance, analyticity, and chiral symmetry. The 
results are compared in \cite{Bennhold_01}. The Giessen model
\cite{Feuster_98}-\cite{Penner_02} 
is based on the Bethe-Salpeter equation in the K-matrix approximation. 
The s-, u-, and t-channel contributions are parameterized via effective 
Lagrangians. Pion and photon induced reactions to the final states $\pi N$, 
$2\pi N$, $\eta N$, $K\Lambda$, $K\Sigma$, and $\omega N$ are included in the 
latest version. The strength of this field theoretical
approach is that constraints from gauge invariance and chiral symmetry can be
included in a natural way. The starting point of the Pitt-ANL model 
\cite{Vrana_00} is an analytic, unitary representation of the s-channel
resonances. This model gives a particularly good description of the inelastic 
threshold openings, but t-channel contributions must be added by hand, and 
gauge invariance is not guaranteed. Finally, the KSU model \cite{Manley_99}
uses a multi-channel Breit-Wigner approach with a phenomenological background
parameterization. The comparison of the results from the three models showed
satisfactory agreement for the first resonance in each partial wave, but 
already for the second resonance, properties like decay branching ratios and
photon couplings varied widely.   

The models discussed above all are reaction models in the sense that they 
try to extract the properties of the excited states of the nucleon from the 
physical observables which then can be compared to the predictions of hadron
models. There is an obvious drawback of this approach. Each nucleon resonance
which is accounted for, introduces parameters into the models, and the
number of parameters grows quickly with increasing incident photon energy.
A direct connection between the quark degrees-of-freedom and photoproduction 
observables, largely reducing the number of parameters, was constructed
by Zhenping Li and collaborators for pseudo-scalar and vector mesons
\cite{Li_95a}-\cite{Zhao_02}. 
Their model for pseudo-scalar 
mesons \cite{Li_97} starts from the effective low energy chirally invariant 
QCD Lagrangian proposed by Manohar and Georgi \cite{Manohar_84}:
\begin{equation}
{\cal L}
=
\bar{\psi}[\gamma_{\mu}(i\partial^{\mu}+V^{\mu}+A^{\mu})-m]\psi+...
\end{equation}
which involves the interaction of the quark field in SU(3) symmetry
$\psi =(\psi(u),\psi (d),\psi(s))^T$
with the field of the pseudo-scalar mesons from the ground state octet 
treated as Goldstone bosons: 
\begin{equation}
\pi=\left| \begin{array}{ccc} -\frac 1{\sqrt {2}} \pi^0-\frac 1{\sqrt{6}}\eta
& \pi^+ & K^+ \\ -\pi^- & \frac 1{\sqrt {2}}\pi^0-\frac 1{\sqrt {6}}\eta &
K^0 \\ K^- & \bar {K}^0 &\sqrt{\frac 23}\eta \end{array}\right|
\end{equation}   
via vector and axial currents:
\begin{equation}
V^{\mu} = \frac{1}{2}(\zeta^{\dagger}\partial_{\mu}\zeta
+\zeta\partial_{\mu}\zeta^{\dagger})\;\;\;\;\;\;\;\;\;\;\;\;\;\;\;\;
A^{\mu} = i\frac{1}{2}(\zeta^{\dagger}\partial_{\mu}\zeta
-\zeta\partial_{\mu}\zeta^{\dagger}),~~~~~~\zeta=e^{i\pi/f}\;
\end{equation}
where $f$ is a decay constant. Then, the CGLN amplitude for the seagull term 
is constructed as well as the u- and s-wave contributions of the 
SU(6)$\otimes$O(3) quark model configurations for $\pi$, $\eta$- and 
$K$-photoproduction from the proton and the neutron in terms of only a few 
constants. Since SU(6)$\otimes$O(3) symmetry is broken, additional 
parameters must account for the configuration mixing in the physically 
observed states. In a similar approach the helicity amplitudes for the 
photoproduction of vector mesons are constructed from a chirally invariant
low energy Lagrangian in \cite{Zhao_98}. 

Finally, we would like to note that a special working group, the Baryon 
Resonance Analysis Group (BRAG \cite{BRAG}), has been formed. The formation of 
the group has been triggered by the large amount of new data from 
electromagnetic and hadronic facilities worldwide; it tries to organize and 
coordinate the efforts. Among the goals of this group are a standardized data 
base for all reactions investigated, a better understanding of the systematic
uncertainties of multipole analysis, and of the extraction of resonance
properties in the framework of the different models. The systematic effects 
are studied via analyses of `benchmark' data sets (see e.g. \cite{Arndt_01}). 

\newpage
\section{The $\Delta$-Resonance Region}
\label{sec:Delta}
\begin{figure}[thb]
\hspace*{0.5cm}
\epsfysize=6.cm \epsffile{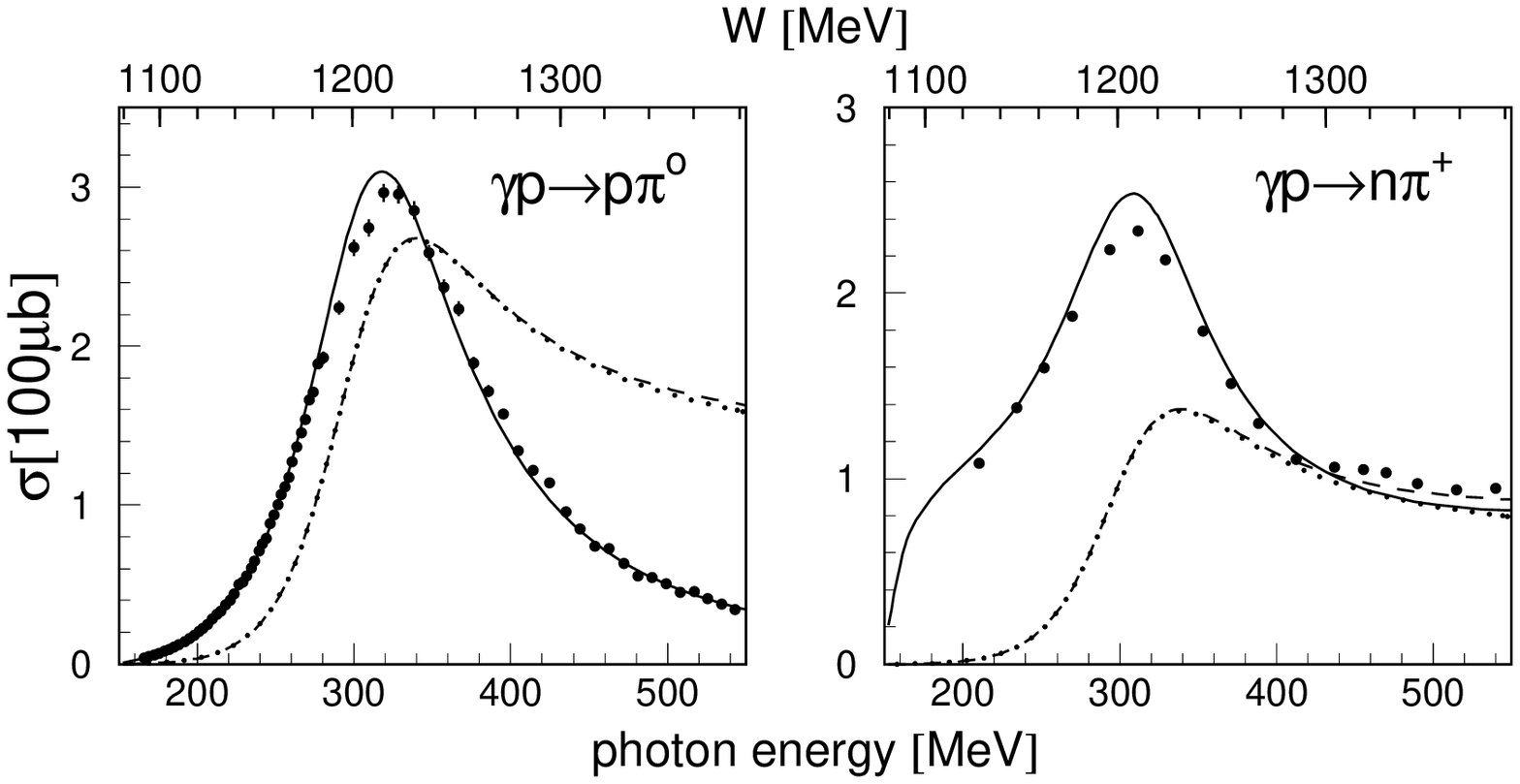}
\hspace*{1.0cm}
\epsfysize=5.5cm \epsffile{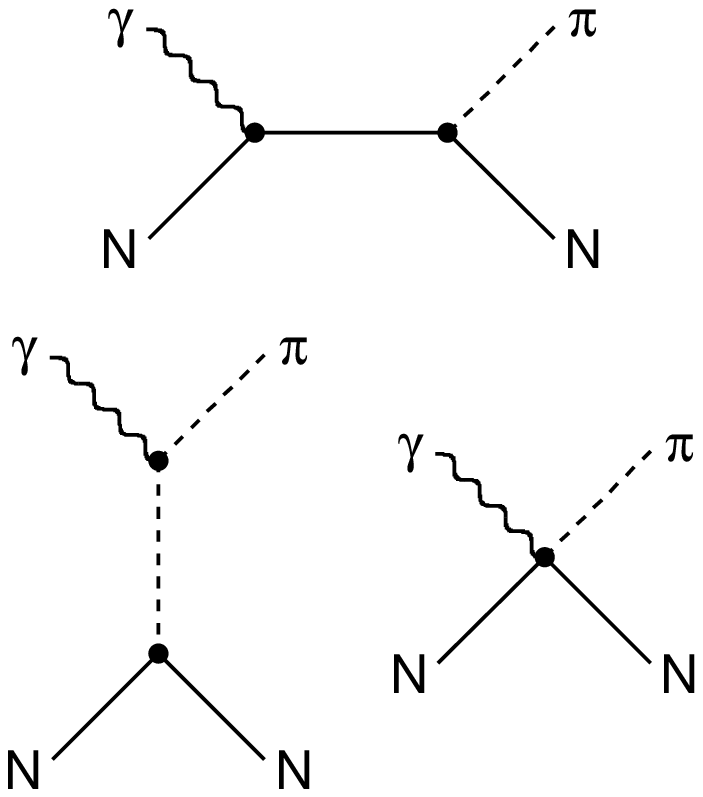}
\caption{Total cross section of neutral (left hand side) and charged 
(middle) pion production from the proton in the $\Delta$-resonance 
region. Data: \cite{Fuchs_96,Haerter_96} ($\pi^o p$) and 
\cite{Buechler_94} ($\pi^+ n$). Curves: MAID2000 \cite{Drechsel_99}, 
Solid: all contributions, dashed: without background but with higher resonances, 
dotted: only P$_{33}$(1232) (dashed and dotted curves almost identical).
Right hand side: background contributions, upper part: nucleon Born term, 
lower part: pion-pole and Kroll-Rudermann (contact) term (only for charged
pions).}
\label{fig_08}       
\end{figure}

\begin{figure}[htb]
\centerline{\epsfysize=4.3cm \epsffile{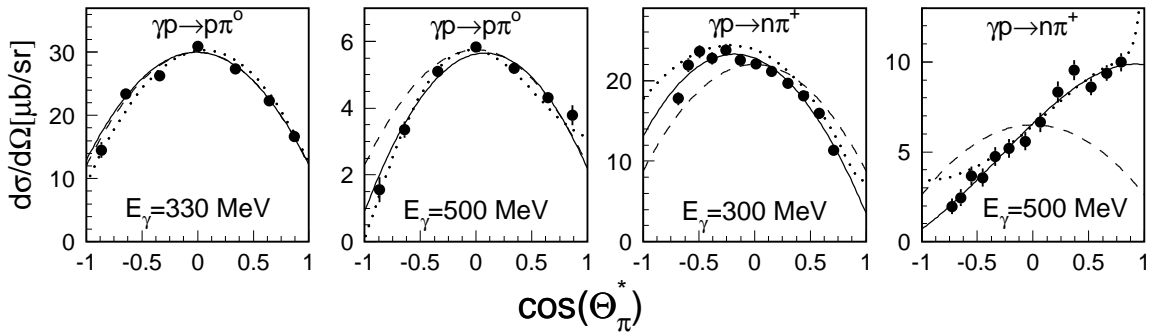}}
\caption{Angular distributions for neutral and charged pion 
photoproduction from the proton. Data are from \cite{Haerter_96,Buechler_94}. 
Solid 
curves: fits to the data, dashed curves: expected distribution for excitation
of the $\Delta$ (see text), dotted curves: MAID2000 model \cite{Drechsel_99} 
(full calculation including all terms).}
\label{fig_09}       
\end{figure}

The $\Delta (1232)$ resonance is the best known exited state of the nucleon.
It has been investigated via many different reactions on the free proton 
and on bound nucleons, in particular in the $\pi N$-final state. The total 
photoproduction cross section and the angular distributions for the reactions 
$\gamma p\rightarrow p\pi^o$ and $\gamma p\rightarrow n\pi^+$ are summarized 
in figs.~\ref{fig_08} and \ref{fig_09}. The angular distributions have been 
fitted with the ansatz
\begin{equation}
\label{eq:angdis}
\frac{d\sigma}{d\Omega}=\frac{q^{\star}}{k^{\star}}
[a+b\;cos(\Theta^{\star})+c\;cos^2(\Theta^{\star})]
\end{equation}
where  $a, b, c$ are free parameters (full curves in fig.~\ref{fig_09}), and 
are compared to the $(5-3cos^2(\Theta^{\star}))$ behavior expected for the 
excitation of the $\Delta$ resonance via the $M_{1+}$-multipole (dashed curves). 
%
%
%
%

The large difference between the neutral and charged channel is demonstrated 
in the figures. The energy dependence of the cross section for 
$\gamma p\rightarrow p\pi^o$ approximates the shape of a Breit-Wigner resonance
more than in the case of charged pions. Furthermore, the angular distribution 
of $\pi^o$-photoproduction close to the resonance position at 
$E_{\gamma}\approx$330 MeV is in excellent agreement with the expectation for 
the $M_{1+}$-multipole transition. 
%
%
%
%
%
%
%
\begin{figure}[hbt]
\begin{minipage}{0.0cm}
{\mbox{\epsfysize=6.cm \epsffile{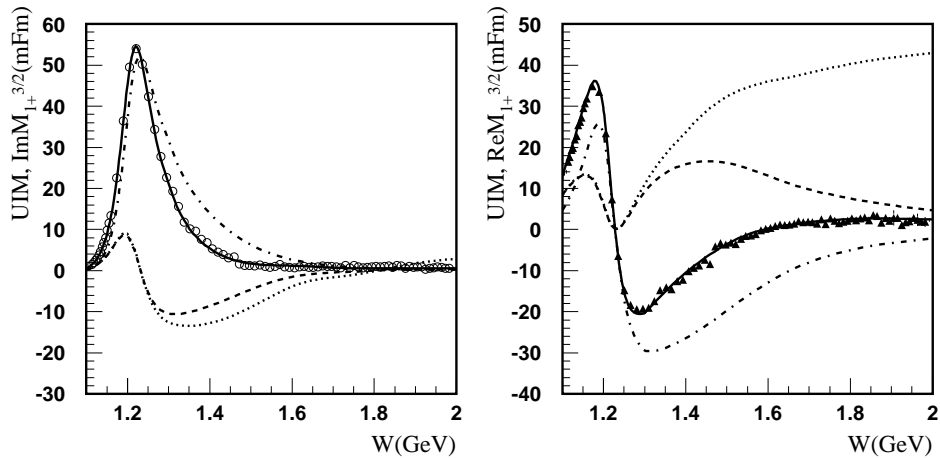}}}
\end{minipage}
\hspace*{12.8cm}
\begin{minipage}{5.3cm}
\caption{Imaginary and real parts of the $M_{1+}$ multipole for $\gamma
p\rightarrow p\pi^o$ in the UIM of
Aznauryan \cite{Aznauryan_02}. Full curves: full model, dashed-dotted:
resonance contribution, dashed: background. Circles and triangles: 
SAID multipole analysis \cite{Arndt_96}. Dotted curves: background in the UIM
of Drechsel et al. \cite{Drechsel_99}}
\label{fig_10}       
\end{minipage}
\end{figure}
%
\hspace*{-0.4cm}
The photoproduction of charged pions 
deviates from this behavior already at the resonance position and has a 
completely different shape at higher incident photon energies. 
The reason is that non-resonant background contributions are 
suppressed for the neutral channel. The Kroll-Rudermann (KR) term
and the pion-pole term cannot contribute since the photon does not couple to 
the neutral pion, so that only nucleon Born terms mix with 
the resonance excitation (see diagrams in fig. \ref{fig_08}). 
On the other hand, photoproduction of charged pions close to threshold is 
dominated by background terms, in particular the Kroll-Rudermann term. 
At higher incident photon energies, the KR-term and the pion-pole term are 
still important.

The relative importance of the background contributions is illustrated by
a comparison of the data to the results of the Unitary Isobar Model (UIM) 
from Drechsel et al. \cite{Drechsel_99}. The full calculation is shown in fig. 
\ref{fig_08} (also in fig.~\ref{fig_09}), along with the calculation with 
non-resonant background excluding other resonances, and the result for the 
$\Delta$ resonance without any other contributions. As expected, contributions 
from resonances other than the $\Delta$ are negligible in this energy range. 
The only eligible candidate would be the tail of the P$_{11}$(1440) `Roper' 
resonance. The cross section ratio of $\approx$2 for the two reactions, when 
only the excitation of the $\Delta$ is considered, simply reflects the isospin 
Clebsch-Gordan coefficients for the strong $\Delta^+\rightarrow\pi^o p$ and 
$\Delta^+\rightarrow\pi^+ n$ decays (see eqs.~\-(\ref{eq:iso_2})). 
Only the amplitude $A^{V3}$ can change the isospin and excite the $\Delta$.  

The non-resonant background is substantial for charged pions. At the resonance 
position the contribution from the $\Delta$ agrees with the data for 
$\pi^o$-production but underestimates it for $\pi^+$-production by roughly a 
factor of two. Unexpectedly, the experimental $\gamma p\rightarrow p\pi^o$ 
cross section deviates significantly at higher incident photon energies from 
the $\Delta$ contribution in the UIM of Drechsel et al. \cite{Drechsel_99} 
since the resonance curve has a very pronounced high energy tail. In the model, 
the large resonance contributions to the total cross section are cancelled by 
background terms. However, the separation of resonance and background terms in 
the models is by no means unique. Due to the unitarization, modifications in 
one sector will also affect the rest. The background in the UIM of ref. 
\cite{Drechsel_99} is constructed from $s$- and $u$-channel nucleon Born terms 
and $t$-channel $\pi$-, $\rho$-, and $\omega$-exchange. Recently, it was argued
by Aznauryan \cite{Aznauryan_02} that this background becomes too large at 
higher incident photon energies, which is artificially compensated in 
\cite{Drechsel_99} by the large high energy tail of the $\Delta$. In a modified 
version of the UIM, Aznauryan included Regge-pole type background amplitudes 
in order to account for $t$-channel contributions of heavier mesons which are 
neglected in \cite{Drechsel_99}. The background contributions to the 
$M_{1+}^{3/2}$ multipole in the UIM's from refs. \cite{Drechsel_99,Aznauryan_02} 
(see fig.~\ref{fig_10}) show indeed a very different high energy behavior which
in turn results in different resonance contributions. It should be noted 
however, that very good agreement is found at energies close to the resonance 
position.
We have discussed this behavior in some detail since the background 
contributions are very important for the following discussion. The message is 
that background contributions are small close to the resonance position for    
$\pi^o$-photoproduction but substantial for $\pi^+$-photoproduction. In both   
cases they are reasonably well under control albeit with some model dependency 
at higher incident photon energies. 

One might wonder what can be learned from photoproduction in the
$\Delta$-resonance region beyond the basic resonance properties like 
excitation energy, width, and electromagnetic coupling which are well known 
\cite{PDG}. In fact, the recent experimental progress has opened the 
possibility for new, detailed studies of this resonance. We 
will discuss four examples: the $E2$ admixture in the 
electromagnetic excitation amplitude, the helicity dependence of the cross 
section, the magnetic moment of the $\Delta$, and the isospin dependence
of the excitation.

\subsection{\it Quadrupole Strength in the $N\rightarrow\Delta$ Transition}
\label{ssec:quadru}

The standard picture for the excitation of the $\Delta$-resonance on the 
nucleon is the spin-flip of a single quark. In the electromagnetic case, the 
spin-flip is induced via the absorption of a magnetic dipole $M1$-photon which 
for pion production corresponds to the $M_{1+}$-multipole. However, from the 
quantum numbers, the excitation can also proceed via the absorption of an 
electric quadrupole photon $E2$, corresponding to the $E_{1+}$ multipole. 
In nuclear physics, electric quadrupole transitions are an important tool for 
the study of quadrupole moments related to deformations of the nuclei. In many 
nucleon models, a tensor force in the hyperfine interaction, first suggested 
by De Rujula, Georgi, and Glashow \cite{Rujula_75}, arises from one gluon exchange
between the quarks, and leads to a d-state admixture in the ground state. This 
is similar to the tensor force in the nucleon - nucleon interaction which gives 
rise to a d-state admixture in the deuteron ground state and results in the 
prolate deformation of the deuteron.
%
%
%
\begin{figure}[hbt]
\begin{minipage}{6.0cm}
{\mbox{\epsfysize=5.0cm \epsffile{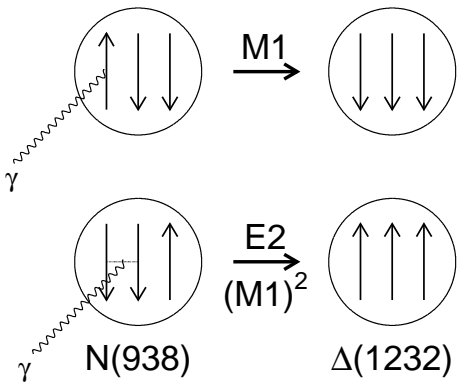}}}
\caption{Upper part: excitation of the $\Delta$-resonance via the spin-flip
of a single quark from absorption of a magnetic dipole photon. Lower part:
absorption of an electric quadrupole photon on a quark pair correlated by 
gluon or meson exchange currents, resulting in a spin-flip of the pair.}
\label{fig_11}       
\end{minipage}
\end{figure}
%
\vspace*{-10.8cm}
\hspace*{6.5cm}
\begin{minipage}{11.9cm}
The situation for the nucleon is different in so far as an intrinsic quadrupole
deformation cannot be directly measured for particles with spin less than one 
but may reveal itself in transitions between the ground and the excited states. 
Therefore, the idea is to look for an $E2$-admixture in the $\Delta$-excitation.
In the quark picture, such an admixture can occur in two different ways. An 
electric  quadrupole photon can induce the spin-flip on a quark, with the 
initial or final state $L=2$, when either the nucleon or the $\Delta$ has a 
d-state admixture in its wave function. This scenario is related to a possible 
deformation of the baryons. However, it was pointed out by Buchmann and 
collaborators 
\cite{Buchmann_97}-\cite{Buchmann_02}
that $E2$ 
admixtures in the absence of d-state components in the wave functions may stem 
from the coupling of the photon to mesonic or gluonic exchange currents. 
Indeed they predict that this process provides a large contribution to the 
observed $E2$ strength. As illustrated in fig.~\ref{fig_11}, it results in the 
simultaneous spin-flip of two quarks which are both in s-states and correlated 
via gluon or meson exchange. Therefore, care has to be taken in the 
interpretation of possible $E2$ admixtures. Nevertheless, it is indisputable 
that the $E2$ admixture in the amplitude is sensitive to the internal structure
of the nucleon, and its measurement provides a stringent test of baryon model 
predictions. The quantity of interest is the ratio:
\end{minipage}

\vspace*{0.5cm}
\begin{equation}
\label{eq:REM}
R_{EM}=\left. \frac{E_{1+}^{(3/2)}}{M_{1+}^{(3/2)}}\right|_{W=M_{\Delta}}
      = \left. \frac{\mbox{Im}E_{1+}^{(3/2)}}{\mbox{Im}M_{1+}^{(3/2)}}
        \right|_{W=M_\Delta} \;,
\end{equation}
where the index (3/2) indicates the isospin $I$ of the $\pi N$-system. The 
Fermi-Watson theorem guarantees that $E_{1+}^{(3/2)}/M_{1+}^{(3/2)}$ is a real 
number, at least up to the two-pion threshold. However, the ratio depends on 
the photon energy, and the resonance position is taken to be the energy where 
the phase is 90$^o$. 
 
Predictions for this ratio are available from many different nucleon models. 
SU(6)-symmetry, as employed in the MIT bag model, requires $R_{EM}$=0. 
Constituent quark models predict values in the range -3.5\%$<R_{EM}<$0\%
\cite{Koniuk_80a},\cite{Gershteyn_81}-\cite{Buchmann_97}
while 
relativized quark models \cite{Capstick_90,Capstick_92} yield small values 
around -0.1\%. The cloudy bag model \cite{Kalbermann_83,Bermuth_88} produces 
results in the range -3.0\%$<R_{EM}<$-2.0\% while Skyrme models \cite{Wirzba_87}
tend to give larger values between -2.5\% and -5\%. First results from QCD 
lattice calculations ($R_{EM}=(+3\pm8)$\% ) \cite{Leinweber_93} presently have 
rather large uncertainties.  

The experimental determination of $R_{EM}$ is difficult. The $E2$ admixture 
is small, and photoproduction reactions are dominated by the magnetic $M_{1+}$ 
multipole. Furthermore, we have already seen that photoproduction in the 
$\Delta$ range has significant contributions from background processes. 
Multipole analyses of data prior to 1999 as in \cite{Arndt_96} made a first 
step towards the extraction of the resonant $E_{1+}^{(3/2)}$-amplitude. As 
pointed out in \cite{Beck_00}, systematic errors in the data base have a large 
influence on the result. This is so because the older experimental results, 
often obtained with untagged bremsstrahlung beams, came from a number of 
experiments with different systematic error sources. When combined, even the 
shapes of angular distributions are affected by different normalization errors. 
Observables, like angular distributions and polarization degrees-of-freedom or 
observables from different isospin channels, were often not consistent. The 
analysis of inconsistent data sets is problematic for weak multipoles which may 
be completely obscured by such effects. Therefore, attempts were made to 
explore new experimental possibilities for a sensitive search for the 
$E_{1+}^{(3/2)}$ admixture. The experiments determined all relevant observables 
simultaneously, so that systematic uncertainties were minimized. Two 
independent measurements were carried out at the MAMI accelerator in Mainz 
\cite{Beck_97,Beck_00} and at the LEGS laser backscattering facility at BNL 
\cite{Blanpied_97,Sandorfi_98}. Both experiments established angular 
distributions and photon beam asymmetries for the reactions 
$\gamma p\rightarrow p\pi^o$ and $\gamma p\rightarrow n\pi^+$. 
In addition, the LEGS experiment determined the same observables for Compton 
scattering. There are two reasons why the choice of reactions and observables 
comes naturally. 

The measurement of the two isospin channels allows the extraction of the 
I=3/2 contribution in the final state. 
From eqs.~\-(\ref{eq:A12},\ref{eq:A14}) follows that all multipole amplitudes 
$M_{1+}$, $E_{1+}$,... denoted as ${\cal{M}}_{l\pm}$ can be decomposed into 
the isospin 1/2 components $_p{\cal{M}}_{l\pm}^{(1/2)}$ and the isospin 3/2 
components ${\cal{M}}_{l\pm}^{(3/2)}$ via:
\begin{equation}
{\cal{M}}_{l\pm}(p\pi^o) =  
_p{\cal{M}}_{l\pm}^{(1/2)}+\frac{2}{3}{\cal{M}}_{l\pm}^{(3/2)}\;\;\;\;\;\;\;\;
{\cal{M}}_{l\pm}(n\pi^+) =  
\sqrt{2}\left(_p{\cal{M}}_{l\pm}^{(1/2)}
-\frac{1}{3}{\cal{M}}_{l\pm}^{(3/2)}\right)\;.  
\end{equation} 
The amplitudes ${\cal{M}}_{l\pm}^{(I)}$ are complex functions of the incident
photon energy. The phases can be related to the corresponding pion - nucleon 
scattering phase shifts $\delta^{(I)}_{l\pm}$ via the Fermi-Watson theorem 
\cite{Watson_54}: 
\begin{equation}
\label{eq:FWTH}
{\cal{M}}_{l\pm}^{(I)} = 
\left|{\cal{M}}_{l\pm}^{(I)}\right|e^{(i\delta^{(I)}_{l\pm}+n\pi)}\;,
\end{equation}
where n is an integer. This relation is strictly valid only below the two-pion 
photoproduction threshold at $E_{\gamma}\approx$310 MeV. It is valid 
approximately well above the two-pion threshold since the $\pi N$
inelasticities in the $P_{33}$ partial wave are small even at energies around 
400 MeV. 

\begin{figure}[hbt]
\centerline{
\epsfysize=4.3cm \epsffile{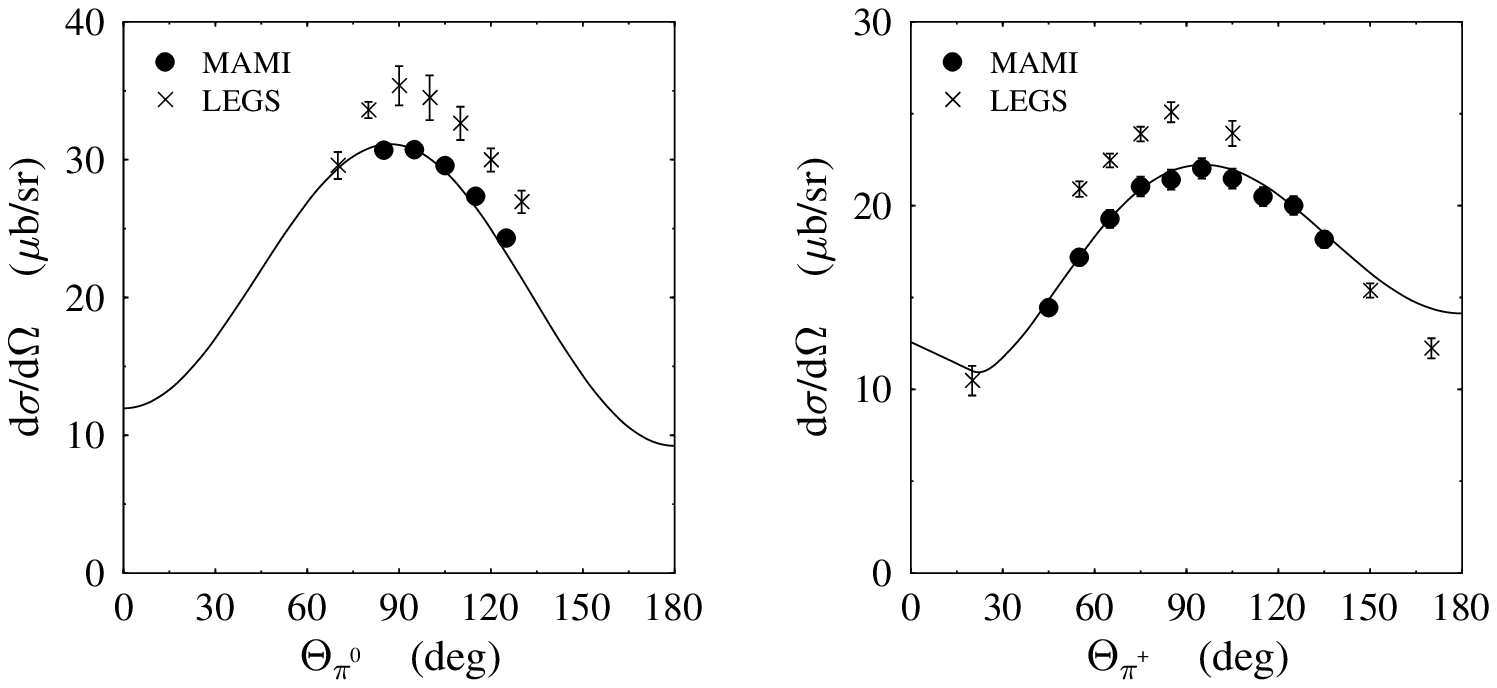}
\epsfysize=4.3cm \epsffile{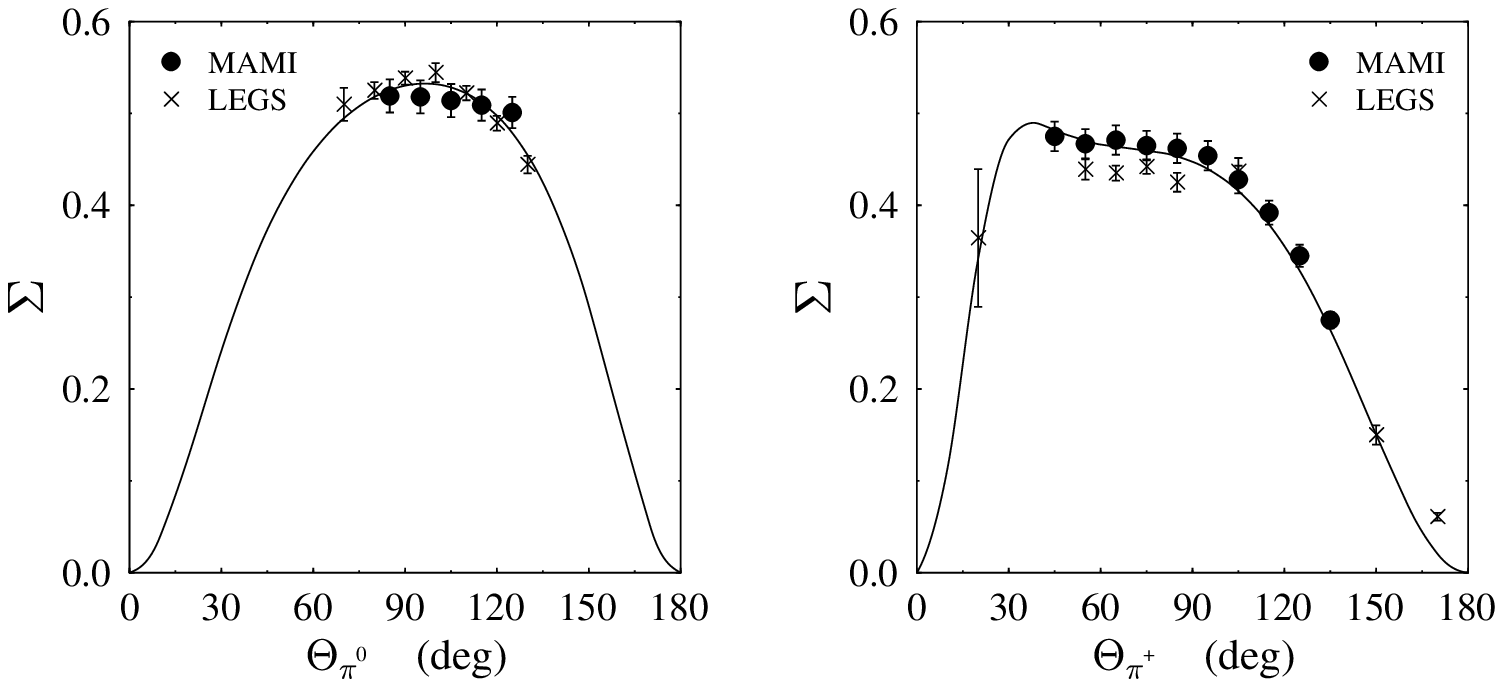}}
\caption{Differential cross sections $d\sigma /d\Omega$ and photon asymmetry
$\Sigma$ for the reactions $p(\vec{\gamma} ,\pi^o)p$ and 
$p(\vec{\gamma} ,\pi^+)n$ for $E_{\gamma}$ = 320 MeV from the LEGS
\cite{Blanpied_97,Sandorfi_98} and the MAMI experiment \cite{Beck_97,Beck_00}.
The curves correspond to the energy dependent analysis of the MAMI data.
(Fig. from \cite{Beck_00}.}
\label{fig_12}       
\end{figure}

The importance of the photon beam asymmetry $\Sigma$ stems from the fact 
that it contains an interference term proportional to the product of the small
$E_{1+}$ term with the leading $M_{1+}$ multipole (see below). The beam
asymmetry is defined by: 
\begin{equation}
\Sigma (\Theta) =
\frac{d\sigma_{\perp}(\Theta)-d\sigma_{\parallel}(\Theta)}
{d\sigma_{\perp}(\Theta)+d\sigma_{\parallel}(\Theta)} \;.
\end{equation}
$\sigma_{\perp}(\Theta)$ and $\sigma_{\parallel}(\Theta)$ are the cross 
sections perpendicular and parallel to the plane defined by the photon
polarization and momentum vectors of a 100\% linearly polarized photon beam. 
It is related to the differential cross section via:
\begin{equation}
\frac{d\sigma (\Theta ,\Phi)}{d\Omega} =
\frac{d\sigma_o (\Theta)}{d\Omega}\left[1-P\Sigma (\Theta) cos(2\Phi)\right]
\end{equation}
where $P$ is the degree of linear polarization of the photon beam and $\Phi$ 
is the angle with respect to the polarization plane. The photon beam at the 
LEGS facility \cite{Thorn_89} is produced via the scattering of laser photons 
at a high energy electron beam. Linear or circular polarization of the laser 
beam is transferred to the backscattered photon beam where the highest transfer 
of linear polarization (100\%) is achieved for photons scattered at 180$^o$, 
corresponding to the highest photon beam energies. The MAMI tagged photon 
facility \cite{Anthony_91} works with bremsstrahlung tagging. Here, linear 
polarization can be produced with coherent bremsstrahlung off a diamond crystal 
\cite{Lohmann_94}. The energy region with the highest degree of polarization 
can be varied with the orientation of the crystal. Reasonable polarization 
cannot be achieved at photon energies larger than $\approx$2/3 of the maximum 
photon energy. This is not a disadvantage for the presently discussed experiment
since the MAMI-B facility provided a photon beam with energies up to 800 MeV 
while the LEGS facility reaches 330 MeV only. The MAMI experiment covered the 
complete range of the $\Delta$-resonance, while the LEGS experiment measured 
only up to the maximum of the $\Delta$-peak. An example for the experimental 
differential cross sections and the photon beam asymmetry is shown in fig. 
\ref{fig_12}.
%
%
%
%

The shape and the absolute scale of the differential cross sections published 
by the two groups do not agree for photon energies close to the resonance 
maximum. The agreement is somewhat better at lower incident photon energies 
\cite{Beck_00}. Close to the resonance peak, a disagreement at the 10\% level 
is observed for both channels, which is well outside the claimed systematic  
uncertainties on the 3\% level. Recently, the same problem has emerged for 
Compton scattering, where an experiment performed at MAMI 
\cite{Galler_01,Wolf_01} finds systematically smaller cross sections than the 
LEGS-experiment. The LEGS-group has investigated this issue in 
\cite{Blanpied_01} where the calibration procedures are discussed in detail. 
The authors emphasize that they do not find any source of additional systematic 
errors. The $p\pi^o$ final channel has been investigated at MAMI using two 
different detection systems, the DAPHNE-detector \cite{Beck_97,Beck_00} as 
well as the TAPS-detector \cite{Haerter_96, Beck_00}. Both results are 
consistent. In addition, the Mainz results are consistent with previous 
measurements at Bonn, for both the $\pi^o p$ \cite{Genzel_74,Genzel_74a} and 
the $\pi^+ n$ \cite{Fischer_72},\cite{Buechler_94} channels. In summary, there 
seems to be a yet unexplained systematic difference in the cross section 
results from Bonn/Mainz versus LEGS on the 10\% level, which is clearly 
unsatisfactory. However, the extraction of the $E2$-admixture involves mainly 
ratios where the results from Mainz and LEGS for $R_{EM}$ are still marginally 
consistent within their uncertainties. 

\begin{figure}[hbt]
\centerline{
\epsfysize=7.85cm \epsffile{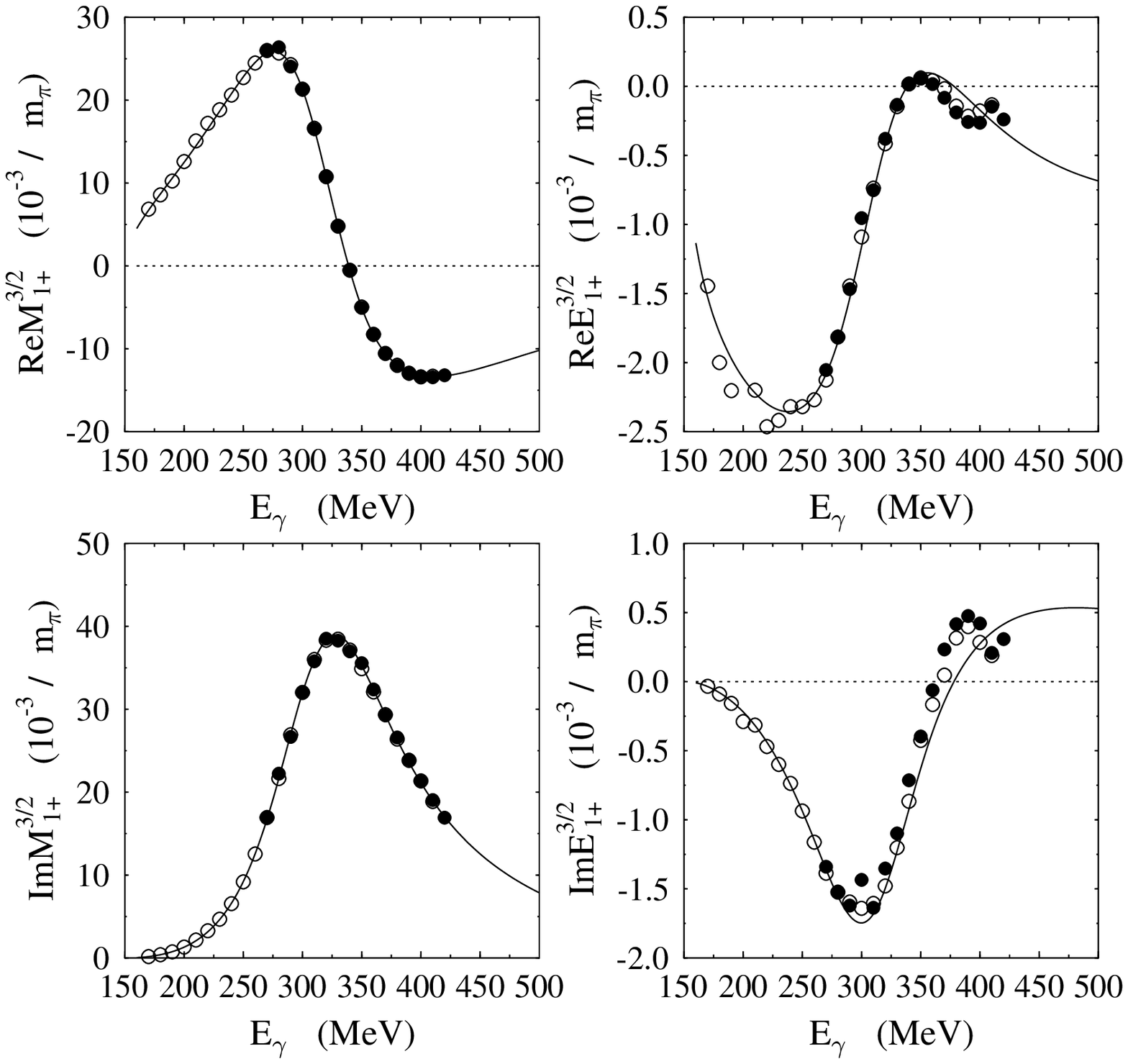}
\epsfysize=8.0cm \epsffile{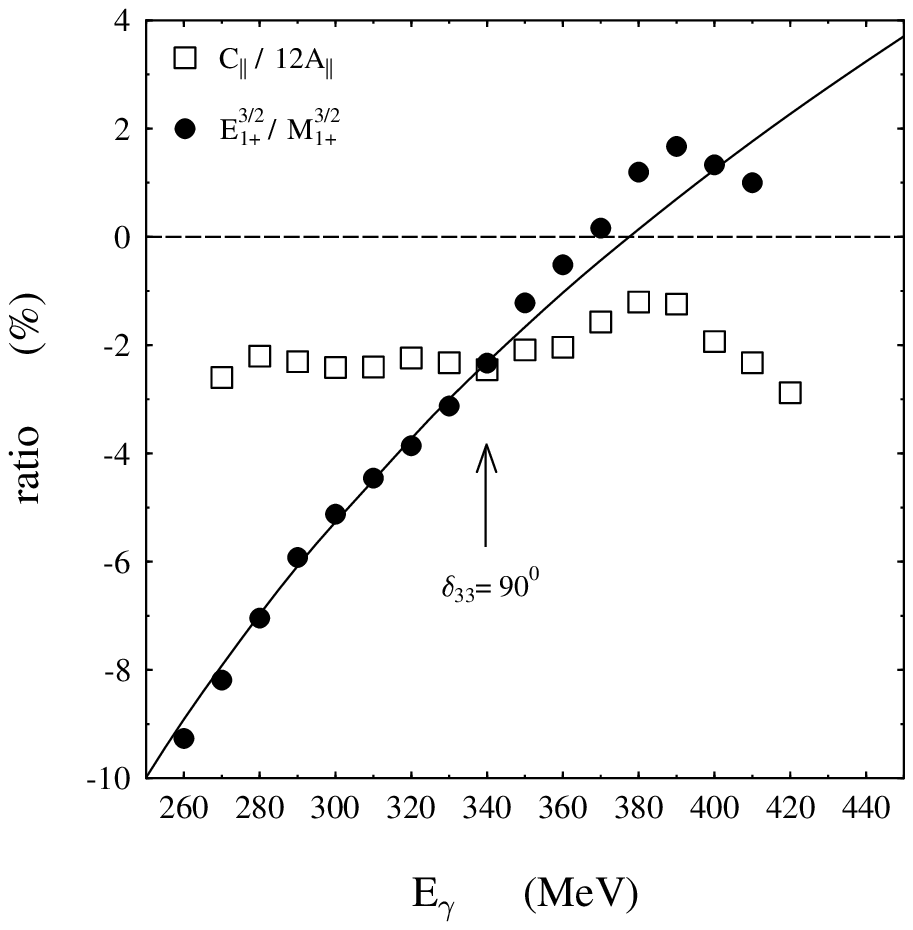}}
\caption{Extraction of $R_{EM}$ \cite{Beck_00}. Left hand side: Real and 
imaginary parts of the $I=3/2$ components of the $M_{1+}$ and $E_{1+}$ 
amplitudes. Solid circles: energy independent fits of the MAMI data alone. 
Line and open circles: energy dependent and energy independent fixed-t 
dispersion analysis of a larger data base \cite{Hanstein_98}. Right hand side: 
energy dependence of the ratio $E_{1+}^{(3/2)}/M_{1+}^{(3/2)}$. Solid circles 
(energy independent) and solid line (energy dependent) results of the fixed-t 
dispersion analysis \cite{Hanstein_98}. Open squares: energy dependence of 
$R\equiv C_{\parallel}/12A_{\parallel}$.
}
\label{fig_13}       
\end{figure}

The first result for $R_{EM}$ from the MAMI data was obtained from an analysis
of the $\pi^o p$ final state \cite{Beck_97} only. We will shortly discuss this
analysis, which is instructive in view of the importance of the polarization 
degree-of-freedom while involving some approximations. As long as only s- and 
p-waves contribute, the differential cross sections can be written as:
\begin{equation}
\label{eq:sp_multi}
\frac{d\sigma_x(\Theta^{\star})}{d\Omega} =
\frac{q^{\star}}{k^{\star}}
\left[A_x + B_xcos(\Theta^{\star})+C_xcos^2(\Theta^{\star})\right]\;,
\end{equation}
where x indicates the unpolarized (0), parallel ($\parallel$) and perpendicular
($\perp$) components. The largest sensitivity to the $E_{1+}$-multipole due to
an interference term with the leading $M_{1+}$ is carried by the parallel
component \cite{Beck_00}:
\begin{eqnarray}
A_{\parallel} & = &
\left|E_{0+}\right|^2+\left|3E_{1+}-M_{1+}+M_{1-}\right|^2\\
B_{\parallel} & = & 
2\mbox{Re}\left[E_{0+}(3E_{1+}+M_{1+}-M_{1-})^{*}\right]\nonumber\\
C_{\parallel} & = & 
12\mbox{Re}\left[E_{1+}(M_{1+}-M_{1-})^{*}\right]\nonumber
\end{eqnarray}  
\begin{equation}
\label{eq:Rrat}
\mbox{with:}\;\;\;\;\;\;\;
R \equiv \frac{1}{12} \frac{C_{\parallel}}{A_{\parallel}} =
\frac{\mbox{Re}\left[E_{1+}(M_{1+}-M_{1-})^{*}\right]}
{\left|E_{0+}\right|^2+\left|3E_{1+}-M_{1+}+M_{1-}\right|^2}\;.
\;\;\;\;\;\;\;\;\;\;\;\;\;\;\;
\end{equation}
It is then argued that at the resonance position 
$\mbox{Re}(M_{1+}-M_{1-})$, $\left|E_{0+}\right|^2$, and 
$9\left|E_{1+}\right|^2$ can be neglected so that \cite{Beck_00}:
\begin{equation}
R = \frac{R_{\pi^o}}{1-6R_{\pi^o}}\;\;\;\;\;\;\;\;
\mbox{with:}\;\;\;\;\;\;\;\;\;\;
R_{\pi^o} =
\frac{\mbox{Im}E_{1+}}{\mbox{Im}M_{1+}-\mbox{Im}M_{1-}}
\approx \frac{\mbox{Im}E_{1+}}{\mbox{Im}M_{1+}^{(3/2)}}\;.
\end{equation}
In the last approximation, the imaginary part of the non-resonant
$M_{1-}$-multipole and the isospin $I=1/2$ component of the $M_{1+}$-multipole,
which resonates in the $I=3/2$ component, are neglected. In this approximation, 
$R_{\pi^o}$ equals $R_{EM}$ (see eq.~\-(\ref{eq:REM})) up to corrections for 
isospin $I=1/2$ contributions to the $E_{1+}$-multipole. In the first analysis 
in \cite{Beck_97}, the $3E_{1+}$-term in the denominator of 
eq.~\-(\ref{eq:Rrat}) was neglected. The above approximations and the neglect 
of higher partial waves are discussed in detail in 
\cite{Davidson_97,Workman_97,Beck_97a,Beck_00}.
The energy dependence of $R\equiv C_{\parallel}/12A_{\parallel}$ is shown
in fig.~\ref{fig_13} (right hand side), and the value obtained for $R$ at 
resonance \cite{Beck_97,Beck_00} is $R = (-2.5\pm0.2_{stat})$\%
which, without corrections for $I=1/2$ contributions to the $E_{1+}$-multipole,
corresponds to $R_{EM} = -2.95$\%.
Including corrections for Born-terms of the order of 10 - 20\% for
$\mbox{Im}E_{1+}^{(1/2)}/\mbox{Im}E_{1+}$, Beck et al. quote \cite{Beck_00}:
\begin{equation}
R_{EM} = (-2.5\pm 0.2_{stat}\pm 0.2_{syst})\%\;. 
\end{equation}
This analysis benefits from the fact that, at the resonance position at 
340 MeV, $\pi^o$-photoproduction is practically free of background 
contributions, as we have already seen in the discussion of the unpolarized 
cross section (see figs.~\ref{fig_08},\ref{fig_09}). Therefore, a full 
multipole analysis is not necessary. On the other hand, the contribution of 
the isospin 1/2 final state to the multipoles is not determined, and can only 
be estimated from model predictions. The clean separation of the isospin 
$I=1/2$ and $I=3/2$ components requires a combined analysis of the $\pi^o p$ 
and $\pi^+ n$ final states. Since the latter is strongly affected by background 
contributions, a much more involved analysis is necessary. Such analyses have 
been performed for the LEGS and the Mainz data. 
 
The LEGS-data were analyzed in an energy dependent way, using a 
parameterization of the energy dependence of the $(\gamma ,\pi)$ multipole 
amplitudes of the following type \cite{Blanpied_97}:
\begin{equation}
\label{eq:legs}
{\cal{M}}_{l\pm}^{i}=\left[{\cal{M}}_{B}^{I}(E_\gamma)+
                   \alpha_1\epsilon_{\pi}+\alpha_2\epsilon^2_{\pi}+
           \alpha_3\Theta_{2\pi}(E_{\gamma}-E_{\gamma}^{2\pi})^2\right]
           \times\left(1+iT_{\pi N}^l\right)+\beta T^l_{\pi N}\;,
\end{equation}
where $E_{\gamma}$ and $\epsilon_{\pi}$ are photon beam energy and pion 
kinetic energy. ${\cal{M}}_{B}^{I}$ are pseudo-vector Born amplitudes 
including $\rho$ and $\omega$ $t$-channel exchange. The $\alpha_i\epsilon^j$ 
terms are a phenomenological parameterization of non-Born background 
contributions, and the 
$\alpha_3\Theta_{2\pi}(E_{\gamma}-E_{\gamma}^{2\pi})^2$ 
term with the unit Heavyside step function 
($\Theta_{2\pi}$ = 1 for $E_{\gamma} > E_{\gamma}$=309 MeV) 
accounts for s-wave double pion production (used only for the $E_{0+}$ 
multipole). The $\pi N$ scattering matrix elements $T^l_{\pi N}$ were taken 
from the SAID multipole analysis \cite{Arndt_96}. The first term of eq. 
(\ref{eq:legs}) parameterizes the background contributions and the 
$\beta T^l_{\pi N}$ term the resonance contributions.
For a single resonance decaying into a single 
channel, the latter has the usual energy dependence of a Breit-Wigner curve. 
The ansatz satisfies the Watson theorem below the 2$\pi$ 
threshold and maintains unitarity at higher energies in a model dependent way. 
The multipoles up to the f-waves were included, and Born terms were kept up to 
$l$=19. The imaginary parts of the Compton amplitudes are connected to the 
$(\gamma ,\pi)$ multipoles via unitarity and their real parts were evaluated 
with dispersion integrals. The calculation of the latter requires 
$(\gamma ,\pi)$ multipoles outside the range of the analysis of the LEGS data, 
which were estimated from other sources (see \cite{Blanpied_97} for details). 
The parameters of the $(\gamma ,\pi)$ multipoles where then fitted to the cross 
section data from the 
$p(\vec{\gamma},\pi^o)p$, $p(\vec{\gamma},\pi^+)n$, and 
$p(\vec{\gamma},\gamma)p$ reactions and additional data for beam- 
\cite{Blanpied_92}, target- , and recoil polarization 
\cite{Dutz_96,Belyaev_83,Getman_81} and for double polarization observables 
\cite{Belyaev_84,Belyaev_86}. Only ratios of the additional polarization data 
were used in order to minimize normalization uncertainties. The $E2$ admixture 
of the $\Delta$ excitation is determined as the ratio of the $\beta$ 
coefficients of the $E_{1+}^{(3/2)}$ and $M_{1+}^{(3/2)}$ multipoles. In the 
fit, this ratio equals the ratio of the imaginary parts of the two multipoles. 
Systematic model uncertainties from higher partial waves, the $\pi N$ phase 
shifts, relative energy calibrations and assumptions used for the calculation 
of the Compton dispersion integrals are included. The authors quote a final 
result of 
\cite{Blanpied_97} $R_{EM} = (-3.0\pm 0.3_{stat+syst}\pm 0.2_{model})$\%. 
After a more refined analysis using an enlarged data base 
the final result is \cite{Blanpied_01}:
\begin{equation}
R_{EM} = (-3.07\pm 0.26_{stat+syst}\pm 0.24_{model})\%\;. 
\end{equation}
The combined results for the differential cross sections and photon beam
asymmetries of the $\pi^o p$ and $\pi^+ n$ channels from the Mainz experiment 
\cite{Beck_00} were analyzed with an energy dependent and an energy independent 
multipole analyses.
%
%
%
%
For the energy independent analysis of the MAMI data alone, 
8 parameters (the s- and p-wave multipole amplitudes 
$E_{0+}^{(I)}$, $M_{1+}^{(I)}$, $E_{1+}^{(I)}$ and $M_{1-}^{(I)}$, $I$=1/2,3/2) 
were fitted to the data independently for 16 energy intervals between 270 
and 420 MeV incident photon energy. Higher partial waves were taken into 
account for the Born-terms. The result for the imaginary and real parts of 
the two multipoles of interest are shown as solid dots in fig.~\ref{fig_13}.
The data were furthermore analyzed with fixed-t dispersion relations based on
Lorentz invariance, isospin symmetry, unitarity, and crossing symmetry
\cite{Hanstein_98}. This second analysis also included more recent Mainz data 
for the differential cross sections of the $\pi^o p$ final state 
\cite{Fuchs_96,Haerter_96}, data from Bonn for the target asymmetry 
\cite{Buechler_94,Dutz_96,Menze_77}, and differential cross sections for
$\pi^-$ photoproduction off the neutron \cite{Rossi_73,Bagheri_88}. The results 
for the multipoles of the two analyses are shown in fig.\ref{fig_13} 
(left hand side, open circles and line).  All three analyses agree very well, 
and the ratio of the $E_{1+}^{(3/2)}$ and $M_{1+}^{(3/2)}$ multipoles is shown 
in fig.~\ref{fig_13} (right hand side). The combined final result from the three 
analyses is quoted as \cite{Beck_00}:
\begin{equation}
R_{EM} = (-2.5\pm 0.1_{stat}\pm 0.2_{syst})\%\;, 
\end{equation}
which is lower than the LEGS result, but still consistent within the combined 
uncertainties.

The data from \cite{Beck_00,Blanpied_97} have also been analyzed by other 
groups. Davidson and Mukhopadhyay \cite{Davidson_97}, and Workman 
\cite{Workman_97} found quite different results for $R_{EM}$. Davidson and 
Mukhopadhyay used their effective Lagrangian model \cite{Davidson_91} for an 
analysis of the data from \cite{Beck_97}, and found a value of 
$R_{EM}$ = ($-$3.19$\pm$0.24)\%. Workman, using the SAID-multipole analysis 
\cite{Arndt_96}, found a much smaller value of $R_{EM}$ = ($-$1.5$\pm$0.5)\% 
\cite{Workman_97}. However, as pointed out in \cite{Beck_00}, these 
discrepancies can be traced to the data bases employed. In case of the 
effective Lagrangian analysis, the inclusion of the Mainz $\pi^+ n$-data in 
the data base lowered the result to $R_{EM}$ = ($-$2.64$\pm$0.25)\%. The small 
value obtained in the VPI multipole analysis is due to the inclusion of older 
data sets. Removal of the $\pi^o p$ data prior to 1980 from the SAID data base 
raises $R_{EM}$ into the range of the other analyses \cite{Beck_00,Workman_98}. 
In summary, the extraction of the $R_{EM}$ value seems to be almost model 
independent as long as the same data basis is used. The main systematic 
discrepancy which remains unresolved is the scale difference between the 
differential cross section data from Mainz and LEGS which results in a 
$R_{EM}$ value close to $-$3\% for LEGS, and close to $-$2.5\% for Mainz. 

Apart from systematic effects in the extraction of $R_{EM}$, another aspect 
is important for the comparison of the experimental result to quark model 
predictions. The very definition of the $E2/M1$ ratio and its relation to 
quantities predicted by models has to be considered. $R_{EM}$ is 
defined to correspond to the  $E2/M1$ ratio at the K-matrix pole on the real 
energy axis. Recently, it has been argued ( see e.g. \cite{Tiator_01}), that 
the T-matrix pole in the complex plane, which can be evaluated with the 
`speed-plot' technique, is a more fundamental quantity. However, the $E2/M1$ 
ratio at the T-matrix pole is a complex quantity and is not easily related to 
the real predictions of the quark models. Finally, as already discussed above, 
the value extracted from the data applies to the `dressed' resonance, while 
models in general will predict it for the `bare' resonance. Kamalov and Yang 
\cite{Kamalov_99} have investigated this problem with a dynamical model using 
a scattering equation of the type shown in fig.~\ref{fig_07}. Fitting the data 
with this model, they indeed find a large difference between the `dressed' 
and `bare' values. The resulting $E2/M1$-ratio for the `dressed' resonance of
($-$2.5$\pm$0.14)\% agrees nicely with the other analyses, but the result for
the `bare' resonance of (+0.25$\pm$0.19)\% is positive and compatible with 
zero. It is thus concluded that the `bare' $\Delta$ is nearly spherical and 
the $E2/M1$ mixing for the `dressed' resonance can be  attributed to the pion 
cloud, i.e. to pion re-scattering effects. In a similar analysis, Sato and 
Lee \cite{Sato_01} also find a significant difference between the `dressed' 
($-$2.7\%) and `bare' ($-$1.3\%) values of  $E2/M1$. However, in their 
dynamical model the `bare' value is negative and has a larger magnitude than 
in the model of Kamalov and Yang. Recently, the predictions of the two models 
for a different observable have been tested experimentally \cite{Bartsch_02}. 
The experiment determined the beam-helicity asymmetry $\rho_{LT'}$ in the 
electroproduction of $\pi^o$-mesons with longitudinally polarized electrons 
($p(\vec{e},e'\pi^o$)p). 
The MAID model \cite{Drechsel_99} as well as the two dynamical models 
correctly predict the negative sign of the measured asymmetry and agree 
fairly well with the angular dependence. However, all fail to reproduce
the absolute magnitude. The MAID model, and the dynamical model of Kamalov and 
Yang {\em overestimate} the magnitude by roughly 30\% while the model of Sato 
and Lee {\em underestimates} it by almost the same value. These results seem 
to indicate that the pion cloud effects are not yet well under control in the 
models. In this sense, any comparison of the experimental value for $R_{EM}$ 
to model predictions must be regarded with care.

For some issues, electron scattering results obtained at finite momentum 
transfer $Q^2$ are closely related to the topic of this review, and we find 
it necessary to include them. In the case of virtual photons, the $C2$ Coulomb 
quadrupole excitation, which corresponds to the $S_{1+}$-multipole,
can additionally contribute. The $C2$ analog to the $R_{EM}$-ratio for the 
$E2$-admixture is the $R_{SM}$-ratio defined via:
\begin{equation} 
R_{SM}=S_{1+}^{(3/2)}/M_{1+}^{(3/2)}\;.
\end{equation}

Perturbative QCD (pQCD) predicts 
\cite{Brodsky_81}-\cite{Carlson_98} 
that, in the limit of very high $Q^2$, only helicity conserving amplitudes 
survive. In particular, the $A_{3/2}$ and $A_{1/2}$ helicity amplitudes 
for the electromagnetic excitation of the $\Delta$ scale like 
\cite{Brodsky_81}:
\begin{equation}  
A_{3/2}  \propto  Q^{-5}\;\;\;\;\;\;\;\;\; 
A_{1/2}  \propto  Q^{-3}
\end{equation}
for large $Q^2$. Consequently, the $A_{1/2}$-amplitude dominates the 
$\Delta$-excitation for large $Q^2$. This means, that pQCD predicts an 
asymptotic $R_{EM}$ value of +100\% (see eqs.~\-(\ref{eq:helrat1},\ref{eq:helrat2})). 
The prediction for $R_{SM}$ at large $Q^2$ is a constant value. Therefore, 
investigating the $R_{EM}$ value as function of $Q^2$, the transition from 
the constituent quark model to the region of pQCD can be studied.

\begin{figure}[hbt]
\centerline{\epsfysize=6.4cm \epsffile{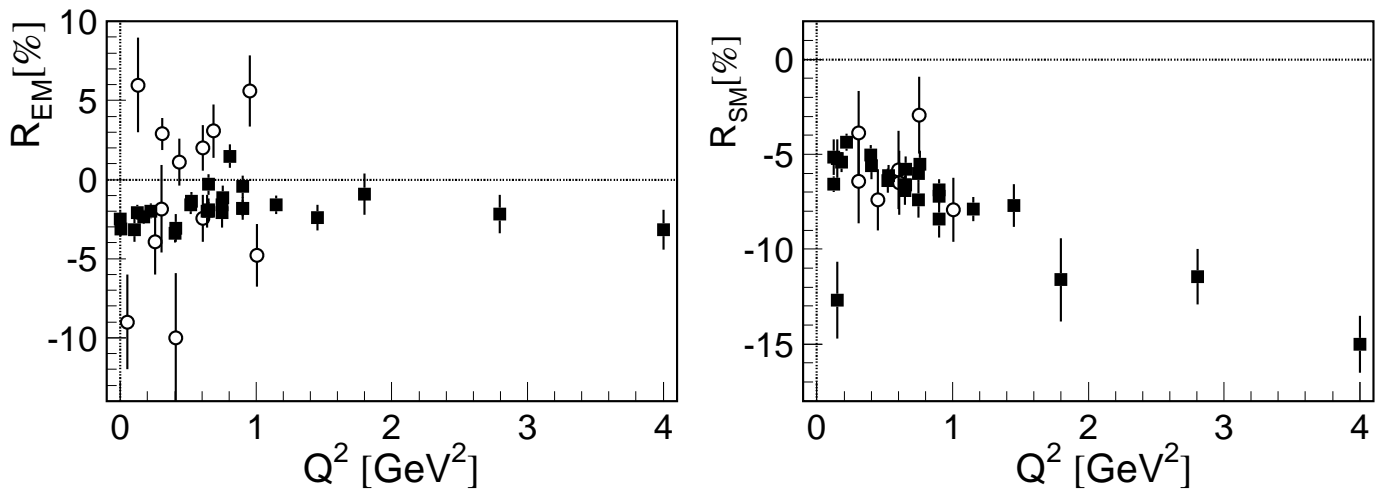}}
\caption{$Q^2$-dependence of $R_{EM}$ (left hand side) and $R_{SM}$ 
(right hand side). Photon point values ($Q^2$=0): refs.  
\cite{Beck_00,Blanpied_01}, values for $Q^2>0$: 
\cite{Mertz_01}-\cite{Papanicolas_01}
and ref. therein. Open symbols: pre-1990 data.}
\label{fig_14}       
\end{figure}

At Bates, Bonn (ELSA), JLab, and Mainz (MAMI), efforts to measure the 
$Q^2$-dependence of $R_{EM}$ and $R_{SM}$ have been undertaken 
\cite{Mertz_01}-\cite{Pospischil_01}.
The results are summarized and compared to older data in fig.~\ref{fig_14}.
It is evident from the figure, that the recent experiments have improved the 
data base significantly. The scattering of the data prior to 1990 was 
substantial. In the case of $R_{EM}$, not even a clear tendency towards 
positive or negative values was visible. Meanwhile, the new data establish 
the $Q^2$-dependence for both ratios more firmly. The result for $R_{EM}$ 
are small, negative values showing no pronounced $Q^2$-dependence up to 
4 GeV$^2$ while $R_{SM}$ seems to drop as function of $Q^2$. Since pQCD 
predicts $R_{EM}$=+100\% and a constant $R_{SM}$-value, it is obvious that up 
to momentum transfers squared of 4 GeV$^2$ an onset of pQCD behavior is not 
visible. 
%
%
%
%
For a more detailed interpretation of the results one should keep
in mind that the extraction of the $R_{EM}$ and $R_{SM}$ ratios is not 
completely model independent. The problem is connected to the non-resonant
background contributions which we have already discussed for the measurements 
at the photon point and which are less well under control for high $Q^2$. The 
measurement of a rather large induced proton polarization $p_n$ in $\pi^o$ 
electroproduction at Bates \cite{Warren_98} was interpreted as evidence for 
background contributions larger than those predicted by models. However, in 
principle this observable and $R_{SM}$ extracted from the double polarization 
measurement $p(\vec{e},e'\vec{p})p$ \cite{Pospischil_01} can involve different 
combinations of multipole amplitudes. In particular, while $p_n$ is sensitive 
to the real part of the background amplitudes, $R_{SM}$ is only sensitive to
the imaginary part \cite{Warren_98,Pospischil_01}, so that it is difficult to 
judge the influence of the background amplitudes on $R_{SM}$.

\subsection{\it Helicity Dependence of Pion Photoproduction in the
$\Delta$-Range}
\label{ssec:delta_gdh}

\begin{figure}[bht]
\begin{minipage}{7.cm}
{\mbox{\epsfysize=2.5cm \epsffile{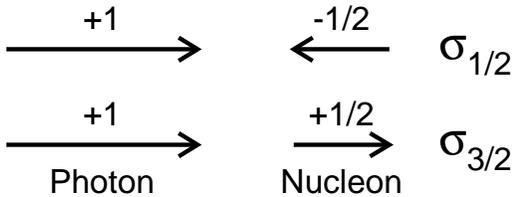}}}
\caption{Definition of $\sigma_{1/2}$, $\sigma_{3/2}$. The arrows symbolize
the spin projections of the photon and nucleon onto the momentum axis of the
incoming photon.}
\label{fig_15}       
\end{minipage}
\end{figure}

The initial photon - nucleon state is characterized by the helicities, i.e. 
the spin projections onto the momentum axis (see app. 6.2).
For a real photon
with $\lambda=\pm 1$ and the nucleon with $\nu_i=\pm 1/2$, two different
possibilities exist which are schematically shown in fig.~\ref{fig_15}.
They correspond to the photoproduction cross sections $\sigma_{1/2}$ and
$\sigma_{3/2}$ with total helicities 1/2 and 3/2. 
%
%
%
%

\vspace*{-5.6cm}
\hspace*{7.cm}
\begin{minipage}{10.7cm}
In 1966, Gerasimov \cite{Gerasimov_66} and independently Drell and Hearn 
\cite{Drell_66} derived the Gerasimov-Drell-Hearn (GDH) sum rule which relates 
the difference of the two helicity components of the total photoabsorption 
cross section to static properties of the nucleon via the GDH-integral:
\begin{equation}
\int_{m_{\pi}}^{\infty}\frac{\sigma_{3/2}(\omega)-\sigma_{1/2}(\omega)}
{\omega}d\omega = 
\frac{\pi e^2}{2m^2_N}\kappa^2\;,
\end{equation}
where $\omega=E_{\gamma}$ is the incident photon energy in the lab frame, 
$m_{\pi}$ the mass of the pion, $m_N$ the mass of the nucleon, and $\kappa$ 
the anomalous magnetic moment of the nucleon. 
\end{minipage}

\vspace*{0.3cm}
\noindent{During} the last few years, the GDH collaboration has undertaken a 
joint effort towards the experimental verification of the sum rule, measuring 
the difference of the helicity components in total photoabsorption. The 
experiment is divided 
into two parts, the first for the energy range from the pion threshold up to 
800 MeV carried out at the Mainz MAMI accelerator and the second in the energy 
range 600 MeV - 3 GeV at the Bonn ELSA accelerator. The experiment requires
a measurement with a circularly polarized photon beam and a longitudinally 
polarized proton target. Circularly polarized photon beams are produced
via backscattering of circularly polarized laser beams (LEGS, GRAAL) or via 
bremsstrahlung of longitudinally polarized electron beams (MAMI, ELSA).
Both are now available at most electron beam facilities. For the longitudinally 
polarized protons, a butanol (C$_4$H$_9$OH) frozen spin target was developed 
\cite{Bradtke_99}. The recent advance in the technology of polarized targets 
(see \cite{Meyer_02} for a review) was essential for the success of the 
experiment. First results for the GDH integral have been published 
\cite{Ahrens_GDH_00,Ahrens_GDH_01}. Here, we will 
not discuss the sum rule itself but emphasize another aspect.

The low energy part of the experiment has not only determined the helicity
dependence of the total absorption cross section, but also provided results
for the helicity difference for exclusive channels
\cite{Ahrens_GDH_01}-\cite{Ahrens_GDH_03}. 
These observables provide valuable new constraints for the multipole analysis 
of photoproduction reactions. As an example, we consider the single pion 
photoproduction reactions $\vec{\gamma}\vec{p}\rightarrow p\pi^o$ and 
$\vec{\gamma}\vec{p}\rightarrow n\pi^+$.
%
%
%
\begin{figure}[thb]
\begin{minipage}{0.0cm}
{\mbox{\epsfysize=7.5cm \epsffile{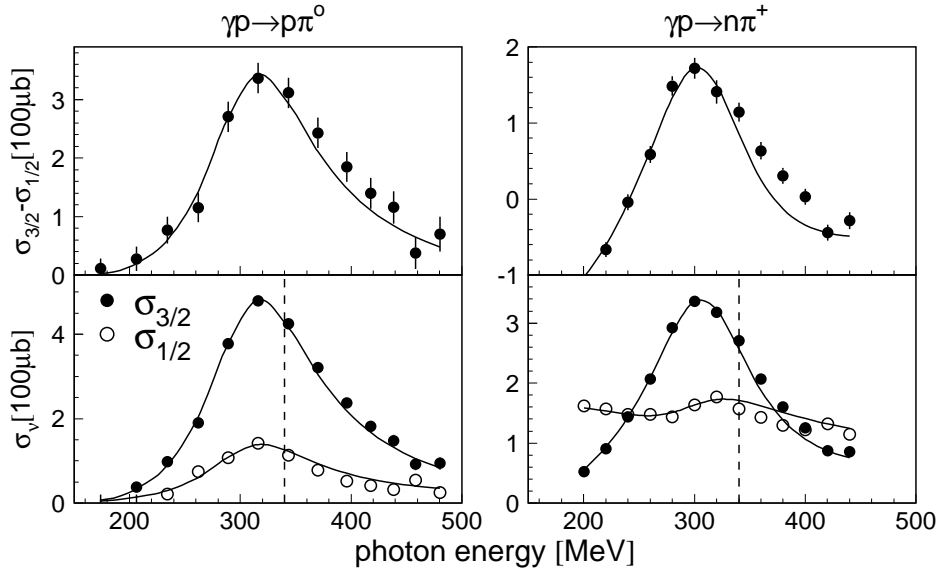}}}
\end{minipage}
\hspace*{12.5cm}
\begin{minipage}{5.5cm}
\vspace*{-0.5cm}
\caption{Helicity dependent pion photoproduction cross sections in the
$\Delta$ range (left hand side: $p \pi^o$, right hand side: $n \pi^+$ final 
states). Top part: cross section difference of the two helicity states 
\cite{Ahrens_GDH_01}. Bottom part: helicity dependent cross sections 
(full symbols: $\sigma_{3/2}$, open symbols: $\sigma_{1/2}$). Curves:  
MAID results \cite{Drechsel_99}. Dashed lines: $\Delta$ resonance position.}
\label{fig_16}       
\end{minipage}
\end{figure}
%
The measured helicity difference $\Delta\sigma = \sigma_{3/2}-\sigma_{1/2}$ 
for these reactions \cite{Ahrens_GDH_00} is shown in fig.~\ref{fig_16}. The 
total cross section for each channel is given by 
$\sigma_{tot}=(\sigma_{3/2}+\sigma_{1/2})/2$ and the cross sections 
corresponding to the helicity 3/2 and 1/2 states can be reconstructed from:
\begin{equation}
\sigma_{3/2} = \sigma_{tot}+\frac{1}{2}\Delta\sigma
\;\;\;\;\;\;\;\;\;\;\;
\sigma_{1/2} = \sigma_{tot}-\frac{1}{2}\Delta\sigma\;.
\end{equation}
The results for the two cross sections are shown in the lower part of 
fig.~\ref{fig_16}. The total cross section for both reactions was taken from 
the MAID parameterization \cite{Drechsel_99} (see also fig.~\ref{fig_08}). 
If only the $\Delta$ excitation were to contribute to the observed cross 
sections, the ratio of the helicity 3/2 and 1/2 components would give directly 
the ratio of the $A_{3/2}$ and $A_{1/2}$ helicity amplitudes of the $\Delta$. 
However, it is obvious from fig.~\ref{fig_16} that the helicity 1/2 
component for the $n\pi^+$ channel has significant background contributions. 
So far, background contributions for all observables investigated have been 
weak for the $p\pi^o$ channel at the $\Delta$ position of 340 MeV incident 
photon energy. It is interesting to check which ratio of the helicity 
amplitudes would follow from the experimental $\pi^o$ cross sections at 
$E_{\gamma}=340$ MeV, assuming background is negligible. 
The result (up to a sign) is:
\begin{equation}
\frac{A_{3/2}}{A_{1/2}}\approx
\sqrt{\frac{\sigma_{3/2}(\gamma p\rightarrow p\pi^0 ,W=M_{\Delta})}
{\sigma_{1/2}(\gamma p\rightarrow p\pi^0 ,W=M_{\Delta})}}
=(1.94\pm 0.15_{stat}\pm 0.10_{syst})\;.
\end{equation}
This value compares very well with recent results for the helicity
amplitudes obtained from the full multipole analyses of the differential cross
sections and photon beam asymmetries as discussed in the previous section:
\begin{eqnarray}
\frac{A_{3/2}}{A_{1/2}} & = & \frac{-(251.0\pm 1.0)(10^{-3}\sqrt{\mbox{GeV}})}
{-(131.0\pm 1.0)(10^{-3}\sqrt{GeV})} = (1.916\pm 0.016)\;, 
~~~~~\mbox{ref. }\cite{Beck_00}\\
\frac{A_{3/2}}{A_{1/2}} & = & \frac{-(266.9\pm 8.0)(10^{-3}\sqrt{\mbox{GeV}})}
{-(135.7\pm 3.9)(10^{-3}\sqrt{\mbox{GeV}})} = (1.967\pm 0.082)\;, 
~~~~~\mbox{ref. }\cite{Blanpied_01}\;.
\end{eqnarray}

The ratio of the helicity amplitudes is connected to the $E2$ admixture in the 
$\Delta$ excitation. For a pure $M1$ transition it would be given by the ratio 
of the Clebsch-Gordan coefficients for the 3/2 and 1/2 states which is 
$\sqrt{3}=1.73$. One can check using the MAID model \cite{Drechsel_99} that 
the cross section ratio $\sigma_{3/2}/\sigma_{1/2}$ for $\pi^o$-photoproduction 
at $E_{\gamma}=340$ MeV is predicted to be very close to the Clebsch-Gordan 
(1.70) when the $E2$-contribution is switched off but all background 
contributions are kept. This is another indication that this ratio seems to be 
almost unaffected by background contributions.

The connection between the helicity and multipole amplitudes for the $\Delta$ 
excitation in the presence of a non-vanishing $E2$ admixture is given by:
\begin{equation}
\label{eq:helrat1}
R_{A}=\frac{A_{3/2}}{A_{1/2}}=
\left.\frac{\sqrt{3}\left(M_{1+}^{\Delta}-E_{1+}^{\Delta}\right)}
{\left(M_{1+}^{\Delta}+3E_{1+}^{\Delta}\right)}
\right|_{W=M_\Delta}\;.
\end{equation}
The ratio of the helicity amplitudes and the $R_{EM}$ value are related by:
\begin{equation}
\label{eq:helrat2}
R_{EM}=\left.\frac{E_{1+}^{\Delta}}{M_{1+}^{\Delta}}\right|_{W=M_\Delta}=
\frac{\sqrt{3}-R_{A}}{3R_{A}+\sqrt{3}}\;,
\end{equation}
which results in $R_{EM}$=$-$2.75\%, in agreement with refs. 
\cite{Beck_00,Blanpied_01}. It must be emphasized that possible background 
contributions have been neglected. The extracted $R_{EM}$ value is therefore 
less well founded than the results obtained from the multipole analysis. 
However, it demonstrates that the results obtained from the differential 
cross sections and photon beam asymmetries are consistent with the outcome 
of the helicity dependent cross sections.  

\subsection{\it The Magnetic Moment of the $\Delta$-Resonance}
The anomalous magnetic moments of proton and neutron \cite{Stern_33} gave the 
first hint for a substructure of the nucleon. As is the case in nuclear physics,
magnetic moments depend sensitively on the details of the wave functions.  
The magnetic moments of the octet baryons ($N$, $\Lambda$, $\Sigma$, $\Xi$) 
are known precisely from spin precession measurements \cite{PDG}. However, 
the lifetimes of the decuplet baryons are so short that a direct measurement 
of the magnetic moment was only possible for the $\Omega^-$. The 
$\Delta$-resonance is a particularly interesting case for which models assume 
a quark structure similar to the nucleon ground state but with spin and 
isospin of the quarks coupled to 3/2 instead of 1/2. 
In the case of SU(3) flavor symmetry, the mass of the nucleon ground state and 
the $\Delta$ would be degenerate, and the magnetic moments would be related 
by $\mu_{\Delta}=Q_{\Delta}\mu_p$. Here, $Q_{\Delta}$ is the charge of the 
$\Delta$ and $\mu_p$ the magnetic moment of the proton. This would imply 
in particular 
that $\mu_{\Delta^+}=\mu_p$, and $\mu_{\Delta^o}=0$. Predictions 
for the $\Delta$-magnetic moments are available from a variety of nucleon 
models (see table \ref{tab_02}). Experimentally, estimates have only been 
obtained for the $\Delta^{++}$.

%
%
%
\begin{figure}[htb]
\begin{minipage}{7.5cm}
{\mbox{\epsfysize=6.cm \epsffile{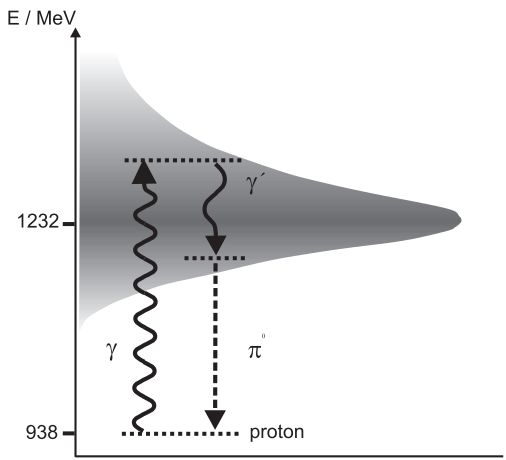}}}
\caption{Principle of the determination of the $\Delta$ magnetic moment
from the reaction $p(\gamma ,\pi^o\gamma ')p$.}
\label{fig_17}       
\end{minipage}
\end{figure}
%

\vspace*{-8.9cm}
\hspace*{8.cm}
\begin{minipage}{9.6cm}
The experimental value for $\mu_{\Delta^{++}}$ is based on two measurements
of the hadron induced reaction $\pi^+ p\rightarrow \pi^+\gamma 'p$ carried
out at the University of California, Los Angeles (UCLA) 
\cite{Arman_72,Nefkens_78} and at the Paul Scherrer Institut, Switzerland 
\cite{Bosshard_91}. The value quoted by the Particle Data Group (PDG) is 
$\mu_{\Delta^{++}}$= (3.7 - 7.5)$\mu_{N}$. This large uncertainty comes mainly 
from a model dependence of the extraction of the magnetic moment from the data. 
The problem is that
$\pi^+$-bremsstrahlung in the initial and final state makes a large 
contribution to the $\pi^+ p\rightarrow \pi^+\gamma 'p$ cross section. 
Although kinematic conditions with maximum destructive interference between
the bremsstrahlung contributions were chosen in the experiments, the model
dependence of the analysis was significant. Different analyses of the UCLA
experiment came up with values ranging from (3.7 - 4.2)$\mu_{N}$ to
(6.9 - 9.8)$\mu_{N}$ 
\cite{Castro_02}-\cite{Heller_87}.
\end{minipage}

%
%
%
\begin{table}[th]
  \caption[Magnetic moments of the $\Delta$]{
    \label{tab_02}
    Model predictions for the magnetic moment of the P$_{33}$(1232) resonance
    in units of the nuclear magneton $\mu_{N}=e\hbar /2m_p$. Models:
    relativistic quark model (RQM), chiral bag model ($\chi$B), chiral
    quark-soliton model ($\chi$QSM), chiral perturbation theory ($\chi$PT),
    QCD sum rules (QCDSR), light cone QCD sum rules (LCQSR), lattice QCD (LQCD).  
}
  \begin{center}
    \begin{tabular}{|l|c|c|c|c|}
      \hline 
        {\bf Method}
      & {\bf $\mu_{\Delta^{++}}/\mu_{N}$ }
      & {\bf $\mu_{\Delta^{+}}/\mu_{N}$ }
      & {\bf $\mu_{\Delta^{o}}/\mu_{N}$ }
      & {\bf $\mu_{\Delta^{-}}/\mu_{N}$ }\\
      \hline \hline
      Experiment \cite{PDG} & 3.7 - 7.5 & & &\\ 
      SU(3) 
      & 5.58 & 2.79 & 0 & -2.79 \\
      RQM \cite{Schlumpf_93} 
      & 4.76 & 2.38 & 0 & -238\\
      $\chi$B \cite{Hong_99} 
      & 3.59 & 0.75 & -2.09 & -1.93\\
      $\chi$QSM \cite{Kim_98} 
      & 4.73 & 2.19 & -0.35 & -2.9\\
      $\chi$PT \cite{Butler_94} 
      & 4.0$\pm$0.4 & 2.1$\pm$0.2 & -0.17$\pm$0.04 & -2.25$\pm$0.25\\
      QCDSR \cite{Lee_98} 
      & 4.13$\pm$1.30 & 2.07$\pm$0.65 & 0 & -2.07$\pm$0.65\\
      LCQSR \cite{Aliev_00} & 4.4$\pm$0.8 & 2.2$\pm$0.4 & 0 & -2.2$\pm$0.4\\
      LQCD \cite{Leinweber_92} 
      & 4.91$\pm$0.61 & 2.46$\pm$0.31 & 0 & -2.46$\pm$0.31\\
      \hline
    \end{tabular}
  \end{center}
\end{table}
%

\vspace*{0.3cm}
\noindent{Recently,} a first attempt was made to measure the magnetic moment 
of the $\Delta^+$ via the photon induced reaction 
$\gamma p\rightarrow \pi^o\gamma 'p$. The principle of the experiment is the 
same as for the pion induced reaction and was first suggested by Kontratyuk 
and Ponomarev \cite{Kontratyuk_68}. It is schematically depicted in fig. 
\ref{fig_17}. The $\Delta$-resonance is excited by a real photon, and decays 
within its final width via an electromagnetic $M1$ realignment transition, 
which is sensitive to the magnetic moment. Finally, the resonance de-excites 
by emission of a $\pi^o$-meson to the nucleon ground state. The small values 
of the $E2/M1$ admixture in the excitation of the $\Delta$, as discussed in 
the previous section, indicate that the quadrupole deformation of the 
$\Delta$ must be small. Furthermore, electric quadrupole realignment 
transitions vanish in the limit of zero photon energy because of time reversal 
symmetry \cite{Drechsel_01}. The next higher magnetic octupole is suppressed 
by two additional powers of photon momenta so that the re-alignment transition 
is dominated by the $M1$ multipole which couples to the magnetic dipole moment.

A pilot experiment \cite{Kotulla_01,Kotulla_02} was performed using the TAPS 
detector \cite{Novotny_91,Gabler_94} at the MAMI-B accelerator. In a fully 
exclusive measurement, the momenta of the recoil proton, the realignment photon, 
and the photons from the $\pi^o\rightarrow 2\gamma$ decay were measured.
The $\pi^o$ mesons were reconstructed via a standard invariant mass analysis. 
Additional kinematic cuts ensured the unique identification of the reaction 
and the rejection of background arising mainly from double $\pi^o$ 
photoproduction events, where one photon had escaped detection due to the 
limited solid angle coverage of the detector.

The contribution of bremsstrahlung processes to the 
$\gamma p\rightarrow \pi^o\gamma 'p$ reaction is less important than for the
$\pi^+ p\rightarrow \pi^+\gamma 'p$ reaction since the latter has the charged
pions in the initial and final state. Nevertheless, the contributions are 
still significant, and a reaction model is necessary for the extraction of 
the magnetic moment. First calculations of the resonant 
$\Delta\rightarrow\Delta\gamma '$ process have been performed in the effective
Lagrangian formalism by Machavariani, Faessler, and Buchmann 
\cite{Machavariani_99} and by Drechsel, Vanderhaeghen, and Giannini
\cite{Drechsel_00}. The experimental data cannot be reproduced since the 
bremsstrahlung contributions are neglected in the calculations, and merely a 
first estimate of the size of the cross section is provided. Recently, Drechsel
and Vanderhaeghen \cite{Drechsel_01} included background diagrams from
bremsstrahlung, non-resonant Born graphs, and vector meson exchange. A useful 
feature of this model is that the $\gamma p\rightarrow p\pi^o$ reaction can be 
calculated as a first step. 
%
%
%
\begin{figure}[hbt]
\begin{minipage}{0.0cm}
{\mbox{\epsfysize=7.cm \epsffile{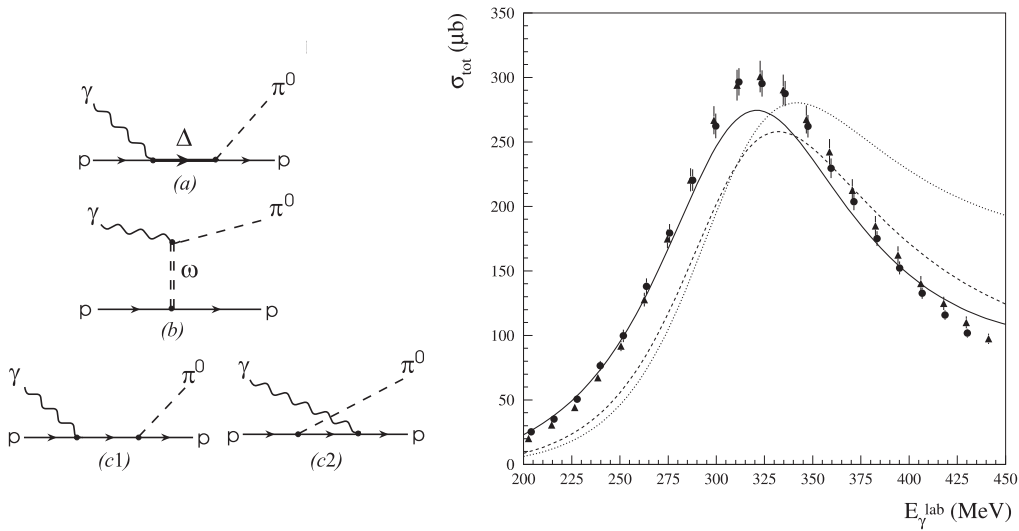}}}
\end{minipage}
\hspace*{14.cm}
\begin{minipage}{4.1cm}
\caption{Left hand side: diagrams included in the model for $p(\gamma ,\pi^o)p$ 
\protect\cite{Drechsel_01}. Right hand side: comparison of the model prediction 
to the measured total cross section of $p(\gamma ,\pi^o)p$. 
Dotted: only $\Delta$ excitation, dashed: $\Delta$ and $\omega$ exchange, 
full: all contributions.
}
\label{fig_18a}       
\end{minipage}
\end{figure}
%
Then, in a second step, the additional photon from 
the realignment transition and from bremsstrahlung contributions can be added.
The diagrams included in the calculation of pion production and the comparison 
of the model result for the total cross section to data for $\pi^o$ 
photoproduction are shown in fig.~\ref{fig_18a}. At energies 
below the maximum of the $\Delta$-resonance, the agreement is good but worsens 
at higher incident photon energies. This discrepancy is probably due to pion 
re-scattering contributions which have not yet been included in the model.

\newpage
%
%
%
\begin{figure}[bht]
\begin{minipage}{9.1cm}
{\mbox{\epsfysize=8.5cm \epsffile{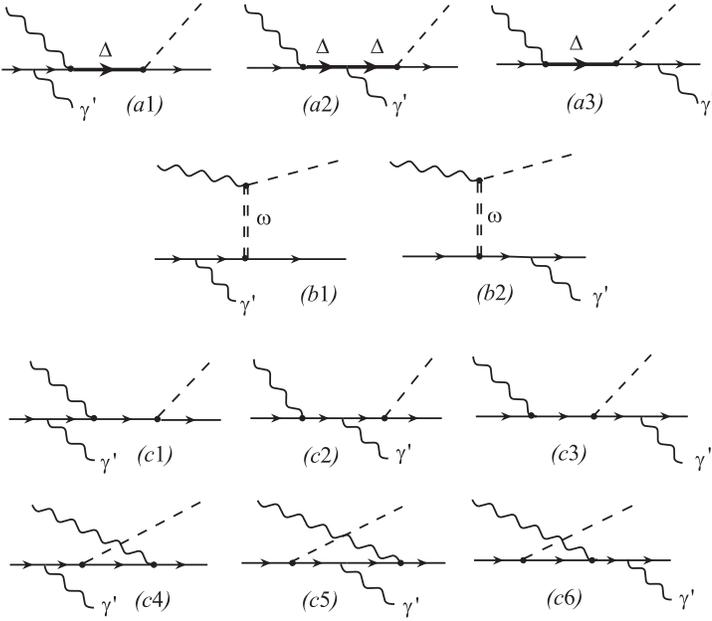}}}
\caption{Diagrams included for the calculation of the 
$\gamma p\rightarrow p\pi^0\gamma '$ reaction \protect\cite{Drechsel_01}.
The process of interest for the $\Delta$-magnetic moment is diagram (a2).
}
\label{fig_18b}       
\end{minipage}
\end{figure}
%

\begin{figure}[hbt]
\begin{center}
\epsfysize=11.5cm \epsffile{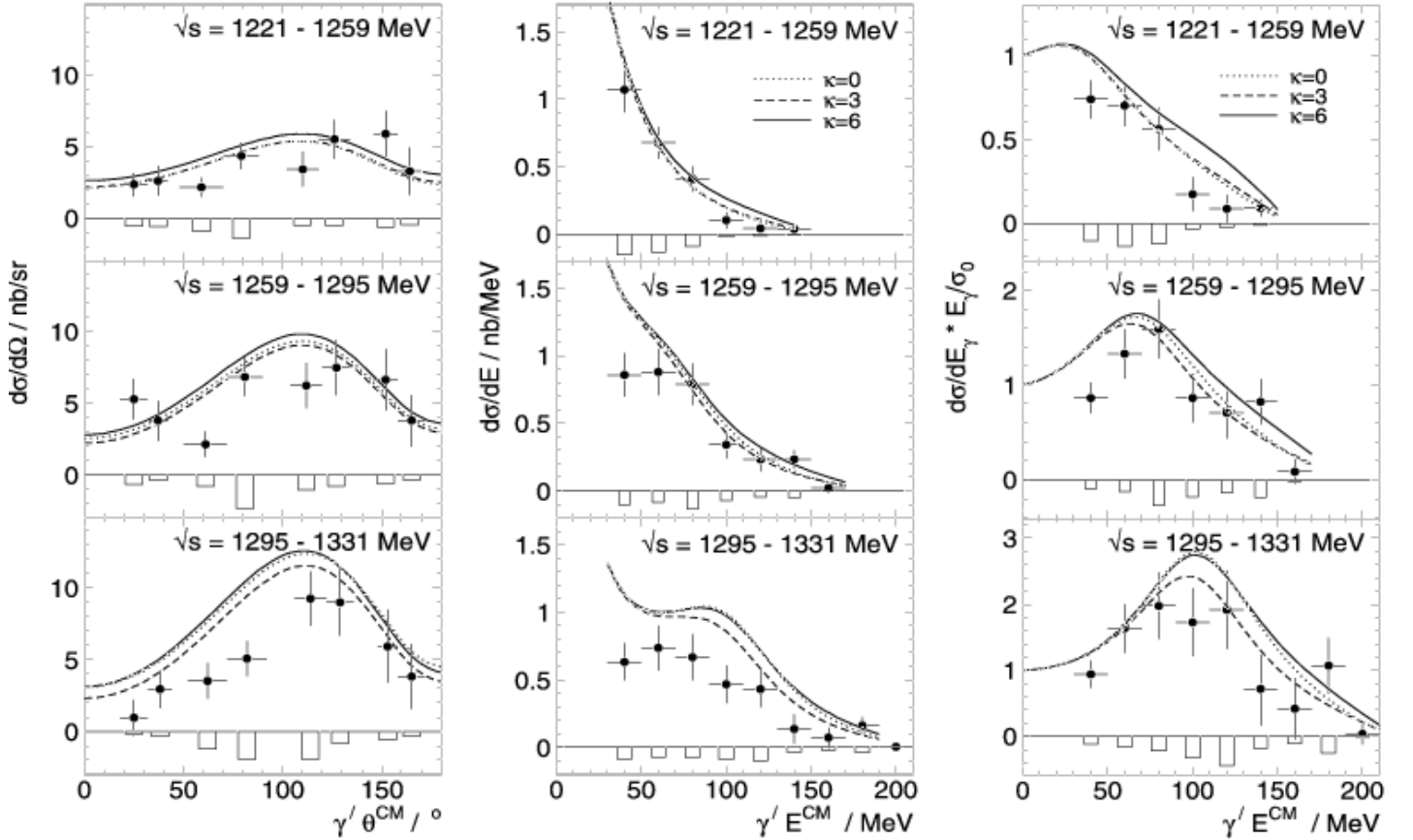}
\caption{Differential cross sections \cite{Kotulla_02} for three ranges of 
incident excitation energies in the cm frame. The systematic errors are shown 
as the bar chart. The left column shows the angular distribution of the 
alignment photon, the middle column its energy. The energy distributions on the 
right hand side have been normalized for both data and calculations to the soft 
photon limit. The calculations correspond to magnetic moments with 
$\kappa_{\Delta^+}$=0,3,6.
}
\label{fig_19}       
\end{center}
\end{figure}

\vspace*{-10.cm}
\hspace*{10.5cm}
\begin{minipage}{7.cm}
The diagrams for the modeling of the $\gamma p\rightarrow p\pi^0\gamma '$ 
reaction are shown in fig.~\ref{fig_18b}. The only new free 
parameter compared to the calculation for pion  photoproduction is the magnetic 
moment $\mu_{\Delta^+}$ of the $\Delta$ resonance in diagram (a2) which is 
expressed in terms of the anomalous magnetic moment $\kappa_{\Delta^+}$ via:
\begin{equation}
\mu_{\Delta^+}=
(1+\kappa_{\Delta^+})\frac{m_N}{m_\Delta} \mu_N
\end{equation}
where $\mu_N$ is the nuclear magneton, and $m_N$, $m_\Delta$ are the nucleon 
and $\Delta$ masses. A comparison of the model prediction for different values
of $\kappa_{\Delta^+}$ to the data, shown in fig.~\ref{fig_19},
can be used for the extraction of the magnetic moment.
\end{minipage}

\vspace*{1cm}
\noindent{The} energy distributions of the $\gamma '$ photon for low energy incident 
photons show the typical $1/E_{\gamma '}$ behavior of bremsstrahlung.
At higher incident photon energies an additional structure develops which is
partly due to the $\Delta$ radiation. The shape of the distributions is rather
well reproduced by the model predictions. However, for the highest incident 
photon energy, the absolute values do not agree for any reasonable value of 
$\kappa_{\Delta^+}$. Since the model already fails for $\pi^o$ photoproduction 
at the high energy side of the $\Delta$, apparently, systematic effects are 
not yet under control. The systematic uncertainty of the data may add to the 
discrepancy.

A rough correction of the effects is attempted in \cite{Kotulla_02} via a
renormalisation of the data and the model predictions to the soft photon limit 
which relates the cross sections for the $\pi^o\gamma 'p$ and $\pi^o p$ final
states in the limit of vanishing $E_{\gamma '}$  \cite{Kotulla_02}:
\begin{equation}
\lim_{E_{\gamma '}\rightarrow 0}\left(\frac{d\sigma}{dE_{\gamma '}}\right)=
\frac{1}{E_{\gamma '}}\sigma_{o}\;,
\end{equation}
where 
\begin{equation}
\sigma_{o}=
\int d\Omega_{\pi^o}\left(\frac{d\sigma}{d\Omega_{\pi^o}}\right)
\frac{\alpha}{2\pi}{\cal F}(t)
\end{equation}
with a kinematic function $\cal{F}$ depending on the four momentum transfer 
$t$ between initial photon and $\pi^o$-meson:
\begin{equation}
{\cal F}(t) = 4\left[-1+\left(\frac{v^2+1}{2v}\right)
ln\left(\frac{v+1}{v-1}\right)\right]\;\;,v=\sqrt{1-\frac{4m_p^2}{t}}\;.
\end{equation}
Model predictions and data were divided by $\sigma_o/E_{\gamma '}$. For the 
model, $\sigma_o$ was taken from the model prediction for neutral pion 
production, and for the data the integral for $\sigma_o$ was evaluated with 
measured angular distributions. In this way, systematic effects are reduced, 
and a value of:
\begin{equation}
\mu_{\Delta^+}=(2.7^{+1.0}_{-1.3}\pm 1.5)\mu_N
\end{equation}
is extracted from the data. The first error is the statistical and the second 
the systematic error which, however, does not include systematic effects in 
the model calculations. The uncertainty of the above result is too severe for 
precise tests of hadron models. A second generation experiment is in 
preparation which will  use a $4\pi$ detector at the MAMI-B accelerator 
promising to improve the statistical quality of the data. In parallel, 
theoretical efforts are continuing to improve the model by including the 
presently neglected re-scattering contributions. Furthermore, the model 
calculations \cite{Drechsel_01} have demonstrated that the 5-fold differential 
cross section $d\sigma/(dE_{\gamma}d\Omega_{\gamma}d\Omega_{\pi})$ has a 
larger sensitivity to $\mu_{\Delta^+}$ for special kinematic regions than the 
distributions measured previously. Also, photon beam asymmetries are predicted 
to have a higher sensitivity to the magnetic moment. These observables will 
be exploited in the follow-up experiment. 

%
%
%
%

The method is in principle not restricted to the $\Delta$. However, the 
reaction $\gamma p\rightarrow p\pi^o\gamma '$ is probably not well suited for 
the extraction of magnetic moments of higher lying resonances. Even if 
background from reactions with higher pion multiplicity could be eliminated, 
it would seem to be almost hopeless to disentangle the contributions from 
different, overlapping resonances in the presence of large non-resonant 
backgrounds. However, at higher energies other meson production reactions 
can be exploited. A measurement of the magnetic moment of the S$_{11}$(1535) 
resonance via the reaction $\gamma p\rightarrow p\eta\gamma '$ seems to be 
promising.  In this case, background from other reactions 
is negligible, and in the relevant range, $\eta$-photoproduction is dominated 
by the excitation of the first S$_{11}$ resonance.    
First model calculations for this reaction have been presented by
Chiang et al. \cite{Chiang_02a}.

\newpage
\subsection{\it The Excitation of the $\Delta$-Resonance on the Neutron}
\label{ssec:delta_neutron}
A full multipole analysis including the isospin structure of the amplitudes
requires the measurement of meson photoproduction from the neutron. Targets of 
free neutrons do not exist, and the cross section must be extracted from 
measurements on neutrons bound in nuclei. One obvious choice of the target 
nucleus is the weakly bound deuteron. A number of experiments using deuteron 
targets have recently been and are currently performed, aiming at the 
investigation of nucleon resonance properties as well as tests of the GDH sum 
rule on the neutron. In the following, we will discuss pion photoproduction 
off the deuteron in the $\Delta$-range. The situation is particularly simple 
when only the $\Delta$-resonance is excited. Only the total isospin changing 
part $A^{V3}$ of the amplitude can contribute, and one can directly read off 
from eqs.~\-(\ref{eq:iso_2}):
\begin{equation}
\label{eq:pion_iso_cross}
\sigma (\gamma p\rightarrow p\pi^o) =  
\sigma (\gamma n\rightarrow n\pi^o) =
2\sigma (\gamma p\rightarrow n\pi^+) =
2\sigma (\gamma n\rightarrow p\pi^-)  
\end{equation}

The contribution of background terms will modify these simple relations. It is 
obvious that experimental information about the reaction on the neutron would 
be very useful for the separation of the background and resonance contributions.
The relations of eq.~\-(\ref{eq:pion_iso_cross}) hold as long as isospin 
violating terms with $\Delta I\geq2$ can be neglected. The investigation of 
pion photoproduction off the neutron thus can be used to search for isotensor 
contributions in the $\Delta$ excitation, provided the background contributions 
can be sufficiently controlled.  

The charged final state $p\pi^-$ had been investigated 30 years ago in a bubble 
chamber measurement of the $\gamma d\rightarrow pp\pi^-$ reaction by the ABHHM 
collaboration \cite{Benz_73}, at Frascati \cite{Chiefari_75}, and later at 
higher energies by the TAGX-collaboration \cite{Asai_90}.
%
%
%
\begin{figure}[hbt]
\centerline{\epsfysize=7.2cm \epsffile{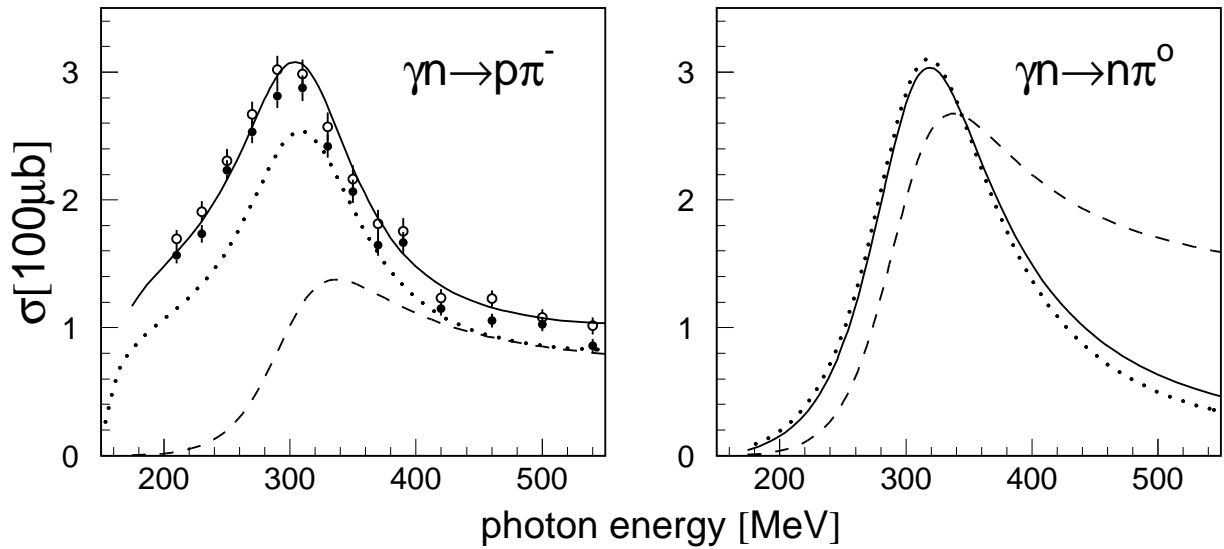}}
\caption{Pion photoproduction in the $\Delta$-region from the neutron.
Left hand side: $p\pi^-$ final state, right hand side: $n\pi^o$ final state. 
Data for $\gamma n\rightarrow p\pi^-$ are from ref. \cite{Benz_73}, full 
and open symbols represent two different extraction procedures of the cross 
section from a measurement of the $\gamma d\rightarrow pp\pi^-$ reaction. The 
full (all contributions) and dashed (only $\Delta$-resonance) curves are the 
MAID predictions \cite{Drechsel_99}. The dotted curves are the MAID results 
for $\gamma p\rightarrow n\pi^+$ (left hand side) and 
$\gamma p\rightarrow p\pi^o$ (right hand side). 
}
\label{fig_20}       
\end{figure}
%
The result for the total cross section is compared to the MAID model in fig.
\ref{fig_20} (left hand side). The agreement is good, and as is the case for 
the $\gamma p\rightarrow n\pi^+$ reaction, the influence of background 
contributions is large. The MAID prediction for the neutral final state is 
shown on the right hand side of the figure. Here, the model predicts that the
cross sections including resonance and background terms are practically 
identical for $\pi^o$-production on the proton and on the neutron. The remarks 
concerning the separation of resonance and background contributions for the 
reaction on the proton apply here, too.

Until very recently, data for $\pi^o$ photoproduction off the deuteron in the 
$\Delta$ range were scarce. Clifft et al. \cite{Clift_73} measured the cross 
section ratio of the $p\pi^o$ and $n\pi^o$ final states in quasifree 
photoproduction off the deuteron, albeit with large statistical uncertainties. 
Some results were obtained for coherent $\pi^o$ photoproduction on the deuteron 
mainly for backward angles 
\cite{Holtey_73}-\cite{Bouquet_74}. 
However, during the last few years, precise data from threshold up to the 
second resonance region have been obtained 
\cite{Bergstrom_98}-\cite{Siodlaczek_01}. 
The experiment at the 
Saskatchewan Accelerator Laboratory (SAL) \cite{Bergstrom_98} covered only 
the threshold region (below 160 MeV incident photon energy), with very good 
statistics. Here, the motivation was to test the predictions of chiral 
perturbation theory. The measurement of ref. \cite{Siodlaczek_01}, which was 
originally motivated as a search for a dibaryon state, covered incident photon 
energies between threshold and 300 MeV, and ref. \cite{Krusche_99} the range 
from 200 - 800 MeV. The latter two experiments which were carried out at 
MAMI with similar detector setups, agree very well in the overlap region from 
200 - 300 MeV. The results from  ref. \cite{Bergstrom_98,Siodlaczek_01} agree 
in the immediate threshold region. The total cross section from the SAL 
experiment seems to be systematically higher by some per cent around 160 MeV, 
although the statistical uncertainties from ref. \cite{Siodlaczek_01} are of 
the same order.

%
%
%
\begin{figure}[hbt]
\centerline{
\epsfysize=2.7cm \epsffile{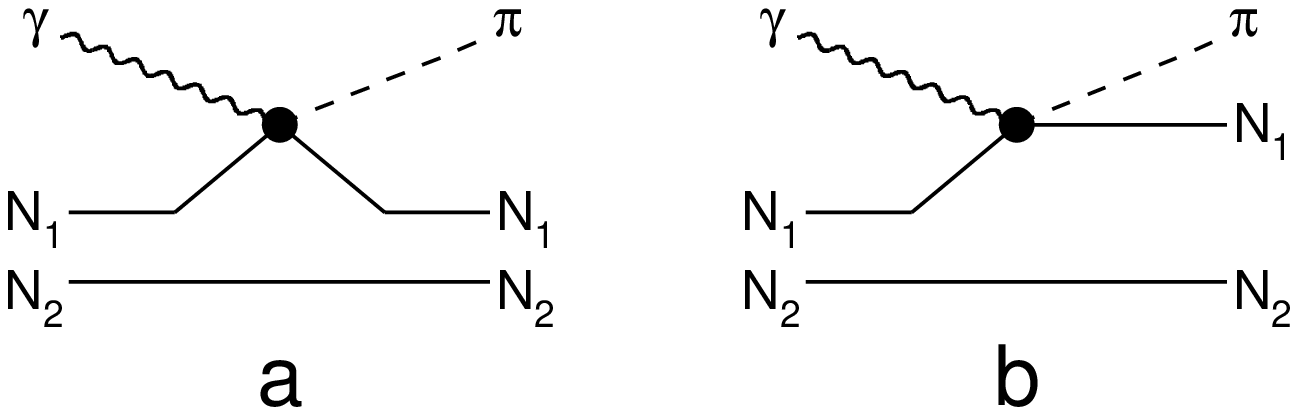}
\hspace*{0.3cm}
\epsfysize=2.7cm \epsffile{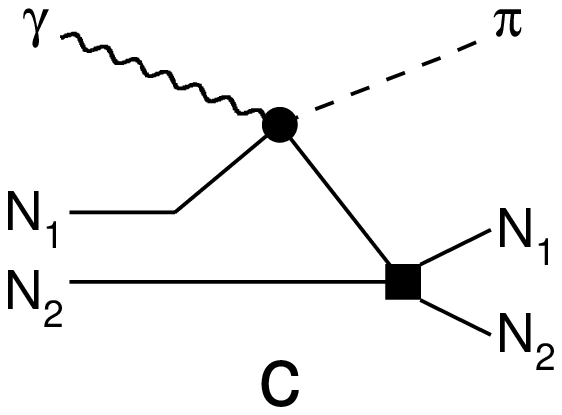}
\hspace*{0.3cm}
\epsfysize=2.7cm \epsffile{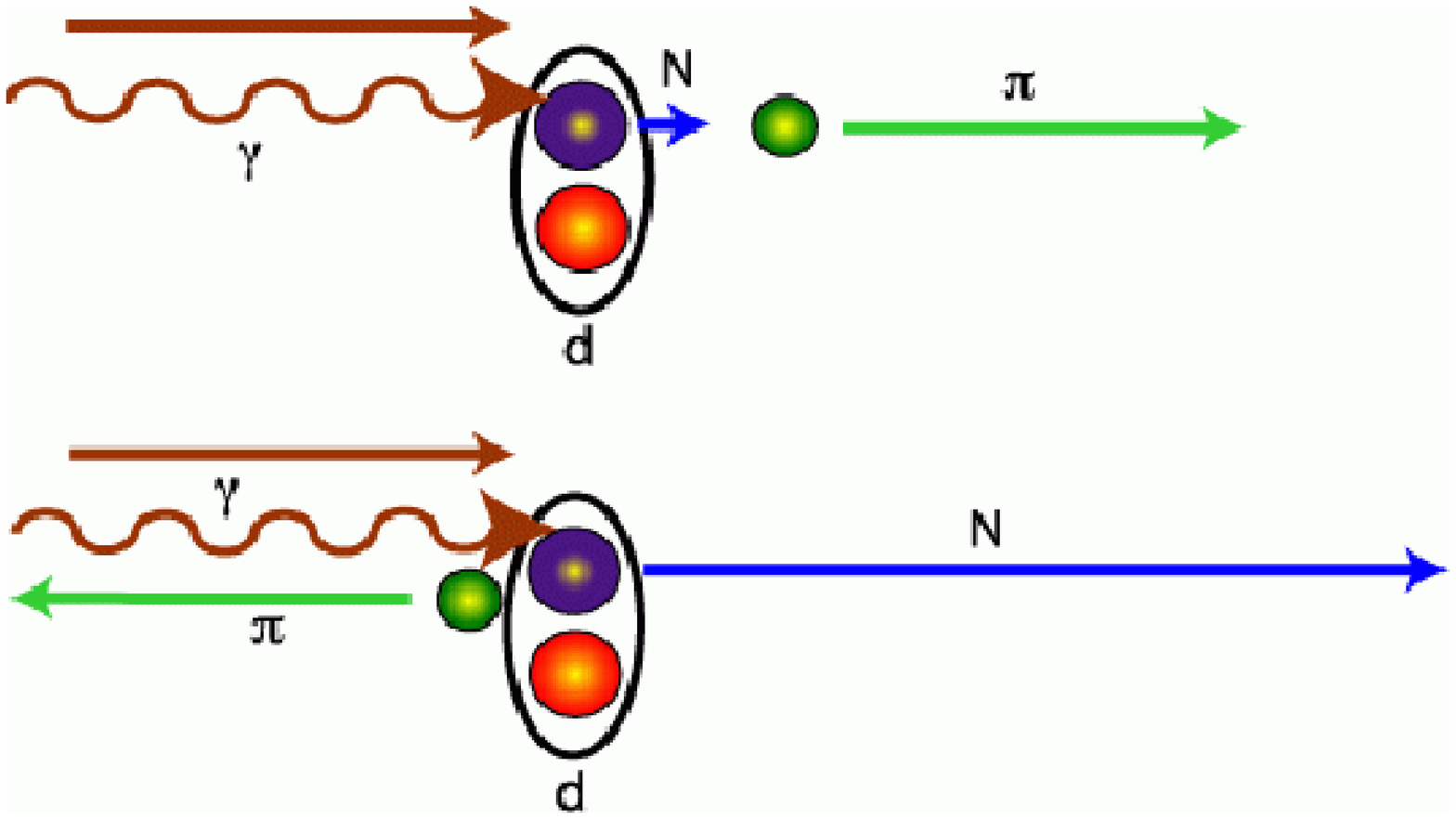}}
\caption{Contributions to neutral pion photoproduction off the deuteron:
(a) coherent, (b) break-up, (c) break-up process followed by 
$NN$ FSI. Right hand side: relative momenta of nucleons.
}
\label{fig_21}       
\end{figure}
%
   
In comparison to photoproduction of charged pions from the deuteron, an 
additional complication arises for neutral pions since, as sketched in fig. 
\ref{fig_21}, the process involves two different reaction mechanisms. In case 
of coherent photoproduction with the $d\pi^o$ final state (diagram a) the 
amplitudes for $\pi^o$-production off both nucleons add coherently. In the 
simplest approximation of the breakup reaction (diagram b) with the $np\pi^o$ 
final state, one nucleon acts as participant, and the other can be regarded 
as spectator. This is the quasifree production, and one might expect that this 
process offers the ideal tool for the investigation of the 
$n(\gamma ,\pi^o)n$ reaction. However, the plane wave impulse approximation 
(PWIA) is not a good approximation for the $d(\gamma ,\pi^o)np$ reaction in 
the $\Delta$ range because the third diagram (c) in fig.~\ref{fig_21}, with 
final state interaction of the two nucleons, gives an important contribution. 
The $NN$ FSI leads to a relation between the coherent and the quasifree process 
since it may (coherent) or may not (breakup) bind the two nucleons in the 
final state. The size of the $NN$ FSI effect depends strongly on the meson 
emission angle. This can be qualitatively understood in the following way. As 
sketched in fig.~\ref{fig_21} the two nucleons in the final state will have a 
large relative momentum when the pion is emitted at backward angles, but 
only a small relative momentum for pions emitted at forward angles. Therefore, 
$NN$ FSI will be important and tend to bind the two nucleons for forward 
emission of the pion leading to an enhancement of the coherent part and a 
suppression of the breakup part. The opposite happens for backward angles.    

Based on the connection between coherent and breakup contributions, Kolybasov 
and Ksenzov \cite{Kolybasov_76}, using the completeness relation, have argued 
that the effect of FSI in the breakup process is just counterbalanced by the 
coherent process so that the sum of the cross sections for the coherent and 
the breakup part with FSI equals the cross section of the pure quasifree 
process without FSI. In this case, the semi-inclusive cross section of the 
$d(\gamma ,\pi^o)X$ reaction, i.e. the sum over coherent and breakup parts, 
is best suited for the extraction of the neutron cross section. This recipe 
has been employed by Siodlaczek et al. \cite{Siodlaczek_01} for a modeling of 
the cross sections with a coalescence model. 
%
%
%
\begin{figure}[thb]
\begin{minipage}{0.cm}
\hspace*{0.2cm}
{\mbox{\epsfysize=10.cm \epsffile{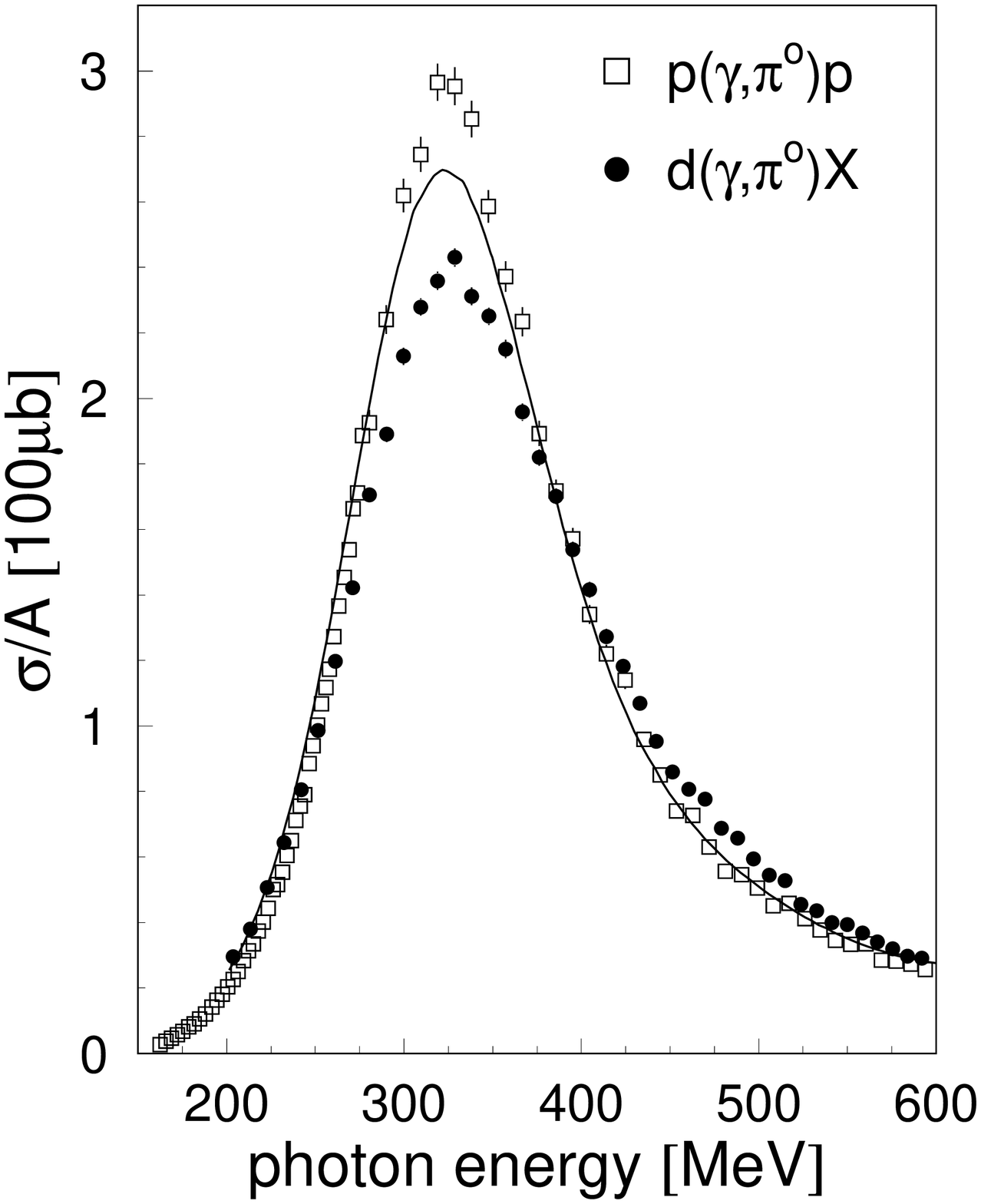}}}
\end{minipage}
\hspace{8.5cm}
\begin{minipage}{0.cm}
\vspace*{.1cm}
{\mbox{\epsfysize=10.5cm \epsffile{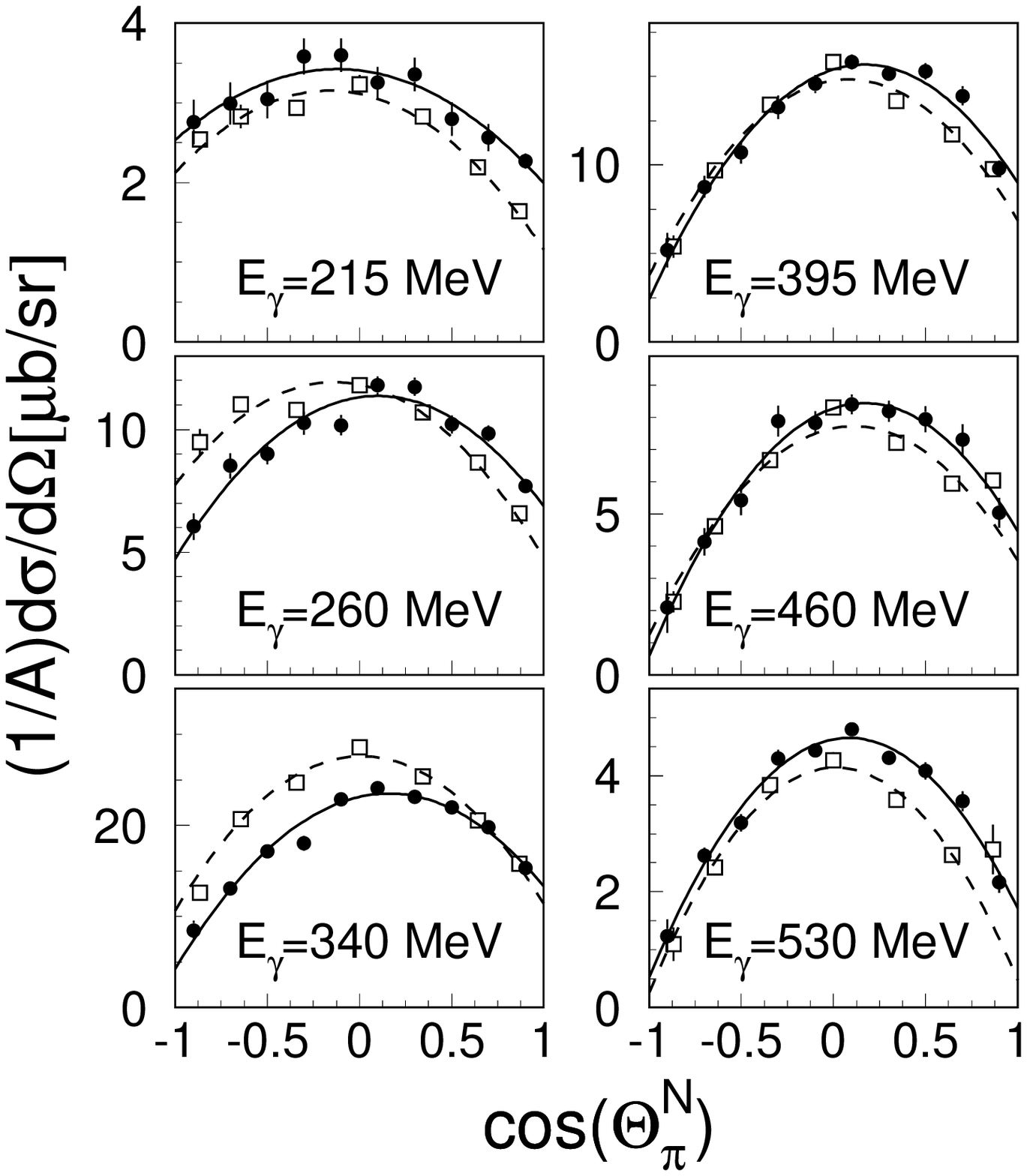}}}
\end{minipage}
\caption{Inclusive $\pi^o$ photoproduction from $^2$H in the $\Delta$ 
range \cite{Krusche_99}. Left: total cross sections of $p(\gamma ,\pi^o)p$
(open squares) and $d(\gamma ,\pi^o)X$ (filled dots) normalized by the mass 
number. Curve: free proton cross section folded with momentum distribution of 
bound nucleons.
Right: angular distributions in the cm frame of the photon and a nucleon 
at rest (symbols as on left side). Curves: fits with eq. (\ref{eq:angdis}).} 
\label{fig_22}       
\end{figure}
%
The relative momentum of the two 
nucleons determines the $NN$ FSI which in turn pushes the reaction into the 
coherent or breakup final state. 
This simple model reproduces the shape of the 
angular distributions quite well, but the absolute scale of the breakup part 
is not in agreement with the data \cite{Siodlaczek_01}.

According to the argument that the inclusive cross section equals the quasifree 
cross section without FSI effects, the neutron cross section would simply 
follow from a comparison of the elementary cross sections folded with the 
nucleon momentum distribution to the measured semi-inclusive deuteron cross 
section. Total cross sections and angular distributions \cite{Krusche_99} of 
the semi-inclusive $d(\gamma, \pi^o)X$ reaction, where both final states 
($d\pi^o$ and $np\pi^o$) were accepted, are compared in fig.~\ref{fig_22} to 
the results for the free proton. The angular distributions are shown for the 
photon - proton cm system for the free proton case, and for the cm system of 
a photon and a nucleon at rest (with zero Fermi momentum) for the deuteron 
case. In comparison with the free nucleon case the angular distributions for 
quasifree pion production from the deuteron should be smeared out only slightly 
by Fermi motion. The distributions for the proton and the deuteron are indeed 
quite similar. As discussed above, the amplitude for the excitation of the 
$\Delta$ on the free proton and on the free neutron must be identical as long 
as isotensor components can be neglected. The MAID model predicts that the 
$\pi^o$-photoproduction cross section including all background terms is nearly 
identical for the proton and the neutron (see fig.~\ref{fig_20}). However, 
fig.~\ref{fig_22} shows that the experimental total cross section close to the 
peak position of the $\Delta$ from the deuteron does not equal twice the Fermi 
smeared proton cross section. Curing this disagreement with a modification of 
the free $\gamma n\rightarrow n\pi^o$ cross section would result in a reduction 
of $\approx$25\% at the peak position \cite{Siodlaczek_01}, while it would 
remain unchanged in the wings of the $\Delta$-peak. Since we know that the 
$\Delta$-excitation should contribute equally for the proton and the neutron, 
this would mean that non-resonant backgrounds contribute very differently for 
proton and neutron. However, this simple model does not allow to draw such 
far-reaching conclusions. A more sophisticated treatment of the FSI effects is 
necessary.   

\begin{figure}[hbt]
\begin{minipage}{0.cm}
\hspace*{1.cm}
{\mbox{\epsfysize=6.0cm \epsffile{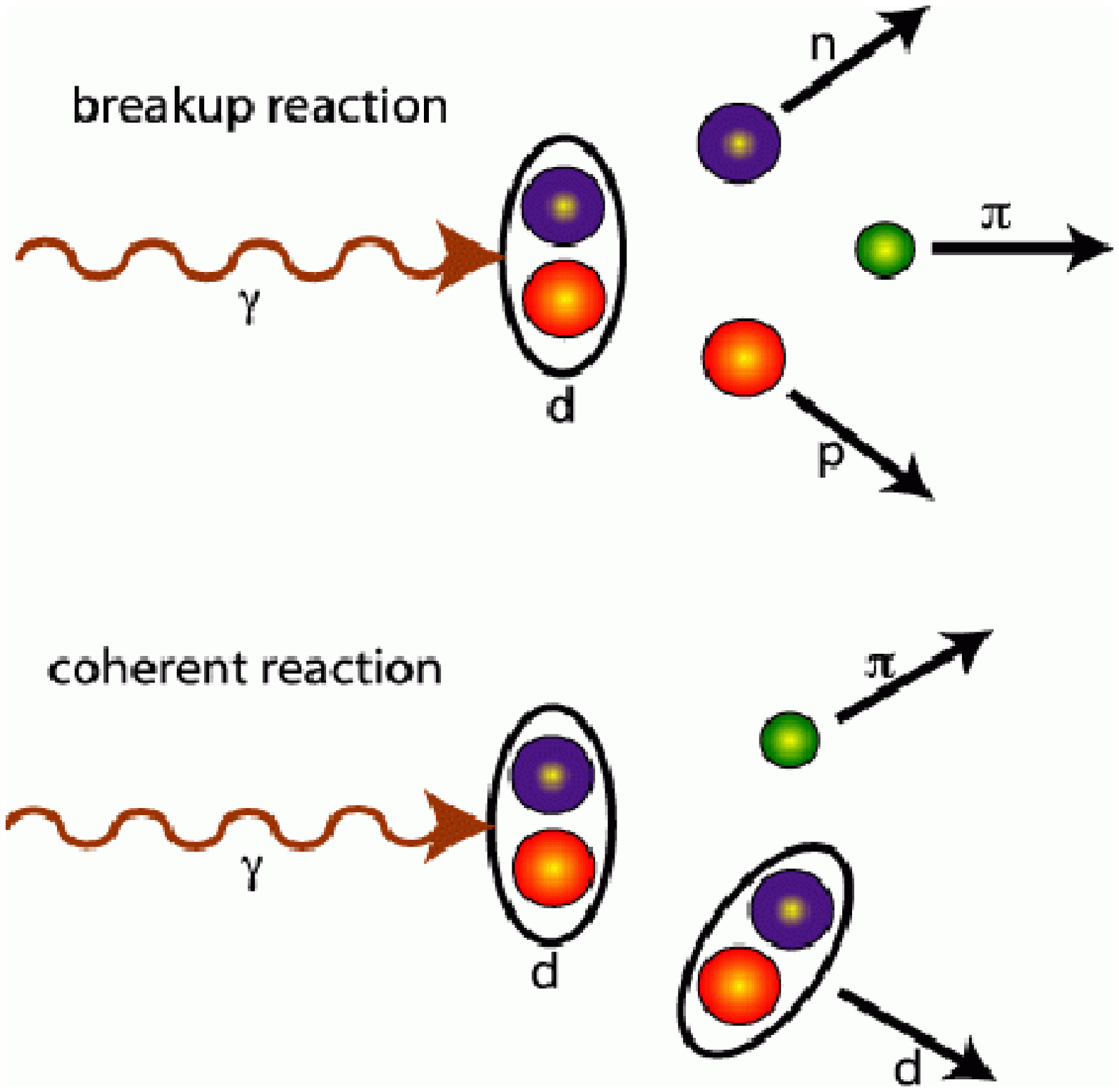}}}
\end{minipage}
\hspace{6.cm}
\begin{minipage}{0.cm}
\hspace*{1.cm}
{\mbox{\epsfysize=7.5cm \epsffile{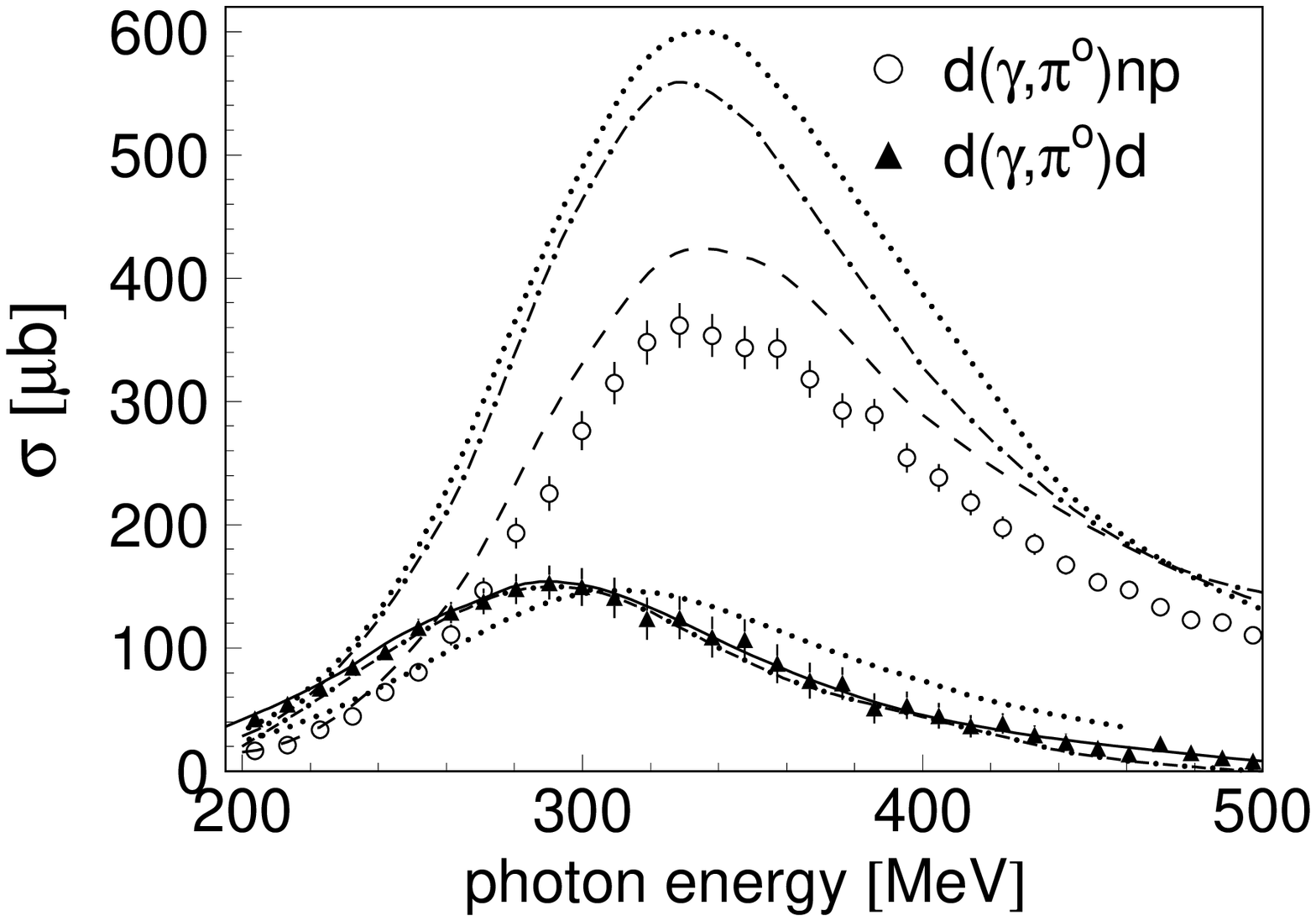}}}
\end{minipage}
\caption{Total cross sections for the reactions $d(\gamma ,\pi^o)np$ (open
circles) and $d(\gamma ,\pi^o)d$ (filled triangles) \cite{Krusche_99}.  
Solid, dash-dotted, and dotted curves for the coherent cross section:  
predictions from models of Kamalov et al. \cite{Kamalov_97}, 
Laget \cite{Laget_81}, and Wilhelm et al. \cite{Wilhelm_96}. 
Break-up reaction: dash-dotted and dashed curves predictions from Laget
\cite{Laget_81} with (dashed) and without (dash-dotted) np FSI, dotted
curve: the PWIA prediction from Schmidt et al. \cite{Schmidt_96}.}
\label{fig_23}       
\end{figure}

\begin{figure}[h]
\hspace{0.1cm}
{\mbox{\epsfysize=10.cm \epsffile{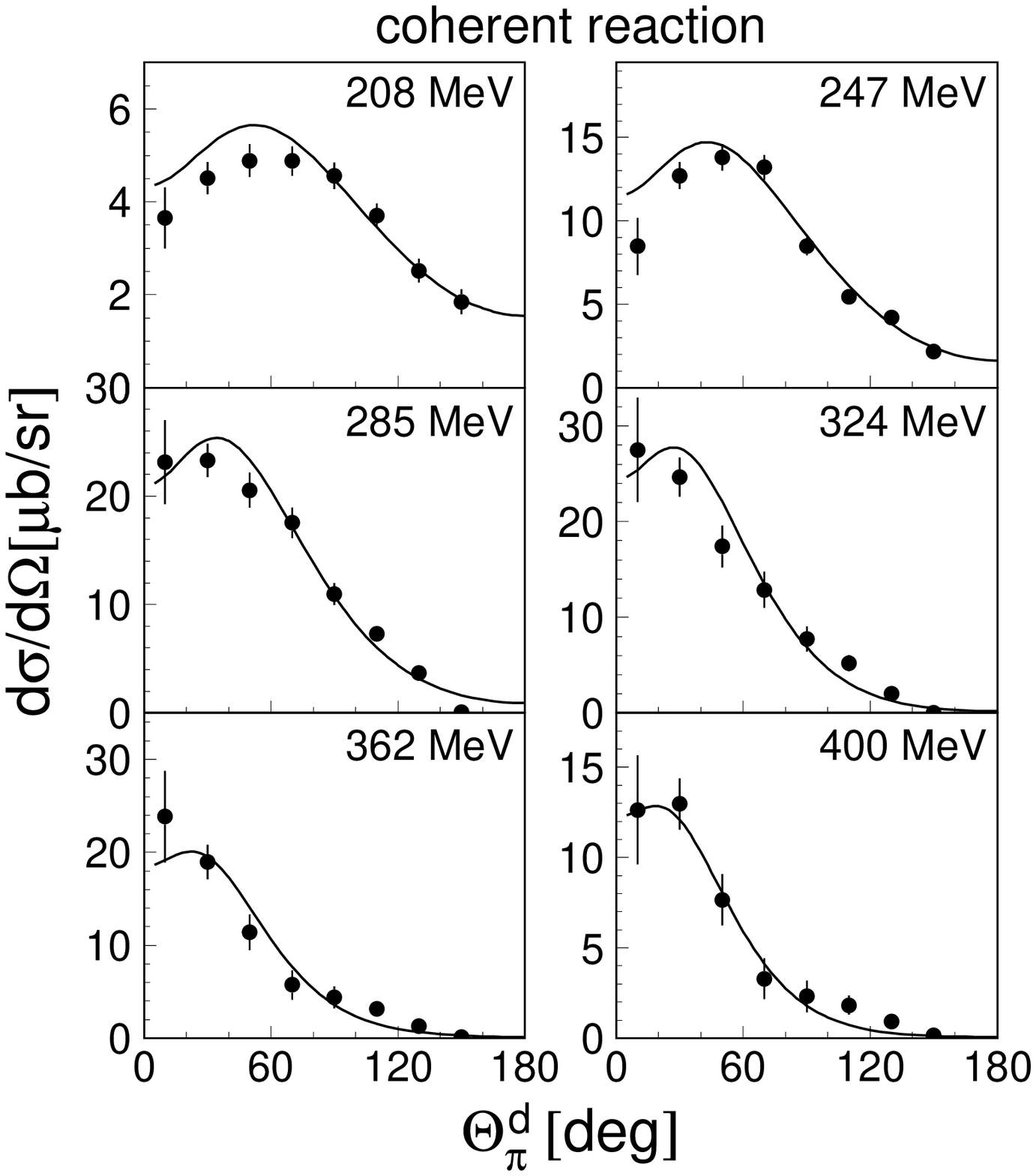}}}
\hspace{0.2cm}
{\mbox{\epsfysize=10.cm \epsffile{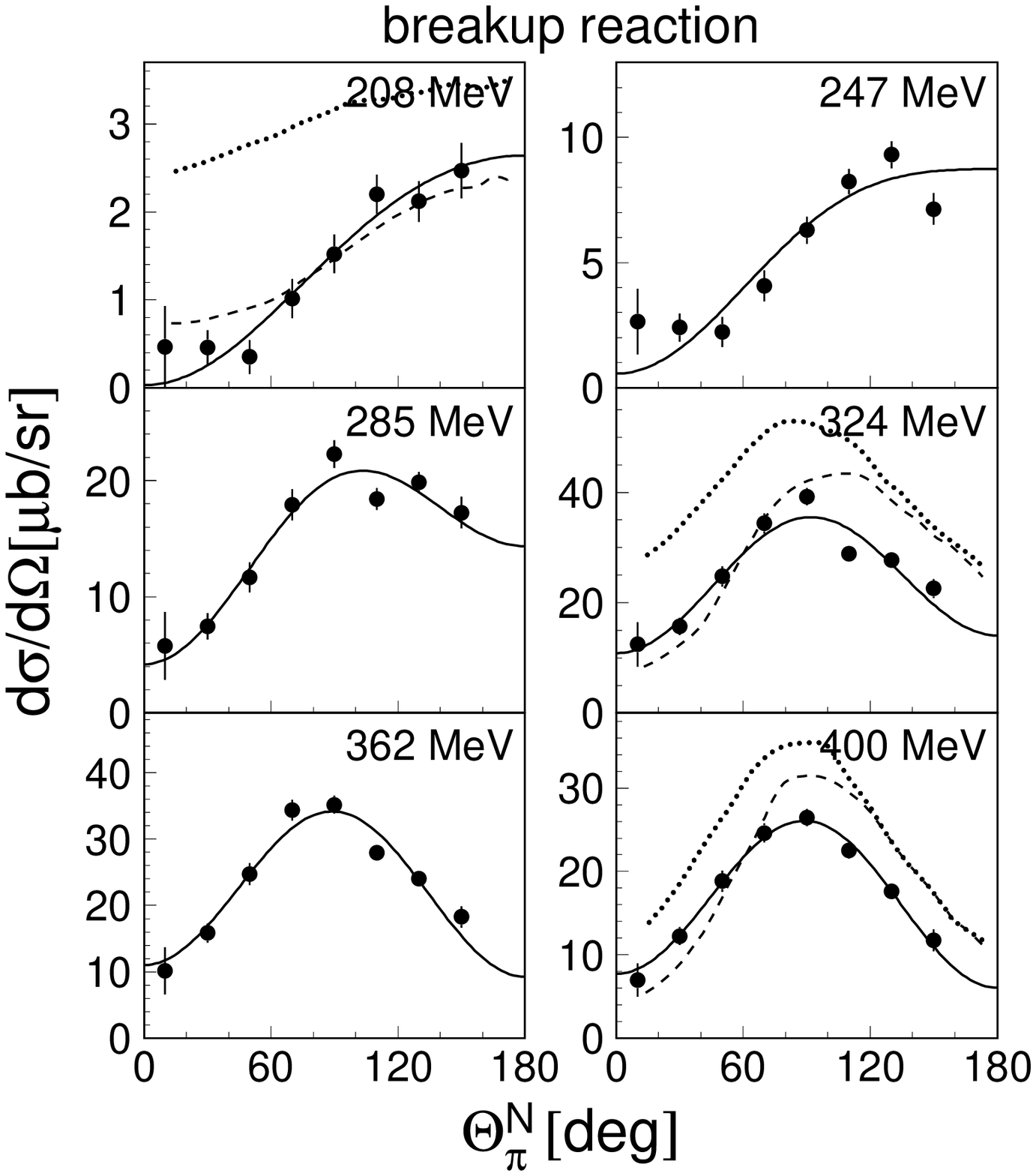}}}
\caption{Angular distributions for the reactions $d(\gamma ,\pi^o)d$ 
(left hand side) and $d(\gamma ,\pi^o)np$ (right hand side) for different bins 
of incident photon energy. The coherent cross sections are shown in the 
photon - deuteron cm system, the break-up cross sections in the 
photon - nucleon cm system. Coherent part, full curves: predictions from 
\cite{Kamalov_97}. Break-up part: full curves: fits to the data with eq. 
(\ref{eq:angdis}), dashed and dotted curves: predictions from ref. 
\cite{Laget_81} with and without $NN$ FSI.}
\label{fig_24}       
\end{figure}

For the extraction of the free neutron cross section, one should ideally have 
reaction models for the coherent and the breakup reaction off the deuteron and 
find agreement with the data for the same free neutron cross section. On the 
experimental side, coherent and breakup contributions have been separated in 
ref. \cite{Krusche_99,Siodlaczek_01} via their different reaction kinematics. 
Total and differential cross sections are summarized in 
figs.~\ref{fig_23},\ref{fig_24} and compared to model predictions. It should 
be noted that the model results are really predictions in the sense that they 
were made before the data were available. In particular, the calculations from 
Laget \cite{Laget_81} predated the experimental results by almost 20 years. 
The experimental angular distributions show the anticipated FSI effects: they 
are peaked at forward angles for the coherent process while the breakup 
distributions are suppressed at forward angles.  

%
%
%
%
%
%

The models for the coherent process are similar as far as the elementary 
production process is concerned, but differ in the treatment of re-scattering 
effects. The calculations by Laget \cite{Laget_81} and Kamalov et al. 
\cite{Kamalov_97} treat the final state interaction in multiple scattering 
theory. Both results are in excellent agreement with the data. Blaazer et al. 
\cite{Blaazer_95} studied re-scattering corrections to all orders by solving 
the Faddeev equations of the $\pi NN$-system. Their results (not shown in the 
figures) are in agreement with the data. Wilhelm and Arenh\"ovel have developed 
a dynamical model \cite{Wilhelm_96} for the coupled $N\Delta$, $NN\pi$, and 
$NN$-systems, which does not reproduce the energy dependence of the total cross 
section as well as the simpler models (see figure \ref{fig_23}). All models 
suggest the presence of pion re-scattering effects but a common conclusion about 
their importance is presently not possible since the various models even 
disagree qualitatively (see \cite{Krusche_99,Schmidt_96} for a detailed 
discussion). Kamalov et al. \cite{Kamalov_97} claim that in the $\Delta$-range 
the main mechanism of FSI is elastic pion scattering, while the contribution 
from charge exchange reactions is small and their final state interaction 
increases the cross section. Wilhelm and Aren\"ovel \cite{Wilhelm_96}, on 
the other hand, argue that charge exchange contributions produce a sizeable 
effect and their FSI lowers the cross section. Unfortunately, the effects in 
the models are most pronounced for the extreme forward angles, where the 
systematic uncertainty of the data is largest (see ref. \cite{Krusche_99} for 
a detailed comparison of the data to the different model predictions). 
Therefore, a systematic uncertainty remains for the extraction of the 
elementary neutron cross section from the data. However, it is fair to say 
that most model predictions are close to the data and that an indication for 
a deviation of the elementary cross section on the neutron from the model 
inputs is not observed. It is interesting to note, that for heavier nuclei 
the DWIA approximations fail, and significant modifications of the $\Delta$ 
excitation appear 
\cite{Rambo_99}-\cite{Krusche_02}.  

The situation is different for the breakup channel. Here, final state 
interaction effects, in this case $NN$ FSI, are more important than for the 
coherent channel. As can be seen in figs.~\ref{fig_23},\ref{fig_24}, the PWIA 
calculations strongly overestimate the data. However, even after inclusion of 
FSI the data are still overestimated. This problem was already apparent in 
\cite{Laget_81} where Laget noted that his calculations reproduced the 
available data for the charged channels and coherent $\pi^o$-production quite 
well but the sum of the cross section from all channels overestimated the 
experimental total photoabsorption cross section. He suggested that the total 
photoabsorption data \cite{Armstrong_72} might suffer from systematic effects, 
but in the meantime those data  have been remeasured \cite{MacCormick_96}, 
yielding the same result. Thus, most of the discrepancy comes from the $\pi^o$ 
breakup channel, which was obviously not well understood in the models. The 
question is whether the failure of the models is connected to the input for the 
elementary $\pi^o$ production off the neutron or to nuclear effects. 

%
%
%
\begin{figure}[hbt]
\begin{minipage}{0.cm}
{\mbox{\epsfysize=6.cm \epsffile{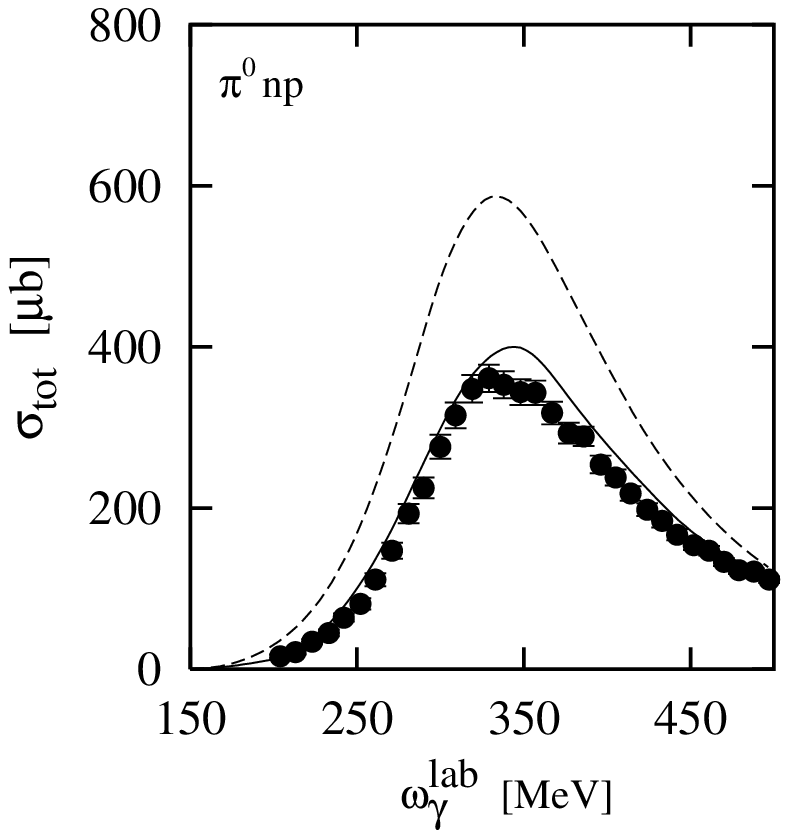}}}
\end{minipage}
\hspace*{5.7cm}
\begin{minipage}{0.cm}
{\mbox{\epsfysize=6.cm \epsffile{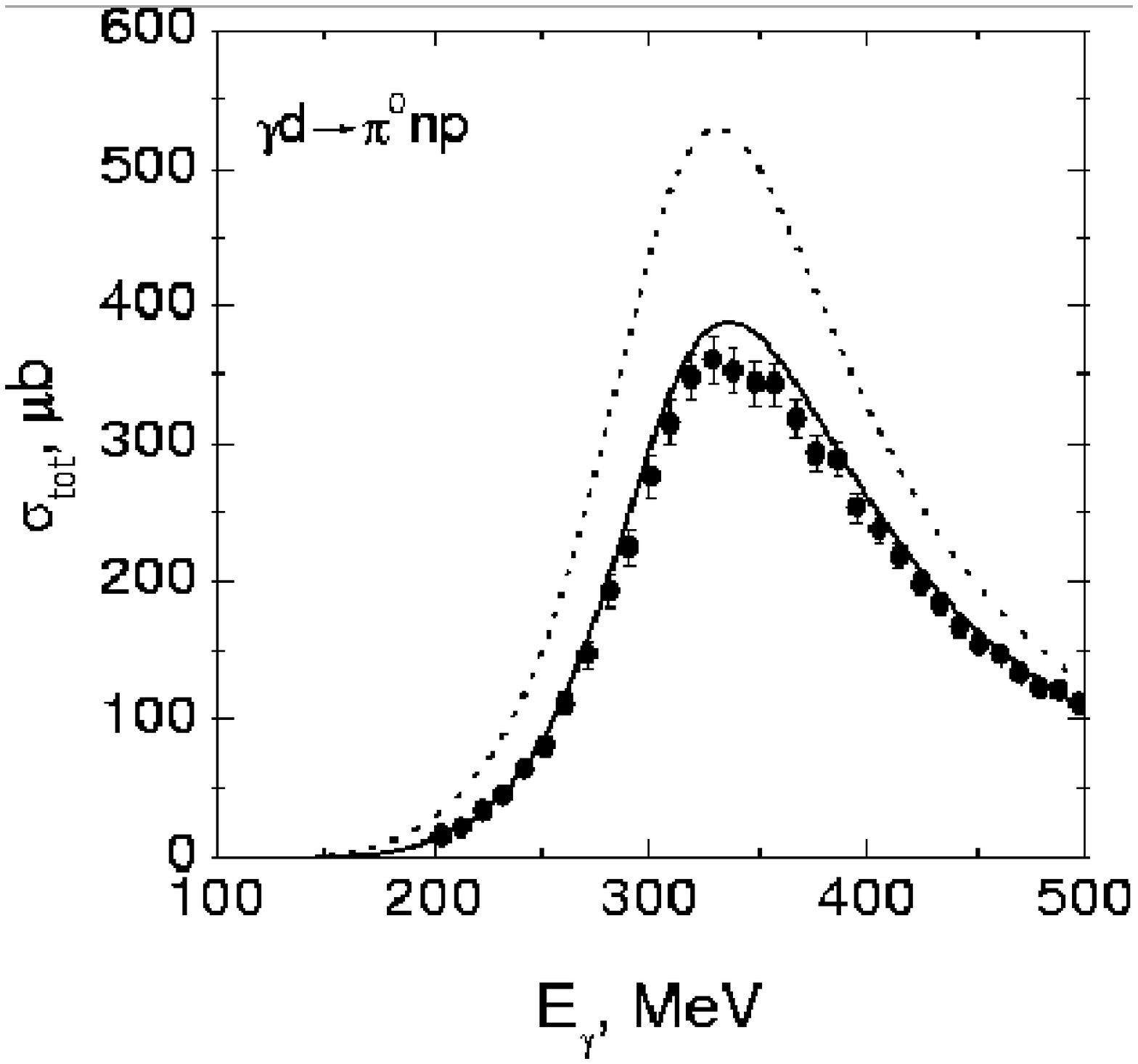}}}
\end{minipage}
\hspace*{6.8cm}
\begin{minipage}{5.cm}
\caption{Total cross section for $\gamma d\rightarrow np\pi^o$, 
compared to the calculation of Darwish et al. \cite{Darwish_02} (left hand side)
and Levchuk et al. \cite{Levchuk_00} (right hand side) (dashed, respectively 
dotted lines: PWIA, solid lines: full calculations). Data are from
\cite{Krusche_99}. 
}
\label{fig_25} 
\end{minipage}
\end{figure}
%

Very recently, new detailed model calculations for the breakup reactions 
$\gamma d\rightarrow np\pi^o$, $\gamma d\rightarrow pp\pi^-$, and
$\gamma d\rightarrow nn\pi^+$ have been presented by Levchuk et al.
\cite{Levchuk_00} and Darwish et al. \cite{Darwish_02}. In both calculations,
care is taken that the elementary cross section on the nucleon is modeled
as realistically as possible. The earlier calculation by Laget \cite{Laget_81}
used the well-known Blomqvist-Laget parameterization \cite{Blomqvist_77} of 
the pion photoproduction amplitude which reproduces the charged channels 
quite well but is known to give a less good description of the 
$\gamma p\rightarrow p\pi^o$ reaction. Levchuk et al. \cite{Levchuk_00}
use CGLN amplitudes taken from the SAID \cite{Arndt_96} and MAID 
\cite{Drechsel_99} multipole analyses. Darwish et al. \cite{Darwish_02} 
use the effective Lagrangian model of Schmidt et al. \cite{Schmidt_96} and 
check with a detailed comparison to cross section data and the SAID and MAID 
multipoles that the elementary reactions are well reproduced. Levchuk et al.
consider only $NN$ FSI, while Darwish et al. include FSI in all two-body
subsystems. They find however, that only $NN$ FSI is important while $N\pi$ 
FSI is negligible. 

The results of the two calculations are similar and in better agreement with 
data than previous models. The predictions for the total cross section of the 
$np\pi^o$ channel are compared in fig.~\ref{fig_25}. Both models almost 
reproduce the data, and the over-prediction of the cross section in the 
$\Delta$ peak is reduced to below 10\%. FSI effects are important in both 
models. However, a closer inspection shows still some difference in the models. 
The PWIA prediction of Darwish et al. is slightly higher than the one of 
Levchuk et al., which seems to indicate a difference in the elementary 
production operator.  
The angular distributions are similar in the models. Typical results from 
\cite{Darwish_02} for the $np\pi^o$ and $pp\pi^-$ final states are summarized 
in figure \ref{fig_26}.
The figure highlights the difficulties of the 
investigation of the $\Delta$ excitation on the neutron from quasifree pion 
production on the deuteron: 
%
%
%
\begin{figure}[hbt]
\hspace*{0.cm}
\begin{minipage}{0.cm}
{\mbox{\epsfysize=4.5cm \epsffile{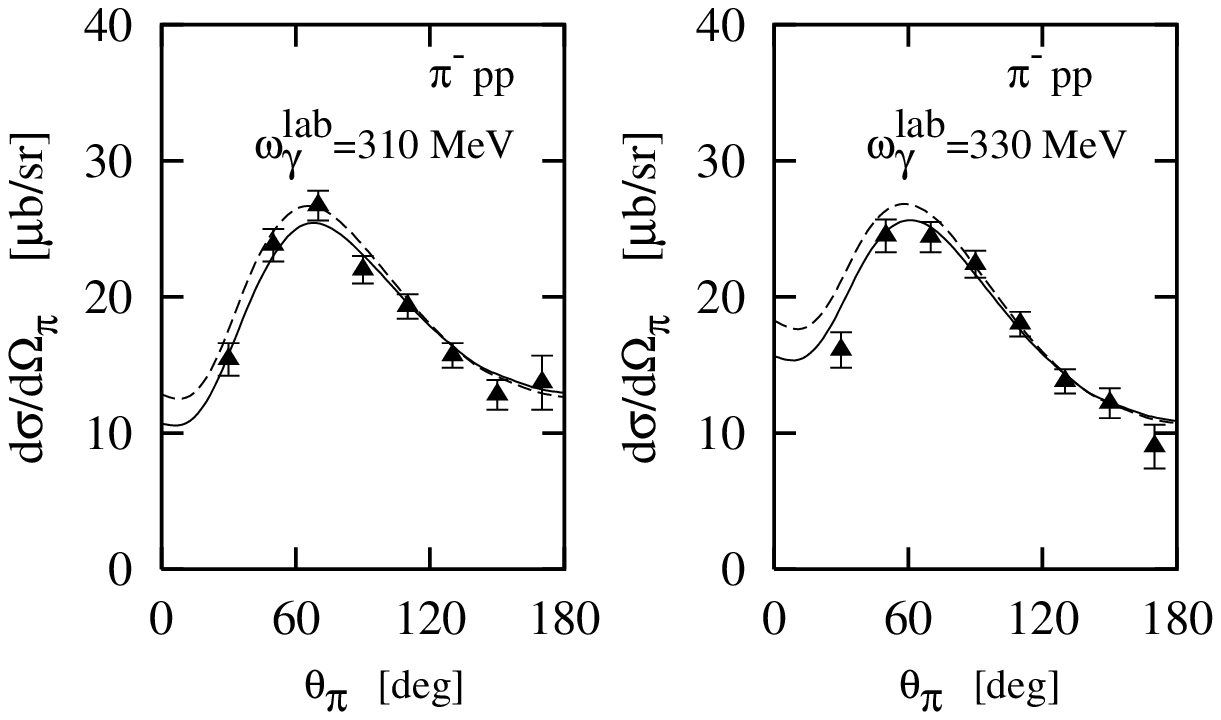}}}
\end{minipage}
\hspace{7.5cm}
\begin{minipage}{0.cm}
{\mbox{\epsfysize=4.5cm \epsffile{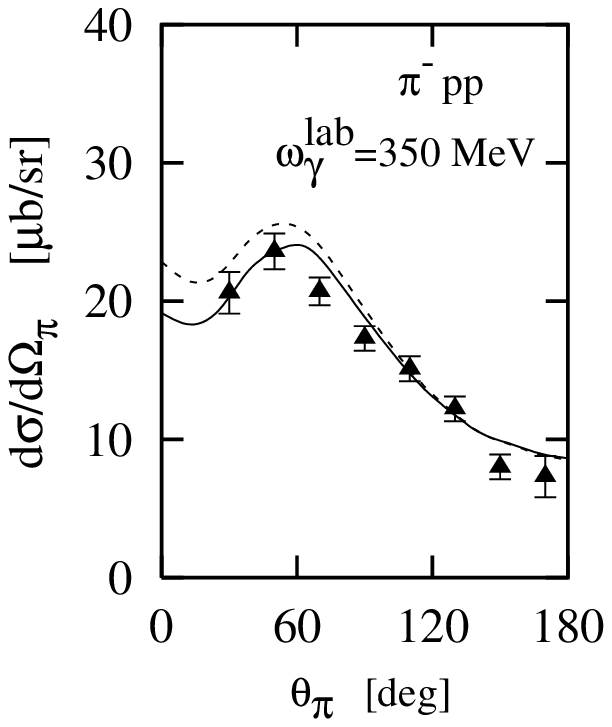}}}
\end{minipage}

\vspace*{-2.0cm}
\hspace*{0.cm}
\begin{minipage}{0.cm}
{\mbox{\epsfysize=4.5cm \epsffile{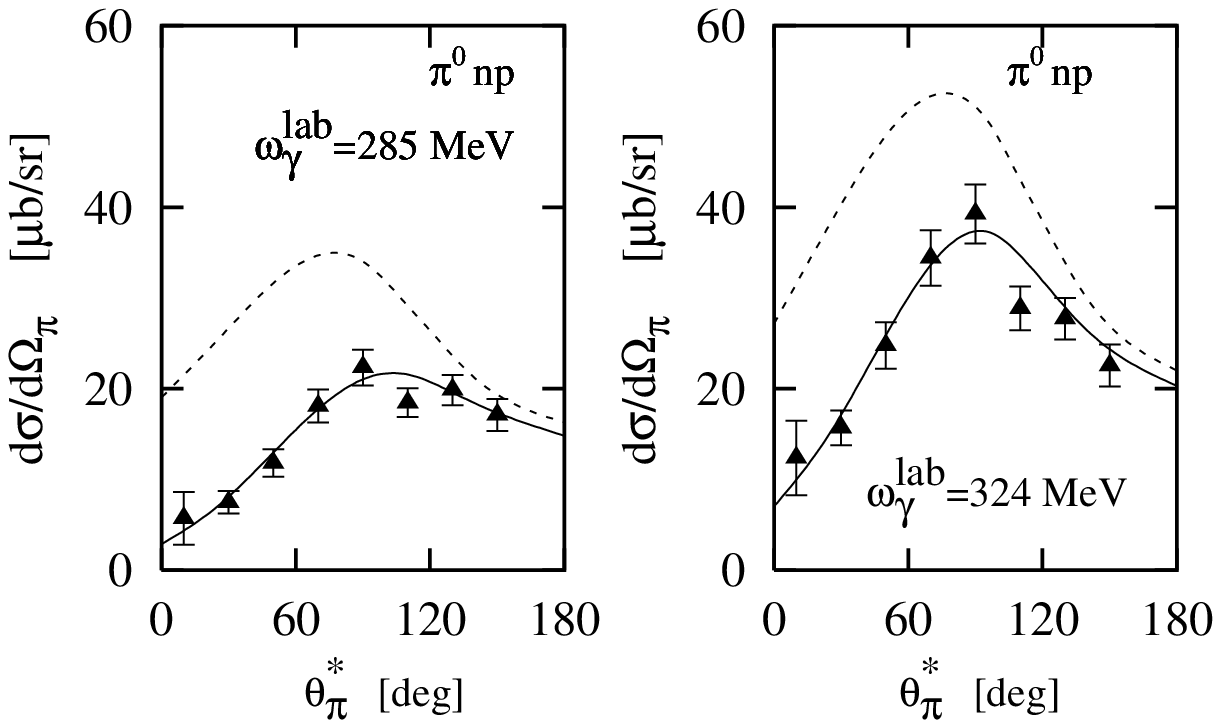}}}
\end{minipage}
\hspace{7.5cm}
\begin{minipage}{0.cm}
{\mbox{\epsfysize=4.5cm \epsffile{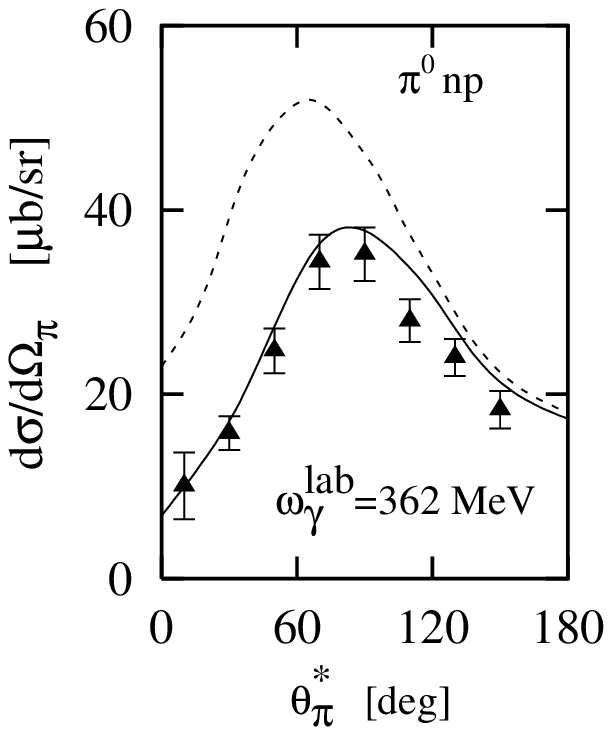}}}
\end{minipage}
\hspace*{-8.3cm}
\begin{minipage}{11.8cm}
\vspace*{6.7cm}
\caption{Angular distributions for the reactions $\gamma d\rightarrow \pi^-pp$
(upper part, data from ref. \cite{Benz_73}) and $\gamma d\rightarrow \pi^onp$
(lower part, data from ref. \cite{Krusche_99}) in the $\Delta$ range compared 
to model calculations from \cite{Darwish_02}. Solid lines: full model,
dashed lines: without FSI.
}
\label{fig_26}  
\end{minipage}
\end{figure}
%

\vspace*{-12.95cm}
\hspace*{11.3cm}
\begin{minipage}{6.4cm}
FSI effects are large for neutral pions, so that substantial model input is 
necessary for the nuclear effects. FSI effects are 
small for the production of $\pi^-$ mesons since the competing coherent channel 
does not exist in this case. However, like in case of $\pi^+$ production from 
the proton, the strong forward peaking of the angular distributions is an 
indication for significant non-resonant background in the elementary process. 
Indeed, Darwish et al. \cite{Darwish_02} find strong contributions of 
nucleon Born terms. 

In summary, the more elaborate model calculations, having become available 
during the last few years, agree with precision data to a degree which
demonstrates that the elementary amplitudes for pion production in the 
$\Delta$ range are reasonably well under control for all isospin channels. 
\end{minipage}

\newpage  
\section{The Second Resonance Region}
\label{sec:secres}

The excitation function of total photoabsorption on the nucleon shows a broad
structure above the energy range of the $\Delta$ resonance at incident photon
energies between 500 and 900 MeV, corresponding to resonance invariant masses
between 1350 and 1600 MeV. This bump in the spectrum is called the `second 
resonance region'. The structure is more complicated than the peak 
corresponding to the $\Delta$ resonance. As discussed below, three overlapping 
nucleon resonances, the tail from the $\Delta$, tails from additional higher 
lying resonances, and background terms contribute. Furthermore, the total 
photoabsorption cross section in the $\Delta$ range stems entirely from single 
pion production. In the second resonance region, the kinematical particle 
thresholds allow  the production of two pions and  $\eta$ mesons.
%
%
%
\begin{figure}[hbt]
\centerline{\epsfysize=9.2cm \epsffile{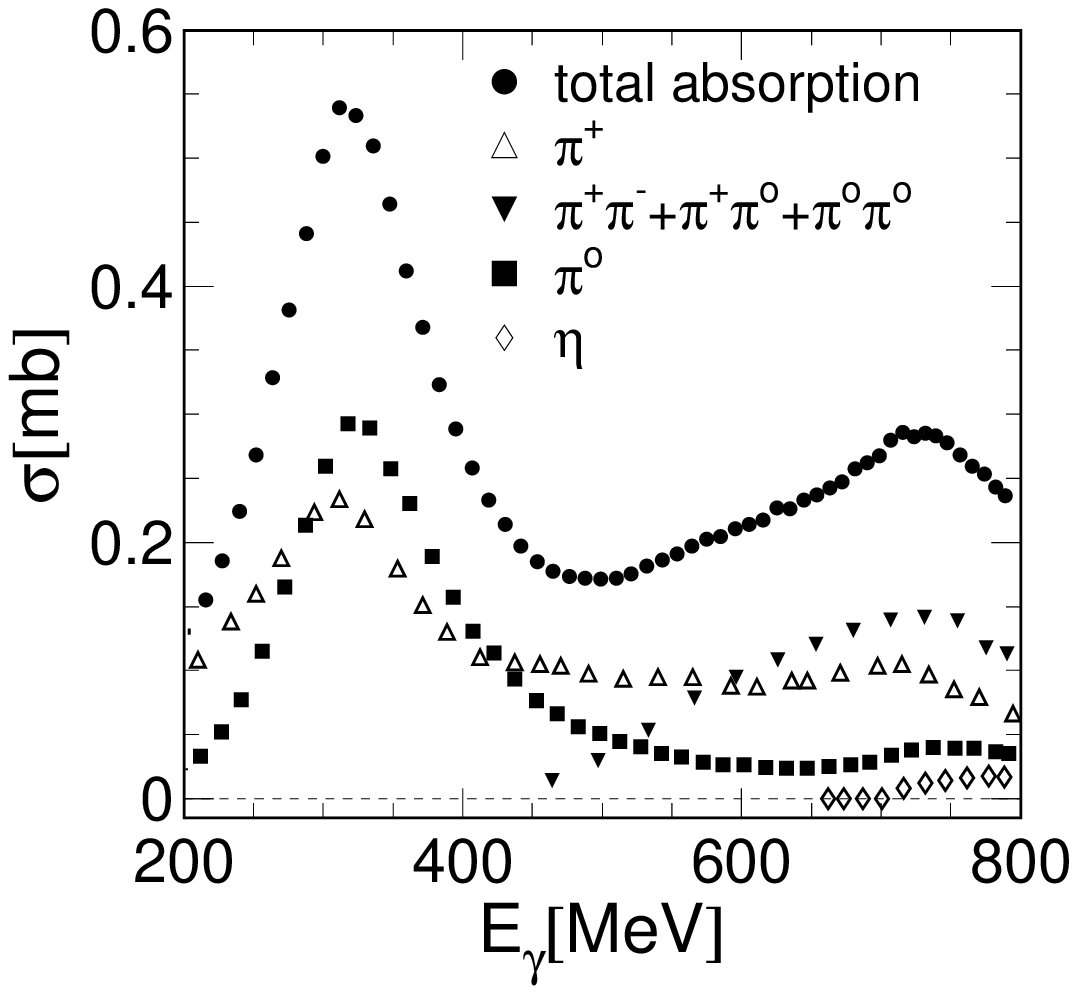}
\epsfysize=9.0cm \epsffile{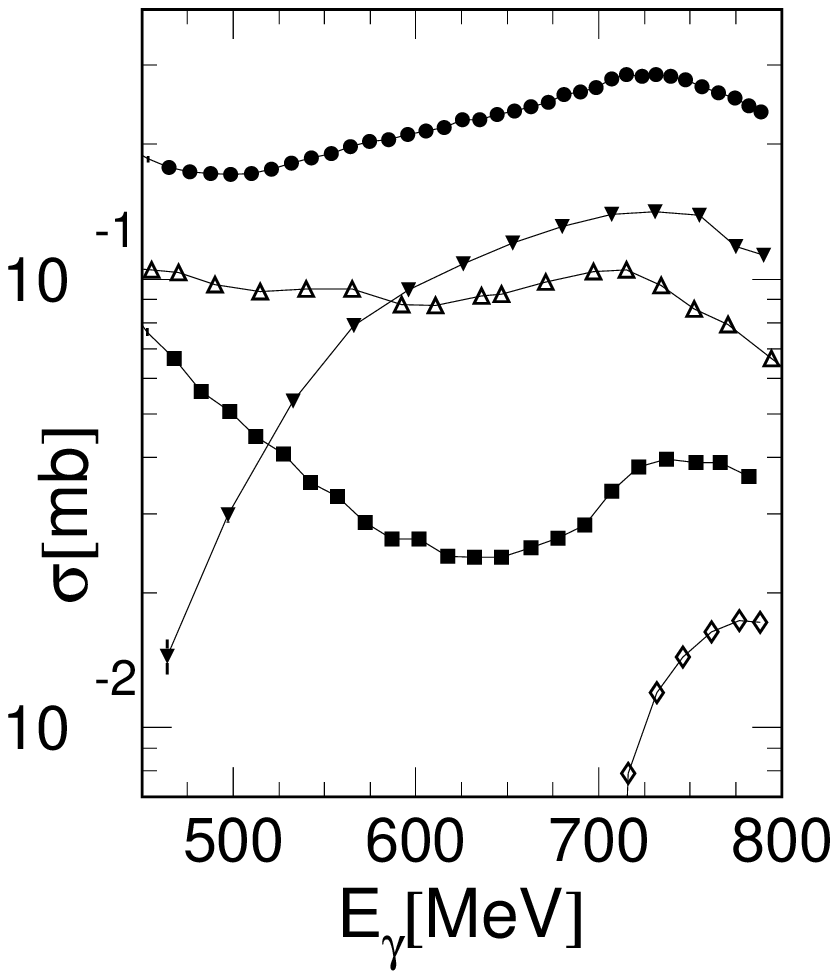}}
\caption{Total photoabsorption and partial cross sections for photoproduction 
off the proton. Data are from \cite{MacCormick_96} (total absorption), 
\cite{ MacCormick_96,Buechler_94} ($\gamma p\rightarrow\pi^+n$), 
\cite{Haerter_96} ($\gamma p\rightarrow\pi^op$), \cite{Braghieri_95,Haerter_97}
(double pion production), and \cite{Krusche_95} ($\gamma p\rightarrow\eta p$). 
Right hand side: second resonance region in logarithmic scale, symbols as on
left side, lines are only to guide the eye.}
\label{fig_27}       
\end{figure}
%
This is demonstrated in fig.~\ref{fig_27}, where the total photoabsorption 
cross section is decomposed into the contributions from the partial channels. 
The cross sections have been measured over the past ten years with the DAPHNE 
\cite{Audit_91} and TAPS detectors \cite{Novotny_91,Gabler_94} at the MAMI 
accelerator. The partial channels add up exactly to the total photoabsorption 
cross section. It is evident, that the resonance bump consists of a complicated 
superposition of the different reaction channels which differ in their energy 
dependence. The neutral and charged single pion production channels behave 
quite differently. Most of the rise of the cross section from the minimum 
around 500 MeV to the maximum around 700 MeV comes from double pion production.
Fitting the total photoabsorption cross sections with Breit-Wigner curves for 
the resonances plus some background as in fig.~\ref{fig_02} would certainly be 
an oversimplification. It should be noted that an understanding 
of this peak structure requires a thorough investigation of the three double 
pion production channels. This is not only important for the discussion of 
resonance excitations on the free nucleon but also forms the basis for 
discussing the experimentally observed strong depletion of the bump structure 
in photoabsorption from nuclei. As we will see in section \ref{ssec:twopi}, 
the interpretation of the double pion production channels is complicated. Only 
the combined progress of experiments and reaction models during the last few 
years has shed light on the dominant reaction mechanisms. The photoproduction 
of $\eta$-mesons contributes only little to the total absorption cross section. 
However, as discussed in detail in sec. \ref{ssec:S11}, it plays a crucial role 
in the investigation of the \s ~resonance.
 
\begin{figure}[htb]
\centerline{\epsfysize=3.72cm \epsffile{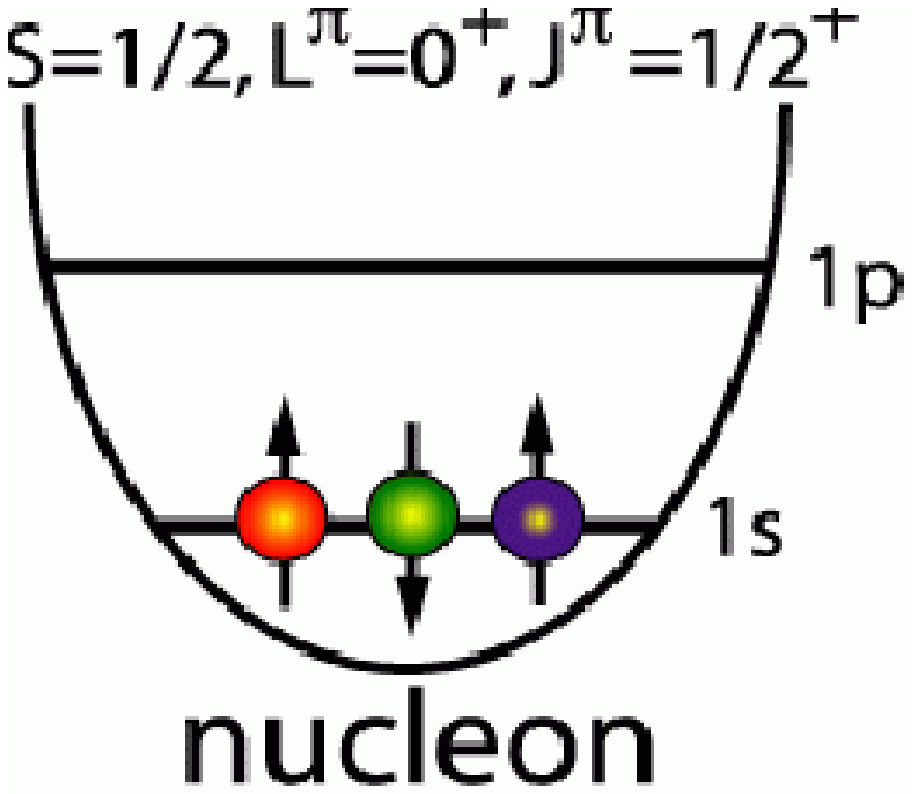}
\epsfysize=3.72cm \epsffile{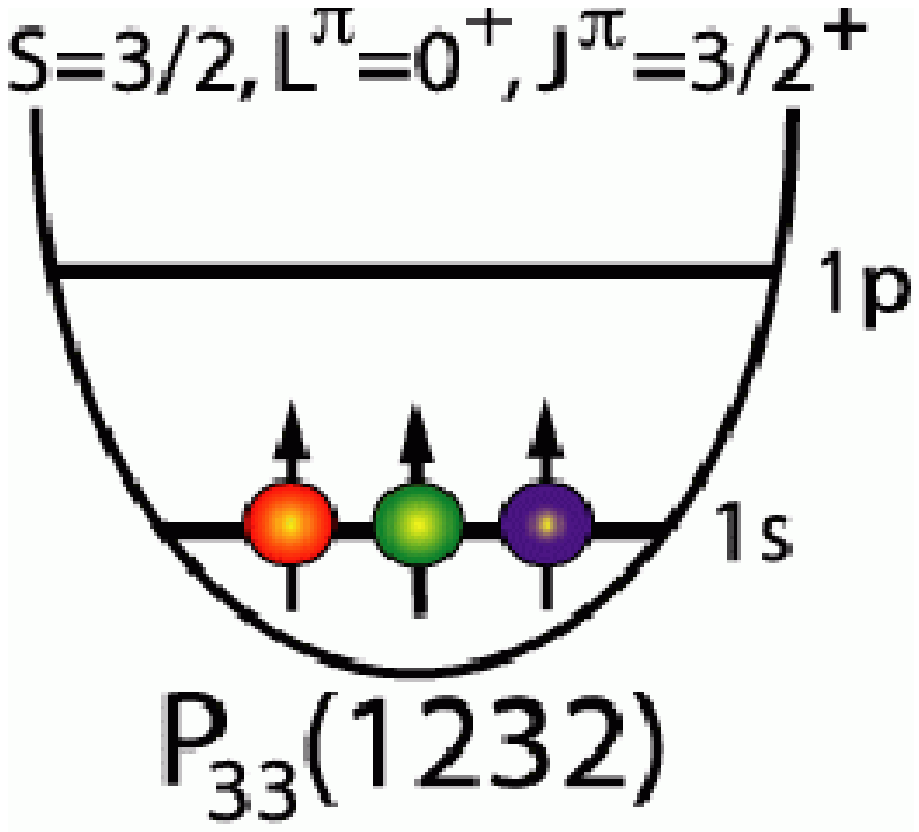}}
\vspace*{0.1cm}
\centerline{
\epsfysize=4.8cm \epsffile{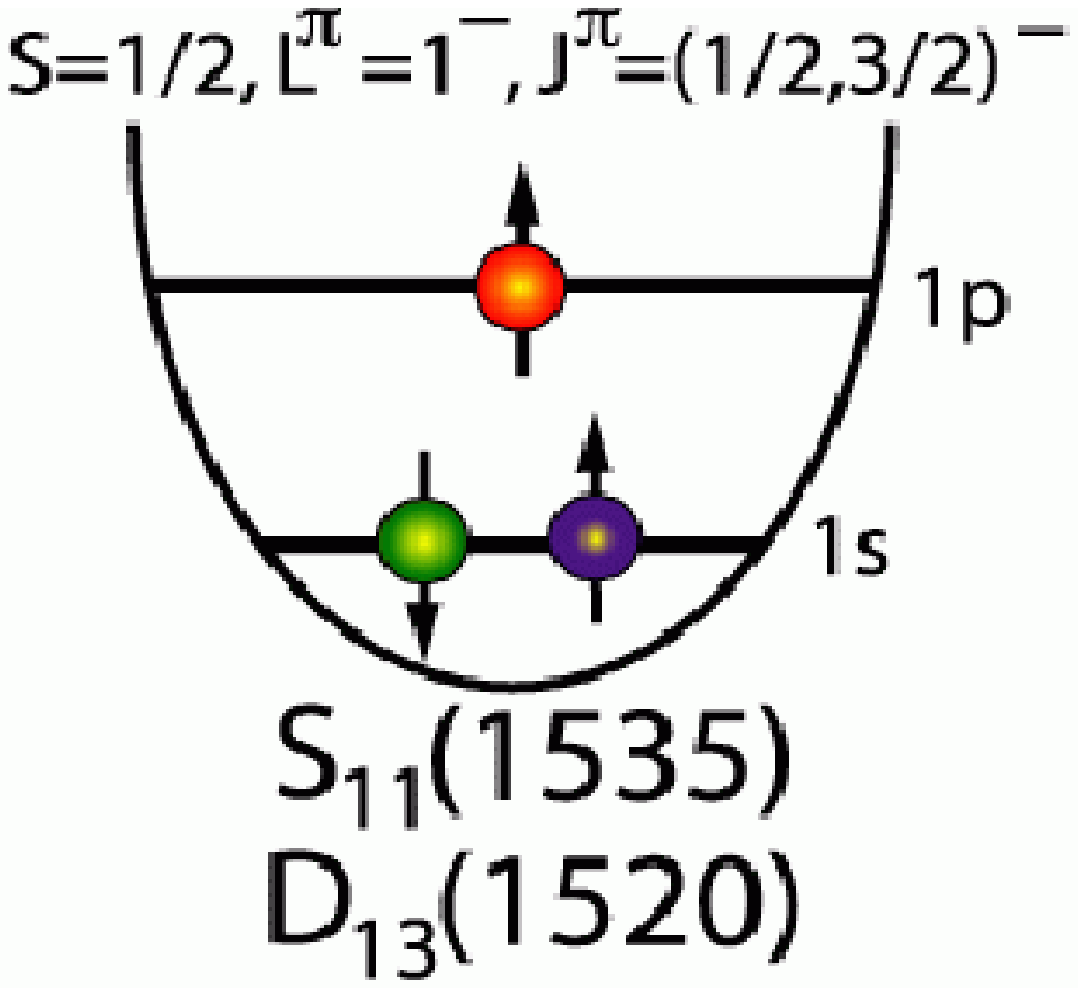}
\epsfysize=4.8cm \epsffile{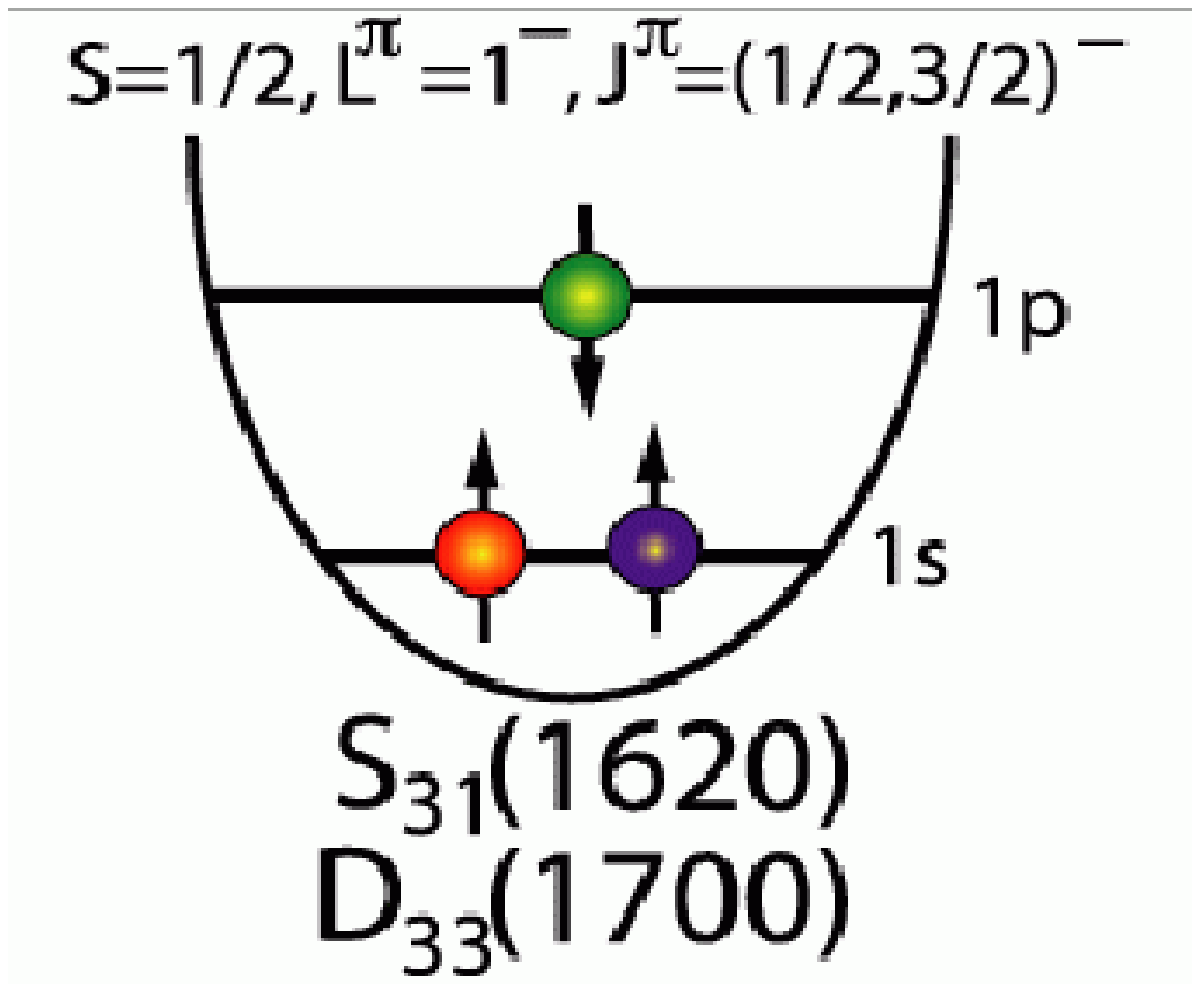}
\epsfysize=4.8cm \epsffile{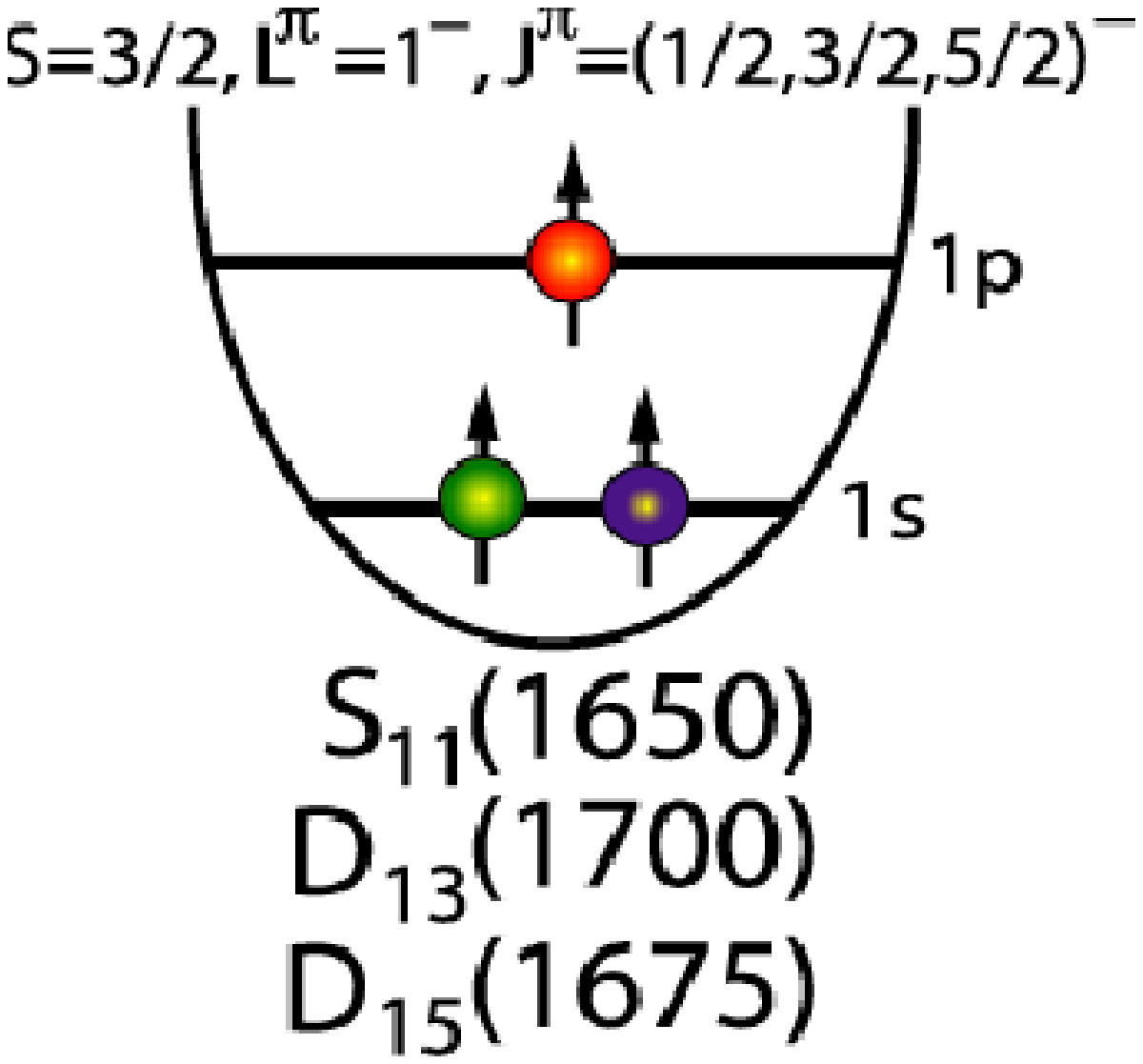}}
\caption{Schematic representation of the  0$\hbar\omega$, 1$\hbar\omega$
excitations of the nucleon in the naive constituent quark model. The labels
correspond to the lowest lying experimentally observed states with the 
respective quantum numbers.  
}
\label{fig_28}       
\end{figure}

In the following, we will discuss briefly  the nucleon resonances expected to 
contribute in the second energy region. A schematic representation of the 
lowest lying nucleon states in the constituent quark model with three quarks 
in a harmonic oscillator potential is given in fig.~\ref{fig_28}. The 
pictorial representation, which neglects the flavor degree of freedom, is a 
strong simplification of the properly anti-symmetrized wave functions
of the model (see \cite{Capstick_00}). Furthermore, mixing of the basis 
states with equal quantum numbers is expected due to spin-orbit interactions. 
An example are the two S$_{11}$ states.
%
%
%

%
%
\begin{figure}[htb]
\centerline{\epsfysize=7.1cm \epsffile{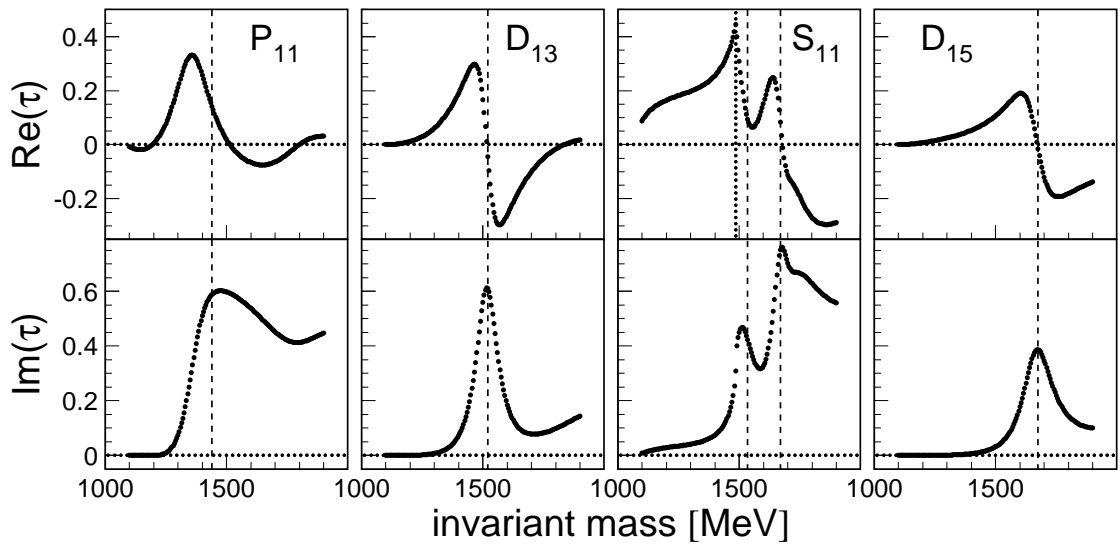}}
\caption{Real and imaginary parts of the $I=1/2$ partial amplitudes for 
$\pi N\rightarrow \pi N$ throughout the second resonance region (taken
from the SAID analysis \cite{Arndt_96,SAID}). Dashed lines: resonance positions,
vertical dotted line: $\eta$ production threshold.
}
\label{fig_29}       
\end{figure}
%

Experimental evidence for the existence of the low-lying excited states is 
provided by pion-nucleon scattering reactions. The relevant partial amplitudes 
from elastic pion scattering, taken from the SAID multipole analysis 
\cite{Arndt_96}, are summarized in fig.~\ref{fig_29}. Three partial waves in 
the isospin 1/2 channel, the P$_{11}$, the S$_{11}$, and the D$_{13}$, show 
clear signals at invariant masses around 1500 MeV. The isospin 3/2 channel is 
not shown but does not contribute to the second resonance region. However, the 
correspondence between the quark model expectations and experimental findings 
is different for various quantum numbers. The lowest lying D$_{13}$, the \d , 
is almost as close to a textbook case as the $\Delta$. The imaginary part shows 
a strong peak while the real part is crossing zero. The same is true for the 
higher-lying lowest D$_{15}$ excitation. 

The S$_{11}$ channel shows two structures related to the \s ~and \ss 
~resonances, which will be discussed further in connection with 
$\eta$-photoproduction. However, the situation for the lowest lying S$_{11}$ 
is less obvious. This resonance is close to the kinematical production 
threshold of the long-lived $\eta$ meson which can be produced off the nucleon 
in an s-wave. The sharp structure in the real part at 1485 MeV corresponds to 
the cusp induced in the pion production amplitude by the $\eta$ threshold. Some 
authors, using the speed plot technique \cite{Hoehler_92,Hoehler_93}, have 
argued that the evidence for the \s ~in pion scattering is not convincing. 
It is claimed that the entire structure could be attributed to the threshold 
cusp. Other authors \cite{Kaiser_95,Kaiser_97} try to explain the unusually 
large branching ratio of the \s ~resonance into $N\eta$ by suggesting that the 
structure in the $\eta$ photoproduction cross section might be attributed to a 
molecular-like $K\Sigma$ state. This scenario would correspond to a quasi-bound 
state rather than to a three-quark excited state of the nucleon. Thus, the 
structure and the very existence of this resonance are subject to debates in 
the literature even though it is assigned a four-star rating in the Particle 
Data Booklet \cite{PDG}. As we will see in sec. \ref{ssec:S11}, recent $\eta$ 
photoproduction experiments have contributed significantly to the discussion.

In addition, the P$_{11}$ partial wave, too, shows a prominent structure close 
to 1500 MeV. This is surprising since a P$_{11}$ excitation corresponds to a 
1s$\rightarrow$2s transition in the constituent quark model. Thus, it would be 
expected at higher excitation energies than the states shown in 
fig.~\ref{fig_28}. Howver, on the contrary, it is centered at a lower energy 
than any of the 
$1\hbar\omega$ states and it has a much larger width. This low-lying P$_{11}$ 
state is the Roper resonance. Non-relativistic constituent quark models do not 
offer a natural description of this state even if the harmonic confinement is 
replaced by a more realistic linear confinement. The state is better described 
in a relativistic quark model with instanton-induced quark forces proposed by 
L\"oring, Metsch, and Petry
\cite{Metsch_00}-\cite{Loering_01c}.
It also decreases in energy in the quark model of Glozman and Riska 
\cite{Glozman_96} 
which adds a chirally invariant meson exchange interaction of the quarks to the 
harmonic confinement. A monopole excitation is found as the 
lowest lying state in the Skyrmion model \cite{Schwesinger_92} which treats 
the nucleon structure as a mesonic field. However, possible `exotic' structures 
of this resonance have also been discussed. Burkert and Li \cite{Li_92} have 
argued that the $Q^2$ dependence of the $A_{1/2}$ helicity coupling in 
electroproduction is in better agreement with a model that treats the Roper as 
a $q^3G$ hybrid, with a gluonic excitation admixed to the three quark state. 
On the other hand, Krehl, Hanhart, Krewald, and Speth have \cite{Krehl_00} 
generated the resonance structure dynamically without a $q^3$ core. In this 
coupled channel meson exchange model for pion-nucleon scattering, the Roper 
resonance appears in the $\sigma N$ channel. Here, $\sigma$ is understood as a 
correlated pion pair in the scalar-isoscalar state. Finally, based on a 
comparison of $\alpha p$ and $\pi N$ scattering experiments, Morsch  and 
Zupranski \cite{Morsch_00} have suggested that the structure in the P$_{11}$ 
channel would be composed of two resonances with different internal structures.
Many aspects of the Roper excitation in hadron induced reactions have been 
summarized in the proceedings of the COSY Workshop on Baryon Excitations
\cite{COSY_00}.  
 
In summary, the second resonance region of the nucleon is a structure formed 
by the three overlapping resonances \pp , \d , \s . Their properties are listed 
in table \ref{tab_03}. Surprisingly, the structure of the \pp ~and \s ~is 
controversial although these are the lowest lying isospin 1/2 excitations of
the nucleon, and all three states hold a four-star status in the 
Particle Data Booklet.

\begin{figure}[htb]
\begin{minipage}{0.0cm}
{\mbox{\epsfysize=5.4cm \epsffile{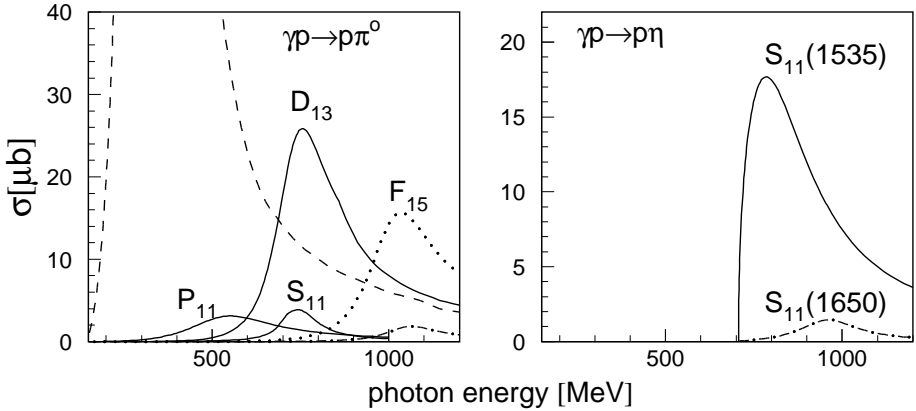}}}
\end{minipage}
\hspace*{12.5cm}
\begin{minipage}{5.8cm}
\vspace*{-0.5cm}
\caption{Contribution of resonances to $\pi^o$ and $\eta$ photoproduction 
(not quantitative). Full curves labelled P$_{11}$, D$_{13}$, and S$_{11}$  
correspond to the \pp, the \d, and the \s ~resonances. The dashed curve 
corresponds to the 
$\Delta$, the dash-dotted curves to the \ss , and the dotted curve to the \f .  
}
\label{fig_30}       
\end{minipage}
\end{figure}

%
%
%
\begin{table}[hbt]
  \caption[Properties of nucleon resonances]{
    \label{tab_03}
Properties of nucleon resonances forming the second resonance region \cite{PDG}. 
(The decay branching ratios into $N\rho$ and $\Delta\pi$ are partial channels
of $N\pi\pi$).}
  \begin{center}
    \begin{tabular}{|c|c|c|c|c|cc|lr|}
      \hline 
      & & 
      & {mass}
      & {width $\Gamma$}
      & {$A_{1/2}^p$}, {$A_{1/2}^n$} 
      &/ {$A_{3/2}^p$}, {$A_{3/2}^n$}
      & \multicolumn{2}{c|}{decays} \\
      state
      & {$I$} 
      & {$J^{P}$} 
      & {[MeV]}   
      & {[MeV]}
      & \multicolumn{2}{c|}{[10$^{-3}$\mbox{GeV}$^{-1/2}$]}
      & \multicolumn{2}{c|}{(\%)} \\
      \hline \hline 
      P$_{11}$(1440) & $\frac{1}{2}$ & $\frac{1}{2}^+$ & 
      1430 - 1470 & 250 - 450 & $-$65$\pm$4 & +40$\pm$10  
      & $N\pi$ & 60 - 70 \\
      & & & & & & & $N\pi\pi$ & 30 - 40\\
      & & & & & & & ~~~~~~~$N\rho$ & $<$8\\
      & & & & & & & ~~~~~~~$\Delta\pi$ & 20 - 30\\
      & & & & & & & $N\eta$ & ?\\
      \hline
      D$_{13}$(1520) & $\frac{1}{2}$ & $\frac{3}{2}^-$ & 
      1515 - 1530 & 110 - 135 & $-$24$\pm$9 & $-$59$\pm$9
      & $N\pi$ & 50 - 60 \\
      & & & & & +166$\pm$5 & $-$139$\pm$11 & $N\pi\pi$ & 40 - 50\\
      & & & & & & & ~~~~~~~$N\rho$ & 15 - 25\\
      & & & & & & & ~~~~~~~$\Delta\pi$ & 15 - 25\\
      & & & & & & & $N\eta$ & 0 - 1\\
      \hline
      S$_{11}$(1535) & $\frac{1}{2}$ & $\frac{1}{2}^-$ & 
      1520 - 1555 & 100 - 200 & +90$\pm$30 & $-$46$\pm$27  
      & $N\pi$ & 35 - 55 \\
      & & & & & & & $N\pi\pi$ & 1 - 10\\
      & & & & & & & ~~~~~~~$N\rho$ & $<$4\\
      & & & & & & & ~~~~~~~$\Delta\pi$ & $<$1\\
      & & & & & & & $N\eta$ & 30 - 55\\
      \hline      
    \end{tabular}
  \end{center}
  \vspace*{-0.7cm}
\end{table}
%
The experimental investigation of the resonance properties, in particular the 
electromagnetic couplings and partial decay widths, is complicated by their 
large overlapping widths. The situation in meson photoproduction reactions is 
sketched in fig.~\ref{fig_30}. The figure shows the expected contribution of 
the three resonances and the tails of other resonances to $\pi^o$ and $\eta$ 
photoproduction off the proton. The contribution is approximated from the 
resonance positions, widths, photon couplings, and decay branching ratios. 
The D$_{13}$ resonance dominates single pion production, and is also important 
for the understanding of double pion production reactions, as we will discuss 
in sec. \ref{ssec:twopi}. The other two resonances contribute only weakly to
single pion production. Photoproduction of the $\eta$ meson is dominated by the 
S$_{11}(1535)$, 
and this fact enables detailed studies of the resonance in the $\eta$ channel. 
In contrast, the P$_{11}$ is not favored in any reaction over the other 
resonances. For the P$_{11}$ resonance, it is mandatory to establish
precise experimental cross sections and polarization observables in order to 
define the relevant multipole amplitudes. The situation is reflected in the 
results discussed in the following sections. Many new results have become 
available for the S$_{11}$ via $\eta$ production. The role of the D$_{13}$ 
not only in single but also in double pion production and even in $\eta$ 
production was investigated in detail. Much less new material is available 
for the P$_{11}$.
%
%
%
%

\newpage
\subsection{\it Single $\pi$ Photoproduction and the \pp ~and \d ~Resonances}
\label{ssec:D13}

Single pion photoproduction is the standard method for the investigation of the
electromagnetic resonance couplings, being one of the most important testing
grounds for hadron models. Partial wave analyses of pion production data up
to 2 GeV incident photon energy are available (see \cite{Arndt_96,Arndt_02}
and ref. therein).
Dispersion relations and unitary isobar models 
\cite{Hanstein_98,Drechsel_99,Aznauryan_02} have been used for the analysis of
the data, and predictions for the photoproduction amplitudes have been made
in the framework of quark models \cite{Zhao_02}. A review of all available
pion photoproduction data and analyses would go far beyond the scope of this 
article. Instead, we will discuss recent advances using the second resonance 
region as an example. 

During the last ten years, the data base for pion photoproduction has been 
improved considerably with respect to two aspects. Many of the older 
bremsstrahlung measurements have been replaced by more precise tagged photon 
beam experiments. An increasing number of polarization observables have been 
measured for the first time. As noted in the most recent pion multipole 
analysis \cite{Arndt_02}, in 1994 bremsstrahlung data comprised still 85\% of 
the then available data. These results were often plagued by poorly understood 
systematic uncertainties causing inconsistencies in the data base. The few data 
sets that were available from tagged photons generally came from low statistics 
measurements. The situation changed due to experimental programs running at 
modern tagged-beam facilities at MAMI-B (Mainz), GRAAL (Grenoble), and LEGS 
(Brookhaven) and will further improve with the upcoming results from CLAS 
(JLab). Progress was fastest in the energy range below 800 MeV incident photon 
energy, since the {MAMI-B} facility produced over 20\% of the $\pi^+n$ data 
after 1995, and almost 90\% of the new $\pi^op$ data \cite{Arndt_02}. In 
parallel with the improvement in data quality, measurements of previously 
unexplored polarization observables produced crucial constraints for the 
analyses. Such results in the $\Delta$ range were already discussed in secs. 
\ref{ssec:quadru}, \ref{ssec:delta_gdh}. In the medium energy range, the 
$\pi^+n$ beam asymmetry ($\Sigma$) was measured at GRAAL 
($E_{\gamma}$=550 - 1500 MeV) \cite{Ajaka_00,Bartalini_02} and the $\pi^op$ 
beam asymmetry ($E_{\gamma}$=500 - 1100 MeV) at Yerevan \cite{Adamian_01}. 
Target asymmetry ($T$) measurements for $\pi^op$ and $\pi^+n$ 
($E_{\gamma}$=220 - 800 MeV) have been reported from ELSA at Bonn
\cite{Dutz_96,Bock_98}. The difference of the helicity components
$\sigma_{3/2}-\sigma_{1/2}$ has been measured for $\pi^op$ with the GDH
experiment in Mainz \cite{Ahrens_GDH_02}. 

%
%
%
\begin{figure}[bht]
\begin{minipage}{0.0cm}
{\mbox{\epsfysize=7.6cm \epsffile{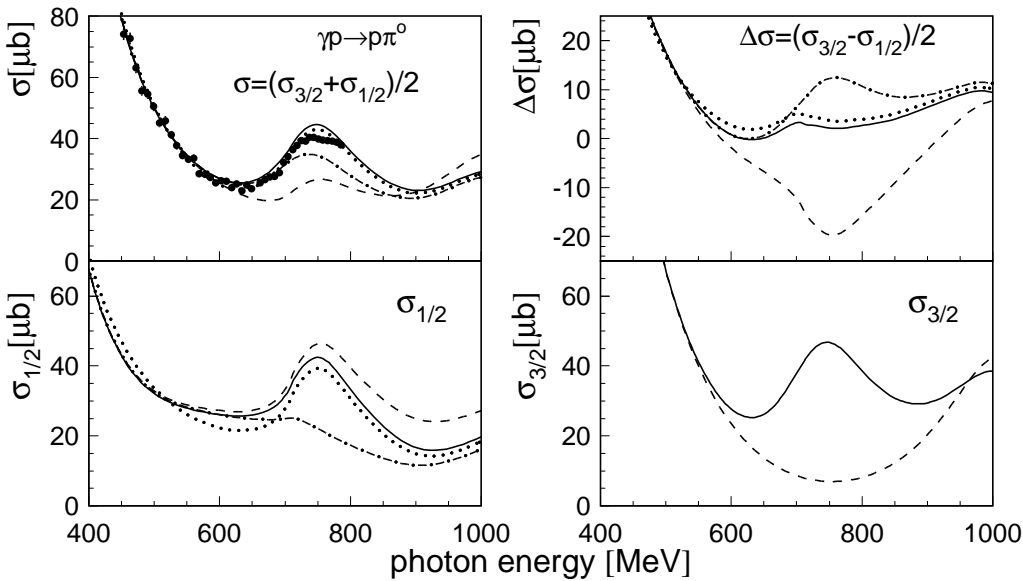}}}
\end{minipage}
\hspace*{14.cm}
\begin{minipage}{4.2cm}
\caption{Unpolarized and helicity dependent total cross sections for 
$p(\gamma ,\pi^o)p$
The curves are the MAID results \cite{Drechsel_99} for the full model (solid),
without D$_{13}$ (dashed), without S$_{11}$ (dash-dotted), and without
P$_{11}$ (dotted). The data for the unpolarized cross section are from
\cite{Krusche_99}.
}
\label{fig_31}       
\end{minipage}
\end{figure}
%

Helicity dependent cross section data are particularly useful for the 
separation of contributions from the $J=1/2$ (P$_{11}$ and S$_{11}$) states 
and the $J=3/2$ D$_{13}$ resonance. This is demonstrated in fig.~\ref{fig_31}
for the reaction $\gamma p\rightarrow \pi^o p$. The figure shows the 
unpolarized cross section, the two helicity components and their difference, 
calculated from the MAID analysis \cite{Drechsel_99} for the full model, and 
for three truncated versions each excluding one resonance. The D$_{13}$ has 
a strong effect on the difference of the helicity components. The resonance 
structure in the helicity $\nu =3/2$ component is entirely due to the D$_{13}$ 
state since the $J=1/2$ resonances cannot contribute. A measurement of 
$\sigma_{3/2}$ would thus put stringent constraints on the $A_{3/2}$ helicity 
amplitude of the D$_{13}$. The S$_{11}$, on the other hand, is expected to 
dominate the resonance structure in the $\nu =1/2$ channel, where the D$_{13}$ 
plays a minor role. The separation of the two resonances in this channel would 
be possible due to the different angular dependence of  s- and d-waves. The 
sensitivity of the unpolarized cross section, the helicity components, as well 
as the photon beam asymmetry to the \pp ~resonance is relatively small.

\begin{figure}[htb]
\begin{minipage}{0.0cm}
{\mbox{\epsfysize=4.cm \epsffile{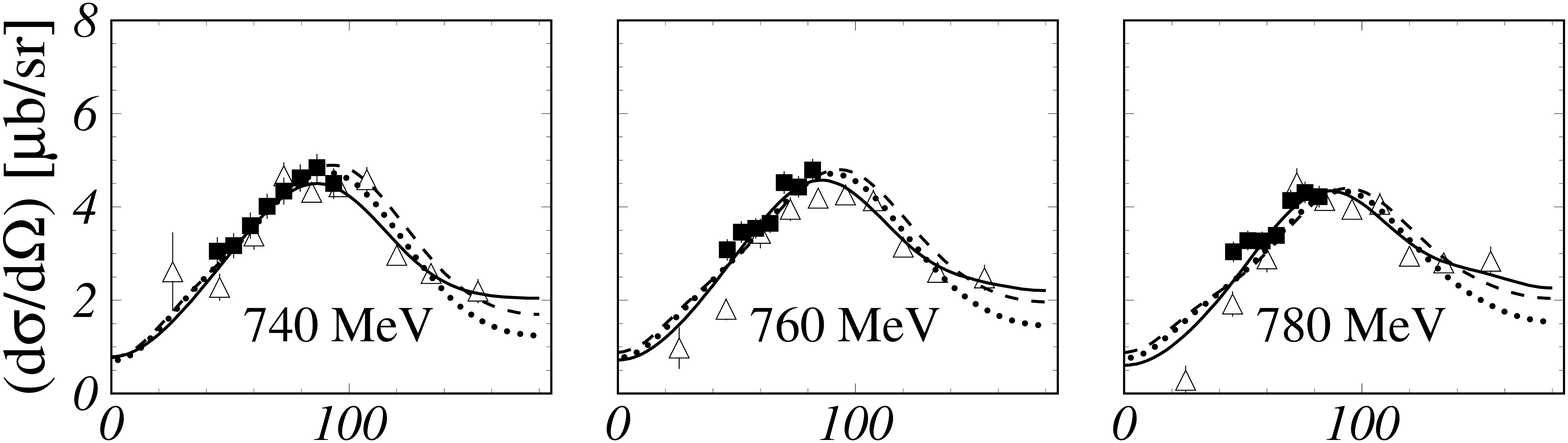}}}
{\mbox{\epsfysize=4.45cm \epsffile{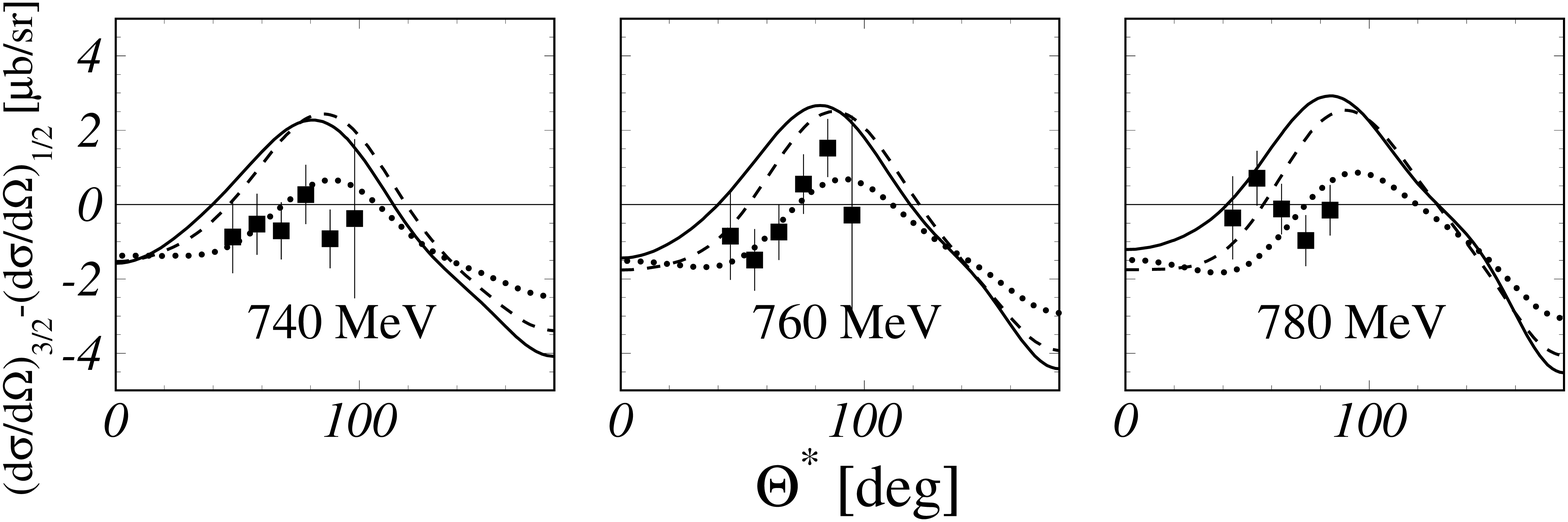}}}
\end{minipage}
\hspace*{13.cm}
\begin{minipage}{5.cm}
\caption{Unpolarized angular distributions (upper part) and difference of the
helicity components (lower part) in the range of the \d
~resonance \cite{Ahrens_GDH_02}. Full squares are from \cite{Ahrens_GDH_02}, 
open triangles from \cite{Krusche_99}. The solid curves correspond to the SAID 
analysis \cite{Arndt_96}, the dashed curve to the MAID solution 
\cite{Drechsel_99}, and the dotted curves to a re-fit of the MAID solution 
to the data.
}
\label{fig_32}       
\end{minipage}
\end{figure}

Figure~\ref{fig_32} shows unpolarized cross sections and the difference of 
the helicity components as a function of the pion polar angle for the reaction 
$\gamma p\rightarrow \pi^o p$ \cite{Krusche_99,Ahrens_GDH_02} throughout the 
excitation range of the \d . 
%
%
%
%
They are compared to the result of the MAID and SAID analyses obtained without 
using the helicity dependent data. Both analyses reproduce the unpolarized 
data. This is to be expected since the angular distributions from 
\cite{Krusche_99} were included in the fits. However, good agreement with the 
helicity difference is not achieved. Therefore, the MAID analysis was re-fitted 
including the polarization data \cite{Ahrens_GDH_02}. The result is indicated 
by the dotted lines. The effect on the unpolarized cross section is small but 
the agreement for the helicity dependence is significantly improved. As 
discussed in \cite {Ahrens_GDH_02}, the modified MAID solution is also in much 
better agreement with the $n\pi^+$ photon asymmetry ($\Sigma$) data but still 
disagrees with the $p\pi^o$ $\Sigma$-measurement. The main difference in the 
amplitudes, extracted from the original and modified MAID fits, apart from 
background contributions, appears in the $E_{2-}^{1/2}$ and $M_{2-}^{1/2}$ 
components which are associated with the D$_{13}$ excitation. In the modified 
fit, the strength of the first is reduced by a factor of (0.81$\pm$0.01) and 
the second is increased by a factor of (1.11$\pm$0.01). These partial 
amplitudes are directly related to the helicity couplings $A_{3/2}$ and 
$A_{1/2}$ of the D$_{13}$ resonance, as long as background  can be neglected.
In \cite{Workman_00}, it can be found that
\begin{eqnarray}
\label{eq:d13_rem}
R_A= \frac{A_{3/2}}{A_{1/2}} & = &
\sqrt{3}\,\frac{E_{2-}+M_{2-}}{E_{2-}-3M_{2-}}
= \sqrt{3}\,\frac{1+R_{M}}{R_{M}-3}\\
R_{M}= \frac{E_{2-}}{M_{2-}} & = & \frac{\sqrt{3}+3R_{A}}{R_{A}-\sqrt{3}}\;.\nonumber
\end{eqnarray}
The change of the multipole amplitudes corresponds to a significant
lowering of the magnitude of the ratio of the helicity couplings 
(see. table \ref{tab_04}) from -9.8 (MAID1998) to -3.8 (MAID2002).
   
\begin{figure}[htb]
\begin{minipage}{6.0cm}
\begin{turn}{90.}
{\mbox{\epsfysize=6.cm \epsffile{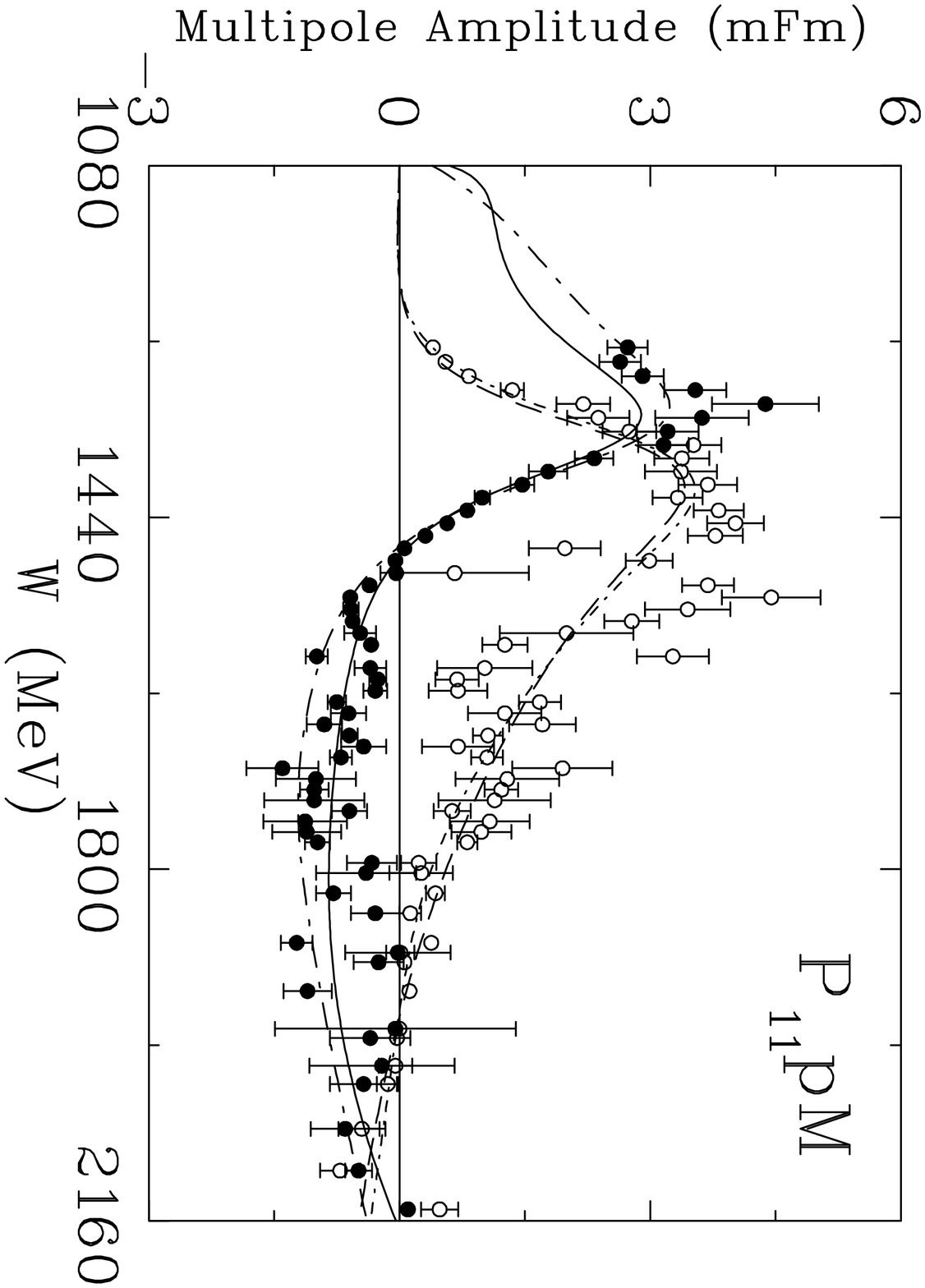}}}
\end{turn}
\end{minipage}
\begin{minipage}{6.0cm}
\begin{turn}{90.}
{\mbox{\epsfysize=6.cm \epsffile{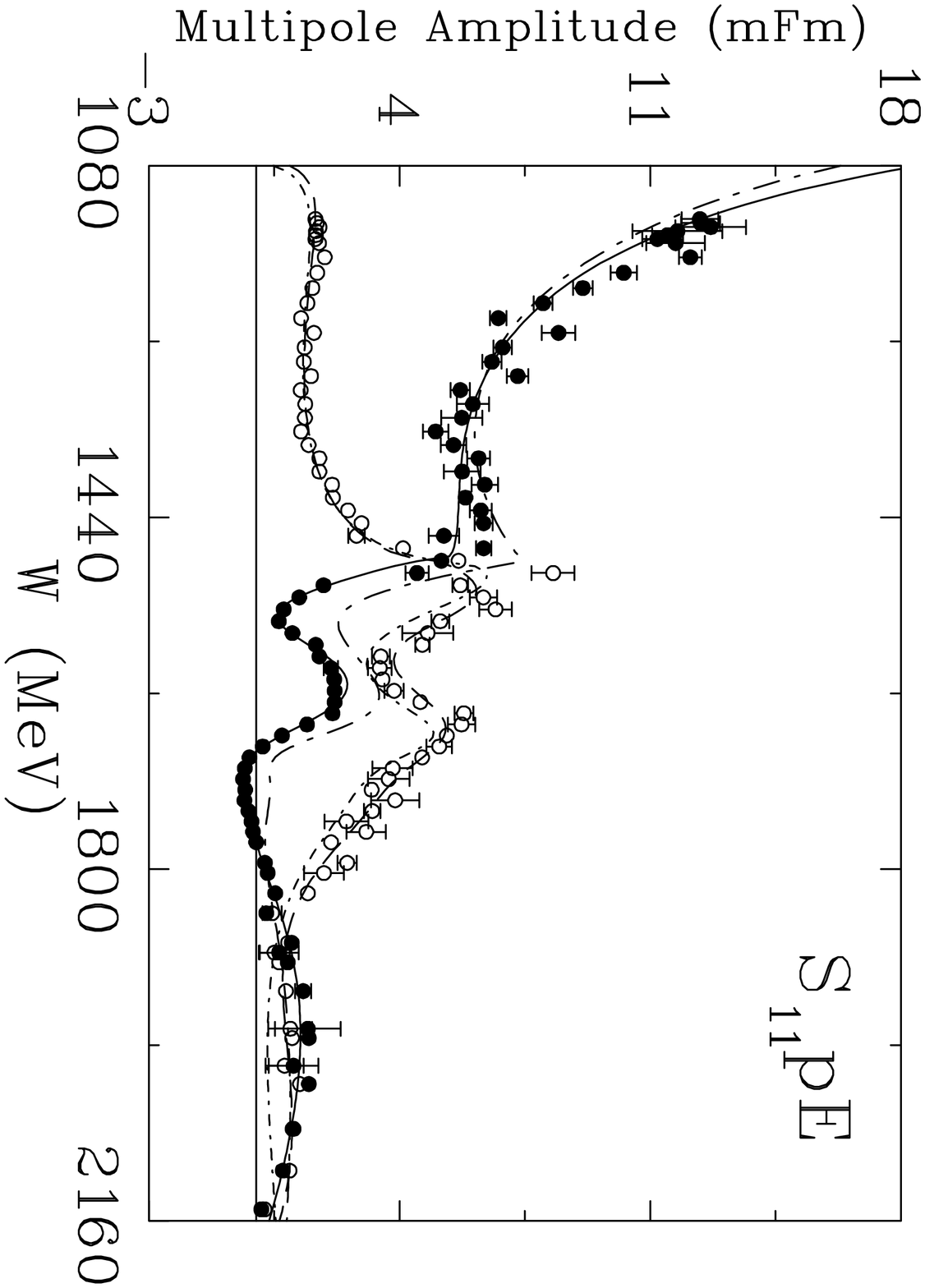}}}
\end{turn}
\end{minipage}\\
\begin{minipage}{6.0cm}
\begin{turn}{90.}
\epsfysize=6.cm \epsffile{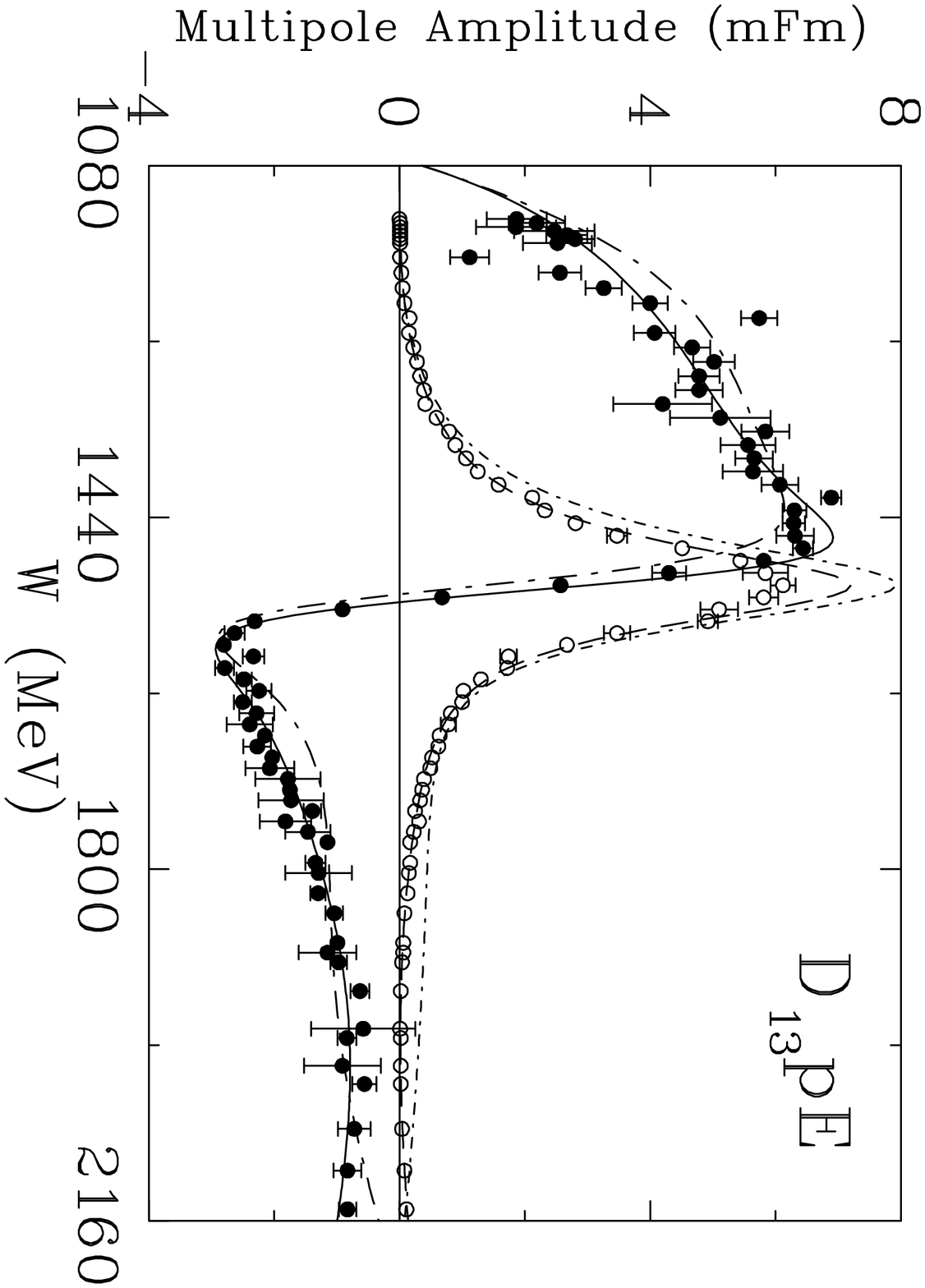}
\end{turn}
\end{minipage}
\begin{minipage}{6.0cm}
\begin{turn}{90.}
\epsfysize=6.cm \epsffile{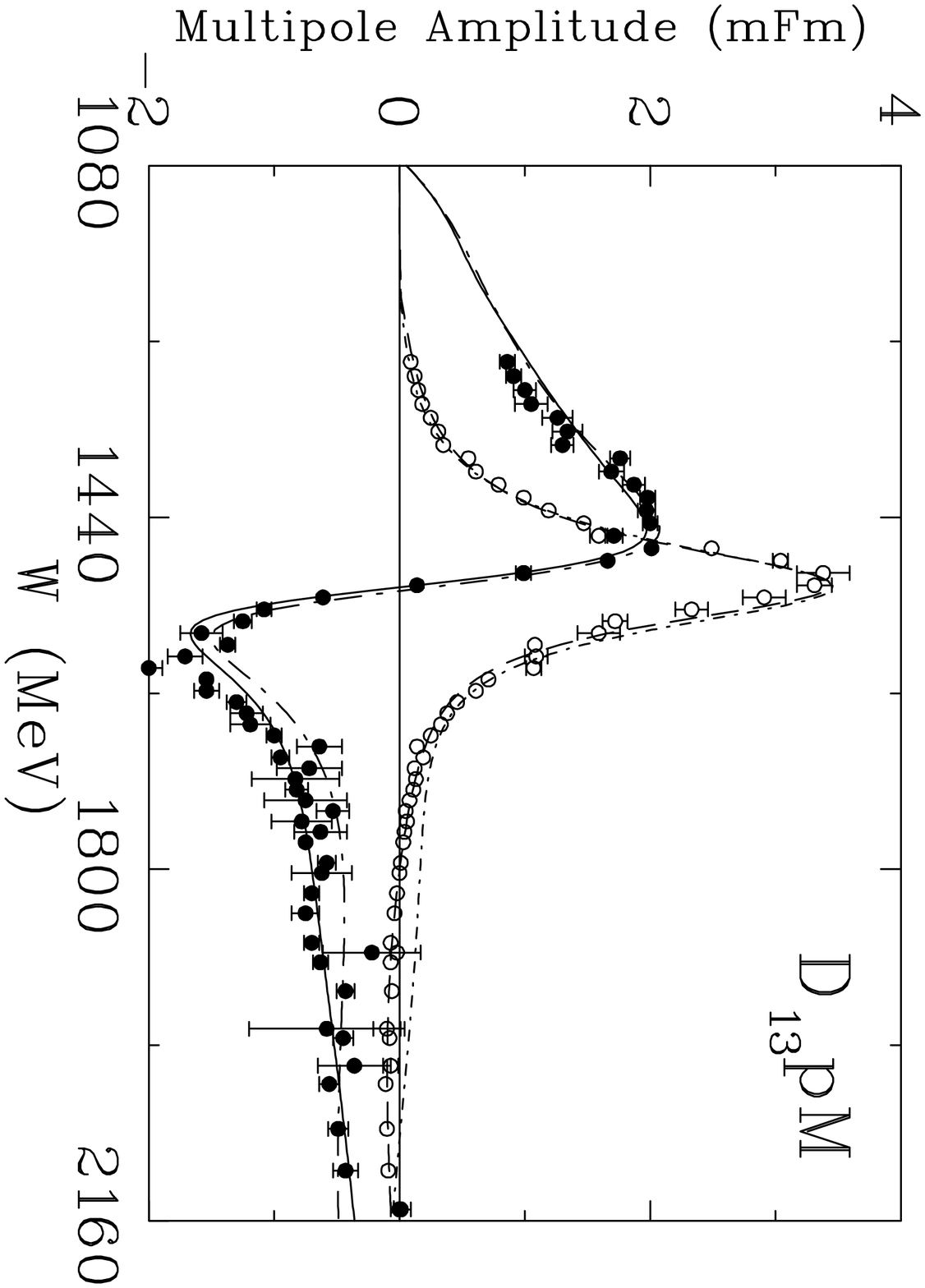}
\end{turn}
\end{minipage}
\hspace*{0.5cm}
\begin{minipage}{5.8cm}
\vspace*{-4.7cm}
\caption{Partial wave amplitudes extracted from pion photoproduction in the 
SAID analysis \cite{Arndt_02}. Solid (dashed) curves: 
energy dependent real (imaginary) parts of the SM02 solution. Filled (open)
circles: real (imaginary) parts of the single energy 
solutions. Long dash-dotted (short dash-dotted) lines correspond to the real
(imaginary) parts of the SM95 solutions \cite{Arndt_96}. Notation: $X_{2I2J}$
denotes orbital angular momentum, isospin and spin, 'p' denotes the
reaction on the proton, and M,E magnetic or electric multipole, e.g. D$_{13}$pM
corresponds to $M_{2-}^{1/2}$ on the proton.
}
\label{fig_33}          
\end{minipage}
\end{figure}

A multipole analysis (SAID) of the full data base has recently been reported 
by Arndt and collaborators \cite{Arndt_02}. The results for the multipoles 
involving the excitation of the S$_{11}$, P$_{11}$, and D$_{13}$ on the proton 
are summarized in fig.~\ref{fig_33}.
%
%
%
%
In the figure, the symbols correspond to the energy independent solutions, i.e.
to multipole fits done independently for each incident photon energy. The 
curves correspond to energy dependent solutions, parameterized in terms of the 
$T$-matrix for $\pi N$-scattering in the appropriate partial wave. The general 
observation is that in all cases the solutions show a structure very similar 
to the corresponding $\pi N$ amplitudes (see fig.~\ref{fig_29}). The 
following remarks can be made to the individual resonances:

\noindent{{\bf $\bullet$ \s}:} 
As has been discussed in the context of pion induced reactions, the structure 
corresponding to the lowest lying S$_{11}$ resonance is obscured by the cusp 
arising from the opening of the $\eta$ production threshold. The determination 
of the helicity couplings of the \s ~has thus always been less precise than 
for the other resonances. The results from different analyses were not in good 
agreement. In their most recent multipole analysis, Arndt et al. \cite{Arndt_02}
find significant sensitivity of the $S_{11}pE$ ($E_{0+}^{1/2}$) multipole to 
details of the data base and of the parameterization. It is concluded that the 
coupling requires a more detailed treatment. The value 
($A^p_{1/2}$ =(30$\pm$3)10$^{-3}\,$\mbox{GeV}$^{-1/2}$) 
that is finally quoted is quite low even when compared to other analyses of 
pion data. It is much lower than results from analyses of $\eta$ 
photoproduction data. We will discuss the S$_{11}$ helicity couplings further 
in the context of $\eta$-photoproduction (see sec. \ref{ssec:S11}).

\begin{figure}[htb]
\begin{minipage}{0.0cm}
\begin{turn}{-90.}
{\mbox{\epsfysize=13.cm \epsffile{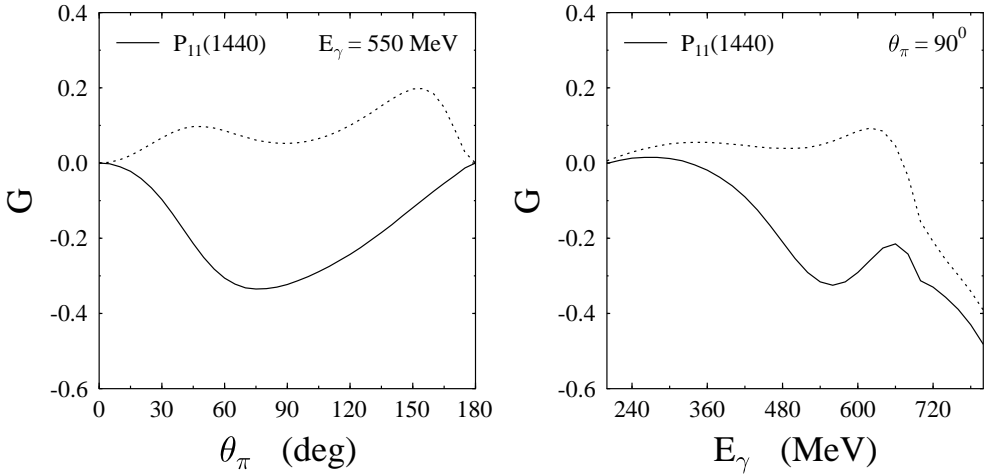}}}
\end{turn}
\end{minipage}
\hspace*{13.2cm}
\begin{minipage}{5.cm}
\caption{Sensitivity of the polarization observable $G$ to the Roper resonance
\cite{Beck_01}. Solid lines: full MAID model, dashed lines: without Roper. 
}
\label{fig_35} 
\end{minipage}
\end{figure}

\noindent{{\bf $\bullet$ \pp}}:
The observables studied so far are marginally sensitive to the P$_{11}$. 
Consequently, the scattering/spread of the energy independent solutions is 
larger than for the other channels. Nevertheless, the helicity couplings 
extracted from the data are given with relatively small uncertainties.
The results of different analyses are in fair agreement 
($A_{1/2}^p$ = $-$67$\pm2$ \cite{Arndt_02}, $-$63$\pm5$ \cite{Arndt_96},
$-$71 \cite{Drechsel_99}). A comparison with the predictions of quark models 
underlines that the Roper resonance does not at all fit into the conventional 
constituent quark picture of the nucleon. Close and Lee \cite{Close_90} predict 
$A_{1/2}^p$=+10, Capstick \cite{Capstick_92} a value of +4, and Bijker et al., 
in their algebraic nucleon model, \cite{Bijker_94} quote a range from 0 to +67. 
(values for $A_{1/2}$ in units of 10$^{-3}$\mbox{GeV}$^{-1/2}$). 
The measurement of the helicity coupling as function of the four-momentum 
transfer can provide important information about the structure of the 
resonance. 
%
%
%
\begin{figure}[tht]
\begin{minipage}{0.0cm}
\begin{turn}{-90.}
{\mbox{\epsfysize=13.cm \epsffile{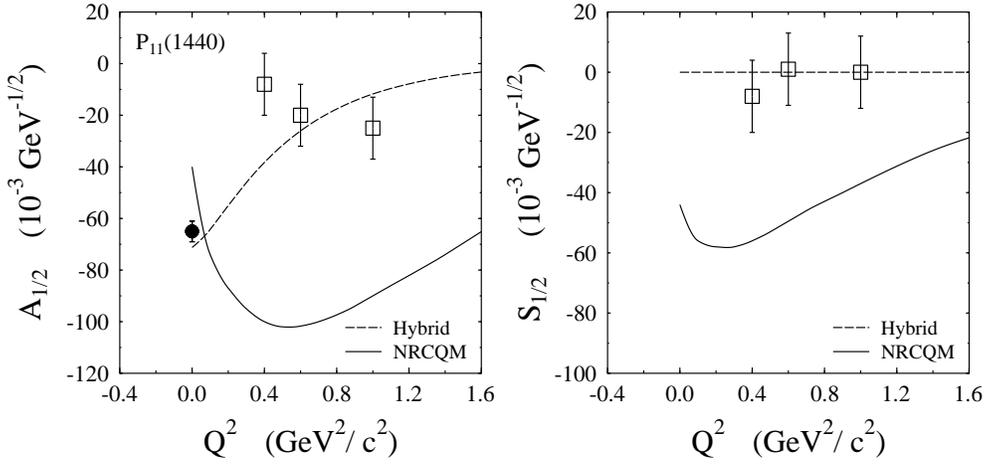}}}
\end{turn}
\end{minipage}
\hspace*{13.2cm}
\begin{minipage}{5.cm}
\caption{Transverse ($A_{1/2}$) and longitudinal ($S_{1/2}$) helicity coupling
of the Roper resonance. Real photon point (filled
circle): PDG \cite{PDG}. Open squares: Gerhardt 
\cite{Gerhardt_80}. Full lines:  $q^3$ quark model prediction (Barnes and
Close) \cite{Barnes_83,Barnes_83a}, dashed lines: $q^3G$ hybrid model (Li,
Burkert and Li) \cite{Li_92}.
}
\label{fig_34}       
\end{minipage}
\end{figure}
%
This is demonstrated in fig.~\ref{fig_34} where the predictions 
for the transverse and longitudinal couplings of a conventional $q^3$ quark 
model \cite{Barnes_83,Barnes_83a} are compared to the prediction of a $q^3G$ 
hybrid model \cite{Li_92}. 
The predictions are quite different but the data at finite four-momentum 
transfers \cite{Gerhardt_80} have large systematic uncertainties from 
incomplete data sets and theoretical assumptions in the analysis. From the 
above discussion, it is evident that it would be helpful to find observables 
more sensitive to the Roper resonance, so that the systematic uncertainties 
could be better controlled. Recently, Beck \cite{Beck_01} has pointed out that
the double polarization observable $G$ (linearly polarized photons and 
longitudinally polarized protons) in the reaction 
$\vec{p}(\vec{\gamma},\pi^o)p$  is ideally suited for this purpose 
(see fig.~\ref{fig_35}). 
%
%
%
%

~\\
\noindent{{\bf $\bullet$ \d}:}
The signal for the \d ~state is by far the clearest. The imaginary part of the 
amplitude displays an almost perfect shape of a Breit-Wigner resonance. Again, 
the properties of interest are the helicity couplings $A_{1/2}$, $A_{3/2}$, 
and their ratio. As discussed in connection with the $\Delta$ resonance 
(see sec. \ref{ssec:quadru}), the onset of perturbative QCD is characterized by 
helicity conservation. For very large $Q^2$, $A_{1/2}\gg A_{3/2}$ is expected. 
At the photon point, the behavior is dictated by the non-perturbative QCD 
effects which may result in a large violation of helicity conservation. 
In fact, in case of the D$_{13}$, the coupling for $Q^2=0$ is dominated by 
$A_{3/2}$. Therefore, the helicity couplings are very sensitive to the internal
structure of the resonances and thus very well suited for model tests. However, 
stringent tests are only possible when the couplings can be determined 
precisely. Usually, the ratio of the couplings can be determined with smaller 
systematic uncertainties than the couplings themselves because resonance 
parameters like width and decay branching ratios cancel. The typical range of 
model predictions for $A_{3/2}^p/A_{1/2}^p$ is indicated in 
table~\ref{tab_04}. Most conventional quark models predict ratios between 
$-$5 and $-$10 but the algebraic model of Bijker et al.~\cite{Bijker_94}
predicts a smaller value of $-$2.5. Typical results extracted from older data
not including the new measurements of polarization observables corresponded
to fairly large values of the ratio (SAID95: $-$8.4, MAID98: $-$9.8). On the 
other hand, values extracted from $\eta$-photoproduction are much smaller, 
between $-$2. and $-$2.5 (see table \ref{tab_04}). The coupling of the D$_{13}$ 
to the $\eta$ channel is weak, the decay branching ratio is smaller than 
0.1 \%. Meanwhile, the pion production is dominated by this resonance, and one 
might wonder whether the extraction of the coupling from the $\eta$ channel is 
possible with reasonable precision, as compared to the pion channel. However, 
as discussed in the following section, the photon beam asymmetry $\Sigma$ in 
$\eta$ photoproduction is extremely sensitive to contributions of the D$_{13}$.

The question is whether a serious discrepancy between the results from pion 
and $\eta$ photoproduction persists. Workman et al. have discussed this problem 
\cite{Workman_00} and find that it is still possible to find a reasonable 
description of the pion data with $R_{A}=-2.5$ although fits to the (old) pion 
data base tend to produce large values of $R_{A}$.
Inclusion of the new polarization observables for pion production into the fits 
has reduced the discrepancy.
We have seen that inclusion of the helicity 
difference $\sigma_{3/2}-\sigma_{1/2}$ into the fit of the MAID model has 
lowered the ratio to only $-$3.8. The same trend is visible for the SAID 
analysis where inclusion of all polarization data has lowered the ratio to 
$-$5.6. The effect of the new data on the fit is clearly visible in 
fig.~\ref{fig_33}. The magnetic multipole is practically unchanged between
the 1995 and 2002 SAID solution but the imaginary part of the electric 
multipole at the resonance position is reduced from about 8 to 6 mFm. According 
to eq.~\-(\ref{eq:d13_rem}) this corresponds to a decrease of $R_{A}$. In fact, 
we could use the same procedure as in \cite{Ahrens_GDH_02}, ignore possible 
background contributions to the imaginary parts and calculate $R_{A}$ from the 
values read-off from fig.~\ref{fig_33} 
(Im($E_{2-}^{1/2})=6$ mFm, Im($M_{2-}^{1/2})=3.3$ mFm) via
eq.~\-(\ref{eq:d13_rem}). Then we obtain $R_{A}=-4.1$ which is even closer 
to the MAID analysis. The discrepancy between pion and $\eta$ results is thus 
reduced from a factor of 4-5 to less than a factor of two.
%
%
%
\begin{table}[bht]
  \caption[D$_{13}$ helicity couplings]{
    \label{tab_04}
Photon couplings of the \d ~resonance. All values in units of
{$10^{-3}\mbox{GeV}^{-1/2}$}. PDG: Review of Particle Properties \cite{PDG},
GW: Breit-Wigner resonance fits to SAID multipole
analysis \cite{Arndt_02}, VPI: 1995 SAID analysis \cite{Arndt_96},
MAID: unitary isobar model \cite{Drechsel_99}, MAID II: re-fit to
helicity cross sections \cite{Ahrens_GDH_02}, ETA-I: analysis of $\eta$
photoproduction with effective Lagrangian model \cite{Mathur_98},
ETA-II: analysis of
$\eta$-production observables with isobar model \cite{Tiator_99}, 
ETA-MAID: isobar model for $\eta$ photoproduction \cite{Chiang_02}. 
QM-I:  non relativistic quark model (Koniuk,Isgur) 
\cite{Koniuk_80}.
QM-II: quark model with relativistic corrections (Close, Li) \cite{Close_90}
(first number: calculation in c.m frame, in brackets: calculation in
Breit-Frame)
QM-III: relativized quark model (Capstick) \cite{Capstick_92}.
$^{\star )}$: re-calculated from $A_{3/2}$, $A_{1/2}$.
AM: algebraic model of hadron structure (Bijker, Iachello, Leviatan)
\cite{Bijker_94}.
}
  \begin{center}
    \begin{tabular}{|c|c|c|c|c|c|}
      \hline 
      Ref. & $A_{1/2}^p$ & $A_{3/2}^p$ & $A_{3/2}^{p}/A_{1/2}^p$ 
      & $A_{1/2}^n$ &  $A_{3/2}^n$\\
      \hline
      PDG (2002) & $-$24$\pm$9 & +166$\pm$5 & $-$6.9$\pm$2.6 $^{\star )}$ 
      & $-$59$\pm$9 & $-$139$\pm$11 \\
      \hline 
      GW (2002) & $-$24$\pm 2$ & +135$\pm 2$ & $-$5.6$\pm$0.5 $^{\star )}$ 
      & $-$67$\pm 4$ & $-$112$\pm 3$ \\
      VPI (1995) & $-$20$\pm 7$ & +167$\pm 5$ & $-$8.4$\pm$3.0 $^{\star )}$ 
      & $-$48$\pm 8$ & $-$140$\pm 10$ \\
      MAID (1998) & $-$17 & +164 & $-$9.8 $^{\star )}$ & $-$40 & $-$135\\ 
      MAID II (2002) & $-$37 & +141& $-$3.8 $^{\star )}$ & &\\ 
      \hline
      ETA-I (1998) & & & $-$2.5$\pm$0.2$\pm$0.4 & & \\
      ETA-II (1999) & $-$79$\pm$9 & & $-$2.1$\pm$0.2 & & \\
      ETA-MAID (2002) & $-$52 & & & & \\      
      \hline 
      QM-I (1980) & $-$23 & +128 & $-$5.56 $^{\star )}$ & $-$45 & $-$122 \\ 
      QM-II (1990) & $-$28($-$30) & +143(+146) & $-$5.1($-$4.9) $^{\star )}$ 
      & $-$46($-$49) & $-$143($-$146) \\  
      QM-III (1992) & $-$15 & +134 & $-$8.9 $^{\star )}$ & $-$38 & $-$114 \\
      AM (1994) & $-$43 & +108 & $-$2.5 $^{\star )}$ & $-$27 & $-$108 \\    
      \hline      
    \end{tabular}
  \end{center}
\end{table}
%
 This is one of the
examples which demonstrates the importance of new precise measurements of 
different observables. When resonance parameters like the helicity ratio are 
uncertain by factors of five, comparisons to model predictions are not useful. 
Often, it is the comparison of results from different channels that reveals
systematic problems, thus underlining the importance of studying resonances 
in other than the pion decay channel alone. In the present case, it has to be 
seen whether the results from pion and $\eta$ photoproduction will eventually 
converge.

\begin{figure}[hbt]
\begin{center}
\begin{turn}{90.}
\epsfysize=7.8cm \epsffile{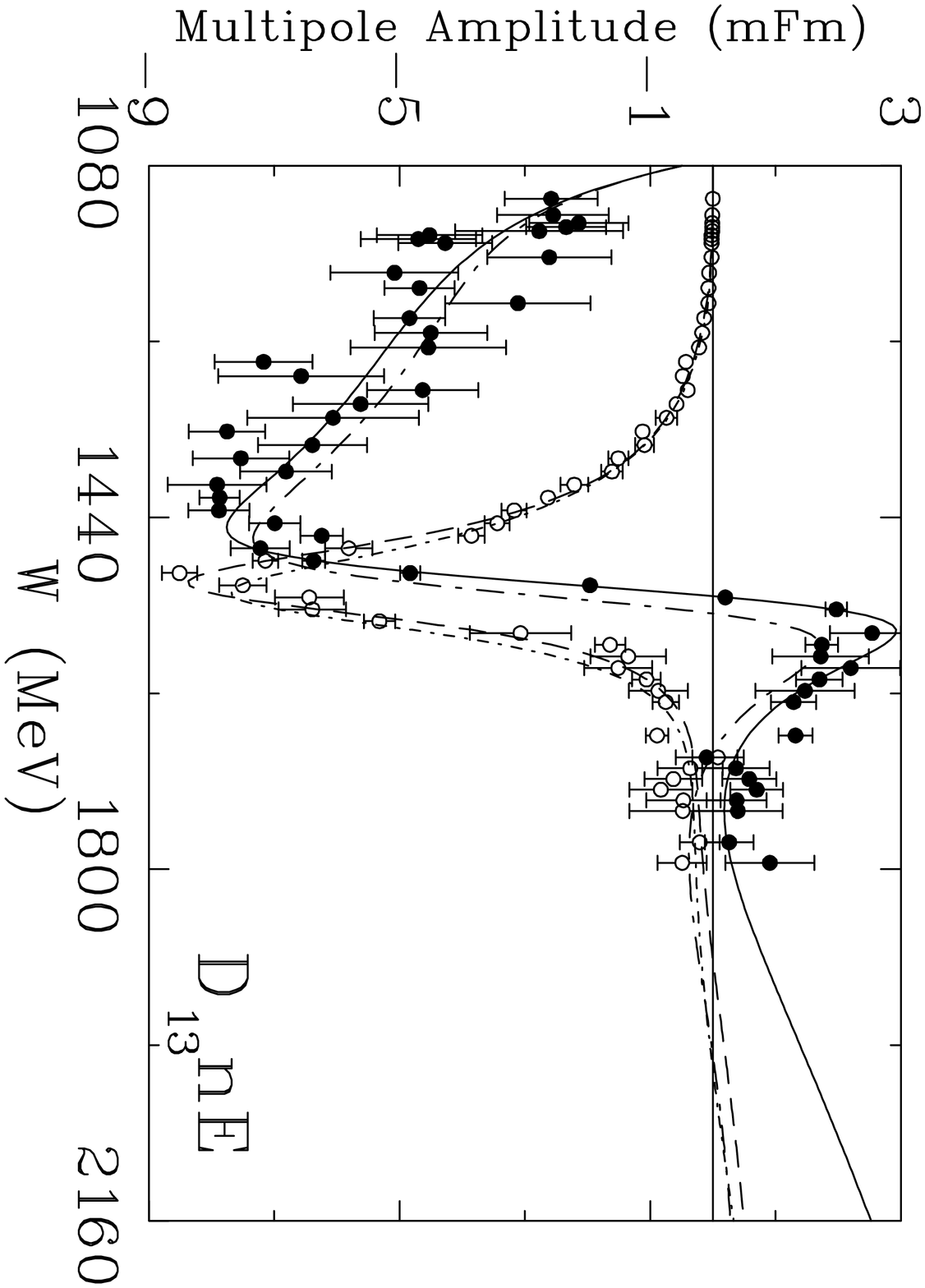}
\end{turn}
\begin{turn}{90.}
\epsfysize=8.0cm \epsffile{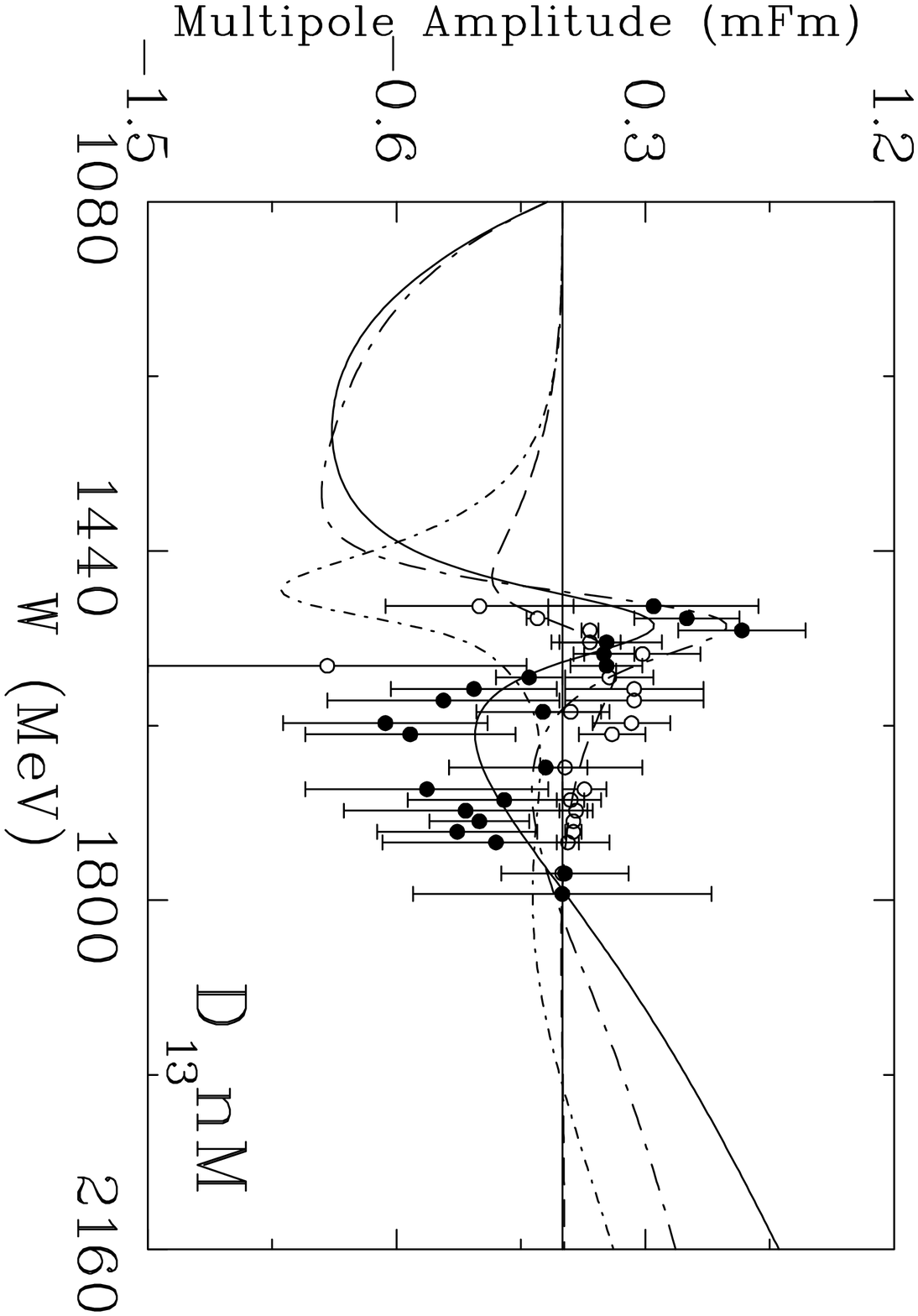}
\end{turn}
\caption{Partial wave amplitudes for the D$_{13}$ excitation on the neutron
from the SAID analysis \cite{Arndt_02}. Caption like fig. 
\ref{fig_33}.
}
\label{fig_36}       
\end{center}
\end{figure}

Until now, we have discussed the second resonance region for the proton. 
Further information about isospin $I=1/2$ $N^{\star}$ resonances is related to 
the isospin structure of their electromagnetic excitation. It involves the two 
independent amplitudes $A^{IS}$ (isoscalar) and $A^{IV}$ (isovector) which are 
related to the proton and neutron amplitudes ($A^p$, $A^n$) via
eq.~\-(\ref{eq:iso_2}). The separation of the isospin components requires
measurements on the neutron which can only be done on neutrons bound in the 
deuteron or other light nuclei. As discussed in sec. \ref{ssec:delta_neutron},
additional systematic problems are introduced which are related to the model 
dependence of the extracted neutron cross section. The data base for the 
reactions off the neutron is thus much more sparse and less reliable than for 
the proton. This is reflected in the results of the multipole analyses.
As an example, the D$_{13}$ multipoles from SAID \cite{Arndt_02} on the neutron 
are shown in fig.~\ref{fig_36}. In particular in the case of the magnetic 
$M_{2-}$ multipole, the situation is unsatisfactory, and a reliable 
determination of the coupling is not possible. The imaginary part at resonance 
position has changed by a factor of four between the 1996 and 2002 analyses. 
Also, the change of the neutron helicity couplings between the two analyses is 
far larger than the quoted uncertainties (see table \ref{tab_04}).

%
%
%
%

\begin{figure}[bht]
\begin{center}
\epsfysize=9.cm \epsffile{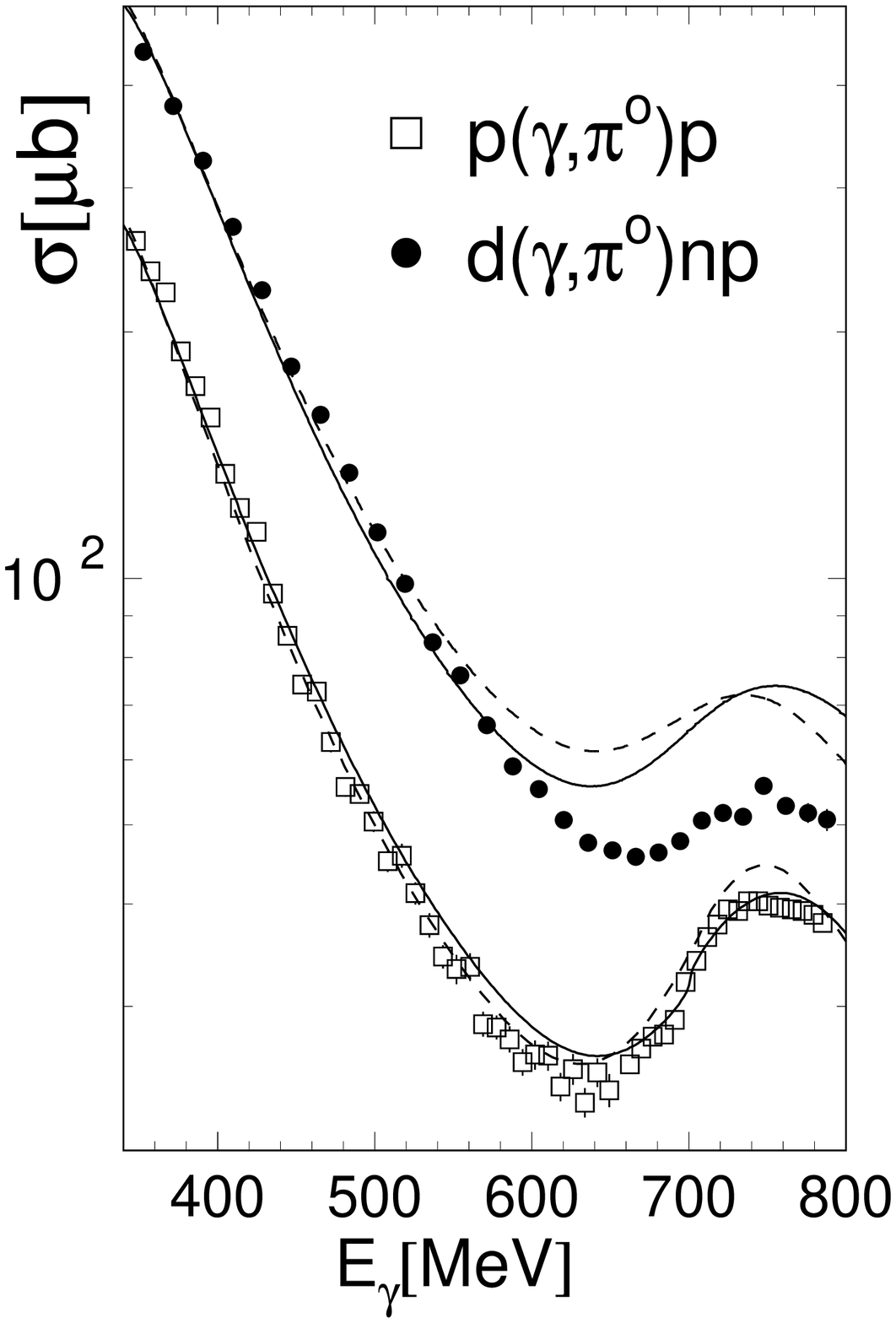}
\epsfysize=9.cm \epsffile{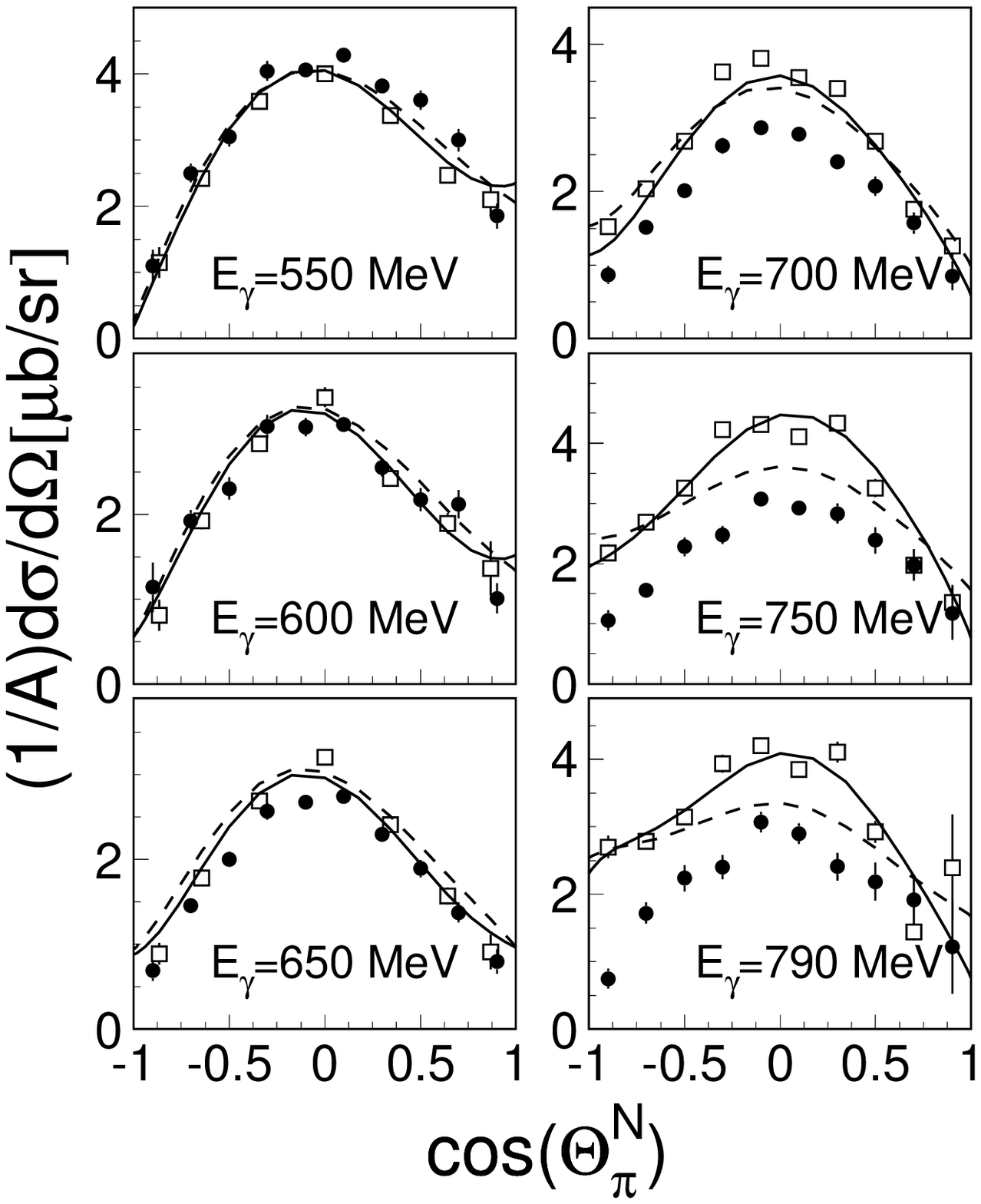}
\hspace*{0.3cm}
\epsfysize=9.cm \epsffile{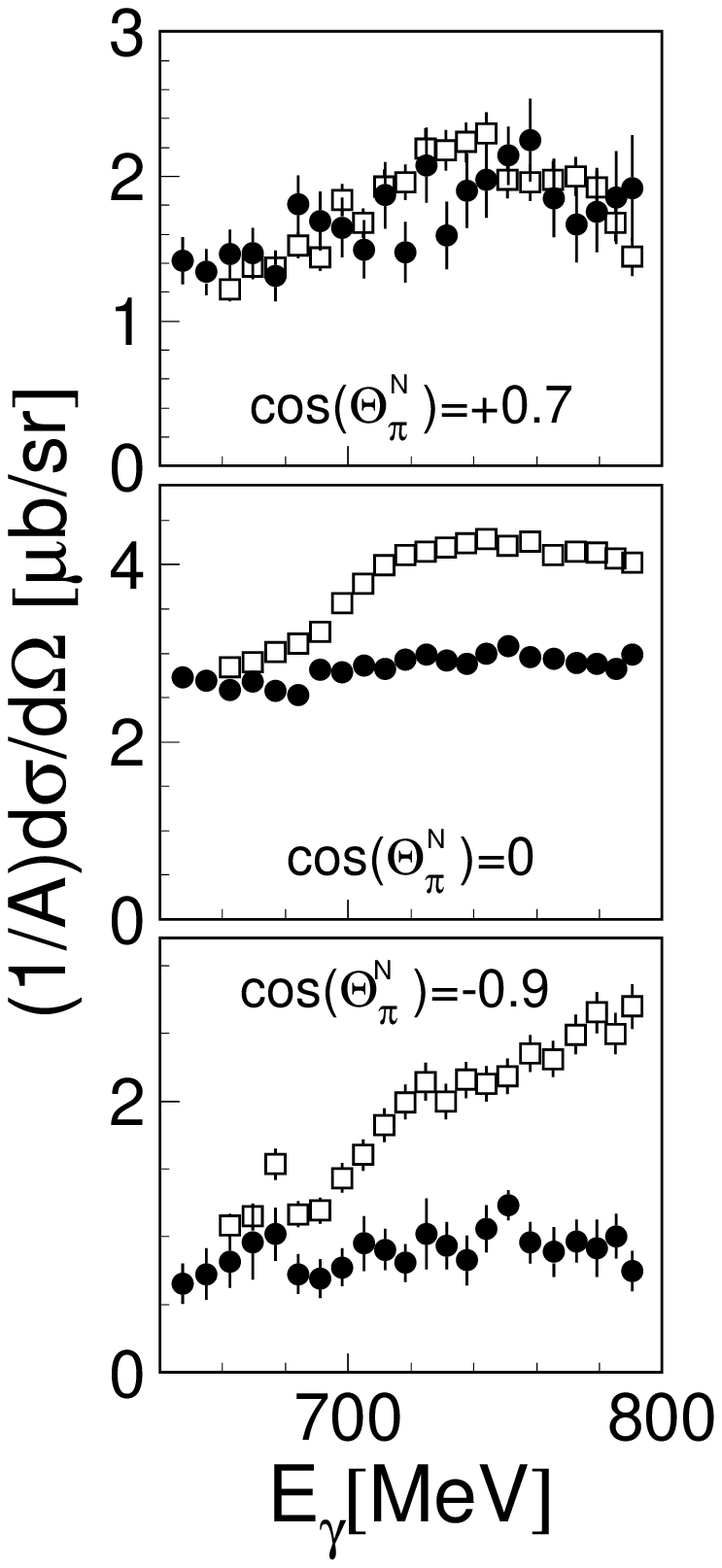}
\caption{Quasifree $\pi^o$ photoproduction off the deuteron (filled circles)
compared to the elementary reaction on the proton (open squares). 
Left hand side: total cross sections. Solid lines: SAID results \cite{Arndt_96}
for the proton cross section and Fermi smeared average of proton and neutron
cross sections. Dashed lines: same for MAID results \cite{Drechsel_99}.
Center: angular distributions normalized to the mass number. Solid lines:
SAID results for the proton, dashed lines: Fermi smeared average over SAID 
proton and neutron cross sections. Right hand side: differential cross section
as function of the incident photon energy.
}
\label{fig_37}  
\end{center}
\end{figure}

Most of the data from deuteron targets still stem from measurements with 
untagged photon beams. Often, only ratios of proton - neutron cross sections 
have been measured with at times insufficient separation of single and double 
pion production channels (see discussion in \cite{Krusche_99}).
The need for better data is obvious. Recently, a new measurement of the 
breakup reaction $\gamma d\rightarrow \pi^o np$ was reported \cite{Krusche_99}. 
The total cross section and typical angular distributions throughout the 
second resonance region are compared in fig.~\ref{fig_37} to the proton data 
and to predictions from the SAID and MAID analyses. 
The deuteron cross section has been modeled in a simple participant - 
spectator approximation where the sum of the proton and neutron cross sections 
predicted by MAID (respectively SAID) was folded with the deuteron Fermi motion 
calculated from the deuteron wave function \cite{Lacombe_81}. 
The proton cross section is well reproduced by both models. 
This is not surprising since the data was included into the fits. 
%
%
%
%
\\
For the deuteron, the PWIA calculations based 
on the MAID and SAID predictions for the neutron cross section agree almost 
perfectly with the data for photon energies up to 550 MeV. The angular 
distributions of the deuteron data are close to two times the proton data 
in this energy range, indicating similar angular distributions for 
$p(\gamma, \pi^o)p$ and $n(\gamma, \pi^o)n$. However,the models significantly 
overestimate the data in the region of the D$_{13}$. In this range, 
the angular distributions for the proton and the deuteron develop 
noticeable differences. This is most clearly seen in the energy dependence of 
the differential cross section for forward, central, and backward pion angles 
(fig. \ref{fig_37}, right side). The cross section is still similar for 
forward angles, but behaves differently for backward angles. The question is 
again whether FSI effects are important as in the $\Delta$ range. The situation 
is different. Coherent contributions do not play any role at the higher 
incident photon energies. The momentum mismatch between participant and 
spectator nucleons is already so large that FSI effects should be much 
reduced. Furthermore, the agreement between the PWIA approximations and the 
data is excellent between 350 and 550 MeV incident photon energy. On the other 
hand, the comparison of the proton and deuteron cross sections at  backward 
angles basically rules out that the effect could be entirely due to the 
$n(\gamma ,\pi^o)n$ reaction. It is not possible to construct a cross section 
for the reaction on the neutron which together with the measured proton cross 
section can reproduce the inclusive deuteron data at backward angles in PWIA.   

The interpretation of the suppression of the structure on the deuteron is 
complicated by the production threshold of the $\eta$ meson at approximately 
705 MeV. It is known \cite{Althoff_79} that the opening of the 
$\eta$-production threshold causes a unitarity cusp at backward angles 
resulting in a pronounced s-shape step in the cross section around the 
threshold at 705 MeV. This cusp structure is superimposed on the rise of the 
cross section towards the D$_{13}$ resonance position. 

In contrast to the $\Delta$ region, model predictions for the nuclear effects 
on the pion production cross sections are not available for this energy region. 
Experimentally, it would be very useful to have measurements where the recoil 
nucleons are detected in coincidence. In that way, one could first investigate 
if the behavior of the $p(\gamma ,\pi^o)p$ reaction is different for free 
protons and protons bound in the deuteron. Overall, the investigation of the 
higher-lying resonances of the neutron is still in an early state. Even results 
for such prominent resonances like the \d ~should be received critically.

\newpage           
\subsection{\it $\eta$-Photoproduction and the S$_{11}$(1535)-Resonance}
\label{ssec:S11}

The lowest-lying S$_{11}$ resonance has two characteristic features which 
distinguish this state from other resonances in this excitation energy range.
They are not easily explained in the framework of nucleon models: the strong 
coupling of the state to the $N\eta$ decay channel and the small slope of its 
electromagnetic transition form factor. The small decay branching ratios of 
the \d ~and \pp ~resonances into $N\eta$ are not surprising since they involve 
higher partial waves close to threshold. However, the decay patterns of the 
first and second S$_{11}$ resonances, both involving $l=0$ transitions, are 
very different. The ratios of the hadronic decay matrix elements 
$\langle N\eta|{\cal{H}}_s|\mbox{S}_{11}\rangle$
and 
$\langle N\pi|{\cal{H}}_s|\mbox{S}_{11}\rangle$ 
follow simply from their partial widths  $\Gamma_{\eta}$, $\Gamma_{\pi}$ via:
\begin{eqnarray}
\frac{\langle N\eta|{\cal{H}}_s|\mbox{S}_{11}\rangle}
{\langle N\pi|{\cal{H}}_s|\mbox{S}_{11}\rangle} & = &
\sqrt{\frac{\Gamma_{\eta}}{\Gamma_{\pi}}
\frac{q_{\pi}^{\star}}{q_{\eta}^{\star}}}\;,
\end{eqnarray}
where $q_{\eta}^{\star}$ and $q_{\pi}^{\star}$ are the meson cm momenta at the
resonance positions. Inserting the nominal resonance masses of 1535 MeV and
1650 MeV, and the partial widths taken from the multi-channel analysis of pion
induced pion and $\eta$ production from Vrana, Dytman, and Lee 
($b_{\eta}$=51\%, $b_{\pi}$=35\% for the \s 
~and 
$b_{\eta}$=6\%, $b_{\pi}$=74\% for the \ss) 
yields:
\begin{equation}
\frac{\langle N\eta|{\cal{H}}_s|\s\rangle}
{\langle N\pi|{\cal{H}}_s|\s\rangle} = 1.9\;\;\;\;\;\;\;\;\;
\frac{\langle N\eta|{\cal{H}}_s|\ss\rangle}
{\langle N\pi|{\cal{H}}_s|\ss \rangle} = 0.35\;.
\end{equation} 
This means more than a factor of five difference for the two resonances.
These decay patterns are very important for the understanding of the underlying 
spin-flavor structure of the two resonances. The only possibility to produce 
such a pattern in the constituent quark model is a fine tuning of the 
configuration mixing of the two S$_{11}$ SU(6)$\otimes$O(3) basis states. 
However, some authors argue (see e.g. \cite{Li_96}), that this mixing is 
probably not sufficient to explain the properties of the two states.

%
%
%
\begin{figure}[thb]
\begin{minipage}{9.1cm}
{\mbox{\epsfysize=6.cm \epsffile{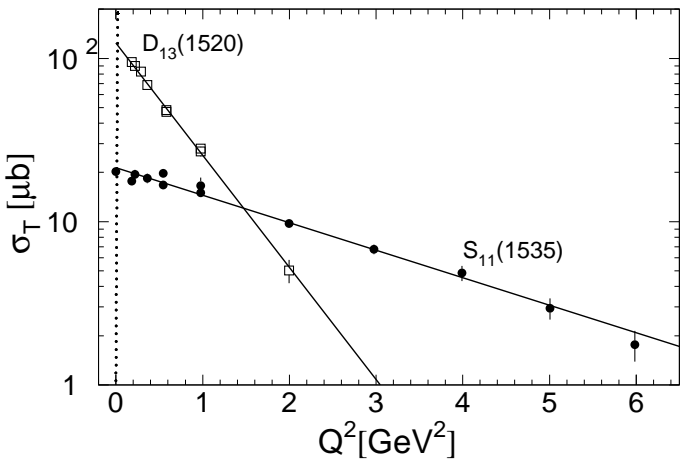}}}
\caption{$Q^2$ dependence of the total transverse cross section for the
excitation of the D$_{13}$ and S$_{11}$ resonances \cite{Brasse_84}.
}
\label{fig_38} 
\end{minipage}
\end{figure}
%

\vspace*{-8.5cm}
\hspace*{9.5cm}
\begin{minipage}{8.2cm}
The other characteristic feature of the \s  ~is the $Q^2$ dependence of its 
electromagnetic transition form factor. It was found in electron scattering 
experiments at Bonn and at DESY in the 1970's
\cite{Kummer_73}-\cite{Brasse_84}
that the decrease of the excitation strengths with four momentum transfer is 
much steeper for the D$_{13}$ resonance than for the S$_{11}$. The effect is so 
large, that although the total cross section for real photons is larger for the 
D$_{13}$ by roughly a factor of 6, the situation is almost reversed at 
$Q^2$=3 GeV$^2$. This behavior is shown in fig.~\ref{fig_38} where the total 
transverse excitation cross sections are plotted versus the momentum transfer. 

\vspace*{1.cm}
\end{minipage}
\noindent{The} separation of the contributions from the two resonances was achieved with 
simplifying assumptions. 
The $\eta$ yield was attributed entirely to the 
S$_{11}$. The cross section difference between inclusive electron scattering 
and $\eta$ production at the $W$ corresponding to the D$_{13}$ excitation was 
attributed completely to the D$_{13}$. Finally, it was assumed that at momentum 
transfers above 4 GeV$^2$ the inclusive cross section is dominated by the 
S$_{11}$. It is difficult to explain this large effect in quark models since 
both resonances belong to the same SU(6) multiplet. It is interesting to note 
that this behavior could be in qualitative agreement with the helicity 
conservation predicted by QCD for large momentum transfers. At the photon 
point, the D$_{13}$ is excited dominantly by the helicity 3/2 amplitude while 
the S$_{11}$ can be only excited by the helicity 1/2 amplitude. Thus, more 
exclusive measurements allowing to extract the precise $Q^2$ dependence of the 
helicity couplings of the two resonances are highly interesting. Such results 
are now becoming available from JLab (see below). 

The unusual decay pattern of the \s ~has been used by several groups as an
argument for very particular structures of this state. One example is the chiral
constituent quark model of Glozman and Riska \cite{Glozman_96,Glozman_96a}. 
In this model,  an interaction of the quarks via exchange of the pseudo-scalar 
octet mesons is introduced in addition to the harmonic confining potential. 
The interaction gives rise to a particular fine structure interaction coming 
from Goldstone boson exchange
%
%
%
\begin{figure}[bht]
\begin{minipage}{0.0cm}
{\mbox{\epsfysize=7.cm \epsffile{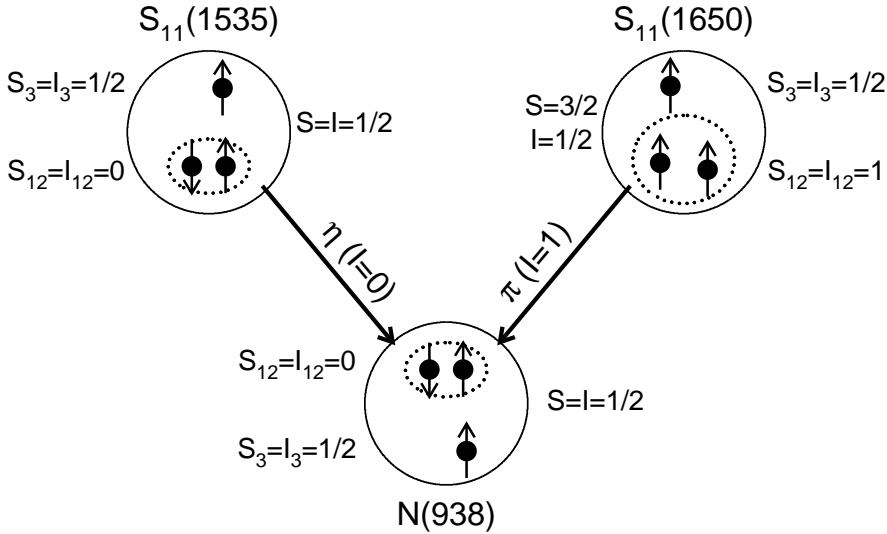}}}
\end{minipage}
\hspace*{13.2cm}
\begin{minipage}{5.cm}
\caption{Quark - diquark clusterization and selection rules in the chiral 
quark model of Glozman and Riska. \cite{Glozman_96,Glozman_96a}. 
$S_{12}$ and $I_{12}$ denote spin and isospin of the diquark, 
$S_{3}$, $I_{3}$ spin and isospin of the third quark and
$S$, $I$ the total spin and isospin of the states.
}
\label{fig_39}       
\end{minipage}
\end{figure}
%
rather than from one-gluon exchange as in most other quark models. As a 
consequence of this interaction, the nucleon states develop a quark-diquark 
structure. In particular, a rather compact diquark with spin/isospin zero 
($S_{12}=I_{12}=0$) appears in the nucleon ground state and the \s , while a 
less closely bound $S_{12}=I_{12}=1$ diquark dominates the wave function of 
the \ss . This leads to the decay selection rules depicted in 
fig.~\ref{fig_39}: the decay of the \ss ~resonance via emission of the 
isoscalar $\eta$ is forbidden since the transition involves an isospin flip of 
the diquark but the decay of the \s ~into $N\eta$ is not hindered. The 
remaining decay strengths of the \ss ~into $N\eta$ is attributed to a small 
admixture of other components to the wave function.

The model discussed above still gives a conventional description of the 
S$_{11}$ in the sense that it is treated as a three-quark configuration. Other 
models, in particular the chiral coupled channel calculations of $\eta$ and 
kaon photoproduction by Kaiser and collaborators \cite{Kaiser_95,Kaiser_97}, 
question this. The model starts from the chiral effective meson - baryon 
Lagrangian so that the only explicit degrees-of-freedom are the baryon and the 
octet mesons. Nucleon resonances are then dynamically generated. As a result 
a strong attraction is found in some of the channels, in particular in the 
$\bar{K}N$ isospin $I=0$ channel and in the $K\Sigma$ isospin $I=1/2$ channel, 
which gives rise to quasi-bound meson-nucleon states. These two quasi-bound 
states show many of the characteristic properties of the $\Lambda$(1405) and 
the \s ~baryon states. The $K\Sigma$ state has a large decay branching ratio 
into $N\eta$, and the cross section for $\eta$ photoproduction on the proton 
is well reproduced by this model (see below). This means that the S$_{11}$ is 
treated as a dynamically generated quasi-bound $(q\bar{q})(qqq)$ state. If this 
interpretation were correct, the immediate question would be: where is the 
$q^3$ S$_{11}$ state predicted by the constituent quark model? A further 
question arises from the $Q^2$ dependence of the form factor. As we will 
discuss below, models of the S$_{11}$ as a $q^3$ configuration tend to predict 
steeper slopes than what is observed. The natural expectation is that a 
molecular-like quasi-bound $K\Sigma$ state should have an even stronger $Q^2$ 
dependence of the transition form factor. Li and Workman \cite{Li_96} have 
argued that the $K\Sigma$ state, if existent, should be strongly mixed with a 
three-quark configuration to account for the behavior of the form factor at 
large $Q^2$. In this case, a third S$_{11}$ resonance with a mass close to the 
two known states should exist. There was circumstantial evidence for a third 
S$_{11}$ resonance close to 1700 MeV in the 1995 VPI analysis of pion elastic 
scattering \cite{Arndt_95}. However, such a state is not seen in the 2000 
PIT-ANL analysis \cite{Vrana_00}. Recently, Saghai and Li \cite{Saghai_01} 
have claimed evidence for a third S$_{11}$ state at 1729 MeV in 
$\eta$-photoproduction. We will see below that this result needs further 
confirmation.

\subsubsection{\it $\eta$-Photoproduction from the Proton}

The investigations of the second resonance region, in particular concerning 
the \s , with $\eta$-photoproduction have intensified since the mid 1990's.
Precise tagged beam experiments accompanied by new theory developments for 
the reaction models have been necessary. Even today, a full multipole analysis 
of the reaction is out of reach. One of the first comprehensive analyses of 
$\eta$ photoproduction was performed by Hicks and collaborators \cite{Hicks_73} 
using an isobar analysis. The analysis parameterized the nucleon resonances in 
Breit-Wigner forms, and included a smooth, phenomenological background 
parameterization. This analysis pointed at the importance of the lowest lying 
S$_{11}$. However, even a refitting of the model 15 years later by Tabakin, 
Dytman and Rosenthal \cite{Tabakin_88} did not reveal much more detail. The 
data base consisted of less than 160 scattered points for differential cross 
sections, mostly from bremsstrahlung experiments which were internally 
inconsistent.
 
\begin{figure}[hbt]
\begin{minipage}{0.0cm}
\epsfysize=10.5cm \epsffile{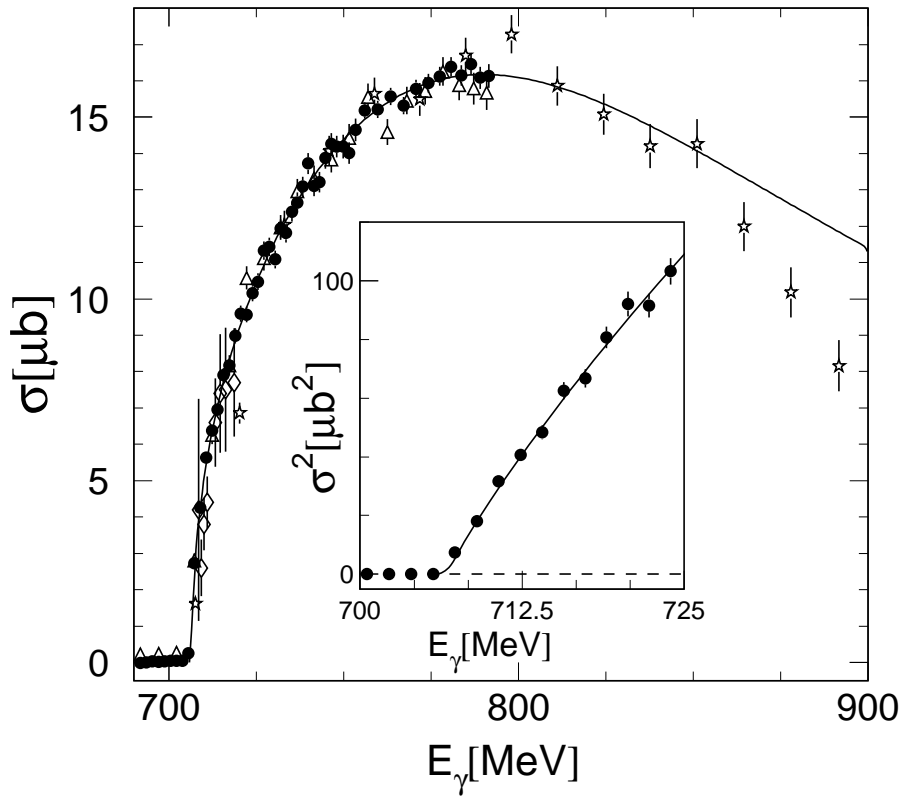}
\end{minipage}
\hspace*{12.5cm}
\begin{minipage}{5.5cm}
\caption{Total cross section for the reaction $p(\gamma ,\eta)p$ in the
threshold region. Open stars: electroproduction close to the photon point
\cite{Schoch_95}, filled circles, (open triangles): photoproduction Mainz, 
two-photon decay channel of the $\eta$ (3$\pi^o$ decay channel)
\cite{Krusche_95}, open diamonds: photoproduction Bonn \cite{Price_95}.
Insert: linear energy dependence of the squared cross section close to 
threshold. The curves are Breit-Wigner fits to the data (see text).
}
\label{fig_40}       
\end{minipage}
\end{figure}

\begin{figure}[hbt]
\centerline{\epsfysize=10.7cm \epsffile{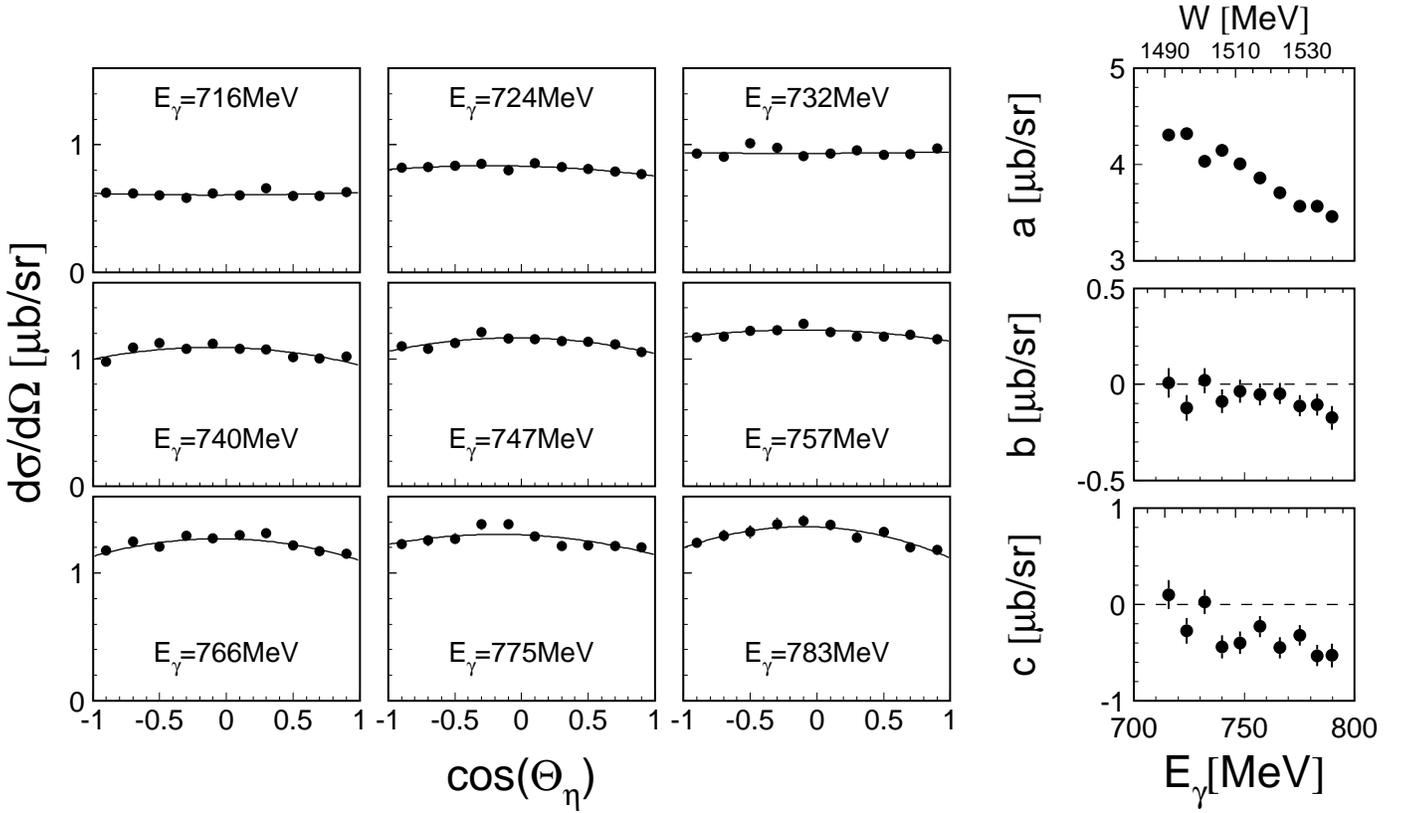}}
\caption{Angular distributions for $p(\gamma ,\eta)p$ \cite{Krusche_95}.
The curves are fits with eq. (\ref{eq:eta_diff}). The pictures on the right
hand side show the energy dependence of the fit coefficients a,b,c.
}
\label{fig_41}       
\end{figure}

First precise measurements of the threshold behavior of the $\eta$ 
photoproduction from tagged beam experiments were reported from Bonn 
\cite{Price_95} and Mainz \cite{Krusche_95} in 1995.  A measurement of 
electroproduction in Bonn \cite{Schoch_95} close to the photon point 
($Q^2$=0.056 GeV$^2$) can be considered as well. The two experiments from Bonn 
reported total cross sections but not angular distributions. The results from 
the three experiments are summarized in figs.~\ref{fig_40},\ref{fig_41}. 
The total cross sections are in good agreement in the overlap region. The Mainz 
experiment used the $\eta\rightarrow 2\gamma$ and $\eta\rightarrow 3\pi^o$ 
decays for the detection of the $\eta$ meson simultaneously. The two channels 
have different instrumental detection efficiencies \cite{Krusche_95}, and the 
agreement indicates small systematic uncertainties. 

%
%
%
%

The typical s-wave behavior of the reaction is apparent from the experimental 
data. The expected energy dependence at threshold is given by:
\begin{equation} \sigma(E_{\gamma})\propto
(E_{\gamma}-E_{thr})^{(l+1/2)}\;\;   l=0,1,2,...\;\; (s,p,d...)\;,
\end{equation}  
where $E_{thr}\approx$707 MeV is the threshold energy. Indeed, the square of 
the cross  section is a linear function of the incident photon energy close to 
threshold, as shown in the insert of fig.~\ref{fig_40}. At the same time, 
the angular distributions are almost isotropic (see fig.~\ref{fig_41}). 
The fit of the total cross section, shown in the figure, was achieved with a 
single Breit-Wigner curve for the S$_{11}$ resonance, neglecting other 
contributions: 
\begin{equation}  \sigma(E_{\gamma})\;\;=\;\;  4\pi
\frac{q_{\eta}^{\star}}{k^{\star}} |E_{0+}|^2 \;\;=\;\; 
\frac{q_{\eta}^{\star}}{k^{\star}} 
\frac{CM_{R}^2\Gamma_{R}^2}{(M_R^2-W^2)^2+M_R^2\Gamma_R^2 x^2} 
\end{equation} 
where the energy dependence of the total width, important due to the proximity 
of the $\eta$ threshold, is parameterized by: 
\begin{equation}
\label{eq:widths}
x=b_{\eta}\frac{q_{\eta}^{\star}}{q_{\eta R}^{\star}} +
b_{\pi}\frac{q_{\pi}^{\star}}{q_{\pi R}^{\star}} +b_{\pi\pi} 
\end{equation}
where $k^{\star}$, $q_{\eta}^{\star}$, $q_{\pi}^{\star}$ are the photon,
$\eta$ and pion cm momenta, $q_{\eta R}^{\star}$, $q_{\pi R}^{\star}$ are the
momenta at resonance position, $M_R$, $\Gamma_R$ are position and width of
the resonance, $b_{\eta}$, $b_{\pi}$, $b_{\pi\pi}$ are the branching ratios
for the indicated decay channels and $W=\sqrt{s(E_{\gamma}})$. Under the 
assumption of S$_{11}$ dominance the electromagnetic helicity coupling follows 
from:
\begin{equation} 
|A_{1/2}^p|=\left[\frac{M_R}{2m_p}
\frac{\Gamma_R}{b_{\eta}}\sigma(M_{R})\right]^{1/2}\;\;
\end{equation}
As discussed below it is assumed that $b_{\eta} = b_{\pi}=0.45$. 
The fit curve corresponds to the resonance parameters 
$M_R =1544\pm 2$ MeV, $\Gamma_R =203\pm 9$ MeV, and 
$A_{1/2}^p =(124\pm3) 10^{-3}$ GeV$^{-1/2}$ \cite{Krusche_97}. 

Contributions from the other resonances in the second resonance region can 
affect the angular distributions, fitted with the ansatz:
\begin{equation}
\label{eq:eta_diff}
\frac{d\sigma}{d\Omega}
=
\frac{q^{\star}}{k^{\star}}
[a+b\;\mbox{cos}(\Theta^{\star})+c\; \mbox{cos}^2(\Theta^{\star})]
\end{equation}
where $\Theta^{\star}$ is the cm polar angle of the $\eta$-mesons.
The $a$-, $b$-, $c$-coefficients can be related to a low energy multipole
expansion of the differential cross sections under the following assumptions: 
\begin{itemize}
\item{The dominance of the \s ~allows to keep only terms proportional to the 
$E_{0+}$ multipole.}
\item{At low incident photon energies in the second resonance region
only $l\leq 2$ multipoles must be accounted for.}
\end{itemize}
In this case the expansion is given by:
\begin{eqnarray}
\label{eq:eta_ang_multi}
a & = & E_{0+}^{2} - Re(E_{0+}^{\star}(E_{2-}-3M_{2-}))\nonumber\\
b & = & 2Re(E_{0+}^{\star}(3E_{1+}+M_{1+}-M_{1-}))\nonumber\\
c & = & 3Re(E_{0+}^{\star}(E_{2-}-3M_{2-}))\;\;\;.
\end{eqnarray}
The $a$ coefficient comes mainly from the S$_{11}$ contribution, the $b$ 
coefficient from a possible interference of the S$_{11}$ with the \pp 
~($M_{1-}$ multipole) and with Born terms and vector meson exchange ($E_{1+}$, 
$M_{1+}$), and the $c$ coefficient from the interference of the S$_{11}$ with 
the D$_{13}$ ($E_{2-}$, $M_{2-}$ multipoles). The results of the fits shown 
in fig.~\ref{fig_41} reflect the dominance of the constant term. The $b$ 
coefficient is small and consistent with zero. Thus, evidence for a 
contribution of the \pp ~resonance was not found. The $c$ coefficient is 
clearly negative which was taken as first evidence for a contribution of the 
D$_{13}$ to $\eta$ photoproduction \cite{Krusche_95} because background 
contributions in this multipole are expected to be negligible. Practically 
identical results were found later in the GRAAL experiment \cite{Renard_02}.

%
%
%
%

\begin{figure}[thb]
\centerline{\epsfysize=8.2cm \epsffile{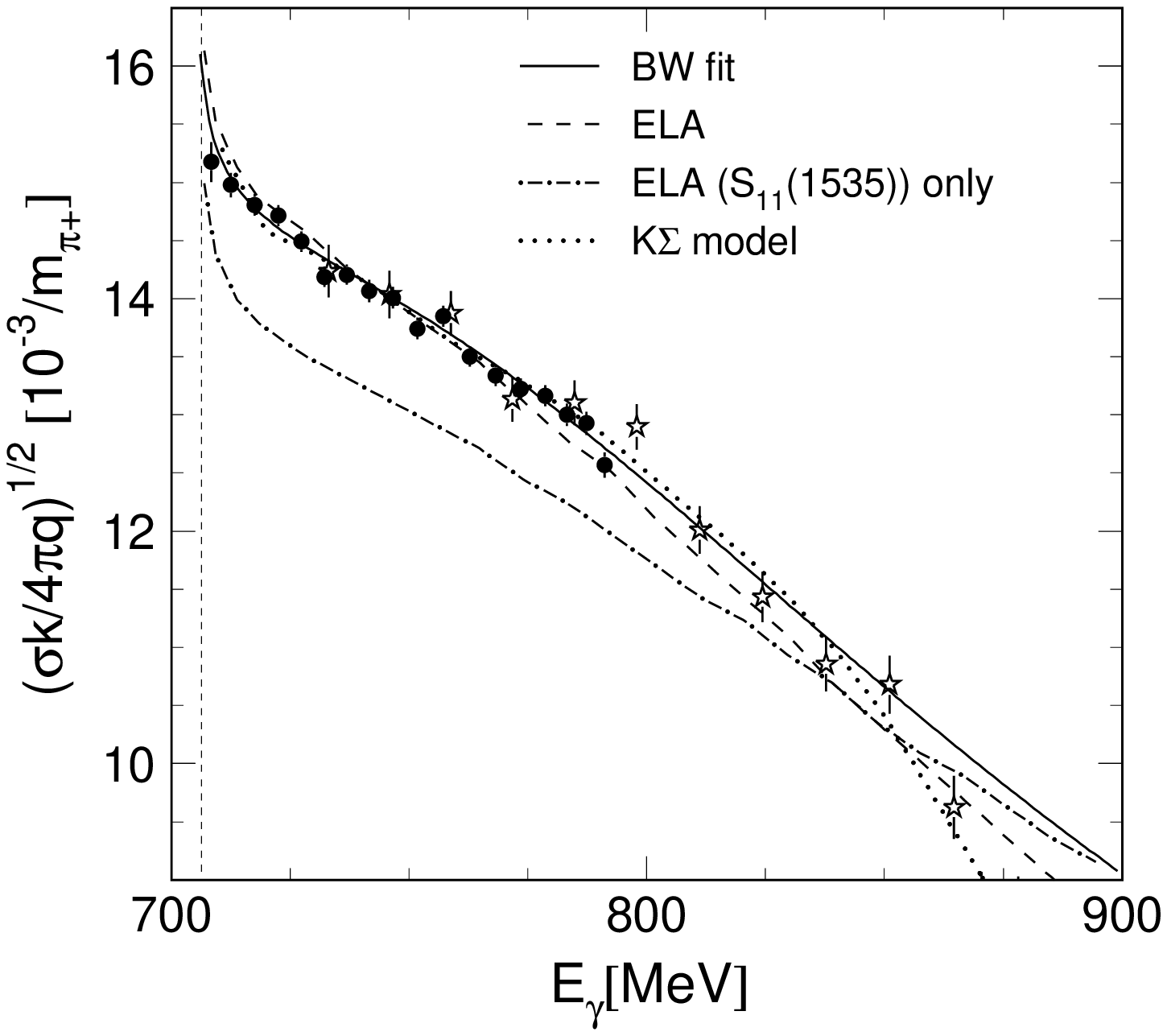}
\epsfysize=8.2cm \epsffile{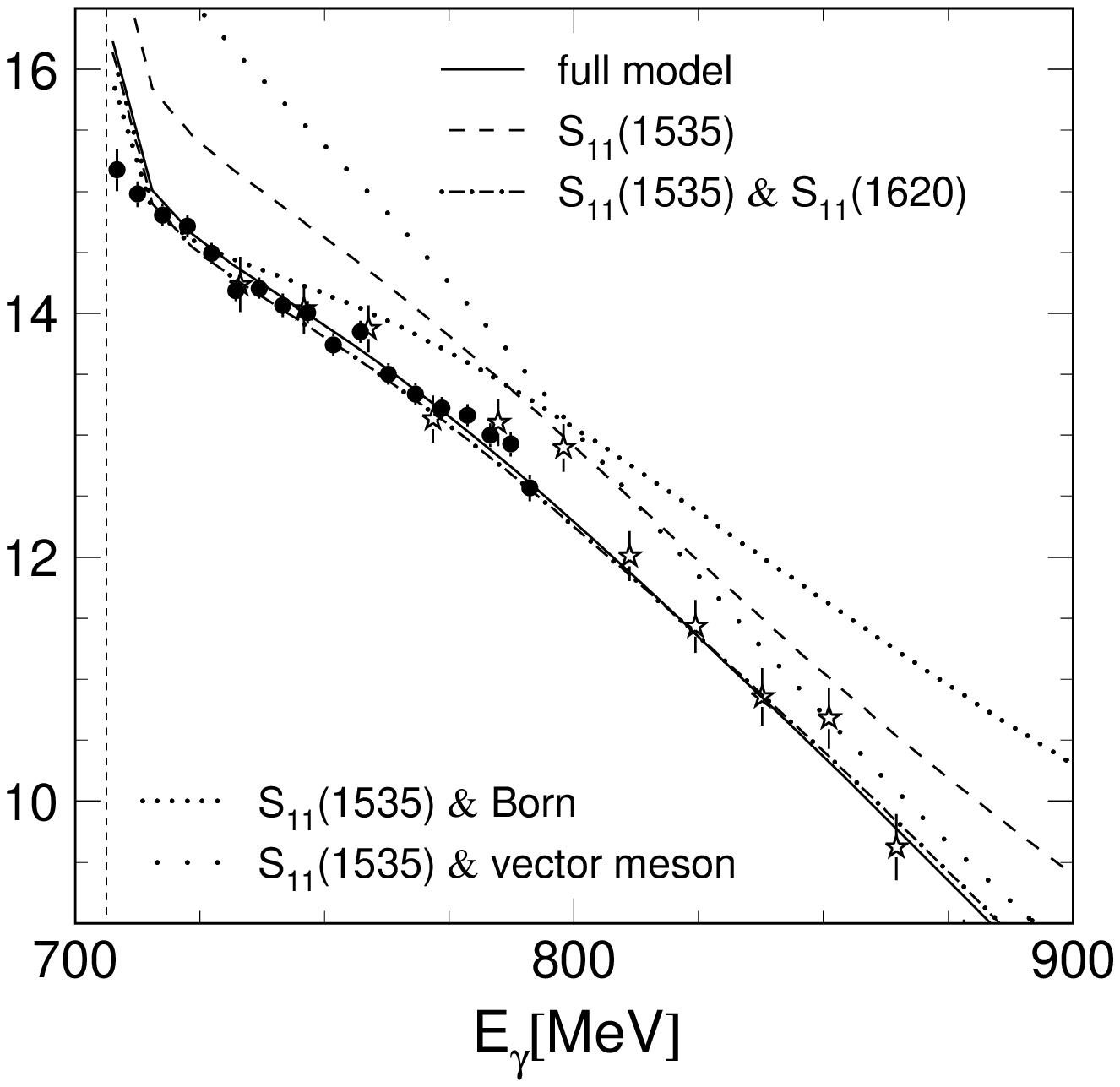}}
\caption{Comparison of the phase space reduced cross section (see text) to 
model calculations. Left hand side: Breit-Wigner fit \cite{Krusche_95}, 
prediction from $K\Sigma$ quasi-bound state \cite{Kaiser_97} and effective 
Lagrangian model (ELA) \cite{Benmerrouche_95,Krusche_97}. For the effective 
Lagrangian model the result of the full model and the result of switching off 
all non-S$_{11}$ contributions is shown. Right hand side: results from ETA-MAID 
\cite{Chiang_02}. Shown is the result of the full calculation (solid line), 
the contribution of the \s ~alone, the contribution of both S$_{11}$ resonances,
and the combination of \s ~with different background contributions.
Data are from \cite{Krusche_95} (filled circles) and \cite{Schoch_95}
(open stars).
}
\label{fig_42}       
\end{figure}

The above analysis seems to account for the main features of the experimental 
results. Obviously, a more detailed analysis is wanted, taking into account 
possible background contributions. The analyses were pushed ahead by the RPI 
group (see e.g. 
\cite{Benmerrouche_91}-\cite{Benmerrouche_96}) 
and the Mainz group (see e.g. \cite{Knochlein_95,Tiator_99,Chiang_02}). 
In both cases, the background contributions from Born terms and vector meson 
exchange are parameterized in an effective Lagrangian formalism. In 
fig.~\ref{fig_42}, results from model calculations are compared to 
experiment. Here, the quantity
$[(\sigma k^{\star})/(4\pi q_{\eta}^{\star})]^{1/2}$ 
is plotted instead of the total cross section. In case of a background-free 
S$_{11}$ excitation it equals $|E_{0+}|$. The first observation is that the 
energy dependence is not intuitive for an amplitude which is resonant at 
$\sqrt(s)\approx 1544$ MeV corresponding to $E_{\gamma}\approx 800$ MeV. 
The amplitude does not peak at this energy but instead rises towards the 
threshold. This behavior stems from the strong energy dependence of the 
resonance width (see eq.~\-(\ref{eq:widths})) due to the phase space opening of 
the $\eta$-channel above threshold. It is apparent that all models can 
reproduce the data at the same level. This holds for the simple Breit-Wigner 
fit \cite{Krusche_95}, the effective Lagrangian model
\cite{Benmerrouche_95,Krusche_97}, the isobar model \cite{Chiang_02}, 
and also for the coupled channel model with the dynamically generated $K\Sigma$ 
quasi-bound state \cite{Kaiser_97}. A comparison of truncated versions of the 
models allows to estimate the background contributions.
This was done in the same way for the ELA \cite{Krusche_97} and the ETA-MAID: 
the full model including all terms was fitted to the data.
Then, different contributions were switched off without refitting any 
parameters. In case of the ELA, the contribution from the \s ~all by itself 
results in a smaller $E_{0+}$ amplitude, roughly 4\% at the resonance position. 
In case of the ETA-MAID, contributions from Born terms and vector meson 
exchange seem to cancel. Only the second S$_{11}$ resonance makes an effect of 
roughly 8\% at the resonance position. 

\newpage
%
%
%
%
Due to the small influence of non-\s ~contributions, one expects that the 
extraction of the electromagnetic coupling of the resonance from $\eta$ 
photoproduction is less prone to systematic uncertainties than the 
extraction from pion data where the \s ~makes only a small contribution.
A comparison of the values of $A_{1/2}^p$ from different analyses of $\eta$
and pion photoproduction is given in table \ref{tab_05}. The extracted 
parameters of the resonances are not independent of each other, and the 
helicity coupling depends on the hadronic widths which are not very well known.
Therefore, we have included renormalized values of the helicity couplings 
assuming a total width of 150 MeV, branching ratios of 
$b_{N\eta}$ =0.5, $b_{N\pi}$ =0.4 and the proportionalities:
\begin{equation} 
A_{1/2}^p\propto \sqrt{\Gamma /b_{Nm}}\;\;\;\;\;  m=\eta ,\pi.
\end{equation}
As seen from the table, the total width of the resonance is not well 
constrained by the analyses of the photoproduction data. Values are ranging 
from 80 MeV to more than 200 MeV where the pion data seem to favor the smaller 
values. After renormalization of the width effects, the scatter of the helicity 
couplings is much reduced.  Almost all the results extracted from $\eta$ 
production fall into the range (90 - 107)$10^{-3}\mbox{GeV}^{-1/2}$ while the 
results from the pion data lie in the range (60 - 80)$10^{-3}\mbox{GeV}^{-1/2}$.
The analysis of $\eta$ photoproduction by Homma et al. \cite{Homma_88} was 
based on a much inferior data base than the other analyses and is not used 
further. The remaining exceptions are the chiral quark model analysis of 
$\eta$ photoproduction by Saghai and Li \cite{Saghai_01} and the most recent 
SAID analysis of pion photoproduction \cite{Arndt_02}. When these two analyses 
are not included, the average values of $A_{1/2}^p$ extracted from the two 
reactions are:
\begin{eqnarray}
\label{eq:a12eta}
(A_{1/2}^p)_{N\eta} = & 100\times 10^{-3}\mbox{GeV}^{-1/2}
& = 182.6\times 10^{-3}\mbox{GeV}^{-1}
\sqrt{\Gamma/b_{N\eta}} \\
\label{eq:a12pi}
(A_{1/2}^p)_{N\pi} = & 78\times 10^{-3}\mbox{GeV}^{-1/2}
& = 127.4 \times 10^{-3}\mbox{GeV}^{-1}
\sqrt{\Gamma/b_{N\pi}}\;\;
\end{eqnarray}
which would be equivalent for $b_{N\eta}/b_{N\pi}\approx 0.5$.

%
%
%
\begin{table}[htb]
  \caption[S$_{11}$ helicity couplings]{
    \label{tab_05}
Photon couplings of the \s ~resonance.  $(A^p_{1/2})_{nor}$: normalized to
$\Gamma =150$ MeV, $b_{N\eta}$ =0.5, $b_{N\pi}$ =0.4, respectively.
Method: $N\eta$: analysis of $\gamma p\rightarrow p\eta$,
$N\pi$: analysis of $\gamma p\rightarrow N\pi$, 
QM: quark model predictions. Sauermann et al. (Sau96) \cite{Sauermann_95}
did a simultaneous analysis of $\eta$ and $\pi$ data.
$b_{N\eta,N\pi}$: $b_{N\eta}$ if Method $N_{\eta}$, $b_{N\pi}$ if 
Method $N_{\pi}$. 
$^{1)}$ error includes uncertainty of partial widths.
$^{2)}$ assumed to be 0.4 for the calculation of $(A^p_{1/2})_{nor}$ and 
        $\xi^p_{\pi}$.
$^{3)}$ not including Hom88 and Sag01.
$^{3)}$ not including Arn02.
See eqs.~\-(\ref{eq:xi},\ref{eq:xi2}) for the definition of the electrostrong 
coupling $\xi$.
}
  \begin{center}
    \begin{tabular}{|c|c|c|c|c|c|c|}
      \hline 
      Ref. & Method & $\Gamma$ & $b_{N\eta,N\pi}$ & $A_{1/2}^p$ & 
      $(A_{1/2}^p)_{nor}$ & $\xi^p_{\eta}$, $\xi,^p_{\pi}$ \\
      & & MeV & & $10^{-3}\mbox{GeV}^{-1/2}$ & $10^{-3}\mbox{GeV}^{-1/2}$ &
      $10^{-4}\mbox{MeV}^{-1}$ \\
      \hline
      Hom88 \cite{Homma_88} & $N\eta$ 
      & 240 & 0.27 & 133$^{+55}_{-39}$ & 77$^{+32}_{-23}$ & 1.7$^{+0.7}_{-0.5}$\\ 
       Kru95 \cite{Krusche_95,Krusche_95a} & $N\eta$      
       & 203 & 0.45 & 125$\pm$25$^{1)}$ & 102$\pm$3 & 2.22$\pm$0.07 \\
      Kno95 \cite{Knochlein_95} & $N\eta$ 
        & 166 & 0.5 & 107 & 102 & 2.25 \\
      Li95 \cite{Li_95a} & $N\eta$ 
      & 198 & 0.5 & 111 & 97 & 2.20 \\ 
      Ben96 \cite{Benmerrouche_95} & $N\eta$
       & 150 & 0.5 & 89$\pm$7 & 89$\pm$7 & 2.04$\pm$0.16 \\
      Sau95 \cite{Sauermann_95} & $N\eta$, ($N\pi$)
      & 162 & 0.55 & 102 & 103 & 2.35 \\ 
      Kru97 \cite{Krusche_97} & $N\eta$
       & 212 & 0.45  & 120$\pm$20$^{1)}$ & 96$\pm$9 & 2.11$\pm$0.20 \\
      Sag01 \cite{Saghai_01} & $N\eta$
      & 162 & 0.55 & 64 & 65 & 1.44 \\
      Chi02 \cite{Chiang_02} & $N\eta$
       & 191 & 0.5 & 118 & 105 & 2.34 \\
      Ren02 \cite{Renard_02} & $N\eta$
       & 150 & 0.55 & 102 & 107 & 2.38 \\
      \hline  
      Average $^{3)}$ & & 180 & & & 100$\pm$3 & 2.24$\pm$0.04 \\ 
      \hline
      Met74 \cite{Metcalf_74} & $N\pi$
      & 100 & 0.34 & 63$\pm$13 & 71$\pm$15 & 0.9$\pm$0.2 \\
      Ara82 \cite{Arai_82} & $N\pi$
      & 173 & 0.38 & 80-83 & 73 - 75 & 0.94 - 0.97 \\
      Cra83 \cite{Crawford_83} & $N\pi$
      & 136 & ? $^{2)}$ & 65$\pm$16 & 68$\pm$17 & 0.9$\pm$0.2 \\
      Arn90 \cite{Arndt_90} & $N\pi$
      & 124 & 0.38 & 78 & 84 & 1.1 \\
      Arn93 \cite{Arndt_93} & $N\pi$
      & 84 & 0.42 & 61$\pm$3 & 84$\pm$4 & 1.10$\pm$0.05 \\
      Arn96 \cite{Arndt_96} & $N\pi$
      & 103 & 0.31 & 60$\pm$15 & 64$\pm$16 & 0.83$\pm$0.21 \\ 
      Arn02 \cite{Arndt_02} & $N\pi$
      & 106 & 0.4 & 30$\pm$3 & 36$\pm$4 & 0.47$\pm$0.05 \\
      Dre99 \cite{Drechsel_99} & $N\pi$
      & 80 & 0.4 & 67 & 92 & 1.2 \\ 
      Che02 \cite{Chen_02} & $N\pi$
      & 95 & 0.4 & 72$\pm$2 & 90$\pm$3 & 1.17$\pm$0.04 \\ 
      \hline  
      Average $^{4)}$ & & 112 & & & 78$\pm$4 & 1.02$\pm$0.05 \\ 
      \hline                      
      Fey71 \cite{Feynman_71} & QM  
      & & & 157 & & \\
      Met74 \cite{Metcalf_74} & QM
      & & & 166 & & \\
      Kon80 \cite{Koniuk_80a} & QM
      & & & 147 & & \\
      Clo90 \cite{Close_90} & QM
      & & & 150 - 160 & & \\
      Cap92 \cite{Capstick_92} & QM
      & & & 76 & & \\
      Bij94 \cite{Bijker_94} & QM
      & & & 126 & & \\                     
      \hline           
    \end{tabular}
  \end{center}
\end{table}
%
%
%
%
\begin{table}[thb]
  \caption[S$_{11}$ hadronic widths]{
    \label{tab_06}
Hadronic widths of the \s ~resonance from analyses of $\pi p\rightarrow \pi p$
and $\pi p\rightarrow \eta p$
}
  \begin{center}
    \begin{tabular}{|c|c|c|c|c|c|}
      \hline 
      Ref. & $\Gamma$ & 
      $b_{\pi}$ & $b_{\eta}$ & $b_{\pi\pi}$ & $b_{\pi}/b_{\eta}$\\ 
      & MeV & & & &\\
      \hline
      KSU92 \cite{Manley_92} & 
      151$\pm$21 & 0.51$\pm$0.05 & 0.43$\pm$0.06 & 0.06$\pm$0.03 & 1.19 \\
      Sau95 \cite{Sauermann_95} & 
      162 & 0.41 & 0.55 & 0.04 & 0.75 \\ 
      Bat95 \cite{Batinic_95} & 
      155$\pm$16 & 0.34$\pm$9 & 0.63$\pm$7 & 0.03$\pm$0.03 & 0.54\\
      Gre97 \cite{Green_97} & 
      167.9$\pm$9.4 & 0.394$\pm$0.009 & 0.568$\pm$0.011 & & 0.69\\  
      Pit-ANL00 \cite{Vrana_00} & 
      112$\pm19$ & 0.35 & 0.51 & & 0.69\\
      \hline           
    \end{tabular}
  \end{center}
\end{table}
%
\clearpage
 
The results for the branching ratios from recent analyses of pion induced
reactions are summarized in table \ref{tab_06}. The more recent results tend 
to smaller values of the ratio, but only one analysis comes close to 0.5 which 
indicates some discrepancy between the results for $A_{1/2}^p$ from pion and 
$\eta$ photoproduction. It is particularly disturbing that the most recent VPI 
multipole analysis of pion photoproduction \cite{Arndt_02} 
(see tab.~\ref{tab_05}) gives such a small value for $A_{1/2}$ (30$\pm$3). 
On the other hand, the authors discuss that the results for the \s ~are 
very unstable and are only given for completeness. This analysis uses the best 
of the available data bases of pion photoproduction making it unlikely that 
earlier analyses of pion photoproduction could be more reliable. The problem 
is that the S$_{11}$ contributes little to pion photoproduction and that the 
close-by threshold cusp obscures the signal.  

\begin{figure}[hbt]
\begin{minipage}{0.0cm}
{\mbox{\epsfysize=6.5cm \epsffile{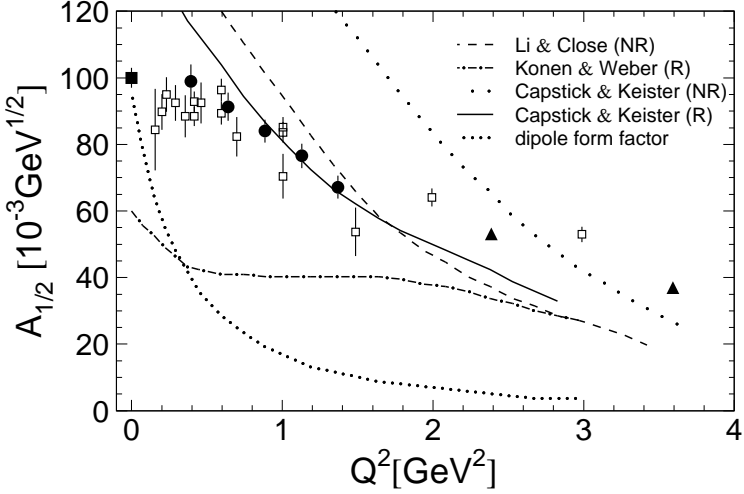}}}
\end{minipage}
\hspace*{11.cm}
\begin{minipage}{6.5cm}
\caption{$Q^2$ dependence of the helicity coupling of the \s ~resonance.
Filled circles: CLAS \cite{Thompson_01}, filled triangles: HMS
\cite{Armstrong_99}, filled square: photon point (see table
\ref{tab_05}), open squares: pre-1985 data  
\cite{Kummer_73}-\cite{Brasse_84}.
All data renormalized to $\Gamma_R$=150 MeV, $b_{\eta}$=0.5.
Quark model calculations: \cite{Close_90,Konen_90,Capstick_95}.
} 
\label{fig_43}       
\end{minipage}
\end{figure}

In this sense, $\eta$ photoproduction is better suited albeit currently lacking 
data for polarization observables in $\eta$ photoproduction. Therefore, a 
complete multipole analysis is impossible, and the results are model dependent.
The dependence seems to be small due to the dominance of the S$_{11}$. 
Interestingly, a coupled channel analysis of $\eta$ {\em and} pion 
photoproduction of Sauermann et al. \cite{Sauermann_95} could describe both 
data sets with a coupling close to the typical results obtained from the 
$\eta$ data alone (see tab.~\ref{tab_05}). Independent of the above 
discussion, the following remarks can be made concerning the electromagnetic 
coupling at the photon point:
\begin{itemize}
\item{Most quark models predict values above 
125$\times 10^{-3}\mbox{GeV}^{-1/2}$, up to almost 
170$\times 10^{-3}\mbox{GeV}^{-1/2}$. 
Even if we adopt the result from $\eta$ photoproduction (eq.~\-(\ref{eq:a12eta}))
such large values require a total width in the range 250 - 400 MeV which is 
unrealistic. Therefore, only the relativized quark model of Capstick 
\cite{Capstick_92} predicts a value in the range of the experimental results. 
However, there could be a further twist. As we have already discussed for the 
$\Delta$, quark models predict the `bare' values of the couplings while the 
results extracted from data usually correspond to the `dressed' vertices in the 
presence of re-scattering terms. Recently, Chen et al. \cite{Chen_02} have 
analyzed pion scattering and pion photoproduction with a coupled channel 
dynamical model. They find not only a value of the `dressed' $A_{1/2}$ which 
lies more in the range of the $\eta$ results than typical results from other 
pion production analyses. They also find a much larger value of the `bare' 
coupling. Rescaled to $\Gamma$=150 MeV, their result corresponds to 
(145$\pm$4)$\times 10^{-3}\mbox{GeV}^{-1/2}$.    
}
\item{
In view of the unexplained discrepancy between the results from pion and $\eta$ 
photoproduction, the value from $\eta$ photoproduction (eq.~\-(\ref{eq:a12eta}))
should be used consistently as the photon point result for the analysis of the 
electromagnetic transition form factor as function of $Q^2$ since the results 
for $Q^2>0$ are all obtained from $\eta$ electroproduction 
(see fig.~\ref{fig_43}).
}
\end{itemize}

Since most of the uncertainty in the electromagnetic coupling comes from
the hadronic widths, Mukhopadhyay and coworkers
\cite{Benmerrouche_91,Nimai_comment,Benmerrouche_96} have suggested to use 
the electrostrong coupling $\xi$ defined via:
\begin{equation}
\label{eq:xi}
\xi= \left(\frac{k^{\star}m_{p}b_{\eta}}
     {q_{\eta}^{\star}M_{R}\Gamma_R}\right)^{1/2}\times A_{1/2}
\end{equation}
for precision tests of models. This quantity is proportional to the product
of the electromagnetic and strong matrix elements:
\begin{equation}
\label{eq:xi2}
\xi \propto \langle N\eta|{\cal{H}}_{s}|S_{11}\rangle
            \langle S_{11}|{\cal{H}}_{em}|\gamma N\rangle
\end{equation}
and the extracted values (see tab.~\ref{tab_05}) indeed agree within a
small band. Our best estimate of this parameter is 
$\xi=(2.24\pm0.05)10^{-4}$ MeV$^{-1}$ which even agrees with 
the original result of Benmerrouche and Mukhopadhyay \cite{Benmerrouche_91} 
(2.2$\pm$0.2) extracted from the sparse old data base. An analogous parameter
can be defined for pion photoproduction, and the experimental results again 
range in a relatively narrow window (see tab.~\ref{tab_05}). It will be 
interesting to get predictions from quark models for these quantities. 

\newpage
The $Q^2$ dependence of the helicity coupling has recently been investigated 
in two experiments at JLab via $\eta$ electroproduction. The measurement with 
the CLAS detector \cite{Thompson_01} covered the $Q^2$ range from 
0.25 - 1.5 GeV$^2$, and the HMS experiment \cite{Armstrong_99} measured two
high $Q^2$ points at 2.4 and 3.6 GeV$^2$.
%
%
%
%
The results are compared in fig.~\ref{fig_43} with older measurements, the 
photon point value and quark model predictions. It should be noted that all 
data are renormalized to $\Gamma_R=$150 MeV and $b_{\eta}$=0.5. In this way, 
the absolute scale of $A_{1/2}$ can vary within the limits of the hadronic 
widths while the shape of the form factor is well determined. The new data 
confirm the small falloff with $Q^2$ which is presently not reproduced by the 
quark models. Unfortunately, a prediction of the form factor from the $K\Sigma$ 
quasi-bound state model of the S$_{11}$ does not exists. Naively, it can be 
expected that the form factor for such a molecular-like system should fall off 
even steeper than the normal quark models.  

\begin{figure}[thb]
\centerline{\epsfysize=4.0cm \epsffile{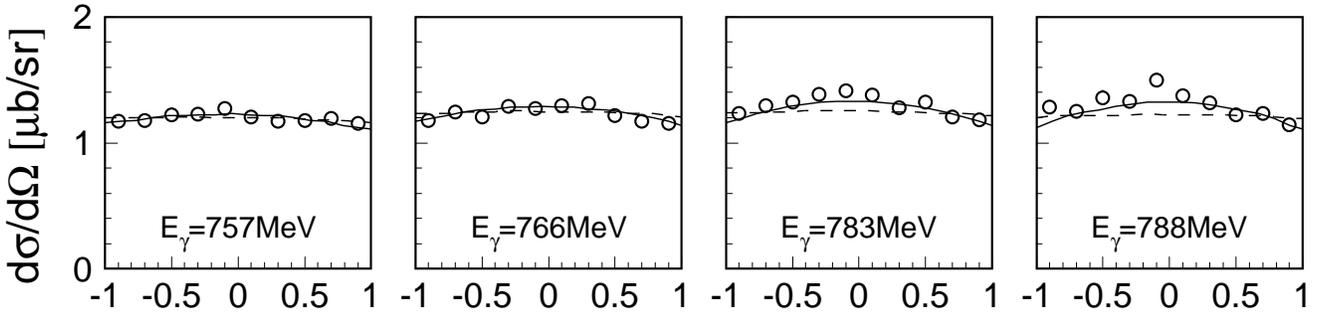}}
\caption{Angular distributions of $\gamma p\rightarrow p\eta$
\cite{Krusche_95} compared to model results of the effective Lagrangian
approach \cite{Benmerrouche_95}. Solid lines: full model, dashed lines:
without D$_{13}$ resonance. 
}
\label{fig_44}       
\end{figure}

The contribution of other resonances to $\eta$ photoproduction in the
threshold region is weak, but in case of the D$_{13}$ it is now well 
established. The 
first indication came from the interference term in the angular distributions 
eq.~\-(\ref{eq:eta_ang_multi}). However, this is a small effect, as can be 
seen in fig.~\ref{fig_44}. Here, calculations in the effective Lagrangian 
approach \cite{Benmerrouche_95} with and without contribution of the D$_{13}$ 
are compared to the data.
%
%
%
%
Polarization observables, in particular the photon beam asymmetry $\Sigma$, 
have a higher sensitivity to the D$_{13}$. A measurement of the photon beam 
asymmetry from threshold up to 1 GeV was reported from the GRAAL experiment 
\cite{Ajaka_98}. A measurement of the target asymmetry $T$ from threshold to
1.15 GeV was done at the Bonn PHOENICS tagged photon facility \cite{Bock_98}. 
Combined analyses of the differential cross sections and the polarization 
observables have been reported by several groups. Mukhopadhyay and Mathur 
\cite{Mathur_98} fitted the data with their ELA. A consistent description of 
all three observables with the exception of the target asymmetry in the 
threshold region was found. At low incident photon energies, the target 
asymmetry has a nodal structure. If enforced in the fit, it spoils the 
agreement with the other observables and with the target asymmetry itself at 
higher incident photon energies. The ratio of the helicity couplings 
$A_{3/2}^p$ and $A_{1/2}^p$ is extracted from the fit with a relatively small 
uncertainty (see table \ref {tab_04}) as well as the electrostrong couplings:
\begin{eqnarray}
\xi_{3/2}(\mbox{D}_{13}) & = & (+0.165\pm0.015\pm0.035)10^{-4}\mbox{MeV}^{-1}\\
\xi_{1/2}(\mbox{D}_{13}) & = &
(-0.065\pm0.010\pm0.015)10^{-4}\mbox{MeV}^{-1}\nonumber
\end{eqnarray}
which also avoid the uncertainty of the hadronic widths. The first error is
statistical, the second reflects the systematic uncertainty from possible
variations of other parameters. The much smaller values of the electrostrong
coupling of the D$_{13}$ as compared to the S$_{11}$ are due to the small
branching ratio of this resonance into $N\eta$. Using eq.~\-(\ref{eq:xi})
and PDG parameters for widths and helicity couplings of the resonance,
the branching ratio $b_{N\eta}(D_{13})$ is estimated in the range 
0.06\% - 0.4\%.
However, this is a rough estimate, more precise values can be obtained when 
all parameters are consistently fitted to the data (see below).

\vspace*{0.5cm}
%
%
%
\begin{figure}[hbt]
\centerline{
\epsfysize=7.5cm \epsffile{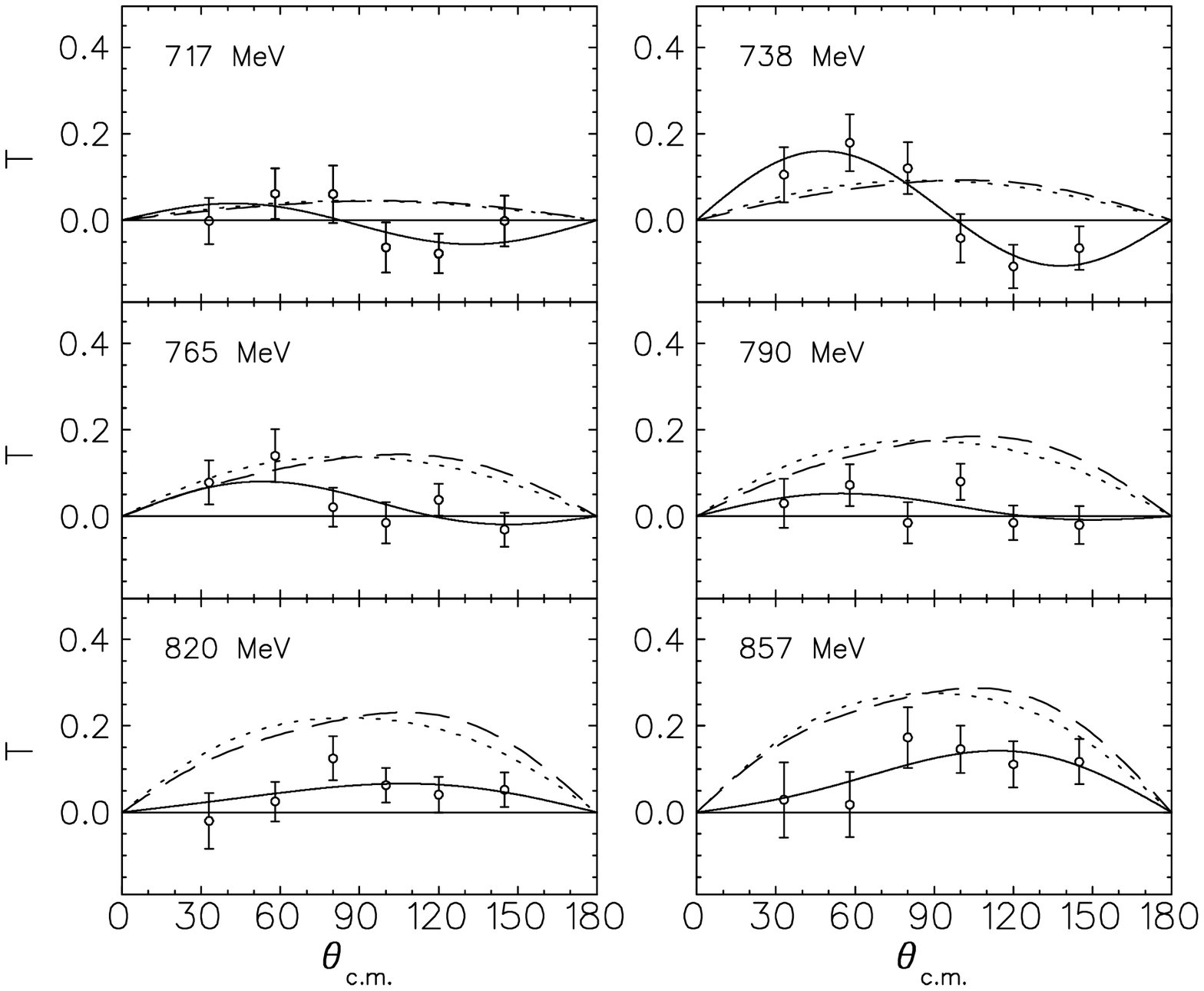}
\epsfysize=7.5cm \epsffile{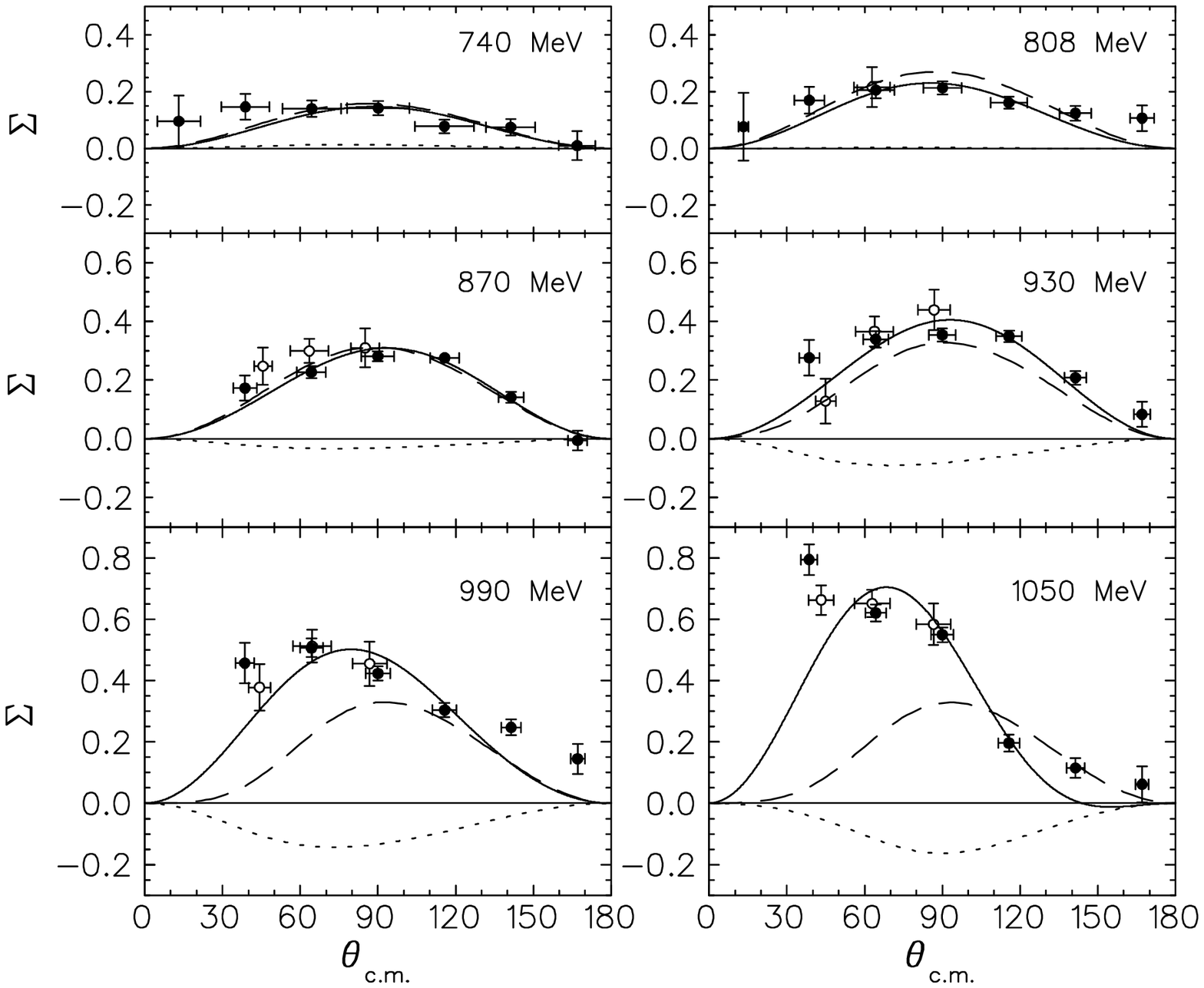}}
\caption{Fit of polarization observables \cite{Tiator_99}. Target asymmetry 
(left side) \cite{Bock_98} and photon beam asymmetry (right side) 
\cite{Ajaka_98}. The curves are from Tiator et al. \cite{Tiator_99}. Solid 
lines: fit to data (see text), dashed lines: isobar model from 
\cite{Knochlein_95}, dotted: isobar model without D$_{13}$. 
}
\label{fig_45}       
\end{figure}
%

Tiator et al. \cite{Tiator_99} have analyzed the data in two different ways.
First they included the polarization observables into the fit of their isobar 
model of ref. \cite{Knochlein_95}. 
The main results are:
\begin{itemize}
\item{The fit simultaneously reproduces all three observables with two
exceptions. In the same way as in the analysis of Mukhopadhyay and Mathur,
the nodal structure of the target asymmetry in the vicinity of the threshold 
is not in agreement with the model (see fig.~\ref{fig_45}, left hand side).
This effect is not yet understood (see discussion below). At the highest 
incident photon energies, the photon beam asymmetry does not agree with the 
data. This is taken as tentative evidence for the contribution of a higher 
lying resonance (see below).
}
\item{The effect of the D$_{13}$ resonance, negligible in case of the 
differential cross sections, is prominent in the beam asymmetry and allows to 
establish the coupling of this resonance to the $N\eta$ channel beyond doubt. 
Figure~\ref{fig_45} (right hand side) compares the model results with 
(dashed) and without (dotted) D$_{13}$ resonance. In the D$_{13}$ range the 
beam asymmetry would basically vanish without this resonance. The large 
positive values found in the experiment agree with the D$_{13}$ contribution 
extracted from the small effect in the angular distributions.
}    
\end{itemize}

The second analysis is a truncated multipole analysis. Differential
cross sections, target, and beam asymmetry do not allow a
fully model independent analysis. However, assuming \s -dominance and 
neglecting partial waves with $l>2$ allows to extend the multipole expansion
for the differential cross section 
(eqs.~\-(\ref{eq:eta_diff}),(\ref{eq:eta_ang_multi})) to the polarization
observables via:
\begin{eqnarray}
\label{eq:texp}
T & = & \mbox{sin}(\Theta^{\star})[d+e\;\mbox{cos}(\Theta^{\star})]\\
\Sigma & = & f\;\mbox{sin}^2(\Theta^{\star})
\label{eq:sigexp}
\end{eqnarray}
where:
\begin{eqnarray}
d & = & \frac{3}{a+c/3}\mbox{Im}[E^{\star}_{0+}(E_{1+}-M_{1+}]\\
e & = & \frac{3}{a+c/3}\mbox{Im}[E^{\star}_{0+}(E_{2-}+M_{2-}]\nonumber\\
f & = & \frac{3}{a+c/3}\mbox{Re}[E^{\star}_{0+}(E_{2-}+M_{2-}]\nonumber\;.
\end{eqnarray}
The magnitudes of the multipoles $E_{0+}$, $E_{2-}$, $M_{2-}$,
$B_{2-}=E_{2-}+M_{2-}$, $A_{2-}=1/2(3M_{2-}-E_{2-})$, and the phase between
the $s$- and $d$-waves are then simple expressions of the fitted coefficients 
$a$,...,$f$ \cite{Tiator_99}:
\begin{eqnarray}
|E_{0+}| & = & \sqrt{a+c/3}\\
|E_{2-}| & = & \frac{1}{4} \sqrt{(a+c/3)(e^2+f^2)}
 \left|1+\frac{c}{3f(a+c/3)}\right|\nonumber\\
|M_{2-}| & = & \frac{1}{12} \sqrt{(a+c/3)(e^2+f^2)}
\left|1-\frac{c}{f(a+c/3)}\right|\nonumber\\
|A_{2-}| & = & -\frac{c}{6f}\sqrt{\frac{e^2+f^2}{a+c/3}}\nonumber\\
|B_{2-}| & = & \frac{1}{3}\sqrt{(e^2+f^2)(a+c/3)}\nonumber
\end{eqnarray}
\begin{equation}
\label{eq:phase}
\mbox{tan}(\Phi_{E_{0+}}-\Phi_{B_{2-}}) =  \frac{e}{f}
\end{equation}
where the definition of the helicity multipoles $A_{2-}$, $B_{2-}$ is given in 
eq.~\-(\ref{eq:A5}), and $\Phi_{E_{0+}}$, $\Phi_{B_{2-}}$ are the phases of the
respective multipoles.

%
%
%
\begin{figure}[hbt]
\begin{minipage}{0.0cm}
\begin{turn}{90.}
{\mbox{\epsfysize=9.3cm \epsffile{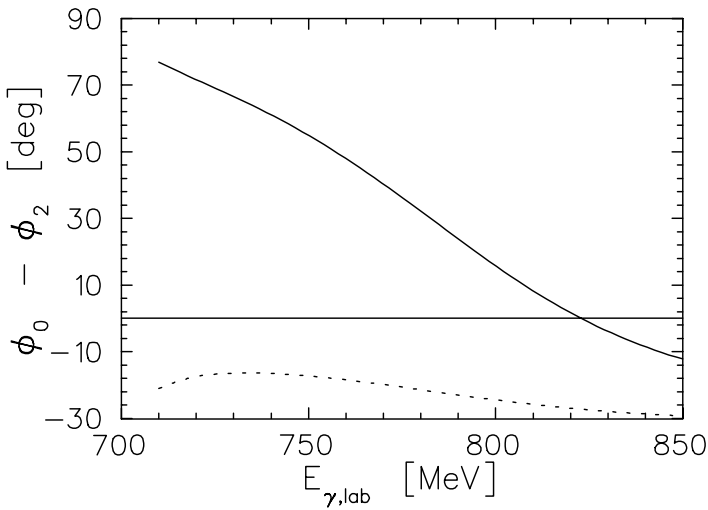}}}
\end{turn}
\end{minipage}
\hspace*{11.0cm}
\begin{minipage}{5.0cm}
\caption{Phase difference between the $B_{2-}$ helicity 3/2 multipole and the
$E_{0+}$ (see eq. (\ref{eq:phase})) \cite{Tiator_99}. Solid line: fit of 
multipole expansion, dotted: Breit-Wigner curves for \s ~ and \d .
}
\label{fig_46}       
\end{minipage}
\end{figure}
%

The multipole expansion allows a simultaneous fit of the differential cross 
sections, beam, and target asymmetry up to incident photon energies of 
$\approx$900 MeV (see fig.~\ref{fig_45}). The low energy $a$, $b$, $c$
coefficients agree with the result of the analysis of the differential
cross section.  This fit sheds light on the problems of the ELA and the isobar 
analyses with the low energy behavior of the target asymmetry. The multipole 
expansion can reproduce the node structure of the target asymmetry. However, a 
rather large, positive, and strongly energy dependent phase difference between 
the $s$- and $d$-waves arises (see fig. (\ref{fig_46})) \cite{Tiator_99}. 

This is unexpected since the $s$-wave is dominated by the \s ~ and the $d$-wave 
by the \d . Both resonances have comparable widths, and the S$_{11}$ lies only 
a little bit higher in energy than the D$_{13}$. Consequently, a Breit-Wigner 
shape of the two resonances results in a small negative phase difference.  At 
least one of the two resonances would not have a Breit-Wigner shape. 
However, the shape of the D$_{13}$ is well established and as discussed above, 
the total cross section of $\eta$ production is dominated by the S$_{11}$ and 
follows a Breit-Wigner curve. It is very desirable to measure the target 
asymmetry more precisely, since the unusual phase stems from the low energy 
behavior of this observable. 

The analysis of the interference terms allows the determination of the decay 
branching ratio of the D$_{13}$ into $N\eta$ and the ratio of the helicity 
couplings of the resonance. This is done in \cite{Tiator_99} where the first 
analysis fits $d\sigma /d\Omega$, $\Sigma$, and $T$ with the model independent 
multipole analysis. In a second scenario, $T$ is ignored, and the relative 
phase between D$_{13}$ and S$_{11}$ is taken from Breit-Wigner shapes. The 
results are summarized in table \ref{tab_07} and compared to other references.  

%
%
%
\begin{table}[bht]
  \caption[D$_{13}$ helicity couplings from $\eta$-production]{
    \label{tab_07}
Properties of the \d ~extracted from $\eta$ photoproduction.
Tia99a and Tia99b \cite{Tiator_99} are the model independent
analysis and the analysis assuming the Breit-Wigner phase.
Muk98 is the ELA analysis of Mukhopadhyay and Mathur, and Chi02 the isobar model
of Chiang et al. \cite{Chiang_02}.
}
  \begin{center}
    \begin{tabular}{|c|c|c|c|c|}
      \hline 
      Ref. & $b_{\eta}$ & $ A_{3/2}/A_{1/2}$ & $\xi_{3/2}$ & $\xi_{1/2}$ \\ 
      & \% & & $10^{-4}$MeV$^{-1}$ & $10^{-4}$MeV$^{-1}$ \\
      \hline
      Tia99a \cite{Tiator_99} & 
      0.08$\pm$0.01 & $-$2.1$\pm$0.2 & 0.185$\pm$0.018 & $-$0.087$\pm$0.013\\
      Tia99b \cite{Tiator_99} &
      0.05$\pm$0.02 & $-$2.1$\pm$0.2 & 0.134$\pm$0.018 & $-$0.087$\pm$0.013\\   
      Muk98 \cite{Mathur_98} &
      & $-$2.5$\pm$0.2$\pm$0.4 
      & 0.165$\pm$0.015$\pm$0.035 &  $-$0.065$\pm$0.010$\pm$0.015 \\ 
      Chi02 \cite{Chiang_02} &
      0.06 & $-$3.2 & 0.155 & $-$0.049 \\
      \hline           
    \end{tabular}
  \end{center}
\end{table}
%

At incident photon energies above $\approx$900 MeV the beam asymmetry is no
longer symmetric around 90$^o$ and cannot be fitted with 
eq.~\-(\ref{eq:sigexp}). The shape requires an additional term in the photon
asymmetry:
\begin{equation}
\Sigma=\mbox{sin}^2(\Theta^{\star})[f+g\;\mbox{cos}(\Theta^{\star})]
\end{equation}
which stems from multipoles with $l\geq 3$ \cite{Tiator_99}. The fitted 
coefficient $g$ rises strongly between 0.9 and 1.1 GeV. This higher partial 
wave is interpreted in \cite{Tiator_99} as a contribution of the F$_{15}(1680)$ 
resonance. A decay branching ratio of this resonance into $N\eta$ of 
$0.15^{+0.35}_{-0.10}$\% is extracted.

In the meantime, the measurement of the angular distributions and the total
cross section has been extended up to incident photon energies of 1.1 GeV
(GRAAL \cite{Renard_02}) respectively 2 GeV (CLAS \cite{Dugger_02}). 
%
%
%
\begin{figure}[hbt]
\begin{minipage}{0.0cm}
{\mbox{\epsfysize=11.cm \epsffile{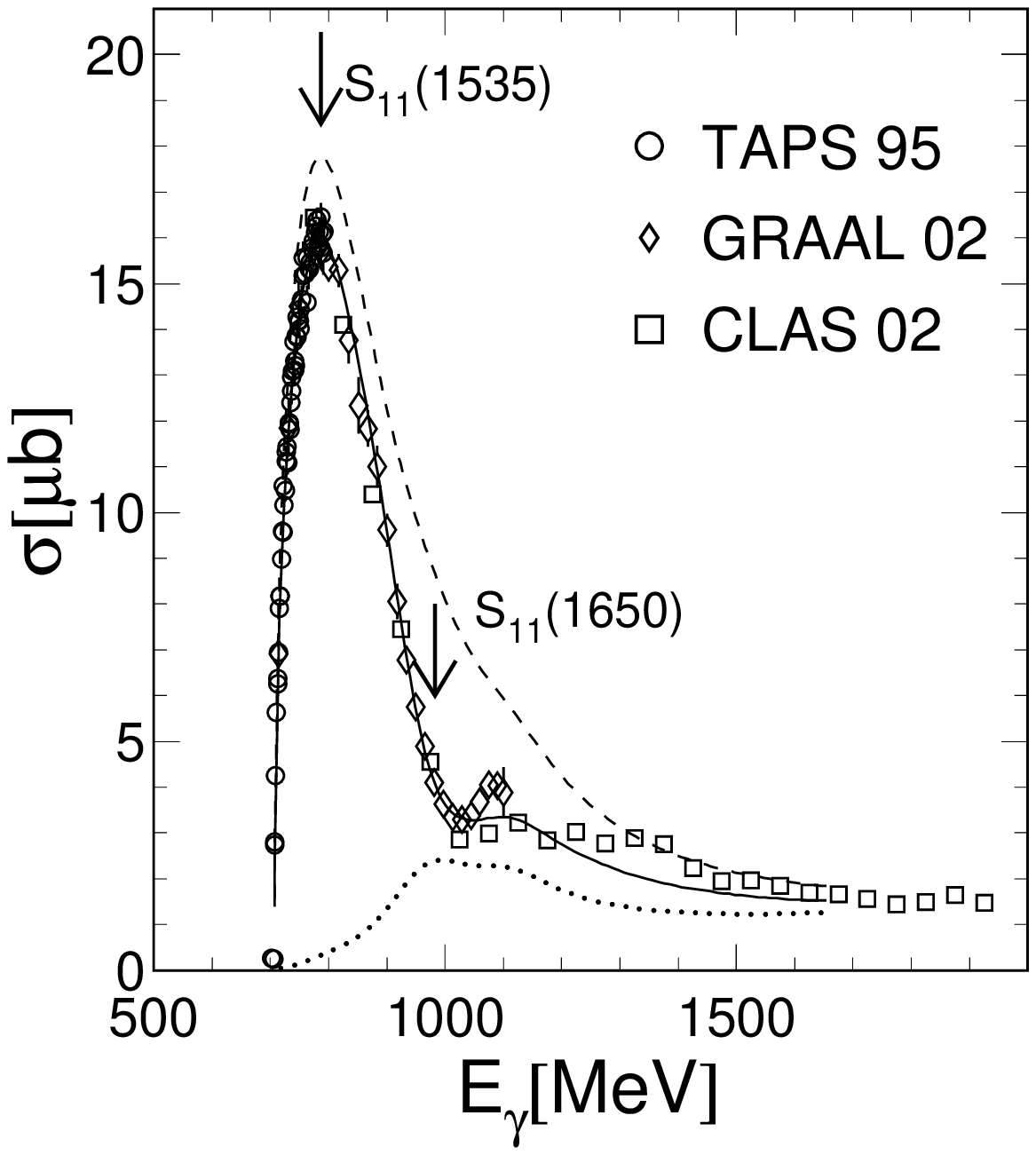}}}
\end{minipage}
\hspace*{9.9cm}
\begin{minipage}{0.0cm}
{\mbox{\epsfysize=5.2cm \epsffile{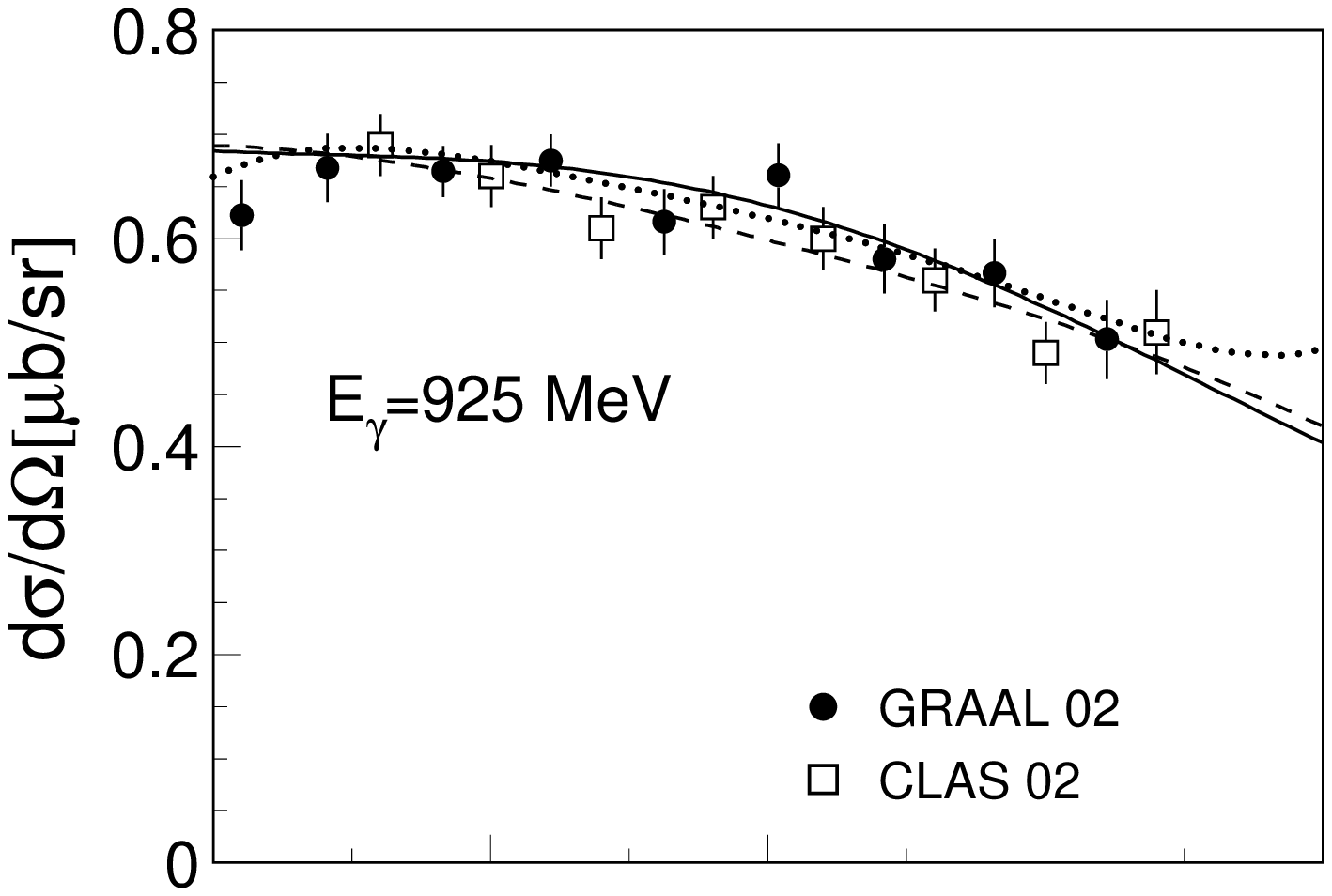}}}
{\mbox{\epsfysize=6.0cm \epsffile{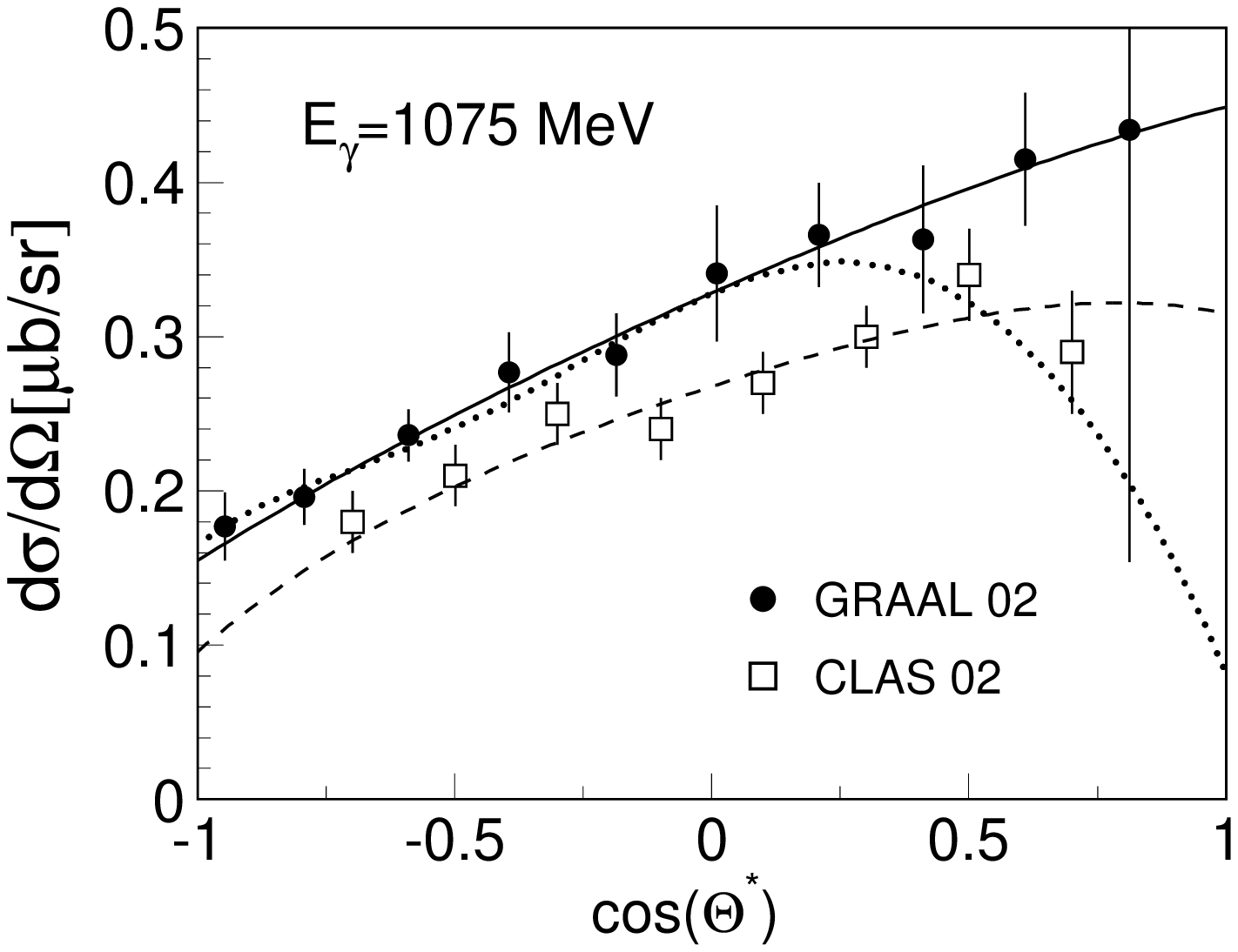}}}
\end{minipage}
\caption{Left: Total cross section for $\gamma p\rightarrow p\eta$. 
Data are from TAPS \cite{Krusche_95} (circles), GRAAL \cite{Renard_02} (diamonds)
and CLAS \cite{Dugger_02} (squares). Curves: ETA-MAID model \cite{Chiang_02},
solid: full model, dashed: without \ss , dotted: without \s . Arrows indicate
positions of the two resonances. 
Right: Selected angular distributions. 
Data: GRAAL \cite{Renard_02} (filled circles), CLAS \cite{Dugger_02}
(open squares). Solid and dashed lines: fits to the data with cubic
polynomials in cos($\Theta$). Dotted: ETA-MAID model \cite{Chiang_02}.
}
\label{fig_47}       
\end{figure}
%
The  results for the total cross section are compared in fig.~\ref{fig_47} 
(left hand side). The data sets are in excellent agreement up to incident 
photon energies of 1.0 GeV. However, between 1.0 and 1.1 GeV a systematic 
difference between the GRAAL and CLAS data is visible. This disagreement is 
important. The small bump in the GRAAL data between 1.0 and 1.1 GeV was 
interpreted in a quark model study of $\eta$ photoproduction by Saghai and Li 
\cite{Saghai_01} as tentative evidence for a third S$_{11}$ resonance with a 
mass of 1712 MeV and a width of 184 MeV. In this model, contributions from all 
known resonances in the relevant energy range are considered. It is argued 
that the fit is considerably improved around $E_{\gamma}=$ 1075 MeV when a 
third S$_{11}$ resonance is introduced. This state could correspond to the 
third S$_{11}$ resonance predicted in \cite{Li_96}. Its inclusion into the fit 
also leads to significantly modified values for widths and electromagnetic 
couplings of the other two S$_{11}$ states (see table~\ref{tab_05}). 
Figure~\ref{fig_47} (right hand side) shows the angular distributions for 
$E_{\gamma}$=925 MeV and $E_{\gamma}$=1075 MeV for a more detailed comparison 
of the two measurements. The total cross section from both experiments agrees 
perfectly at the lower energy while disagreeing for the second energy. Also, 
the angular distributions agree within the statistical uncertainty for the 
lower photon energy, and polynomial fits to the data give identical results 
for the total cross section. However, the GRAAL data at 1075 MeV are 
systematically higher for all angles. Furthermore, large uncertainties arise 
from the extrapolation of the angular distribution to forward angles. In 
contrast to the lower photon energy, polynomial fits to the data differ 
strongly from the fit of the ETA-MAID model \cite{Chiang_02}. Further 
experimental effort should clarify the situation in this energy region. 
It is of particular importance to measure the angular distributions also for 
the extreme forward angles apart from a careful absolute normalization of the 
data.

At still higher incident photon energies the CLAS data \cite{Dugger_02} show 
a tendency for strong forward peaking of the angular distributions. This might 
be an indication for growing t-channel contributions. However, a detailed 
analysis of the reaction in this energy range is presently not available.

\newpage
\subsubsection{\it $\eta$-Photoproduction from Light Nuclei}

As already discussed in the case of pion photoproduction, information about 
the isospin structure of the electromagnetic excitation can only be gained 
from measurements on nucleons bound in (light) nuclei. This introduces 
additional systematic uncertainties due to the nuclear effects. On the other 
hand, ratios of helicity couplings on the proton and neutron are free of the 
large systematic uncertainties of the hadronic widths of the resonances. In 
case of the isoscalar $\eta$ mesons, it is possible to extract the isospin 
structure from a comparison of the cross section on the proton, the neutron 
and coherent production from the deuteron via:
\begin{equation}
\label{eq:s11_iso}
  \sigma_p \sim |A^{IS}_{1/2} + A^{IV}_{1/2}|^2,\;\;\;\;\;
  \sigma_n  \sim  |A^{IS}_{1/2} - A^{IV}_{1/2}|^2,\;\;\;\;\;
  \sigma_d  \sim  |A^{IS}_{1/2}|^2
\end{equation}
where $A^{IS}_{1/2}$ and $A^{IV}_{1/2}$ are the isoscalar and isovector parts 
of the helicity amplitude. In case of the neutron cross section, nuclear 
effects in the quasifree measurement must be considered. In case of the 
deuteron, the proportionality stands for a reaction model for coherent $\eta$ 
photoproduction. The first attempts along these lines of investigating the 
isospin structure of the \s ~were undertaken in the late 1960's with 
bremsstrahlung. Bacci \cite{Bacci_69} and  collaborators measured the breakup 
reaction and found that the inclusive cross section from the deuteron is 
roughly twice as large as for the proton at an incident photon energy of 
850 MeV. The conclusion was, that $\sigma_{n}\approx \sigma_{p}$. 
According to eqs.~\-(\ref{eq:s11_iso})), either the isoscalar or the isovector 
part would dominate, and the other would vanish. In the same year, Anderson 
and Preprost \cite{Anderson_69} measured the coherent cross section from the 
$I=0$ deuteron and found a rather large cross section which is only possible 
when the isoscalar part is large. Consequently, both experiments together 
seemed to indicate that $A_{1/2}^n/A_{1/2}^p\approx$+1 and thus the isovector 
part is {\it negligible}. This however, was in sharp contrast to the results 
from pion photoproduction and to model predictions which both favored ratios 
between $-$0.6 and $-$1, indicating that the isovector part is {\it dominant}. 
This problem was solved during the last few years with a series of precise 
measurements of breakup and coherent $\eta$ photoproduction from the deuteron 
and from He-isotopes 
\cite{Krusche_95c}-\cite{Pfeiffer_02}.

%
%
%
\begin{figure}[thb]
\centerline{
\epsfysize=8.0cm \epsffile{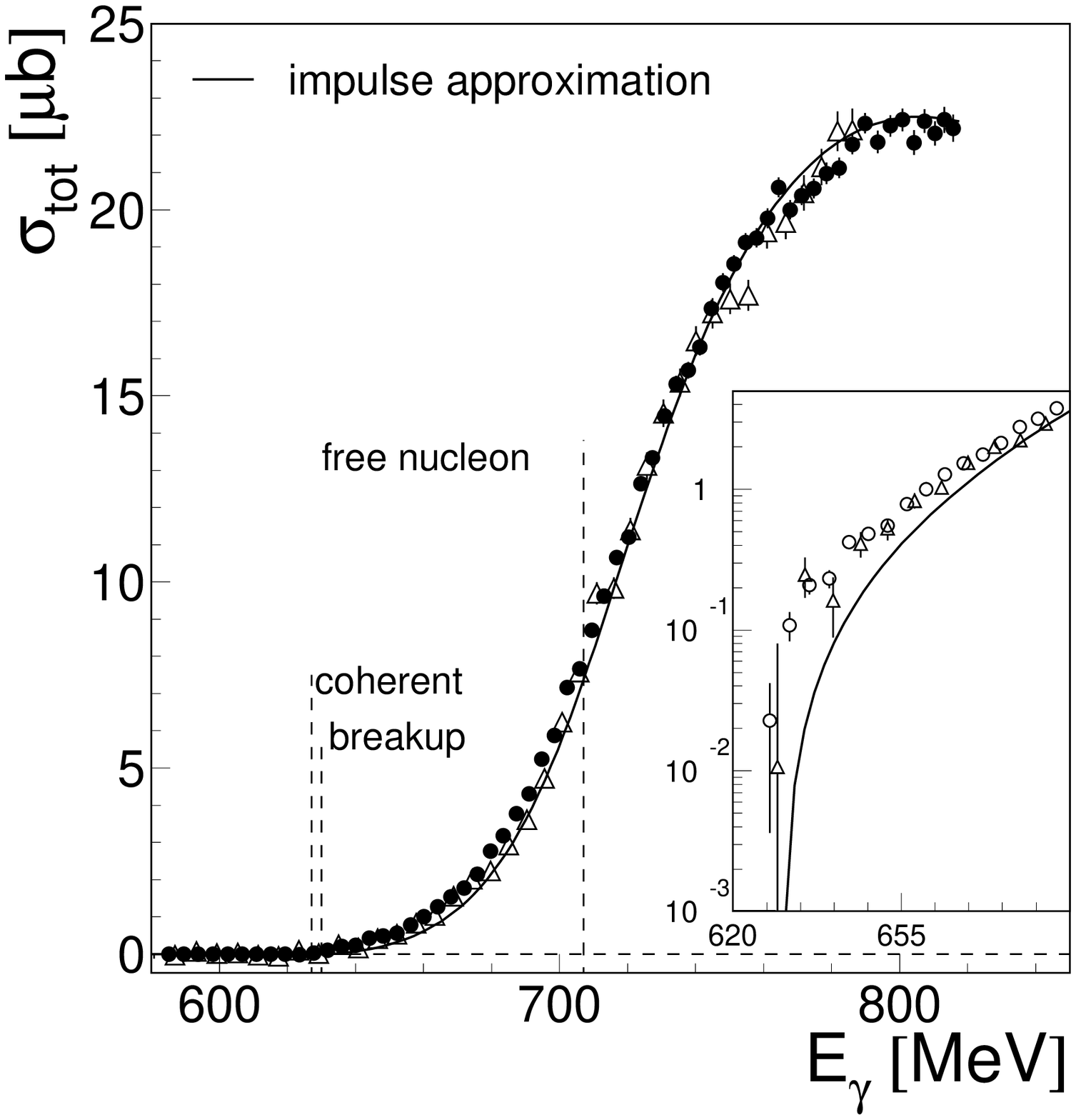}
\epsfysize=8.0cm \epsffile{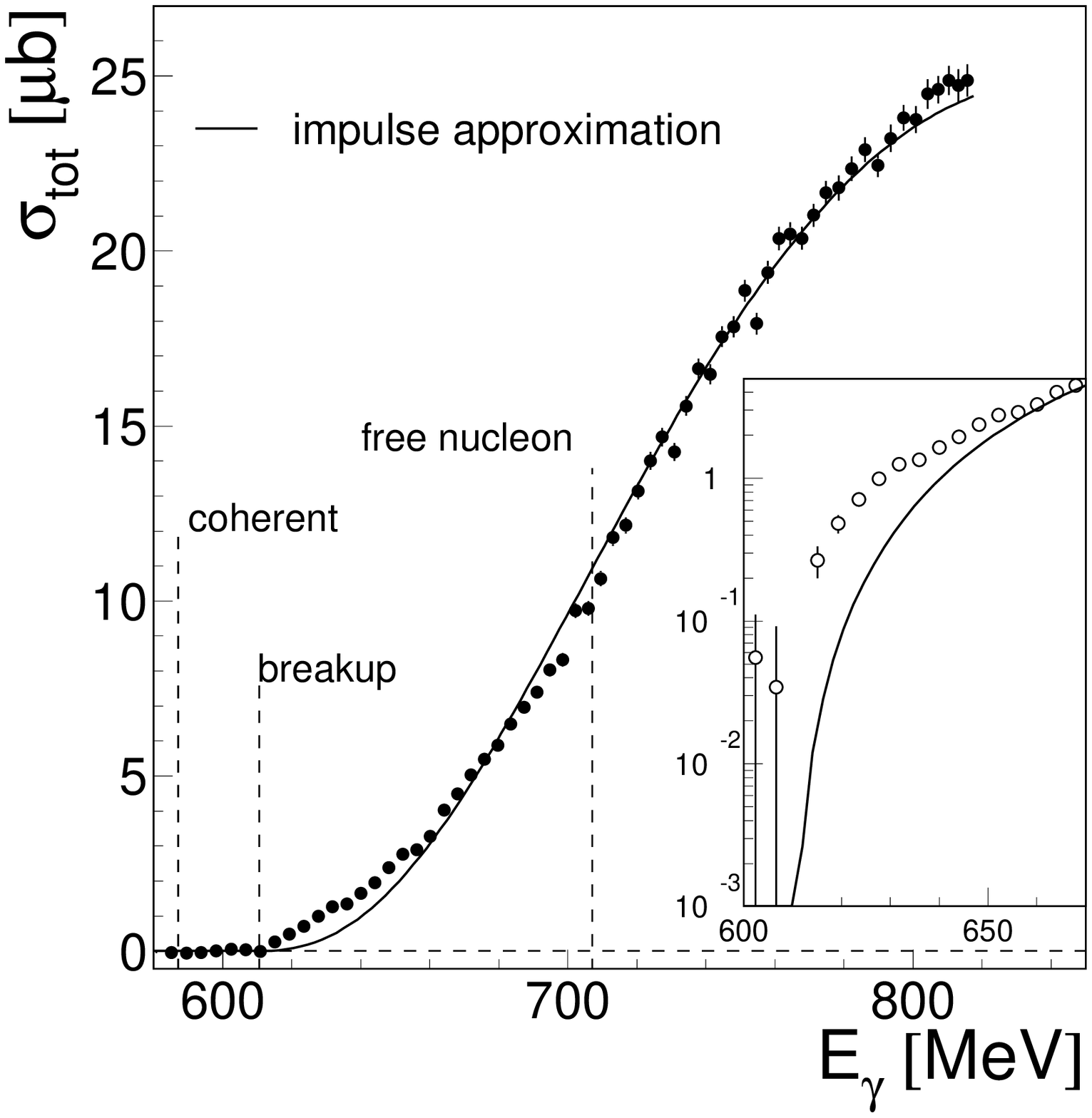}}
\caption{Inclusive $\eta$ photoproduction cross section. Left hand side:
from the deuteron. Circles: ref. \cite{Weiss_02}, triangles: 
ref.~\-\cite{Krusche_95c}. Right hand side: from $^4$He \cite{Hejny_99}. 
For both pictures the dashed lines indicate the coherent, the breakup, and the 
free nucleon production thresholds. The solid curves are the result of the 
impulse approximation model under the assumption of a constant
$\sigma_n /\sigma_p$=2/3 ratio (see text). Inserts: threshold region. 
}
\label{fig_48}       
\end{figure}
%

The first group of experiments aimed at the extraction of the cross section
ratio from the neutron and the proton via quasifree $\eta$-photoproduction
from light nuclei. Different experimental concepts were exploited:
\begin{itemize}
\item{Measurement of the inclusive $A(\gamma ,\eta)X$ reaction. Comparison of 
the sum of the Fermi-smeared free proton cross section and a Fermi-smeared 
ansatz for the free neutron cross section to the inclusive nuclear cross 
section in PWIA. Variation of the ansatz for $\sigma_n$ until agreement is
achieved.}
\item{Coincident measurement of $\eta$-mesons and recoil nucleons. Extraction
of the ratio of $\sigma_n /\sigma_p$ of the quasifree cross sections measured
under identical conditions as a function of the incident photon energy 
$E_{\gamma}$.}
\item{Coincident measurement of $\eta$-mesons and recoil nucleons and 
reconstruction of the effective $\sqrt{s^{\star}}$ and effective incident 
photon energy $E_{\gamma}^{\star}$ from the final state kinematics
($E_{\eta}$, $E_{R}$, $\vec{p}_{\eta}$, $\vec{p}_{R}$:
energy and momentum of meson and recoil nucleon).
Extraction of $\sigma_n /\sigma_p$ as a function of $E_{\gamma}^{\star}$:
\begin{equation}
\label{eq:e_eff_1}
s^{\star}=(E_{\eta}+E_R)^2-(\vec{p}_{\eta}+\vec{p}_R)^2
\end{equation}
\begin{equation}
\label{eq:e_eff_2}
E_{\gamma}^{\star}=(s^{\star}-m_R^2)/2m_R\;\;.
\end{equation}
}
\end{itemize}

Two examples for the inclusive method are summarized in fig.~\ref{fig_48}.
Shown is the total, inclusive cross section of the reactions 
$d(\gamma ,\eta)X$ \cite{Krusche_95c,Weiss_02} and $^4$He$(\gamma ,\eta)X$
\cite{Hejny_99}. The curves correspond to PWIA calculations. Here, the sum of
the elementary cross sections off the proton and off the neutron were folded
with the momentum distributions of the bound nucleons. For the free neutron
cross section an ansatz was taken to be proportional to the free proton cross
section. Agreement is obtained in both cases for $\sigma_n/\sigma_p\approx$ 2/3.
The influence of the different nuclear momentum distributions is severe. The 
total cross sections from $^4$He and from the deuteron are almost equal at an 
incident photon energy of 800 MeV although twice the number of nucleons is 
involved in the He case. FSI effects are not included, and the effect stems 
from the nucleon momentum distributions. The good agreement of both data sets 
with impulse approximations using the same neutron - proton ratio is reassuring 
for the application of this method. 

%
%
%
\begin{figure}[hbt]
\begin{minipage}{9.75cm}
{\mbox{\epsfysize=7.0cm \epsffile{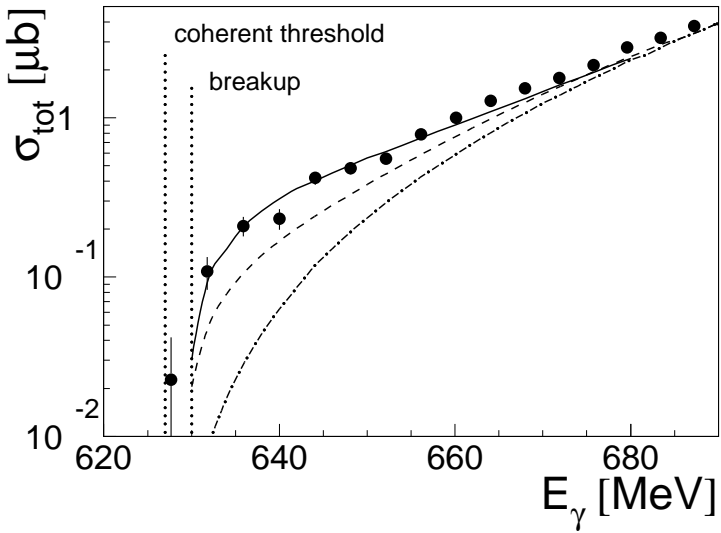}}}
\caption{Threshold behavior of $d(\gamma ,\eta)X$. Data from \cite{Weiss_02}.
Model calculations
from Sibirtsev et al. \cite{Sibirtsev_02}. Dotted: PWIA, dashed: PWIA and
$NN$ FSI, full: PWIA and $NN$ FSI and $N\eta$ FSI.
}
\label{fig_49}       
\end{minipage}
\vspace*{-10.6cm}
\end{figure}
%

\hspace*{10.0cm}
\begin{minipage}{7.8cm}
As shown in the inserts of 
fig.~\ref{fig_48}, and in more detail in fig. \ref{fig_49}, the agreement with 
the PWIA calculations suffers in the 
vicinity of the absolute thresholds. The threshold behavior is further 
discussed in \cite{Hejny_02}. At least in case of the deuteron, where model 
calculations are available, it is understood via final state interaction 
effects which are large close to threshold. Fix and Arenh{\"o}vel have studied 
these effects 
\cite{Fix_all} 
and found that the data can only be 
reproduced with a three body (Faddeev-type) calculation of the $NN\eta$ system 
although the main contribution comes from $NN$ FSI. Similarly, Sibirtsev et al.
\cite{Sibirtsev_01,Sibirtsev_02} found a dominant contribution from
$NN$ FSI and some importance of the interference between $NN$ and $N\eta$
FSI (see fig.~\ref{fig_49}).
These effects are interesting in themselves for the study of the $\eta$-nucleon 
interaction. Here, it is only important, that they are negligible at incident 
\end{minipage}

 \noindent{photon} energies above the free nucleon threshold because the momentum mismatch 
between participant and spectator nucleon is always large at these energies. 
Then, the extraction of the neutron-proton cross section ratio is not spoiled . 

The exclusive reaction with detection of the recoil nucleons was also
investigated for deuteron and $^4$He targets
\cite{Hoffmann-Rothe_97,Hejny_99,Weiss_02}.
%
%
%
\begin{figure}[hbt]
\centerline{
\epsfysize=6.5cm \epsffile{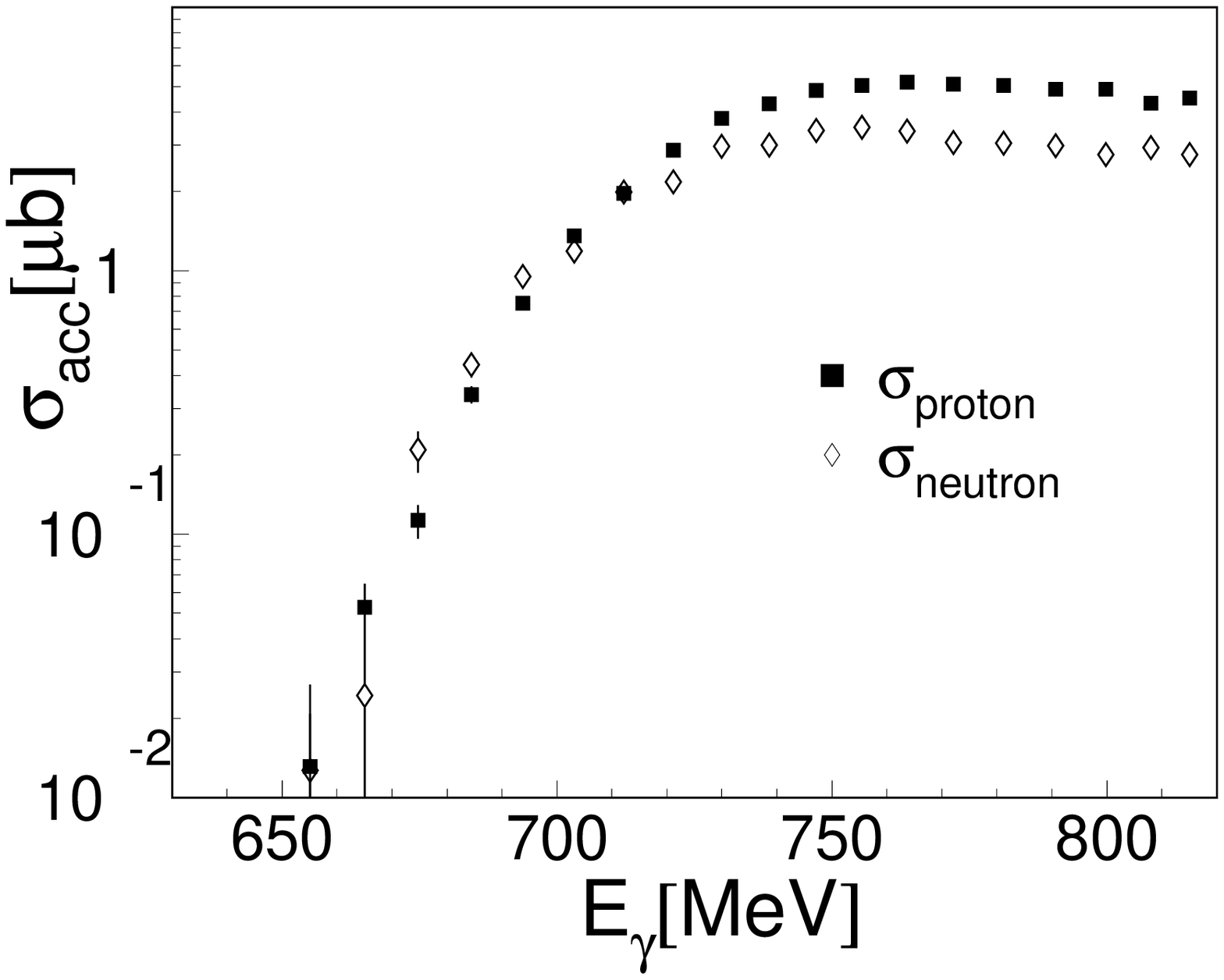}
\epsfysize=6.5cm \epsffile{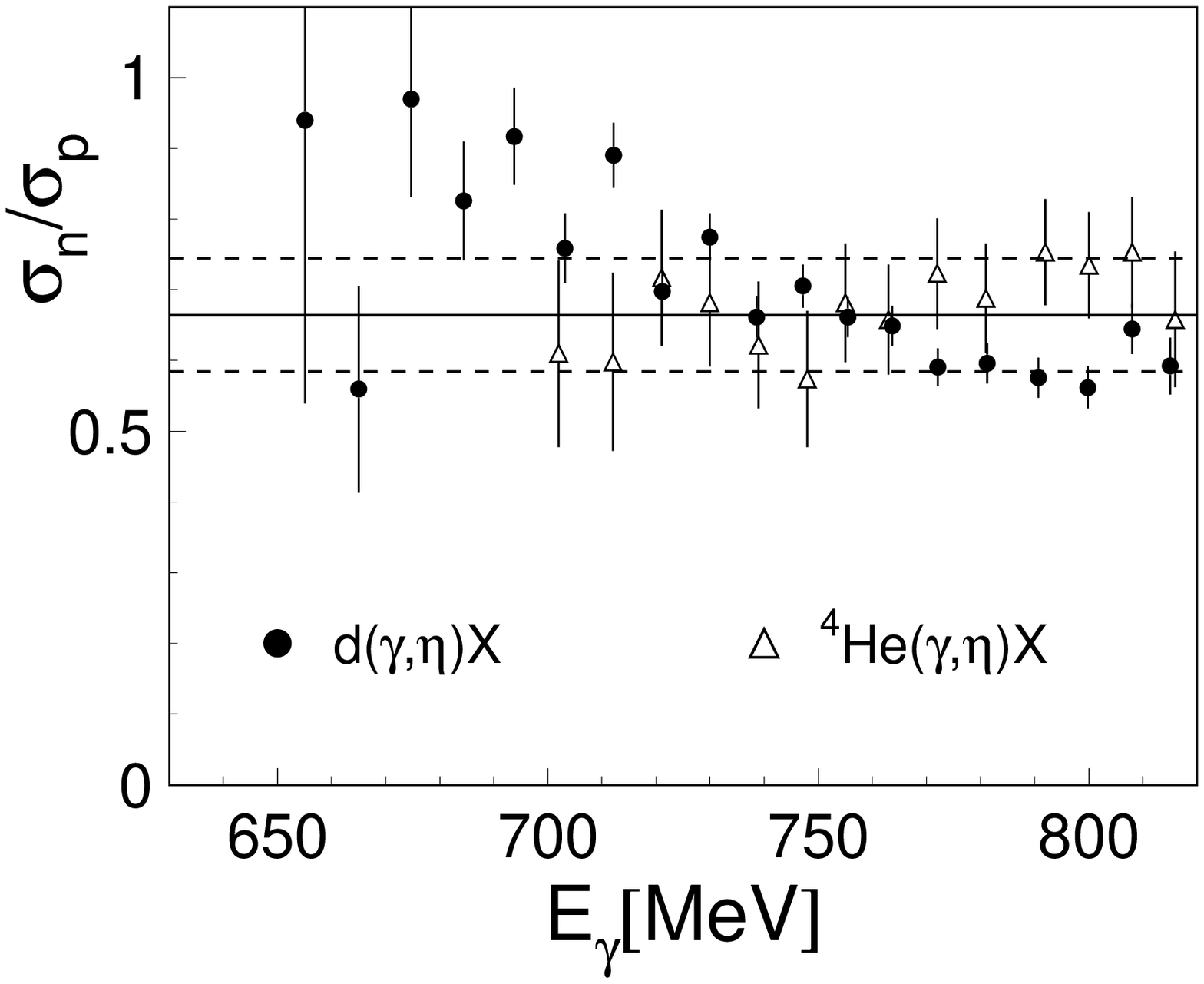}}
\centerline{
\epsfysize=6.5cm \epsffile{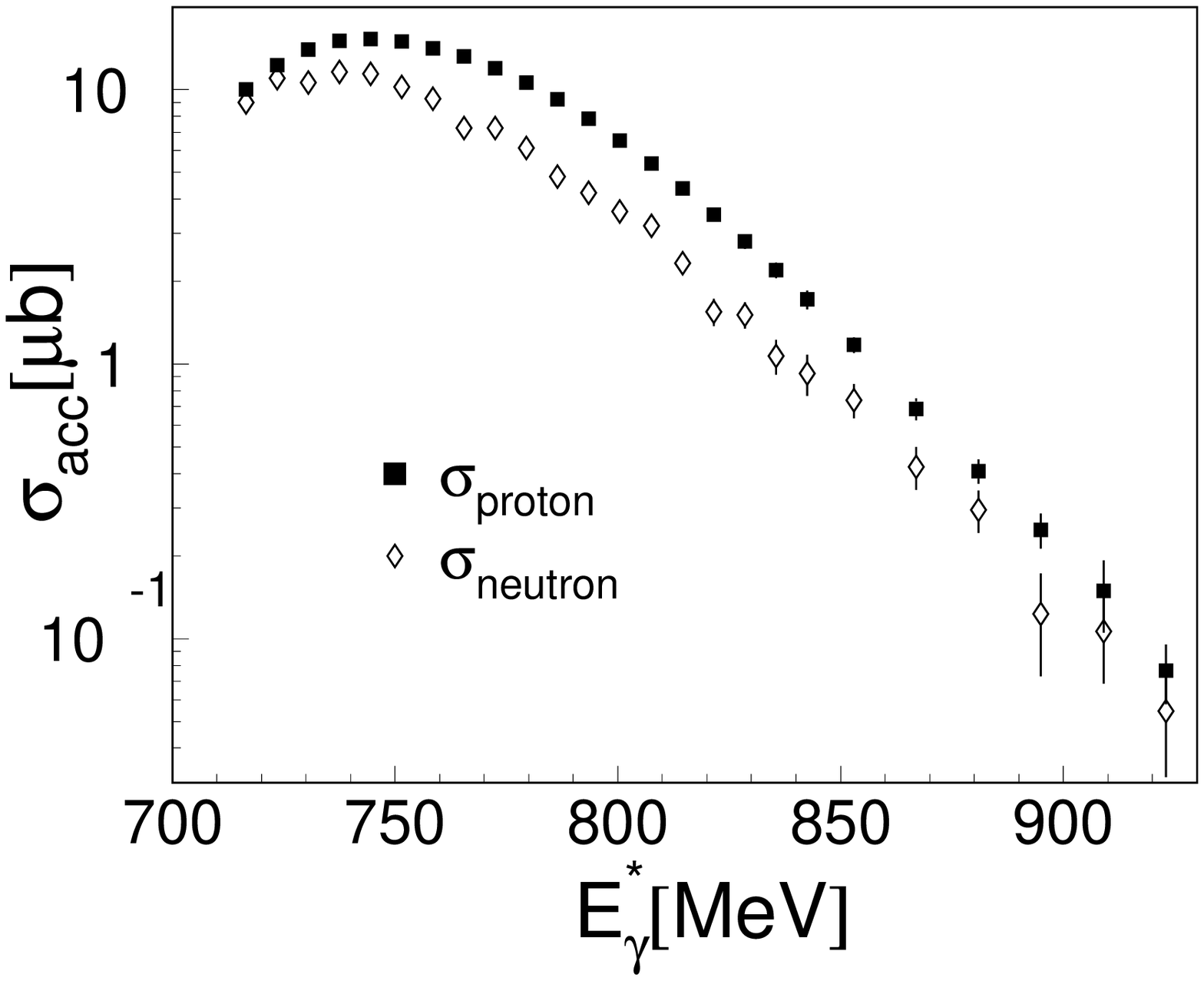}
\epsfysize=6.5cm \epsffile{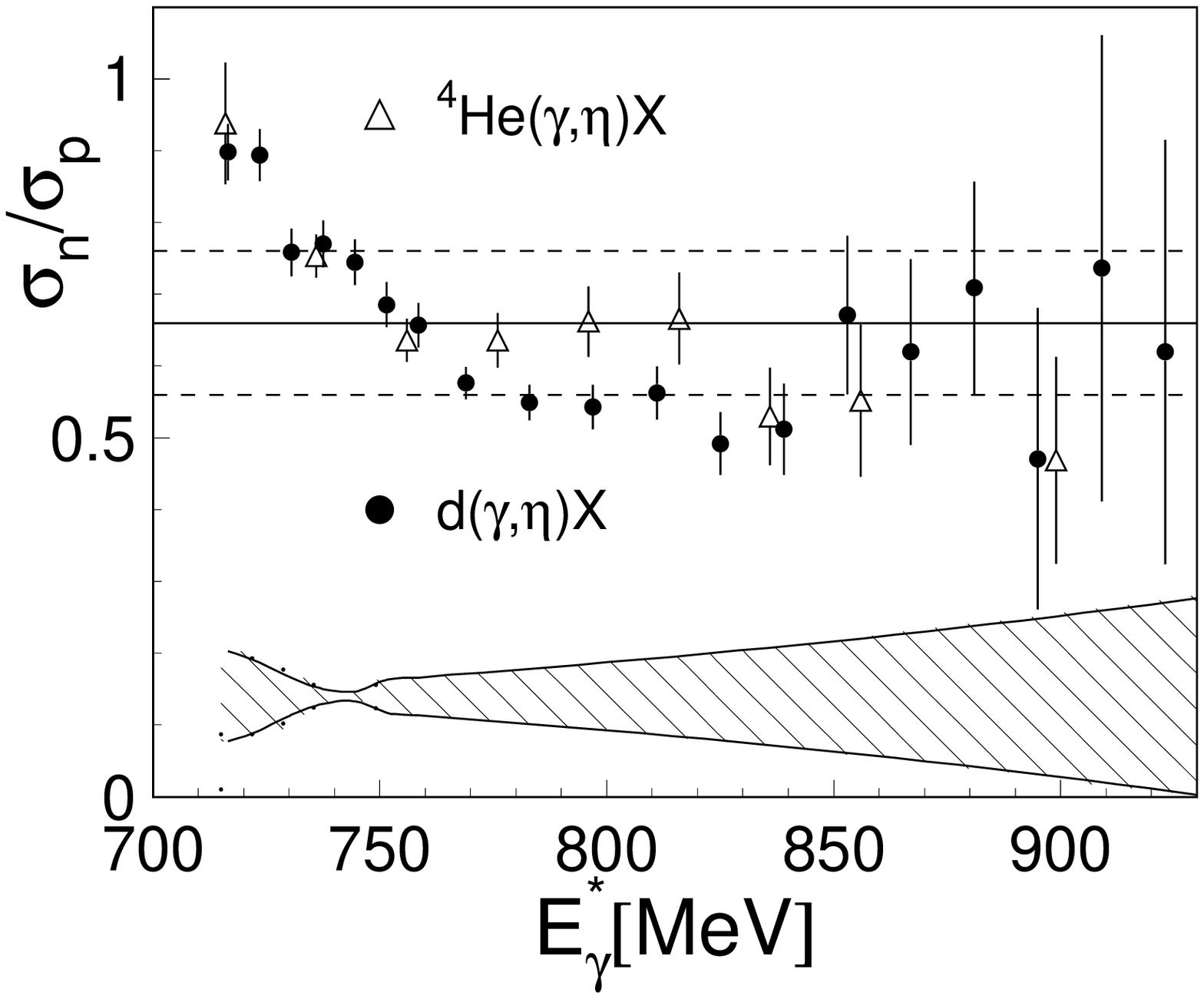}}
\caption{Exclusive $\eta$ photoproduction off the deuteron \cite{Weiss_02}
and $^4$He \cite{Hejny_99}. 
Left hand side: measured cross section within the detector acceptance for the
proton and the neutron from the deuteron target. Right hand side: neutron/proton 
cross section ratios $\sigma_n/\sigma_p$ obtained for $d$ and $^4$He tagets. 
Upper parts: versus incident photon energy, bottom part: versus effective 
photon energy. 
}
\label{fig_50}       
\end{figure}
%
Typical results are summarized in fig.~\ref{fig_50}. The measured cross
sections are given within the detector acceptance \cite{Hejny_99,Weiss_02} 
since a small fraction of the phase space of the recoil nucleons was not 
covered in the experiments. Meanwhile, this should not significantly influence
the ratios (see discussion in \cite{Weiss_02}). The ratios are shown as a 
function of the incident photon energy and as function of the effective photon 
energy. The latter is defined (see eqs.~\-(\ref{eq:e_eff_1},\ref{eq:e_eff_2})) 
as the laboratory energy of a photon which produces the same $\sqrt{s}$ on a 
{\it nucleon at rest} as was reconstructed from the reaction kinematics for 
the incident nucleon with Fermi motion. This representation avoids the 
averaging over the Fermi momenta. It is thus better suited for the comparison 
to model predictions, and extends to higher effective photon energies for 
reactions with Fermi momenta anti-parallel to the incident photon. However, it 
is subject to an additional systematic uncertainty, shown as a shaded band in 
the lower right hand corner of the figure, which arises from the relative 
detector energy calibration for protons and neutrons. The results obtained 
from the exclusive measurements are in agreement with a neutron/proton ratio 
close to 2/3. The rise of the ratios close to the thresholds is due to the FSI 
effects discussed above.

\begin{figure}[hbt]
\begin{minipage}{0.0cm}
{\mbox{\epsfysize=10.5cm \epsffile{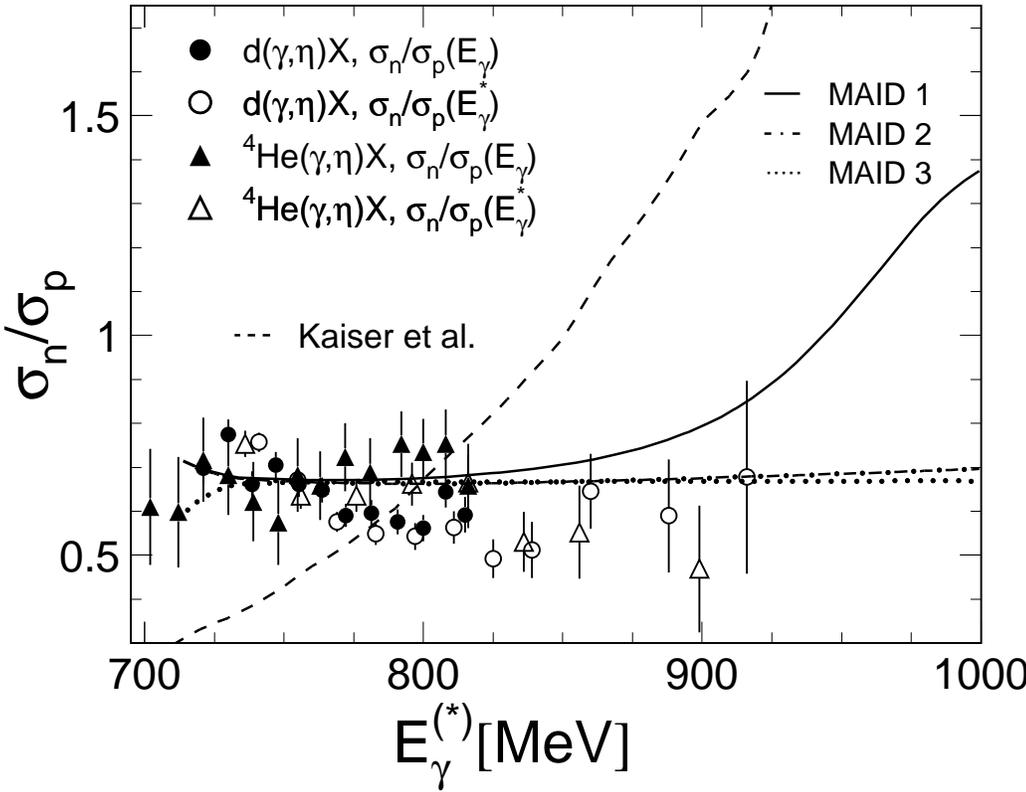}}}
\end{minipage}
\hspace*{13.5cm}
\begin{minipage}{4.7cm}
\caption{Ratio of exclusive neutron - proton cross sections for the deuteron
and for $^4$He \cite{Weiss_02,Hejny_99}. Dashed curve: prediction from the 
$K\Sigma$ model of the \s ~(Kaiser et al.
\cite{Kaiser_95,Kaiser_97}). The curves labelled MAID are the 
predictions from the MAID model \cite{Chiang_02} for the full model (MAID 1), 
the S$_{11}$(1535)-resonance, Born terms and vector meson exchange (MAID 2),
and for the \s ~alone (MAID 3).
}
\label{fig_51}       
\end{minipage}
\end{figure}

The results from the different exclusive measurements are compared in fig.
\ref{fig_51} to the predictions from the $K\Sigma$ model of the
\s ~\cite{Kaiser_95,Kaiser_97} and to the results of the ETA-MAID model
\cite{Chiang_02}. The $K\Sigma$ model is not favored by the data which are in 
good agreement with the ETA-MAID model.
%
%
%
%
The MAID predictions in the range used for the extraction of the ratio of the 
helicity couplings of the \s ~($E_{\gamma}\approx$ 750 - 800 MeV) are
practically identical for the full model and for truncations of the model up to
the extreme case where only the \s ~is considered. This is further evidence 
that the helicity ratio extracted from the data is not influenced by background
contributions. At higher photon energies a strong rise of the ratio is 
predicted which comes from the contribution of higher lying resonances. The 
presently available data do not extend to sufficiently high energies for a test 
of these predictions. Measurements on the deuteron at higher incident photon 
energies are necessary for the search for contributions from other resonances 
to the $N\eta$ channel. 

The results for the neutron/proton cross section ratio in the S$_{11}$ range 
from the different experiments are summarized in table \ref{tab_08} along with 
the ratios of the helicity couplings resulting under the assumption that 
background contributions are negligible. The uncertainty of the ratio of the 
couplings has much smaller uncertainties than the couplings themselves since 
the poorly known hadronic widths cancel. The results from all experiments are 
in good agreement. The systematic effects (which dominate the uncertainties) 
come from different sources (different targets, different extraction methods). 
It is therefore reasonable to average over the results which yields a precise 
result for 
the comparison to quark model predictions. In addition, the table includes two 
results for the analysis of the data from \cite{Krusche_95,Krusche_95c} in the 
framework of models which explicitly included background terms. One of them 
lies a little above and one a little below the other results. In contrast, the 
results from multipole analyses of pion photoproduction are more scattered due 
to the small contribution of the S$_{11}$ to this channel. The result indicates 
that either the isoscalar or the isovector part is dominant. Which one it is 
depends on the relative sign of the couplings. The results from pion 
photoproduction agree on a negative sign, corresponding to a dominant isovector 
part. 
%
%
%
\begin{table}[thb]
  \caption[S$_{11}$ neutron - proton ratio]{
    \label{tab_08}
Cross section ratio of $n(\gamma ,\eta)n$ and $p(\gamma ,\eta)p$ and the
corresponding helicity ratios. Muk95: effective Lagrangian analysis of 
proton and neutron cross sections, Sau95: coupled channel analysis.\\
$^{1)}$ the result from Bac69 \cite{Bacci_69} (bremsstrahlung beam, no invariant
mass analysis for $\eta$, Fermi motion neglected) is not included.
}
  \begin{center}
    \begin{tabular}{|c|c|c|c|c|}
      \hline 
      Ref. & target & method & 
      $\sigma_n/\sigma_p$ & $|A_{1/2}^n|/|A_{1/2}^P|$ \\      
      \hline
      Bac69 \cite{Bacci_69} & $^2$H & $\eta$ detected
      & $\approx$1 & $\approx$1\\
      Kru95 \cite{Krusche_95} & $^2$H & $\eta$ detected 
      & 0.66$\pm$0.07 & 0.81$\pm$0.04 \\
      Hof97 \cite{Hoffmann-Rothe_97} & $^2$H & recoil nucleons 
      & 0.68$\pm$0.06 & 0.83$\pm$0.04 \\
      Hej99 \cite{Hejny_99} & $^4$He & $\eta$ detected  
      & 0.67$\pm$0.07 & 0.82$\pm$0.04 \\ 
      Hej99 \cite{Hejny_99} & $^4$He & $\eta$ and recoil nucleons  
      & 0.68$\pm$0.09 & 0.83$\pm$0.05 \\
      Wei02 \cite{Weiss_02} & $^2$H & $\eta$ detected 
      & 0.67$\pm$0.07 & 0.82$\pm$0.04 \\
      Wei02 \cite{Weiss_02} & $^2$H & $\eta$ and recoil nucleons
      & 0.66$\pm$0.10 & 0.81$\pm$0.06 \\         
      \hline
      & & average${^1)}$ & 0.67$\pm$0.03 & 0.82$\pm$0.02 \\
      \hline 
      Muk95 \cite{Nimai_95} & & ELA, data from \cite{Krusche_95,Krusche_95c}
      & & 0.84$\pm$0.15 \\
      Sau95 \cite{Sauermann_95} & & CC, data from \cite{Krusche_95,Krusche_95c}
      & & 0.80 \\
      \hline         
      Met74 \cite{Metcalf_74} & & pion production & & 0.8\\
      Ara82 \cite{Arai_82} & & pion production & & 0.9\\
      Cra83 \cite{Crawford_83} & & pion production & & 1.5\\
      Arn90 \cite{Arndt_90} & & pion production & & 0.74\\
      Arn93 \cite{Arndt_93} & & pion production & & 0.75\\
      Arn94 \cite{Arndt_96} & & pion production & & 0.33\\
      Arn02 \cite{Arndt_02} & & pion production & & 0.53\\
      \hline
    \end{tabular}
  \end{center}
\end{table}
%

\begin{figure}[thb]
\begin{minipage}{0.0cm}
{\mbox{\epsfysize=7.cm \epsffile{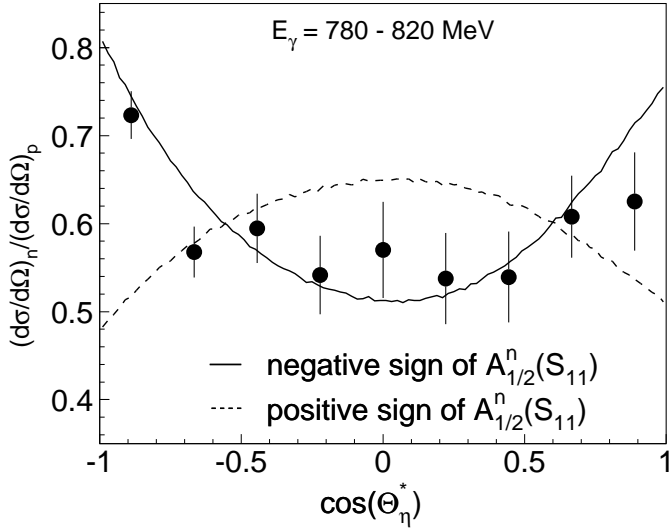}}}
\end{minipage}
\hspace*{11.0cm}
\begin{minipage}{6.0cm}
\caption{Ratio of the differential neutron/proton cross sections in the range of
the \s ~compared to predictions (eq. (\ref{eq:dsapp})) with negative and 
positive relative sign of the S$_{11}$ couplings \cite{Weiss_02}.
}
\label{fig_52}       
\end{minipage}
\end{figure}

\newpage
A possibility to access the relative sign in $\eta$ photoproduction is the 
interference term of the S$_{11}$ with the D$_{13}$ in the angular 
distributions (coefficient $c$ in eq.~\-(\ref{eq:eta_ang_multi})). Relying on 
the dominance of the S$_{11}$ and Breit-Wigner shapes of the resonances, the
angular distributions on the proton ($N=p$) and the neutron
($N=n$)  can be approximated in the form \cite{Weiss_02}:
\begin{eqnarray}
\label{eq:dsapp}
\left(\frac{d\sigma}{d\Omega}\right)_{N} & \propto & (A_{1/2}^N(S_{11}))^2 +
{\cal{G}}(E_{\gamma})A_{1/2}^N(S_{11})A_{1/2}^N(D_{13})(3cos^2(\Theta)-1)
\;\;,
\end{eqnarray}
where the function $\cal{G}$ is identical for proton and neutron and can be
determined from the angular distributions of $p(\gamma ,\eta)p$. Inserting
$A_{1/2}^p(D_{13})$=($-$24$\pm$2) and $A_{1/2}^n(D_{13})$=($-$67$\pm$3)
\cite{Arndt_02} and folding with the deuteron Fermi momenta results in the
curves compared to the data in fig.~\ref{fig_52}. Obviously, the negative 
sign for $A_{1/2}^n$ is favored. 

%
%
%
\begin{figure}[hbt]
\begin{minipage}{0.0cm}
{\mbox{\epsfysize=7.5cm \epsffile{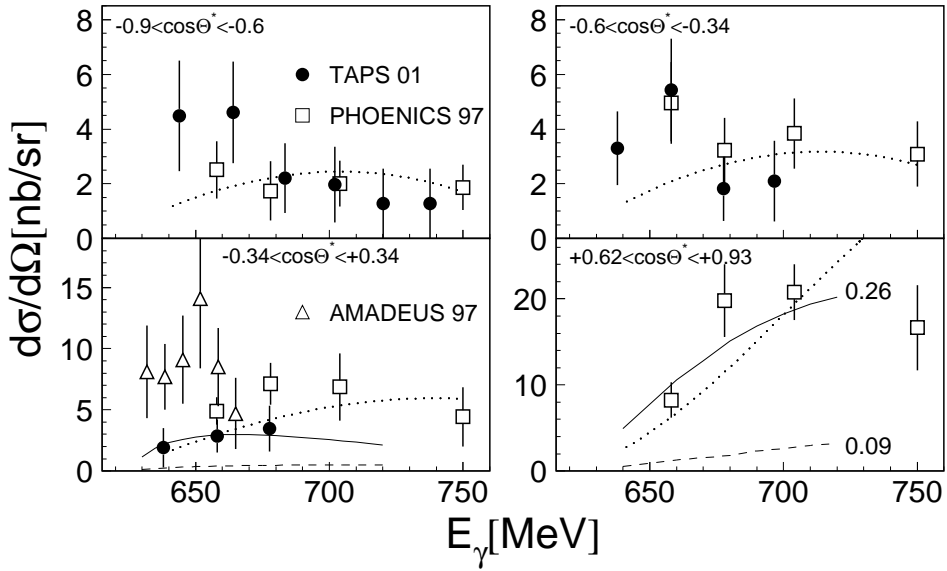}}}
\end{minipage}
\hspace*{13.0cm}
\begin{minipage}{5.0cm}
\caption{Coherent $\eta$ photoproduction off the deuteron. The data are from
TAPS \cite{Weiss_01} (full circles),  PHOENICS \cite{Hoffmann-Rothe_97}
(open squares), and AMADEUS \cite{Hoffmann-Rothe_97} (open triangles). 
Calculations by Fix and Arenh\"ovel \cite{Fix_all} for various values of 
$|A_{1/2}^s|/|A_{1/2}^p|$ (solid: 0.26, dashed: 0.09), and Ritz  and Arenh\"ovel 
\cite{Ritz_01} (dotted) are compared to the data.
}
\label{fig_53}       
\end{minipage}
\end{figure}

%
Consequently, only the measurement of coherent $\eta$ photoproduction off 
the deuteron by Anderson and Preprost \cite{Anderson_69} favored a dominant 
isoscalar contribution. In the meantime, this result has been ruled out.
The reported cross sections at $\Theta_{\eta}^{\star}\approx$ 90$^o$
ranged from 40 nb/sr to 27 nb/sr in the energy range from 665 to 705 MeV,
but in \cite{Krusche_95c}, an upper limit (without detection of actual
signal) of 10 nb/sr was found. The results from two recent measurements are
summarized in fig.~\ref{fig_53}. The measurements with the PHOENICS and 
AMADEUS detectors in Bonn \cite{Hoffmann-Rothe_97} detected the recoil 
deuterons.
The measurement with the TAPS detector in MAINZ \cite{Weiss_01} detected 
deuterons and $\eta$-mesons in coincidence. The results at 90$^o$ are much 
smaller than reported earlier in \cite{Anderson_69}, in case of the TAPS 
measurement more than an order of magnitude, so that they support a dominant 
isovector part of the amplitude. However, there is one last twist. When the 
signs of the helicity amplitudes and the ratio of their magnitudes are known, 
the isoscalar contribution to the amplitude can be calculated
from eqs.
(\ref{eq:s11_iso}): $A_{1/2}^{IS}/A_{1/2}^p$=(0.09$\pm$0.01). 
However, models for the coherent process using this ratio 
(see fig.~\ref{fig_53})
under predict the coherent cross section, which is better described with
$A_{1/2}^{IS}/A_{1/2}^p\approx$0.25. Recently, Ritz and Arenh{\"o}vel 
\cite{Ritz_99,Ritz_01} have argued that there is an unexpected large complex 
phase between $A_{1/2}^p$ and $A_{1/2}^n$ due to re-scattering contributions. 
In their model they find: 
\begin{eqnarray}
A_{1/2}^p & = & (+120.9-\mbox{i}\;66.1)\times 10^{-3}\mbox{GeV}^{-1/2}\\
A_{1/2}^n & = & (-114.0-\mbox{i}\;1.7)\times 10^{-3}\mbox{GeV}^{-1/2}\nonumber
\end{eqnarray} 
which allows to reproduce the coherent cross section (see fig.~\ref{fig_53}) 
and the ratio of the couplings determined from the breakup reactions 
($|A_{1/2}^n|^2/|A_{1/2}^p|^2$=0.68). However, the magnitude of the coupling 
$A_{1/2}^p$ (138$\times 10^{-3}$GeV$^{-1/2}$) is quite large compared to other 
results. 

A further check of the isospin structure can be obtained from coherent $\eta$ 
photoproduction off He-isotopes which have different quantum numbers. The 
photoexcitation of the S$_{11}$ resonance involves the $E_{0+}$ spin-flip 
amplitude. According to the above discussion, the excitation is dominantly 
isovector. Consequently, the coherent reaction should be almost forbidden for 
the $I=J=0$ $^4$He nucleus but comparatively large for the $I=J=1/2$ $^3$He 
nucleus. In agreement with this expectation, an investigation of 
$^4$He$(\gamma ,\eta)^4$He did not find a signal but only upper limits for 
the reaction \cite{Hejny_99}. Meanwhile, total cross sections up to 250 nb at 
photon energies between 650 - 720 MeV were found for $^3$He \cite{Pfeiffer_02}.  
%
%
%
%

\newpage                   
\subsection{\it Double Pion-Photoproduction}
\label{ssec:twopi}

\begin{figure}[bht]
\begin{minipage}{0.0cm}
{\mbox{\epsfysize=7.8cm \epsffile{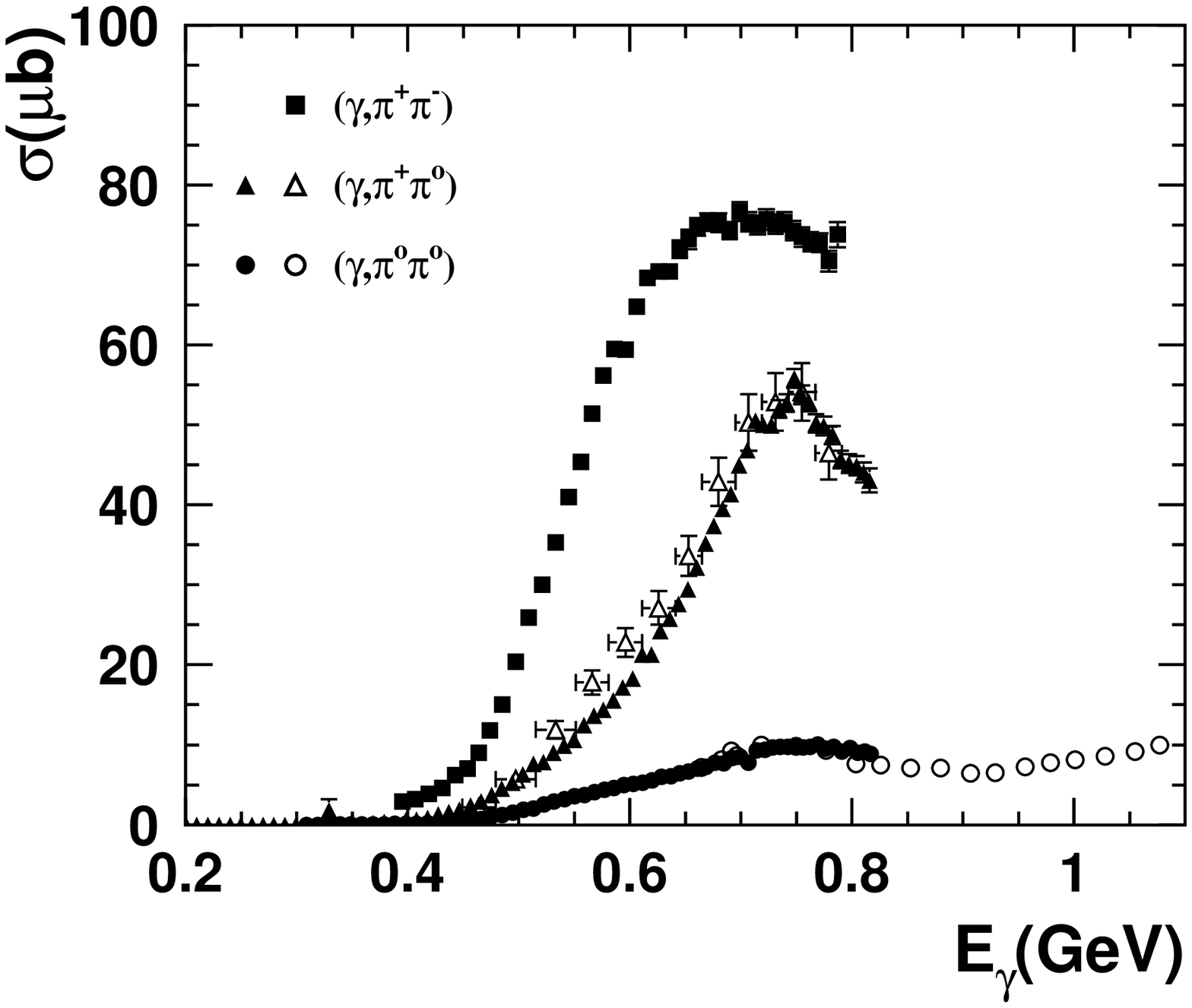}}}
\end{minipage}
\hspace*{9.3cm}
\begin{minipage}{0.0cm}
{\mbox{\epsfysize=7.8cm \epsffile{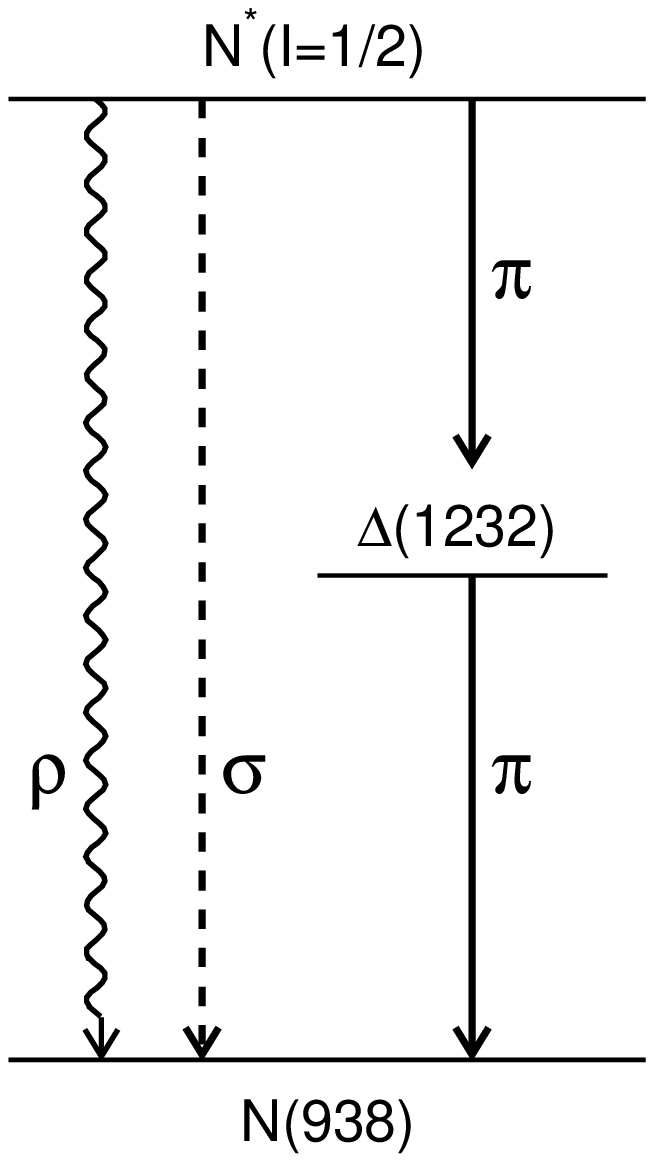}}}
\end{minipage}
\hspace*{4.2cm}
\begin{minipage}{4.5cm}
\caption{Left hand side: Total cross section of the three isospin channels of 
double pion production on the proton. Data are from \cite{Braghieri_95,
Langgaertner_01,Kotulla_01,Hourany_01}. Right hand side: possible resonance
contributions to double pion production in the second resonance region.}
\label{fig_54}       
\end{minipage}
\end{figure}

The production of pion pairs is important for the study of the second resonance 
region of the nucleon. We have already discussed (see fig.~\ref{fig_27}) that 
almost 50\% of the total photoabsorption cross section at incident photon 
energies of $\approx$800 MeV can be attributed to double pion production.
This is responsible for the bump structure in the total cross section of 
figs.~\ref{fig_27},\ref{fig_54}. The much discussed suppression of the peak 
for photoproduction on nuclei will not be understood without a thorough 
investigation of double pion  production. Furthermore, double pion production 
gives access to interesting decay properties of the resonances.
%
%
%
%
Three different resonance decay processes can contribute in the double pion 
channel (see fig.~\ref{fig_54}, right hand side): the decays via emission of 
the $\rho$ and $\sigma$ mesons to the nucleon ground state with subsequent 
decays of the mesons into pion pairs and the cascade decay of the resonances 
via the intermediate $\Delta$ resonance. The sequential decays allow the study 
of resonance - resonance transitions. The decay via the $\rho$ meson is 
suppressed since its nominal mass ($m_{\rho})\approx$771 MeV corresponds to a 
kinematical production threshold on the free nucleon of 1086 MeV, well above 
the second resonance region. However, the large width of $\approx$150 MeV 
allows contributions from its low energy tail. The existence of the 
scalar-isoscalar $\sigma$ meson is still under dispute 
(PDG: $f_{0}$(600) with $m=$ 400 - 1200 MeV, $\Gamma$ = 600 - 1000 MeV). 
In the present context, it is not important whether we call such a broad state 
a meson or understand it as a correlated pion pair in the scalar-isoscalar 
state. What we have discussed so far corresponds to resonance decays to the 
nucleon ground state. In analogy to nuclear physics, we expect valuable 
structure information from transitions between higher lying states. 
In addition, like in single meson production, non-resonant background 
contributions complicate the picture and require detailed reaction models for 
the extraction of the resonance contributions.  

Until very recently, data for double pion production reactions came mostly
from bubble chamber experiments. Therefore, $\gamma p\rightarrow p\pi^+\pi^-$ 
had been the only isospin channel measured with reasonable precision. 
Total cross sections and invariant mass distributions of the 
$\pi^+\pi^-$-, $p\pi^+$-, and $p\pi^-$-pairs are available in the literature 
\cite{ABBHHM_68}-\cite{Hilpert_66}. 
New interest arose during the last few years when the isospin channels 
with neutral particles became accessible. In a series of experiments with the 
DAPHNE \cite{Audit_91} and TAPS \cite{Novotny_91,Gabler_94} detectors at the 
Mainz accelerator all isospin channels 
\cite{Braghieri_95}-\cite{Langgaertner_01}
except $\gamma n\rightarrow n\pi^+\pi^-$ were measured up to the second 
resonance region. The two detectors are complementary in the sense that DAPHNE 
has advantages for reactions with many charged particles in the final state 
while TAPS is optimized for the 2$\gamma$ decay of neutral pions. At higher 
incident photon energies the $2\pi^o$ final state became available at GRAAL 
in Grenoble \cite{Hourany_01} and the $\pi^+\pi^-$ final state at SAPHIR in 
Bonn \cite{Klein_96}.
 
\begin{figure}[hbt]
\begin{minipage}{0.cm}
{\mbox{\epsfysize=7.5cm \epsffile{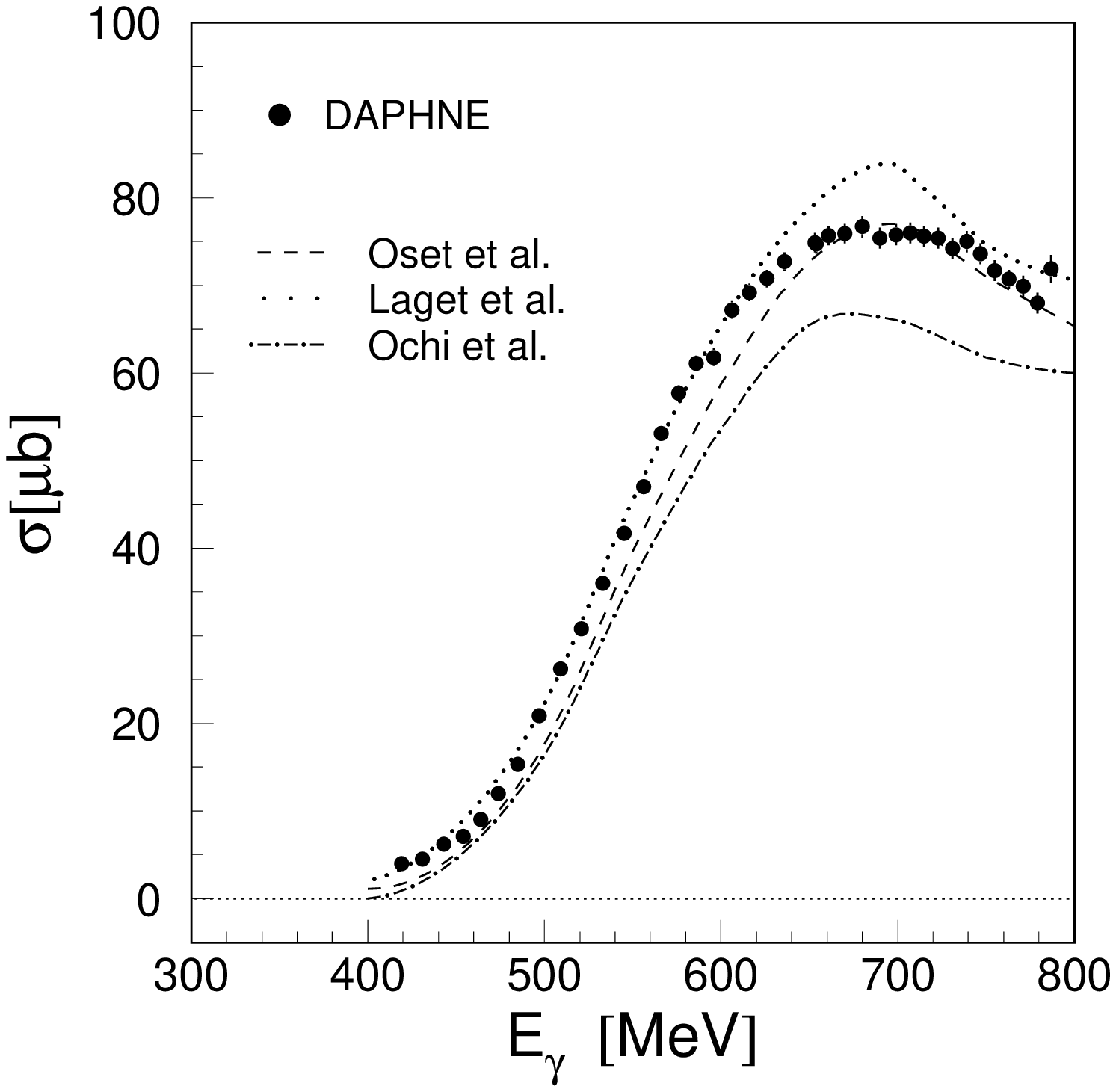}}}
\end{minipage}
\hspace*{7.3cm}
\begin{minipage}{0.cm}
\vspace*{-1.cm}
{\mbox{\epsfysize=2.2cm \epsffile{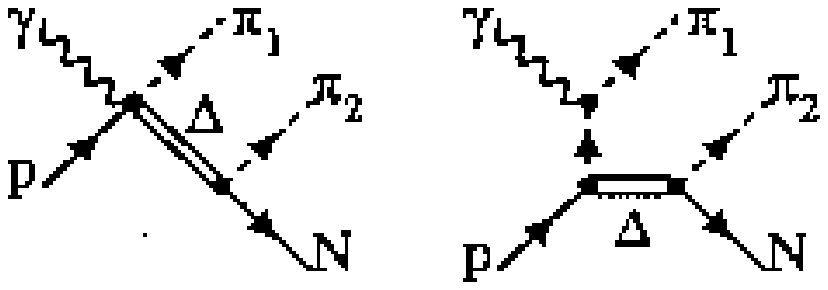}}}\\
\vspace*{0.5cm}
\hspace*{2cm}
{\mbox{\epsfysize=2.3cm \epsffile{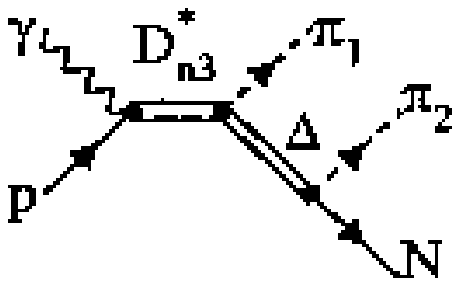}}}
\end{minipage}
\hspace*{6.5cm}
\begin{minipage}{4.3cm}
\caption{Left hand side: Total cross section for the reaction 
$\gamma p\rightarrow p\pi^+\pi^-$. Data are from  \cite{Braghieri_95}. 
Dashed, dotted, and dash-dotted curves: calculations 
from 
\cite{Tejedor_96}-\cite{Ochi_97}.
Right hand side: important contributions to the reaction (many more diagrams
give small contributions, see e.g. \cite{Tejedor_96}).
}
\label{fig_55}       
\end{minipage}
\end{figure}

The $\gamma p\rightarrow p\pi^+\pi^-$ reaction was analyzed in an early 
attempt to extract the dominant production mechanisms by L\"uke and S\"oding 
\cite{Lueke_71}.
The total cross section is small between threshold at $\approx$310 MeV and 
$\approx$400 MeV. It then rises from $\approx$400 MeV to a maximum at 
$\approx$650 MeV (see fig.~\ref{fig_55}). This rise reflects the 
$\gamma p\rightarrow\pi\Delta$ threshold smeared by the width of the 
$\Delta$-resonance. It is accompanied by a strong peak at the mass of the 
$\Delta$ in the invariant mass distribution of the $p\pi^+$-pair which is 
absent in the $p\pi^-$-invariant mass. An important contribution is assigned 
to the $\gamma p\rightarrow\Delta^{++}\pi^-$ channel while the 
$\gamma p\rightarrow\Delta^{o}\pi^+$ channel is negligible. The $\Delta\pi$ 
intermediate state could be populated by the decay of a resonance. However, 
a more detailed analysis \cite{Lueke_71} showed that the reaction is dominated 
by the $\Delta$-Kroll-Ruderman term and the pion pole term 
(see fig.~\ref{fig_55}, diagrams upper right corner). 
%
%
%
%

More recent analyses of this reaction 
\cite{Tejedor_96}-\cite{Ochi_97}, 
taking into account the new precise data from the DAPHNE-detector 
\cite{Braghieri_95}, have solidified this picture. Although the level of 
agreement between predictions and data is somewhat different (see fig.
\ref{fig_55}), all models find the reaction dominated by the  
$\Delta$-Kroll-Ruderman term. However, even though the direct contribution 
from higher resonances is negligible, it was pointed out by Oset and 
coworkers \cite{Tejedor_96} that the peak-like structure between 600 and 
800 MeV is due to an interference of the $\Delta$-Kroll-Ruderman term with 
the sequential decay of the D$_{13}$(1520)-resonance via 
$\gamma p\rightarrow \mbox{D}_{13}\rightarrow\Delta\pi$
(diagram lower right corner in fig.~\ref{fig_55}). This allows the extraction 
of the coupling constant of the D$_{13}$-decay into 
$\Delta\pi$ \cite{Tejedor_96b}. At higher photon energies (up to 2 GeV) 
total cross sections and invariant mass distributions were obtained with the 
SAPHIR-detector at ELSA \cite{Klein_96}. The invariant mass distributions 
show clear signals for $\Delta\rightarrow N\pi$ and $\rho\rightarrow\pi^+\pi^-$ 
contributions. 

\begin{figure}[thb]
\hspace*{1.cm}
\begin{minipage}{0.0cm}
{\mbox{\epsfysize=15.8cm \epsffile{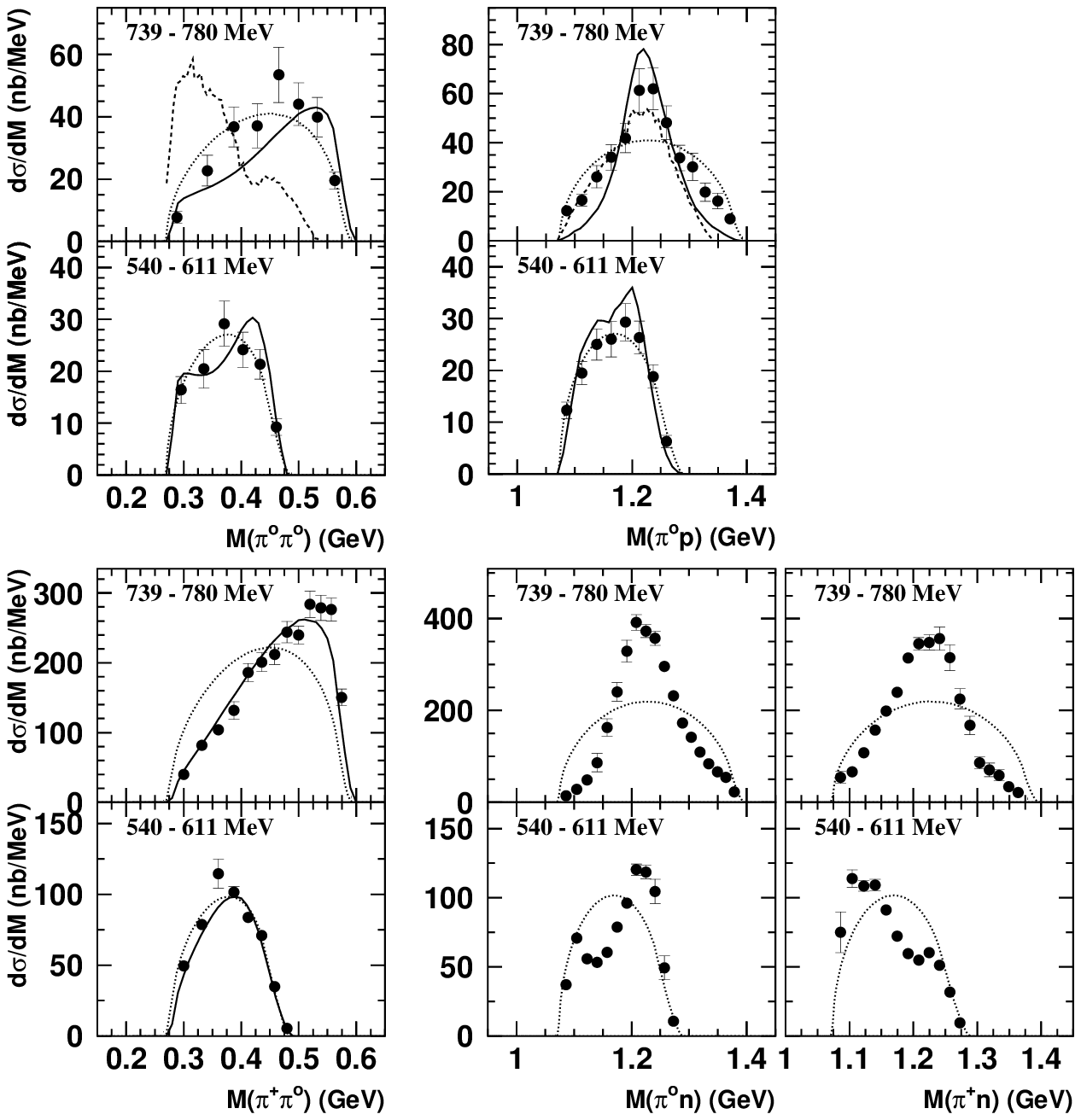}}}
\end{minipage}
\hspace*{11.2cm}
\begin{minipage}{0.cm}
\vspace*{-8.35cm}
{\mbox{\epsfysize=7.7cm \epsffile{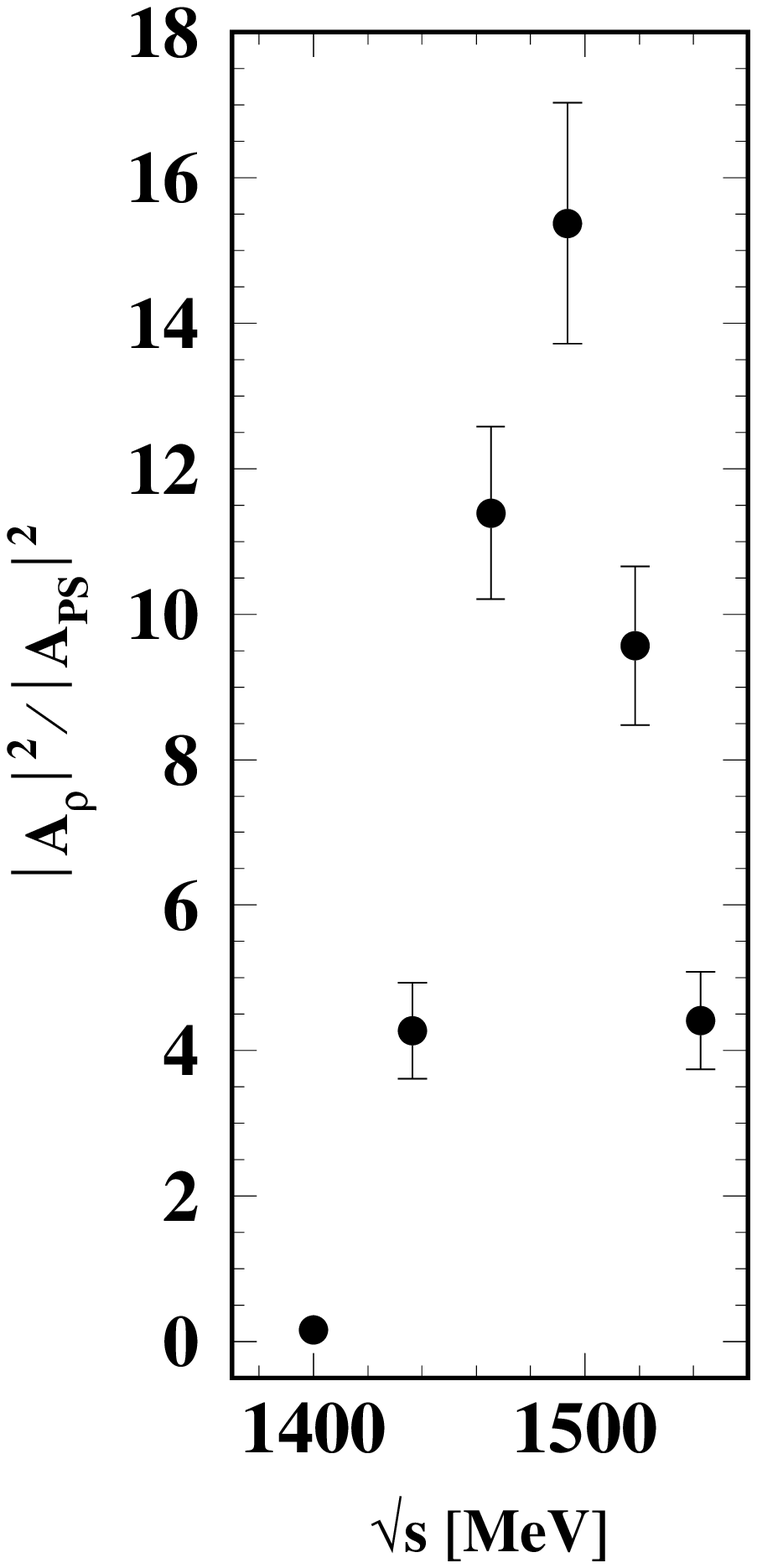}}}
\end{minipage}
\caption{Invariant mass distributions for the reactions 
$\gamma p\rightarrow p\pi^o\pi^o$ and $\gamma p\rightarrow n\pi^+\pi^o$.
Upper part (left and middle): double $\pi^o$ production, data from 
\cite{Wolf_01}. Dotted curves: phase space, dashed: model with dominant 
$\mbox{P}_{11}\rightarrow N\sigma$ contribution \cite{Murphy_96}, solid: 
model with dominant D$_{11}\rightarrow \Delta\pi$ decay \cite{Tejedor_96}. 
Lower part: $n\pi^+\pi^o$ final state, data from \cite{Langgaertner_01}. 
Dotted curves: phase space, full: model including $\rho$ mesons 
\cite{Nacher_01}. Upper right corner: fitted amplitude ratio for $\rho$ 
production and phase space for $n\pi^+\pi^o$ \cite{Langgaertner_01}.   
}
\label{fig_58}       
\end{figure}

The final state with two neutral pions is particularly well suited for the
study of resonance decays. Most background terms are excluded since the photon 
does not couple to the neutral pion and the $\rho$-meson does not decay into a 
pair of neutral pions.
%
%
%
\begin{figure}[hbt]
\begin{minipage}{0.cm}
{\mbox{\epsfysize=8.cm \epsffile{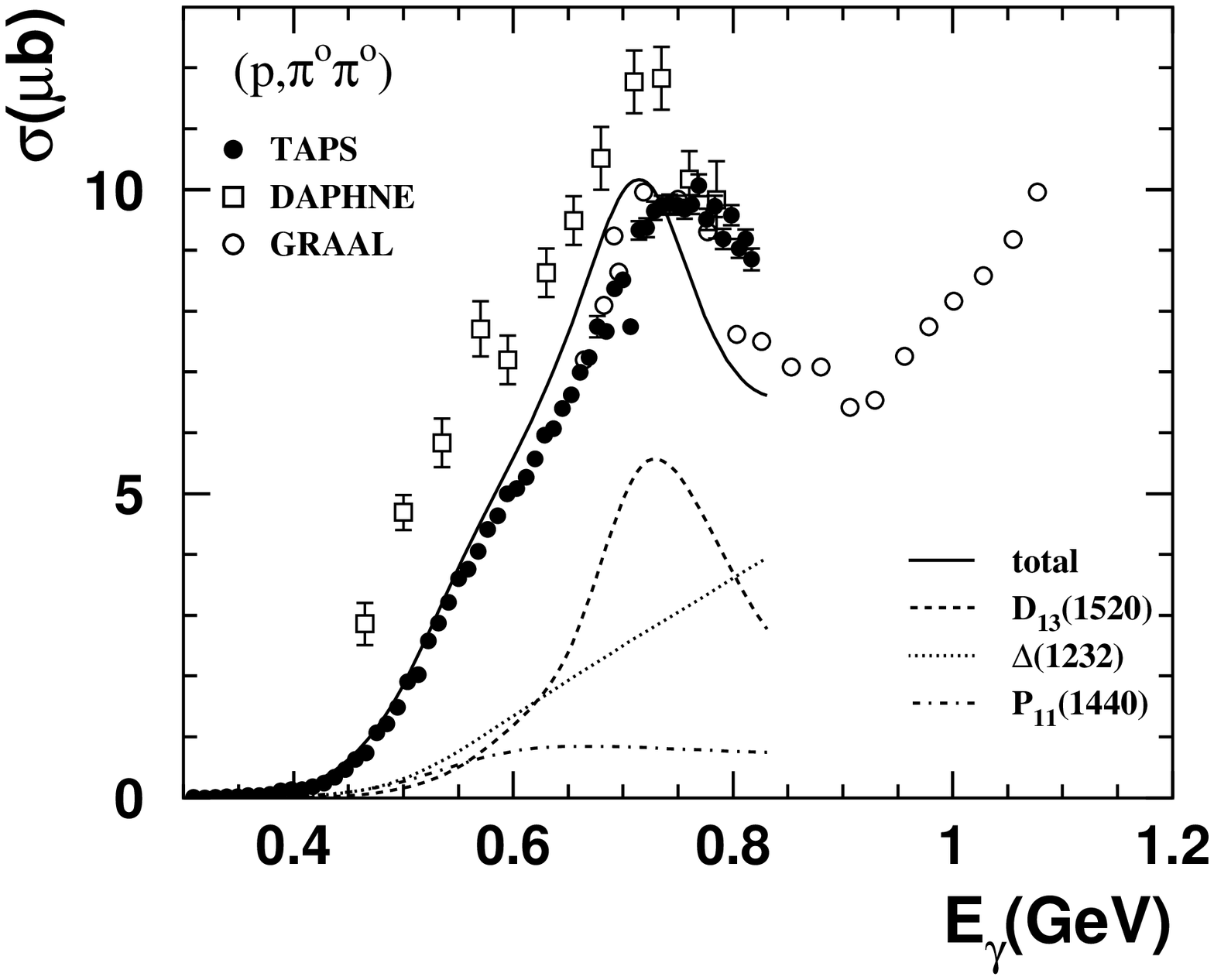}}}
\end{minipage}
\hspace*{10.cm}
\begin{minipage}{0.cm}
\vspace*{-1.5cm}
{\mbox{\epsfysize=2.5cm \epsffile{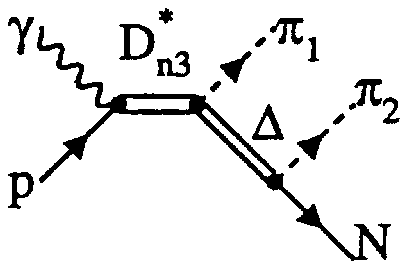}}}\\
\vspace*{0.5cm}
{\mbox{\epsfysize=2.5cm \epsffile{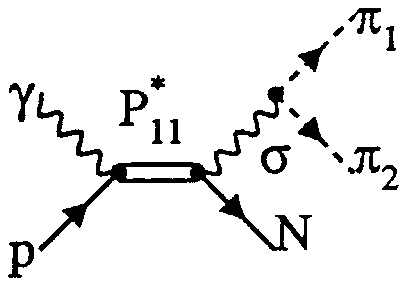}}}
\end{minipage}
\hspace*{4.3cm}
\begin{minipage}{3.55cm}
\caption{Total cross section for $\gamma p\rightarrow p\pi^o\pi^o$.
Data are from references \cite{Braghieri_95,Kotulla_01,Hourany_01}. Curves
are from the Oset model
\cite{Tejedor_96}. Right hand side: dominant reaction mechanisms predicted by
different models (see text).
}
\label{fig_56}       
\end{minipage}
\end{figure}
%
Surprisingly, the two models from refs. \cite{Tejedor_96,Murphy_96} made very 
different predictions. One of them \cite{Tejedor_96} predicted that the 
sequential decay of the D$_{13}$(1520) resonance via 
D$_{13}\rightarrow\pi\Delta\rightarrow\pi\pi N$, the other \cite{Murphy_96}, 
that the decay of the P$_{11}$(1440) resonance via a correlated pair of pions 
in a relative s-wave (P$_{11}\rightarrow\sigma N$) is the dominant process 
(see diagrams in fig.~\ref{fig_56}, right hand side). The total cross section 
was 
measured in several experiments with the DAPHNE \cite{Braghieri_95} and TAPS 
\cite{Haerter_97,Wolf_00,Kotulla_01} detectors in Mainz and at higher incident 
photon energies at GRAAL \cite{Hourany_01}. The total cross section agrees 
better \cite{Braghieri_95,Haerter_96} with the result from the Oset model 
\cite{Tejedor_96} in spite of a systematic discrepancy between the DAPHNE and 
TAPS data. The question was finally solved by the invariant mass distributions 
measured with TAPS (see fig.~\ref{fig_58}). While the $\pi^o\pi^o$-mass 
distribution is explained by phase space, the $p\pi^o$-distribution shows a 
peak at the mass of the $\Delta$(1232). This favors the sequential D$_{13}$ 
decay over the correlated P$_{11}$ decay which predicted a strong shift of the 
$\pi^o\pi^o$ invariant mass towards small values. The contribution of the 
partial channels to the total cross section in the Oset model is shown in 
fig.~\ref{fig_56}. The resonance-like contribution from the sequential decay 
of the D$_{13}$ is dominant.

\begin{figure}[hbh]
\hspace*{0.5cm}
\begin{minipage}{0.cm}
\vspace*{0.9cm}
{\mbox{\epsfysize=8.cm \epsffile{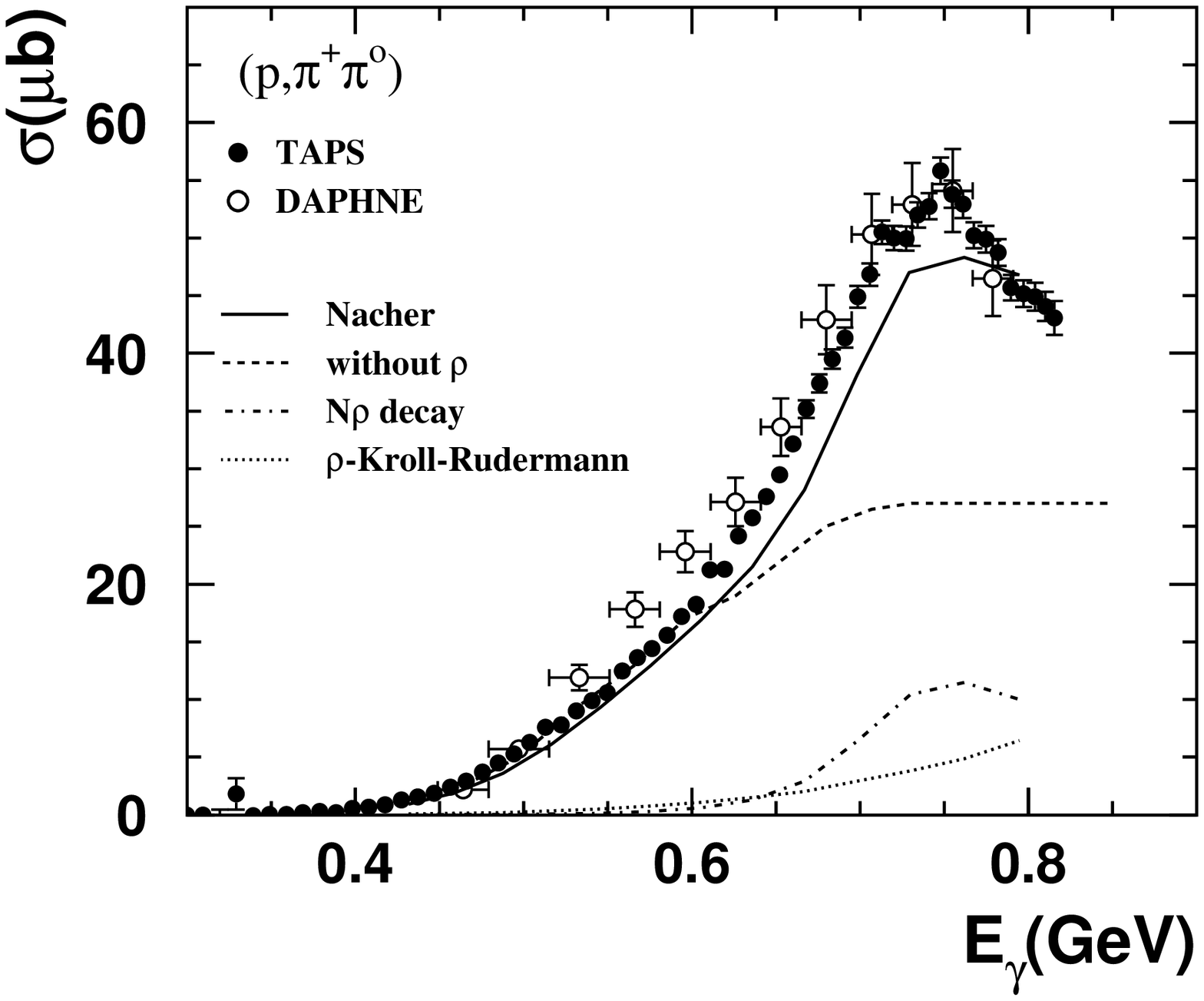}}}
\end{minipage}
\hspace*{10.5cm}
\begin{minipage}{0.cm}
\vspace*{-0.5cm}
{\mbox{\epsfysize=2.4cm \epsffile{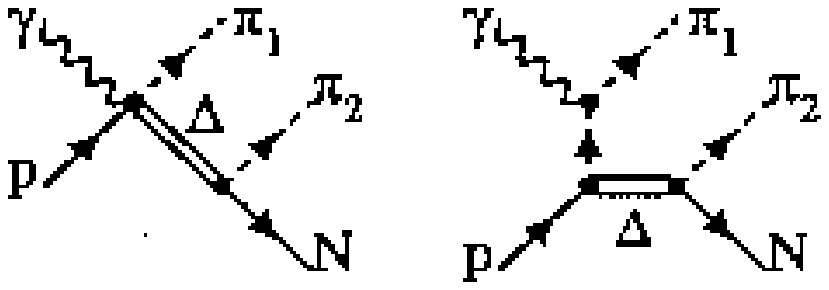}}}\\
\vspace*{0.5cm}
{\mbox{\epsfysize=2.4cm \epsffile{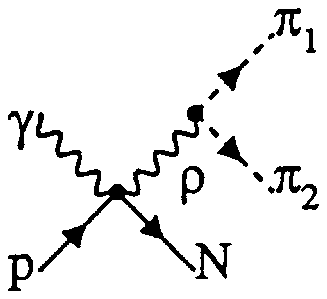}
\hspace*{0.5cm}
\epsfysize=2.4cm \epsffile{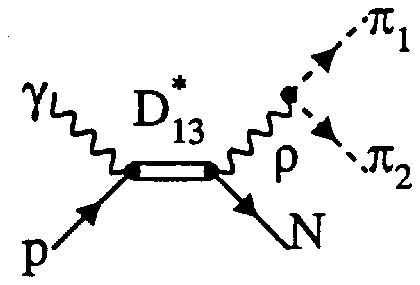}}}
\end{minipage}
\caption{Left: Total cross section of $\gamma p\rightarrow n\pi^o\pi^+$.
Data from: \cite{Braghieri_95,Langgaertner_01}. Curves from: \cite{Nacher_01}.
Right: upper part: leading diagrams, lower part: contribution of the
$\rho$-meson.
}
\label{fig_57}       
\end{figure}

\begin{figure}[hbt]
\begin{center}
\epsfysize=6.5cm \epsffile{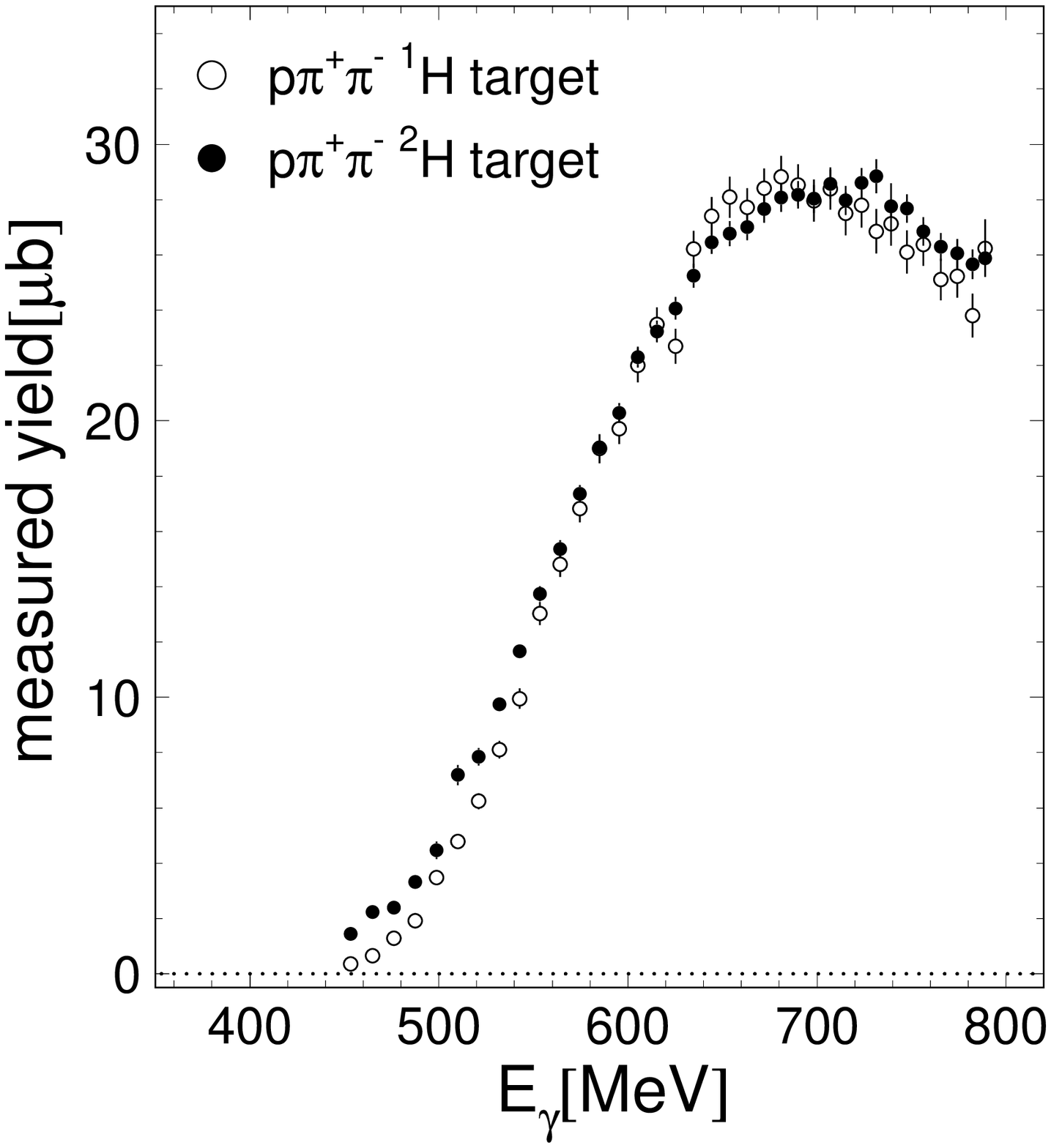}
\epsfysize=6.5cm \epsffile{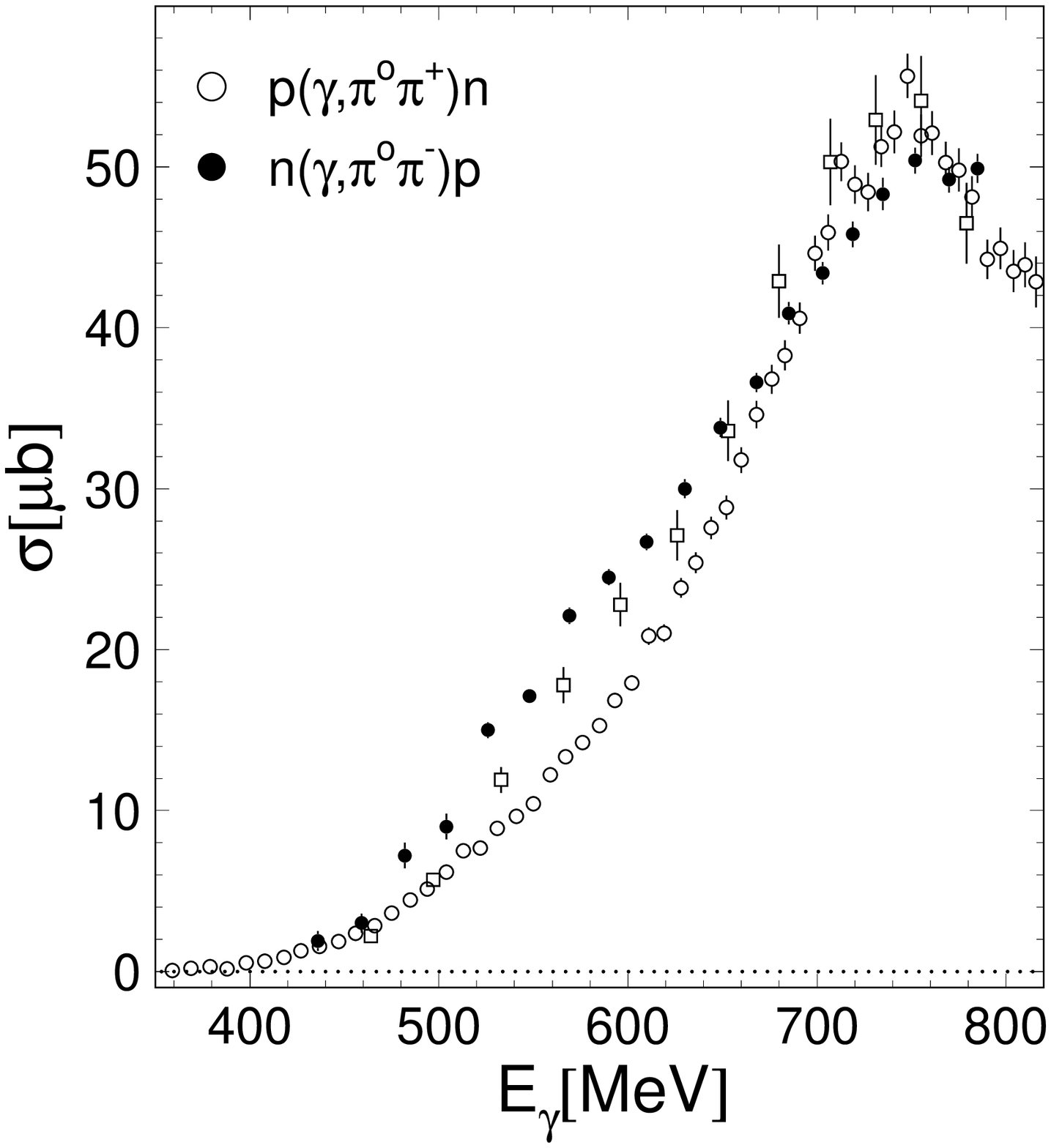}
\epsfysize=6.5cm \epsffile{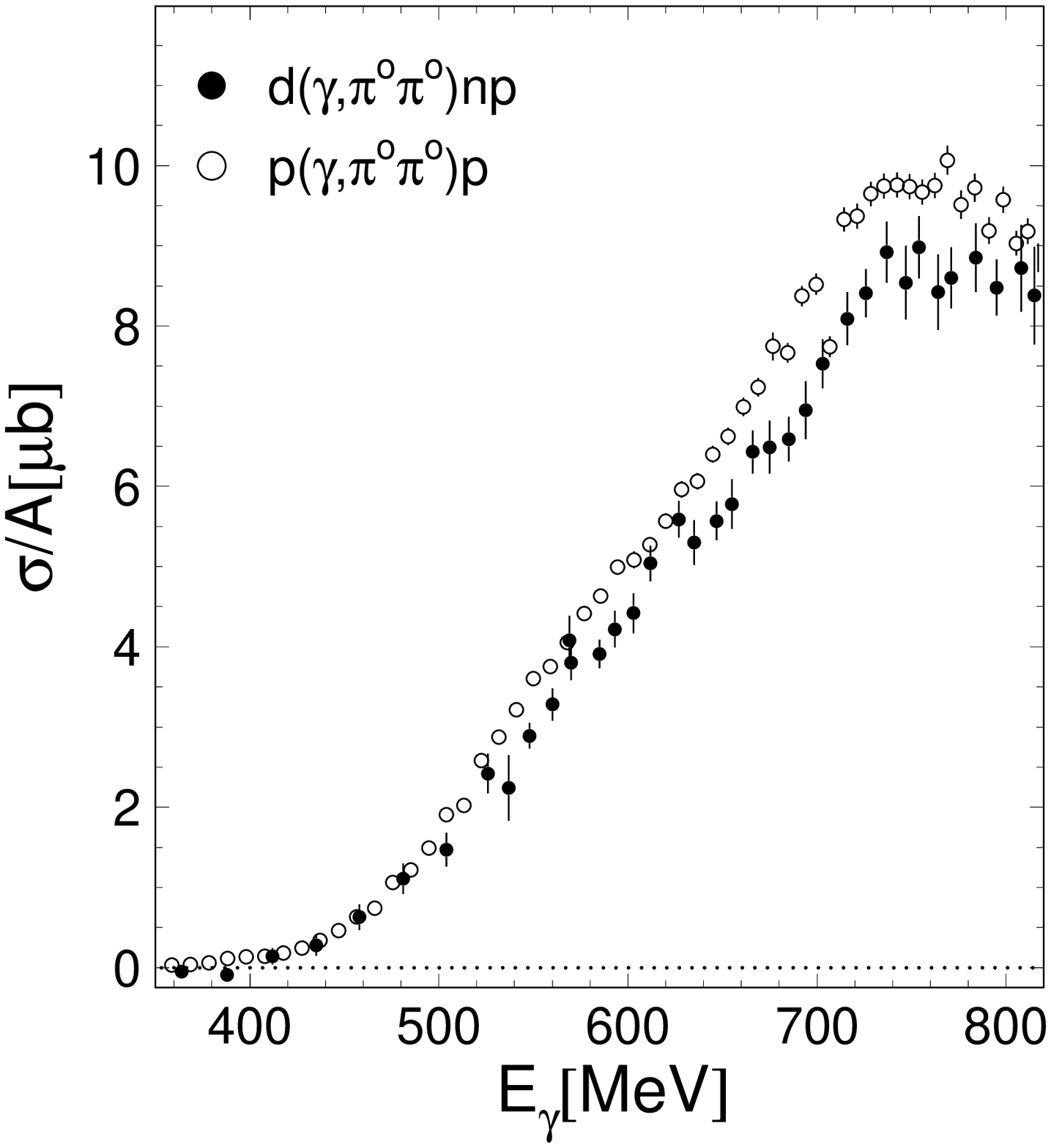}
\caption{Double pion production from the neutron. Left: comparison of
$p\pi^+\pi^-$ from proton and deuteron target (only the yield within the
acceptance of the DAPHNE detector) \cite{Zabrodin_97}. Middle: comparison of
$n\pi^o\pi^+$ \cite{Braghieri_95,Langgaertner_01} with $p\pi^o\pi^-$ 
\cite{Zabrodin_97}. Right: comparison of 2$\pi^o$ from the proton and deuteron. 
}
\label{fig_60}       
\end{center}
\end{figure}

It came as a surprise when the first measurement of the 
$\gamma p\rightarrow n\pi^+\pi^o$ reaction \cite{Braghieri_95} produced a total 
cross section that was strongly underestimated by all model predictions. 
Typical results from the models \cite{Tejedor_96,Murphy_96} predicted a 
behavior of the total cross section corresponding to the dashed curve in 
fig.~\ref{fig_57}. A measurement of the reaction \cite{Langgaertner_01} with 
the TAPS detector confirmed the experimental result. The same problem was found 
for the $\gamma n\rightarrow p\pi^-\pi^o$ reaction which was measured in 
quasifree kinematics from a deuteron target \cite{Zabrodin_97} 
(see fig.~\ref{fig_60}). An important piece was obviously missing in the 
models which predicted that the dominant contribution comes from the 
$\Delta$-Kroll-Rudermann term and the pion pole term, as is the case in the 
double charged channel. 

%
%
%
%
%
%
%
%

Ochi et al. \cite{Ochi_97} suggested a strong contribution of off-shell 
$\rho$-meson decays even at low incident photon energies. This proposal was 
attractive since $\rho$-mesons would mainly contribute to the $\pi^+\pi^o$ and 
$\pi^-\pi^o$ channels. The $\rho^o\rightarrow \pi^o\pi^o$ decay is forbidden, 
and the contribution to $\pi^+\pi^-$ is suppressed since the neutral $\rho$ is 
not produced via the Kroll-Rudermann term. Indeed, the model predicted a 
substantially larger cross section for the mixed charge channels. In other 
aspects it used more simplifications than other models and described the other 
charge states less well. However, this suggestion motivated a careful study of 
the invariant mass distributions of the pion - pion and pion - nucleon pairs 
from this reactions. Zabrodin et al. \cite{Zabrodin_99} measured the 
distributions from the $\gamma n\rightarrow p\pi^-\pi^o$ reaction and found a 
deviation of the pion - pion invariant mass from phase space, showing an 
enhancement at large invariant masses. The analysis was complicated by effects 
of the bound neutron. Subsequently, invariant mass distributions of the 
$\gamma p\rightarrow n\pi^+\pi^o$ reaction were obtained by Langg\"artner et 
al. \cite{Langgaertner_01}. Typical examples are compared in fig.~\ref{fig_58}
to the double $\pi^o$ channel. The comparison is particularly instructive since 
the double $\pi^o$ channel does not have a contribution from the $\rho$.

The $\pi^o n$ invariant mass peaks already at low incident photon energies at 
the $\Delta$ mass. This signal does not appear in the $\pi^+ n$ invariant 
mass at low photon energies. This behavior is expected if, as predicted in 
the models, the process is dominated at low photon energies by the 
$\Delta$-KR- and $\Delta$-pion-pole terms. Since the photon does not couple to 
the neutral pion, the charged pion is produced at the first vertex and the 
neutral pion stems from the subsequent $\Delta^o\rightarrow N\pi^o$ decay, 
giving rise to the $\pi^o n$ invariant mass correlation. At higher incident 
photon energies, sequential decays of $N^{\star}$ resonances may contribute. 
It can be seen from the relevant Clebsch-Gordan coefficients that for the 
D$_{13}\rightarrow\Delta\pi\rightarrow\ N\pi\pi$ reaction both sequences of 
the charged and neutral pion are equally probable, so that the $\Delta$-signal 
may also appear in the $\pi^+ n$ invariant mass. The double $\pi^o$ channel 
has the $\Delta$ signal only at higher incident photon energies from the 
sequential D$_{13}$ decay, since the KR-term is forbidden.

The $\pi^o\pi^o$ invariant mass is similar to phase space behavior, but at the 
higher incident photon energies the $\pi^o\pi^+$ invariant mass is clearly 
shifted to large masses as expected for a contribution of the $\rho^+$ meson. 
The $\pi^o\pi^+$-data were fitted with a simple model assuming only phase space 
and $\rho$-decay contributions \cite{Langgaertner_01}
via:
\begin{equation}
\frac{d\sigma}{dm}\propto
|a(\sqrt{s})+b(\sqrt{s})p_{\pi}(m_{\pi\pi})D_{\rho}(m_{\pi\pi})|^2
ps_{\sqrt{s}\rightarrow\pi\pi N}
\end{equation}
where $p_{\pi}$ is the momentum of the $\pi$ in the $\rho$ rest frame, 
$ps_{\sqrt{s}\rightarrow\pi\pi N}$ is the three body phase space factor, and 
$D_{\rho}$ represents the $\rho$-meson propagator. 

The constants $a$ and $b$ were fitted to the data. The ratio of the matrix 
elements for $\rho$-meson decays and phase space components was calculated via:
\begin{equation}
\frac{|A_{\rho}|^2}{|A_{ps}|^2}=
\frac{\int|b(\sqrt{s}) p_{\pi}(m_{\pi\pi})D_{\rho}(m_{\pi\pi})|^2dm_{\pi\pi}}
{\int|a(\sqrt{s})|^2dm_{\pi\pi}}\;.
\end{equation}

The result of the ratio of the matrix elements is shown in fig. \ref{fig_58} 
(upper right corner). This is the ratio of the matrix elements without phase 
space factors. The relative contribution of the $\rho$-decay matrix element 
peaks near the position of the D$_{13}$ resonance hinting at a significant
D$_{13}\rightarrow N\rho$ contribution to $\pi^o\pi^+$-photoproduction.
Such a contribution would be interesting in view of the unexplained suppression
of the second resonance bump in the total photoabsorption on nuclei. Since in 
this region the $D_{13}$ resonance has the largest coupling to the initial 
photon - nucleon state, it has been argued that an in-medium broadening of 
this resonance is a likely cause for the observations. The broadening could 
arise from a coupling of the resonance to the $N\rho$ final state since it 
is predicted \cite{Klingl_97} that the  $\rho$-meson itself is broadened 
appreciably in the nuclear medium. The extraction of a possible 
D$_{13}\rightarrow N\rho$ contribution requires a detailed model for the 
reaction $\gamma p\rightarrow n\pi^o\pi^+$ taking into account interference 
effects between all contributions. In view of the new experimental results, 
the Oset group has updated the model \cite{Nacher_01}, now  including $\rho$ 
contributions correctly. The new calculation is compared to the total cross 
section and the invariant mass distributions demonstrating that the inclusion 
of the $\rho$ is essential.

%
%
%
\begin{figure}[thb]
\begin{minipage}{0.0cm}
\epsfysize=9.cm \epsffile{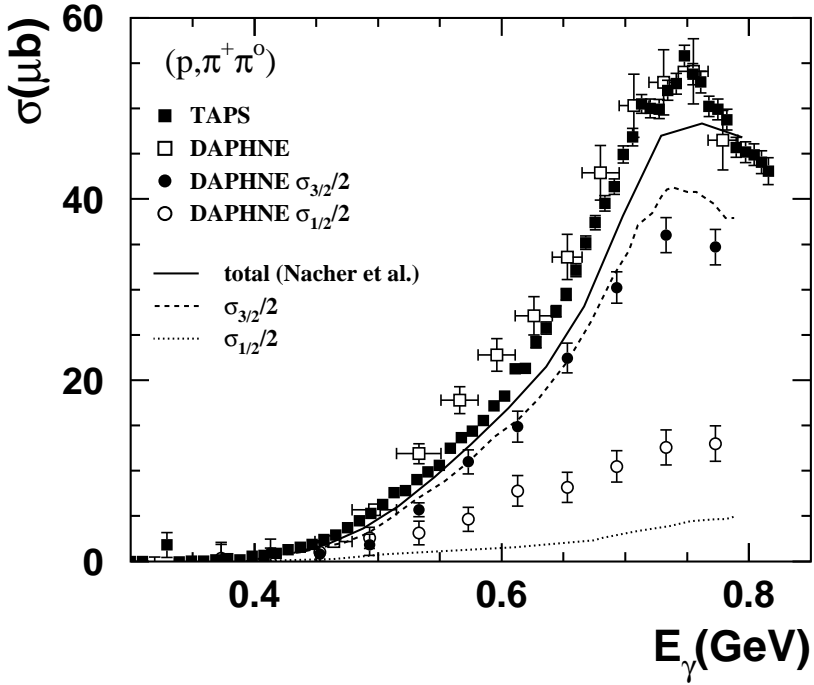}
\end{minipage}
\hspace*{12.0cm}
\begin{minipage}{5.5cm}
\caption{Helicity dependence of the $\gamma p\rightarrow n\pi^o\pi^+$
reaction. Shown are the total cross section and the $\sigma_{3/2}$ and
$\sigma_{1/2}$ partial cross sections. Data from are from 
\cite{Zabrodin_97,Langgaertner_01,Ahrens_GDH_03}. 
Curves (model predictions) are from Nacher et al. \cite{Nacher_01,Nacher_02}.
}
\label{fig_59}       
\end{minipage}
\end{figure}
%
The invariant mass distributions have clarified the relative importance of
the different reaction mechanisms involving resonance decays (sequential decay 
with an intermediate $\Delta\pi$ state, emission of $\rho$ meson). However, 
they cannot assign these reactions to a specific resonance. For example, the 
large importance of the D$_{13}$ and the negligible contribution of the 
S$_{11}$ in the models results from the photon couplings and decay widths 
which are input parameters to the calculations. However, recently the GDH 
collaboration has measured the helicity dependence of the cross section for 
the $n\pi^o\pi^+$ final state \cite{Ahrens_GDH_03}. The result 
(see fig.~\ref{fig_59}) shows that 
most of the resonance structure occurs in the helicity $\nu$=3/2 channel.
The $\nu$={1/2} channel, where a contribution from the S$_{11}$ would show up, 
has a flatter energy dependence and contributes less than 30\% to the total 
cross section. The model predictions of Nacher et al. \cite{Nacher_02} agree 
qualitatively with the distribution of the strength on helicity 3/2 and 1/2
while underestimating the $\nu$=1/2 contribution.
  
The isospin channels involving neutron targets have been studied in quasifree 
kinematics on the deuteron (see fig.~\ref{fig_60})
\cite{Haerter_97}-\cite{Kleber_00}. 
A comparison of the yields 
for the $p\pi^+\pi^-$ final state from the free proton and quasifree proton 
bound in the deuteron \cite{Zabrodin_97} demonstrates that nuclear effects are 
negligible except for Fermi smearing. The total cross sections for the 
$p(\gamma, \pi^+\pi^o)n$ \cite{Braghieri_95,Langgaertner_01} and 
$n(\gamma, \pi^-\pi^o)p$ reactions are almost equal, and the cross section of 
$d(\gamma, \pi^0\pi^o)np$ is almost twice as large as for 
$p(\gamma, \pi^o\pi^o)p$. For double $\pi^o$ production, the cross sections 
for protons and neutrons must be similar. Altogether, a large isospin 
dependence is not found which is important for the interpretation of the 
nuclear data.
%
%
%
%

\section{Conclusion and Perspectives}

The use of meson photoproduction reactions for the investigation of baryon
resonances aims at a better understanding of non-perturbative QCD and has 
developed rapidly over the last decade. In this review, we have concentrated 
on the lowest-lying states of the nucleon which have been studied in much 
detail. The experimental precision that has been achieved recently has been 
possible due to the impressive progress in accelerator and detector technology.
Among these are the measurement of the $E2$ admixture in the excitation of the 
$\Delta$ resonance and the investigation of the magnetic moment of the $\Delta$.
The detailed investigation of $\eta$ photoproduction resulted not only in 
precise coupling constants for the \s ~resonance but also allowed to extract a 
decay branching ratio of the \d ~as small as 0.06\%. Finally, the understanding 
of the reaction mechanisms for double pion production saw significant progress 
due to precise measurements of the different isospin channels. 
 
What can we expect next? The next generation of experiments concentrating on 
the low-lying resonances will see a further advance in sensitivity. The advance 
will mainly be due to the use of 4$\pi$ detection systems like the combinations 
of the TAPS detector with the Crystal Barrel in Bonn or with the Crystal Ball 
in Mainz. The combination of 4$\pi$ detectors with linearly and circularly
polarized photon beams as well as polarized targets will provide access to a 
wealth of new observables comprising a powerful tool for the extraction of 
specific resonance properties. The measurement of the magnetic moment of the
$\Delta$ and also of the \s ~will certainly profit. Up to now, model independent
multipole analyses have only been possible for pion photoproduction. Within the 
foreseeable future, $\eta$ photoproduction should reach the same stage as 
more and more polarization observables are determined.

We have only briefly touched upon the problem of `missing' resonances. The 
search for the many predicted but yet unobserved nucleon states will gain 
increasing importance as the modern facilities can operate in the necessary 
energy range. 
%
%
%
\begin{figure}[hbt]
\centerline{\epsfysize=7.0cm \epsffile{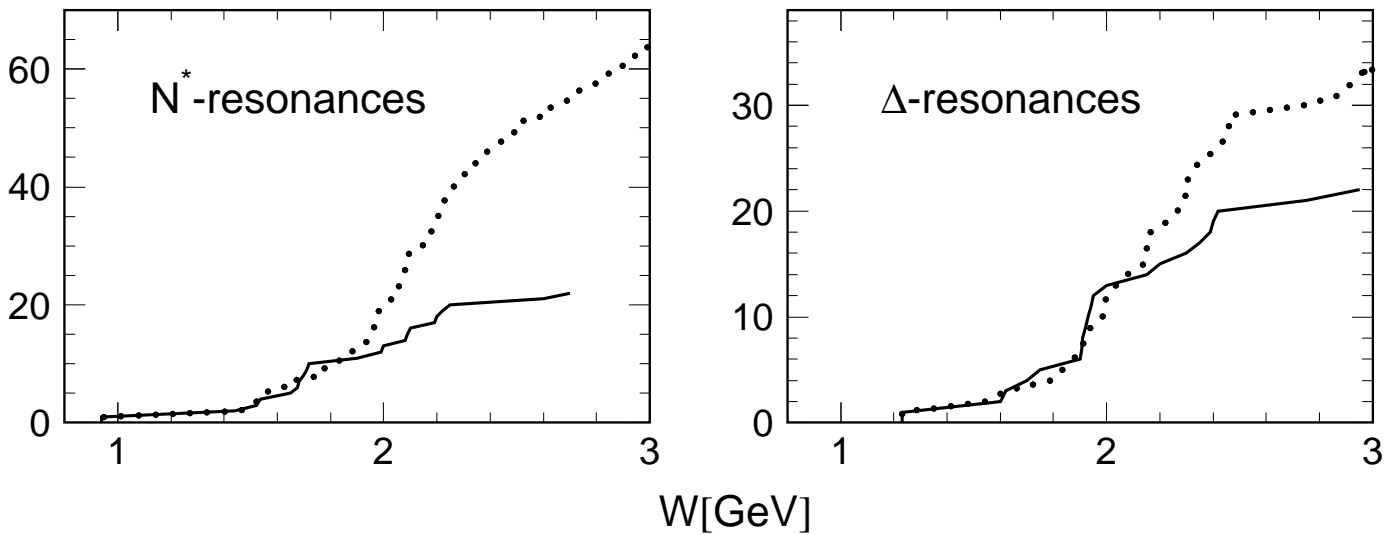}}
\caption{Comparison of predicted \cite{Capstick_94} and observed \cite{PDG}
numbers of nucleon resonances.
}
\label{fig_61}       
\end{figure}
%
As schematically shown in fig.~\ref{fig_61}, the region of interest starts at 
resonance masses of 2 GeV corresponding to photon energies of roughly 1.7 GeV 
which are available at CLAS and at ELSA. It is in this energy regime that the 
discrepancy between the number of observed states and the number of predicted 
states sets in. What is the best strategy to identify new states? We have 
discussed in detail that the extraction of resonance contributions is 
problematic even in the second resonance region. The problems at higher 
energies are expected to be yet more severe in spite of the availability of 
high precision experiments. The development of reaction models, allowing the 
well-defined extraction of resonance contributions, will prove to be a 
challenge. However, it seems neither necessary nor feasible to provide a 
`complete' experimental level scheme of the nucleon. It would already add to 
the success of the quark model if some of those resonances could be identified 
which are predicted to have large decay branching ratios into channels other 
than $N\pi$. 

\begin{figure}[thb]
\centerline{
\epsfysize=8.5cm \epsffile{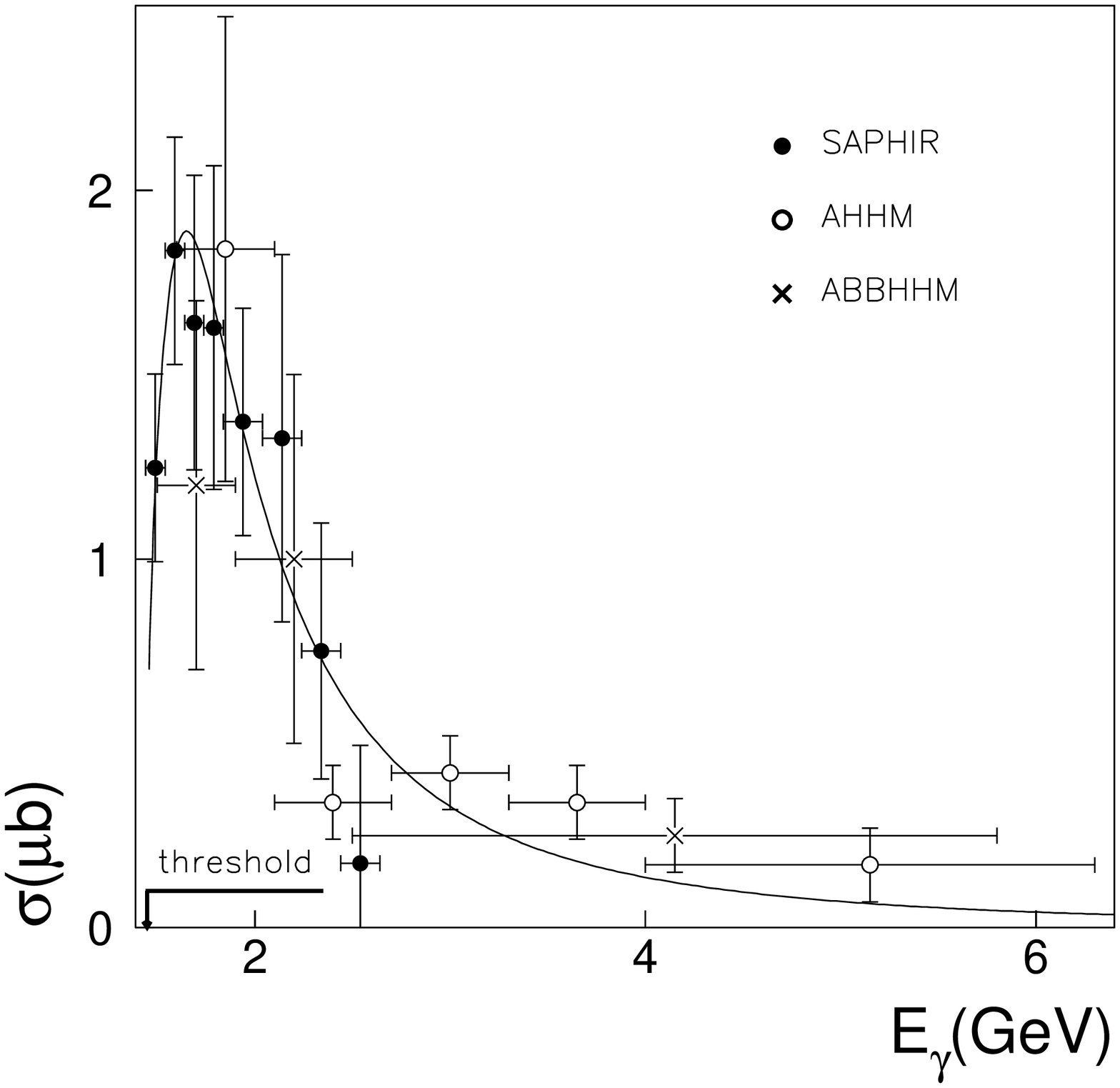}
\epsfysize=8.5cm \epsffile{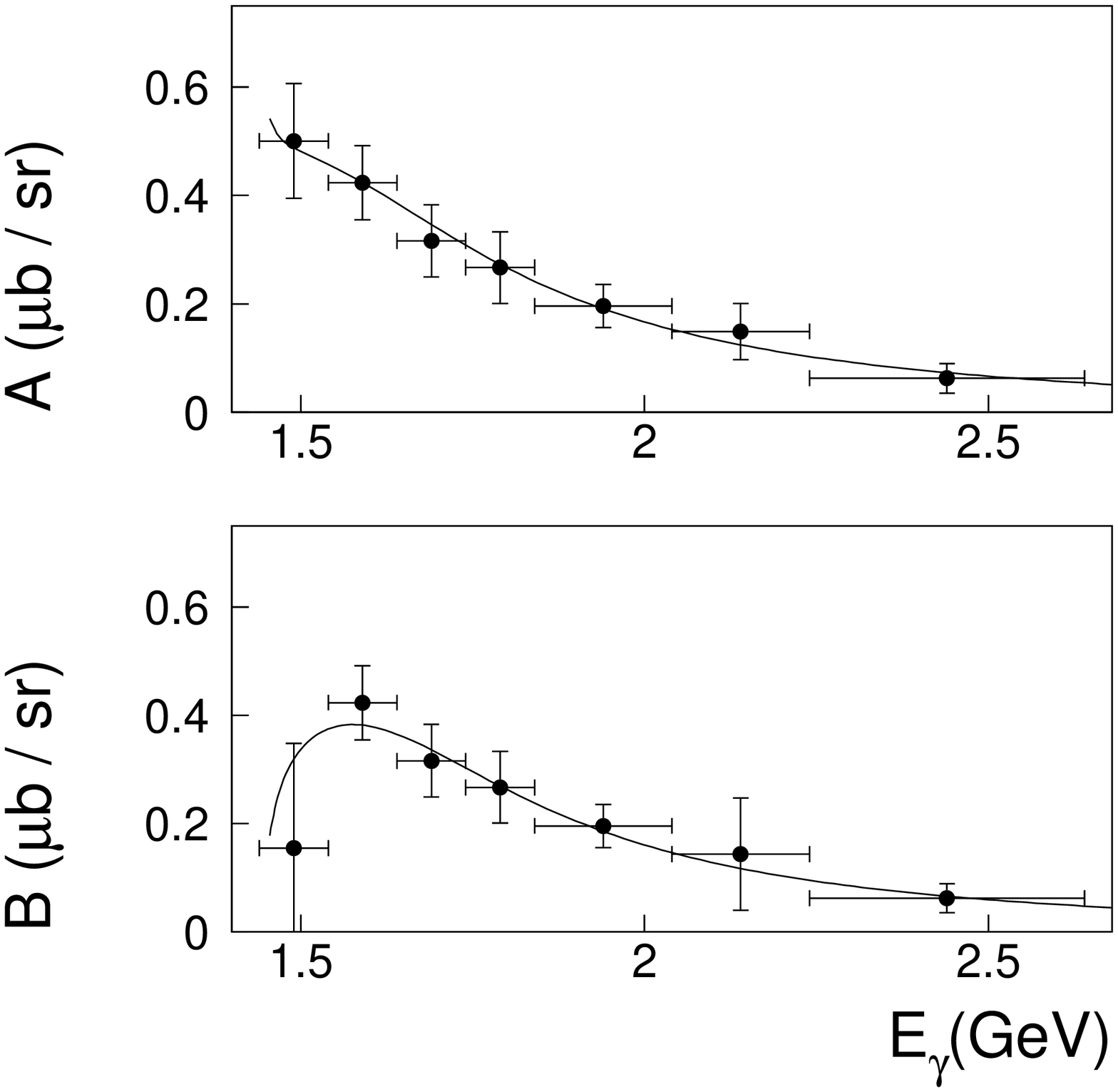}}
\caption{The reaction $p(\gamma ,\eta ')p$ at threshold \cite{Ploetzke_98}. 
Left hand side: total cross section. Data are from SAPHIR \cite{Ploetzke_98} 
(filled circles) and older data from \cite{ABBHHM_68,Struczinski_76}. The curve 
is a fit to the data. Right hand side: fitted coefficients $a, b$ of the 
angular distributions. Curves are fits with Breit-Wigner resonances. 
}
\label{fig_62}       
\end{figure}

One strategy is to concentrate on the threshold regions of heavier mesons. 
Close to threshold, only few partial waves contribute, reducing the necessary 
complexity of the models. There has already been an attempt to exploit the 
$\eta'$ threshold region \cite{Ploetzke_98}. The cross section for 
$\gamma p\rightarrow p\eta'$ rises rapidly from threshold at 
$E_{\gamma}$=1.45 GeV to a maximum at $\approx$1.8 GeV
(see fig.~\ref{fig_62}, left hand side). This might indicate a situation 
similar to $\eta$ photoproduction at threshold. 
%
%
%
%
However, the angular distributions close to threshold are not isotropic but 
show a strong forward - backward asymmetry. This is shown at the right hand 
side of fig.~\ref{fig_62} where the coefficients $a, b$ of fits of the angular 
distribution with the ansatz eq.~\-(\ref{eq:eta_diff}) are plotted versus the 
incident photon energy (the coefficient $c$ is consistent with zero within the 
statistical accuracy). In contrast to $\eta$-photoproduction, the 
$b\times \mbox{cos}(\Theta)$-term is comparable to the constant term $a$. The 
simplest interpretation given in \cite{Ploetzke_98} in terms of resonance 
contributions is an interference between an S$_{11}$ and a P$_{11}$ resonance. 
A fit of the data with Breit-Wigner parameterizations of the two resonances is 
shown in the figure. Parameters of these resonances are quoted in 
\cite{Ploetzke_98}. However, the data have limited statistical accuracy and 
the systematic uncertainty, in particular of the absolute normalization, is 
substantial (other analyses of the SAPHIR data gave a peak cross section 
around 0.5$\mu b$ \cite{Link_00}). More precise  measurements of the angular 
distributions and of polarization observables will be soon available.
 
Similar studies are underway for the $\omega$ and the $\Phi$ meson.
In the first case, evidence for a strong contribution of the P$_{11}$(1720)
resonance was found \cite{Barth_03}, while in the second case resonance
contributions could not be identified \cite{Barth_03a}. 

\begin{figure}[thb]
\centerline{\epsfysize=7.5cm \epsffile{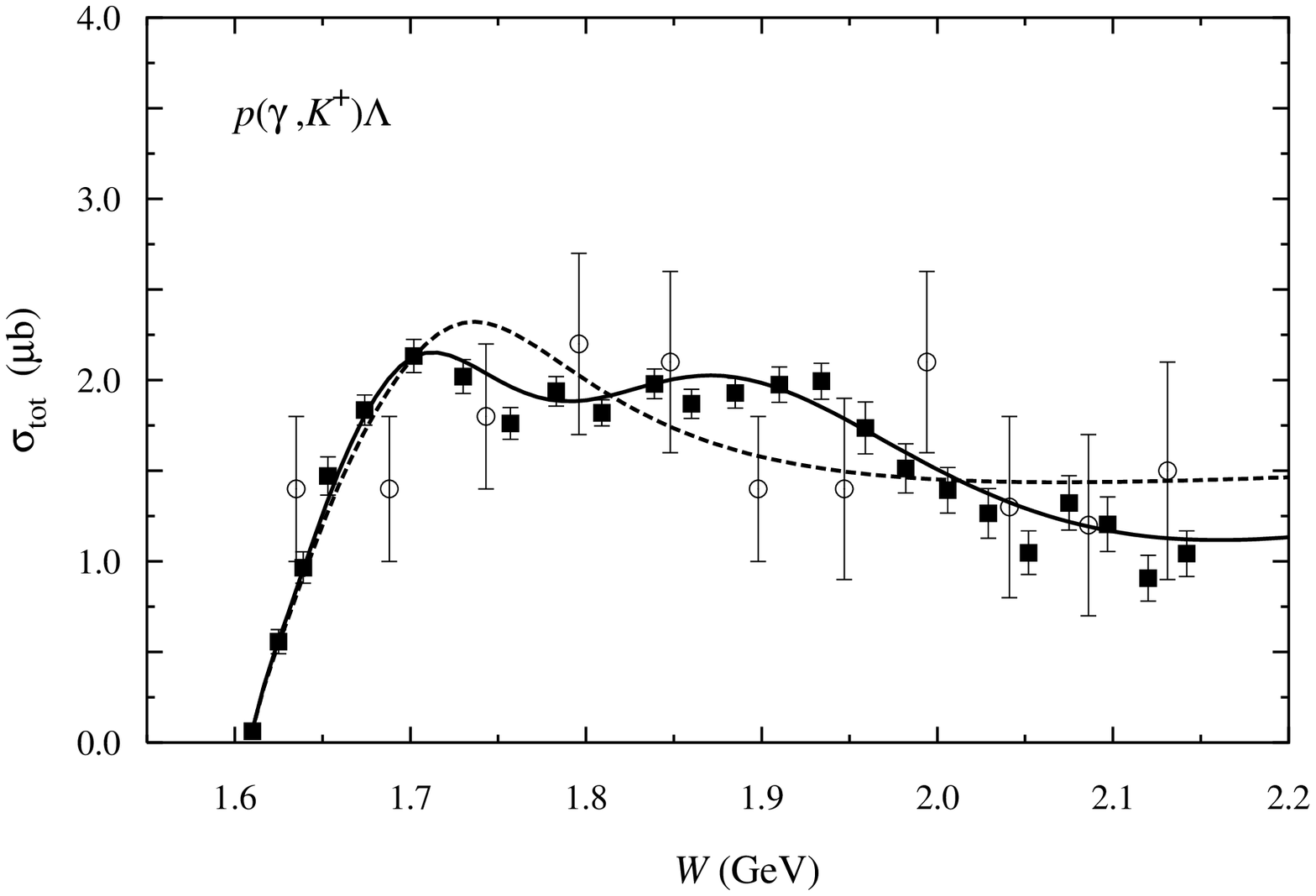}
\epsfysize=7.8cm \epsffile{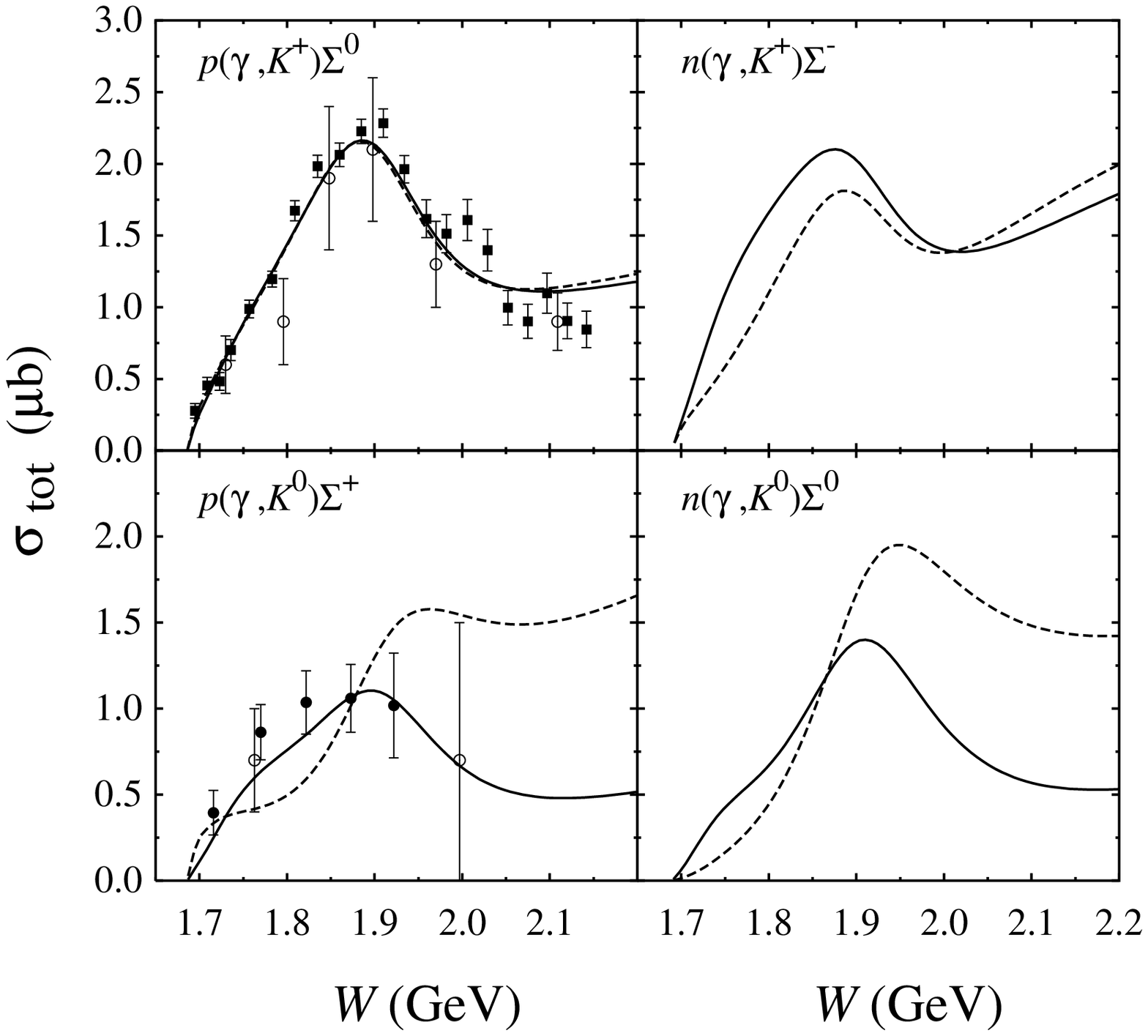}}
\vspace*{-0.7cm}
\caption{Final states with open strangeness \cite{Mart_00a,Mart_00b}.
Left hand side: Total cross section for $p(\gamma ,K^+)\Lambda$. 
Data: SAPHIR \cite{Tran_98} (solid squares),
ABBHHM collaboration \cite{ABBHHM_69a} (open squares). 
Model fits: Mart et al.\cite{Mart_00a}, full and dashed lines: with and without
D$_{13}$(1960) resonance. Right hand side: Total cross section for $K\Sigma$
photoproduction. Data: SAPHIR \cite{Tran_98,Goers_99} (solid symbols), 
ABBHHM \cite{ABBHHM_69a} (open symbols). Model fits: Mart
\cite{Mart_00b}, full and dashed lines: with and without P$_{13}$(1720)
resonance.
}
\label{fig_63}       
\end{figure}

Also, there are recent advances in photoproduction of non-strange nucleon 
resonances in reactions with open strangeness.  The final states $K\Lambda$ 
and $K\Sigma$ have been studied at SAPHIR in Bonn 
\cite{Tran_98,Goers_99,Glander_03}. Further results are expected from CLAS at 
JLab, from LEPS at SPring8, and from the Crystal Barrel and TAPS at ELSA. 
Due to isospin selection, $K\Lambda$ can only couple to $I=1/2$ $N^{\star}$ 
resonances while $N^{\star}$ and $\Delta$ resonances can decay into $K\Sigma$. 
In the quark model, such decays have been investigated by Capstick and Roberts 
\cite{Capstick_98}. They predict that both channels have at least the potential
to confirm weakly established resonances around excitation energies of 2 GeV. 
The data from SAPHIR are compared in fig.~\ref{fig_63} to calculations
by Bennhold and Mart \cite{Mart_00a,Mart_00b} in the framework of an isobar
model. In both cases, the reactions are dominated by resonance contributions.
The total cross section for the $K\Lambda$ channel exhibits a structure around 
resonance masses of 1.9 GeV, which was not obvious in older data with inferior
statistical quality. This structure is not reproduced by the model when 
only well established nucleon resonances are included 
(dashed line in fig.~\ref{fig_63}, left hand side). Much better agreement 
is achieved when a further D$_{13}$ resonance is included (solid line). A 
D$_{13}$ resonance with the necessary properties (relative strong coupling to 
the $K\Lambda$ channel and to the photon) is predicted by the quark model 
\cite{Capstick_98} at an excitation energy of 1960 MeV. Furthermore, there is 
circumstantial experimental evidence for such a state from pion induced 
$K\Lambda$ production \cite{Saxon_80,Bell_83}. Nevertheless, more data, in 
particular polarization observables, will be necessary to establish the 
contribution of this state.
%
%
%
\vspace*{-0.5cm}
%

In case of the $K\Sigma$ channel, cross section data with reasonable statistical
quality is presently only available for the $K^+\Sigma^o$ final state 
(see fig.~\ref{fig_63}, right hand side). An interpretation in terms
of resonance contributions with only one of the four possible isospin channels
is certainly ambiguous. As an example fig.~\ref{fig_63} shows two model fits
of the reaction \cite{Mart_00b} which do include (solid lines), or do not 
include (dashed lines) a contribution of the P$_{13}$(1720) four-star resonance.
The fit quality for the  $K^+\Sigma^o$ final state is almost the same while 
large differences are predicted for the other channels. The available data for 
the $K^o\Sigma^+$ channel seem to favor a coupling of this resonance to the
$K\Sigma$ channel, but are not yet sufficiently precise to warrant a final 
conclusion. An investigation of the threshold behavior of this channel could 
shed new light on the structure of the low-lying S$_{11}$ resonances. We have 
discussed the peculiar decay pattern of the S$_{11}$(1535) as compared to the 
S$_{11}$(1650) in the context of $\eta$-photoproduction. Kaiser et al. 
\cite{Kaiser_95,Kaiser_97} suggested a $K\Sigma$ quasibound state as 
possible explanation for the properties of the S$_{11}$(1535). Li and 
Workman \cite{Li_96} have argued that such a configuration must be strongly 
mixed with a three-quark state in order to explain properties like the $Q^2$
dependence of the helicity amplitude. If this mixing occurs, a third S$_{11}$
resonance should exist close to the other two. Due to its structure it
should couple strongly to the $K\Sigma$ channel. The final states with
neutral kaons would be most sensitive to a resonant s-wave behavior at 
threshold since, as discussed for pion photoproduction, the leading Born terms 
are suppressed. Attempts to study these reactions are under way at ELSA and
SPring8.

\newpage
%
%
%
\begin{figure}[thb]
\begin{minipage}{6.0cm}
\epsfysize=6.cm \epsffile{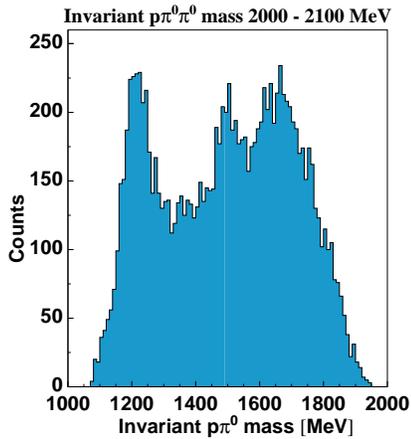}
\caption{Invariant mass distribution of the $p\pi^o$ pairs observed in double
$\pi^o$ production from initial states in the range $\sqrt{s}\approx$ 2000
-2100 MeV \cite{Junkersfeld_02}. 
}
\label{fig_64}       
\end{minipage}
\end{figure}
%

\begin{figure}[hbt]
\centerline{
\epsfysize=5.0cm \epsffile{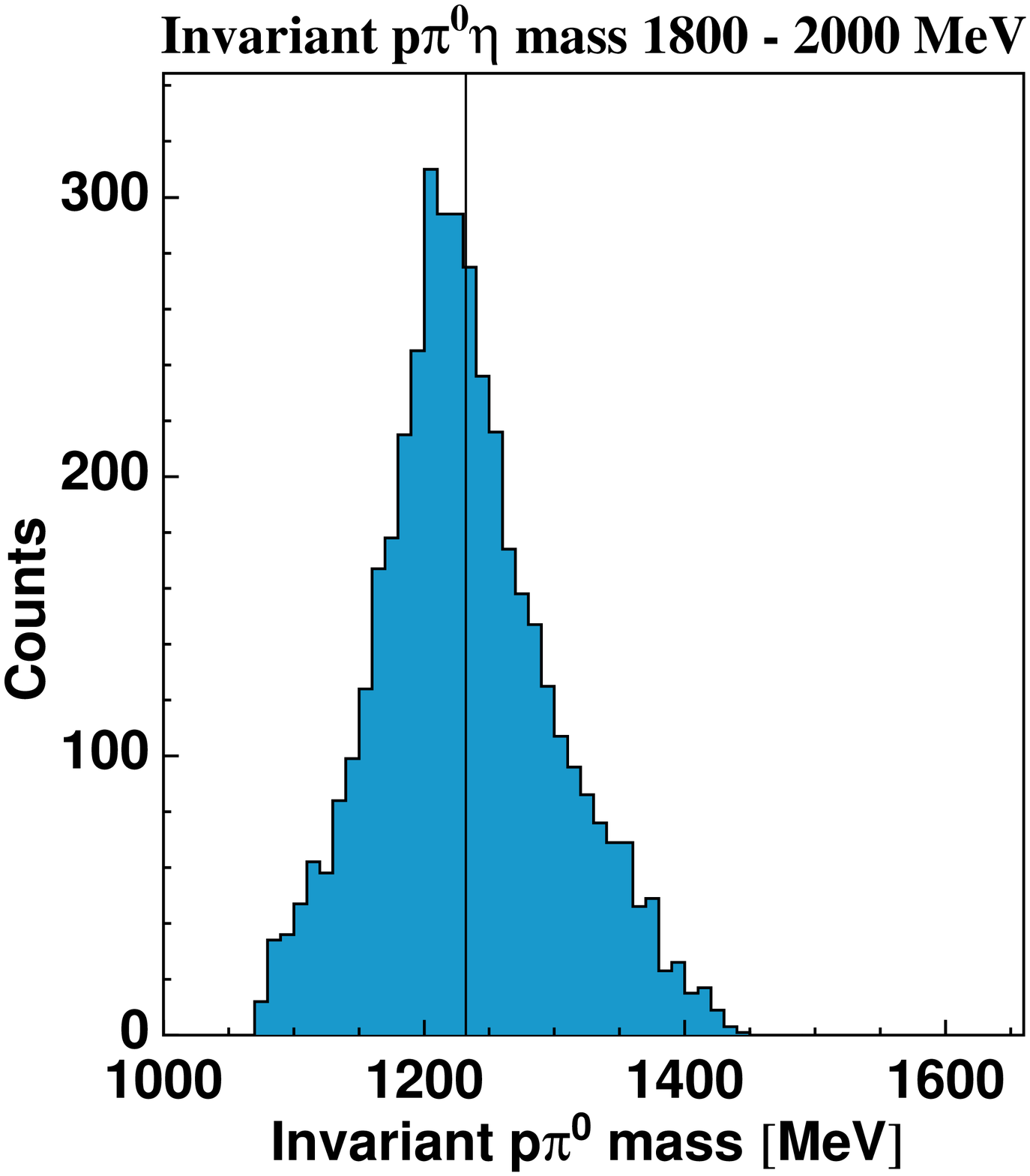}
\epsfysize=5.0cm \epsffile{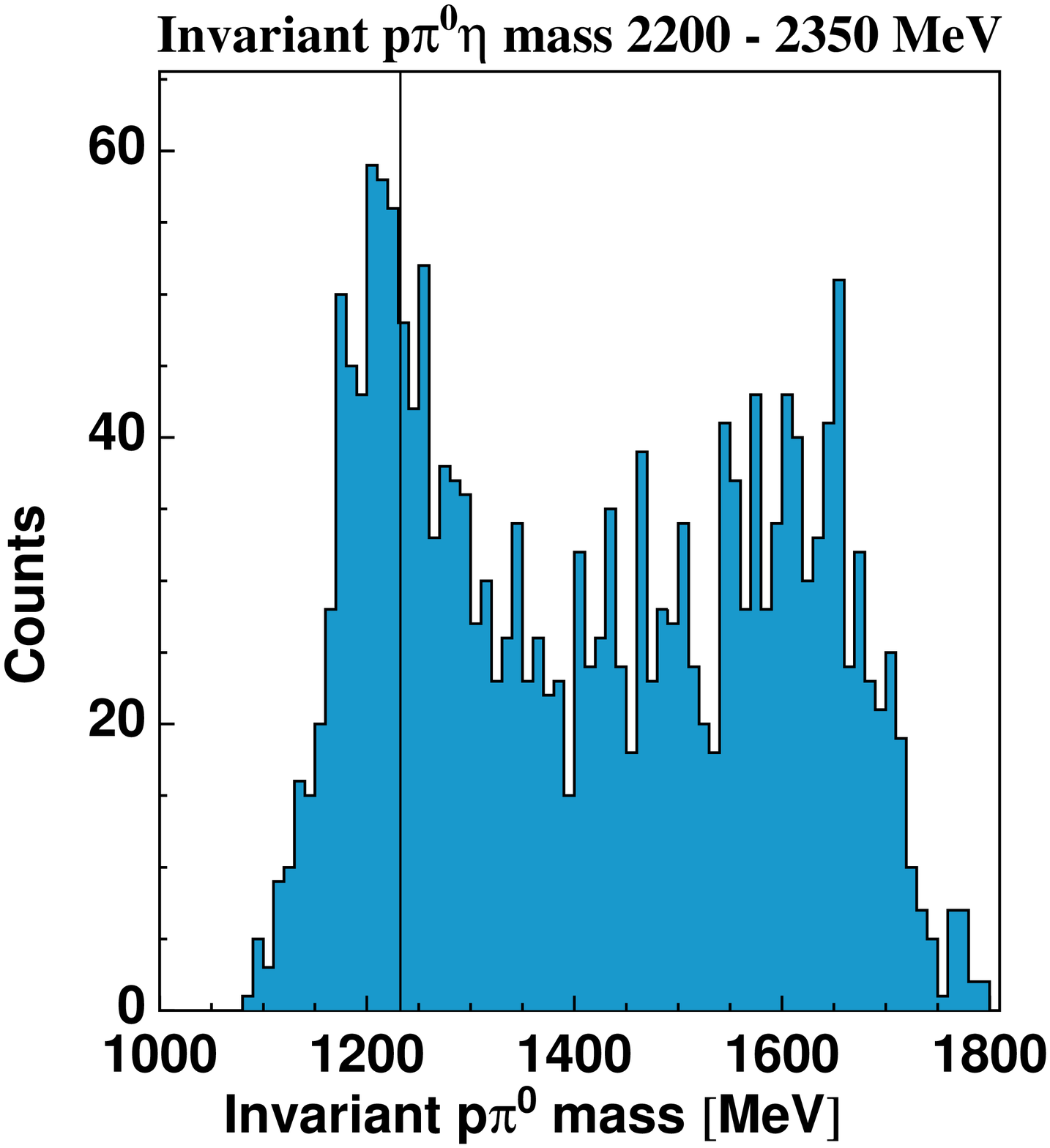}
\epsfysize=5.0cm \epsffile{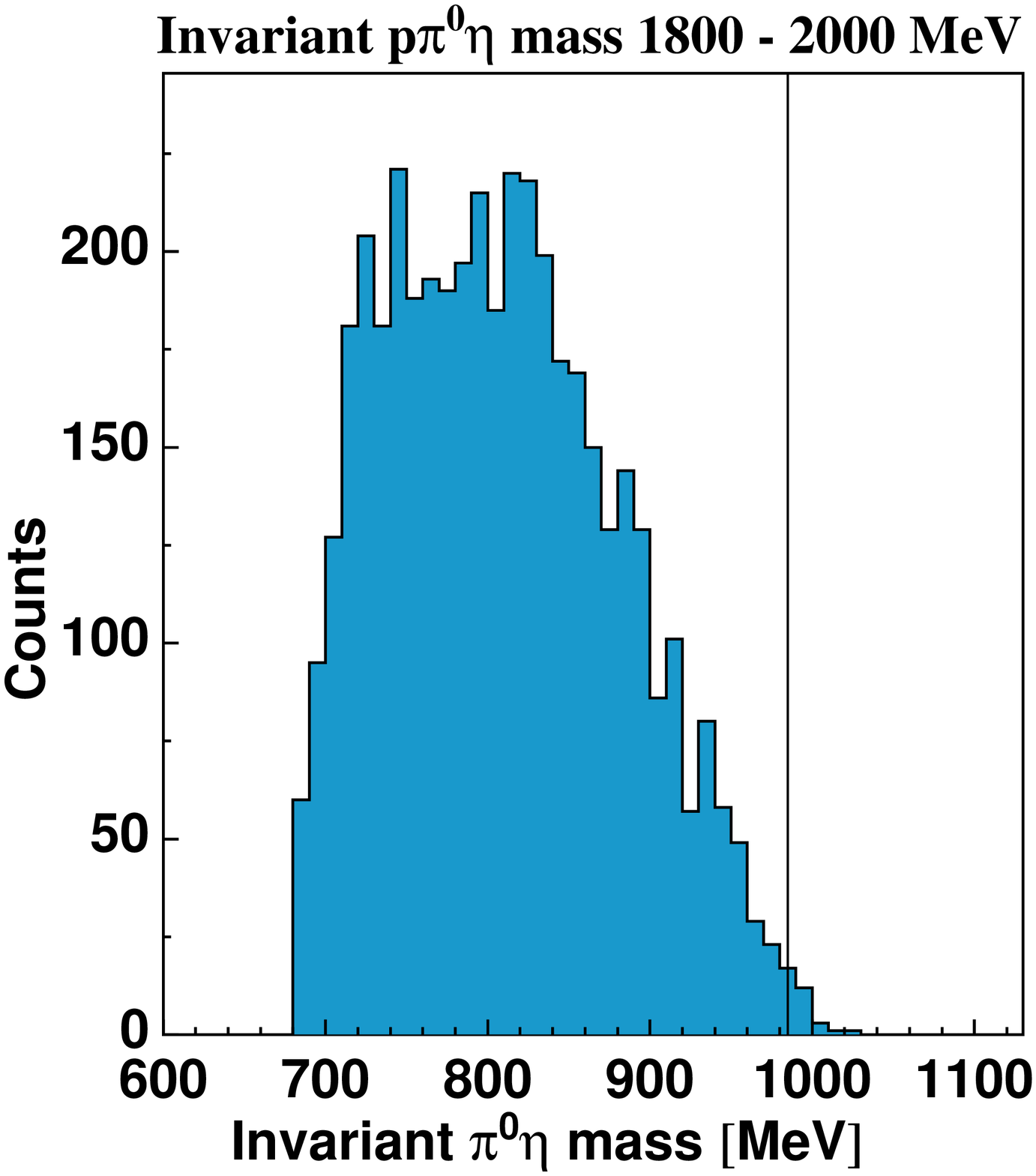}
\epsfysize=5.0cm \epsffile{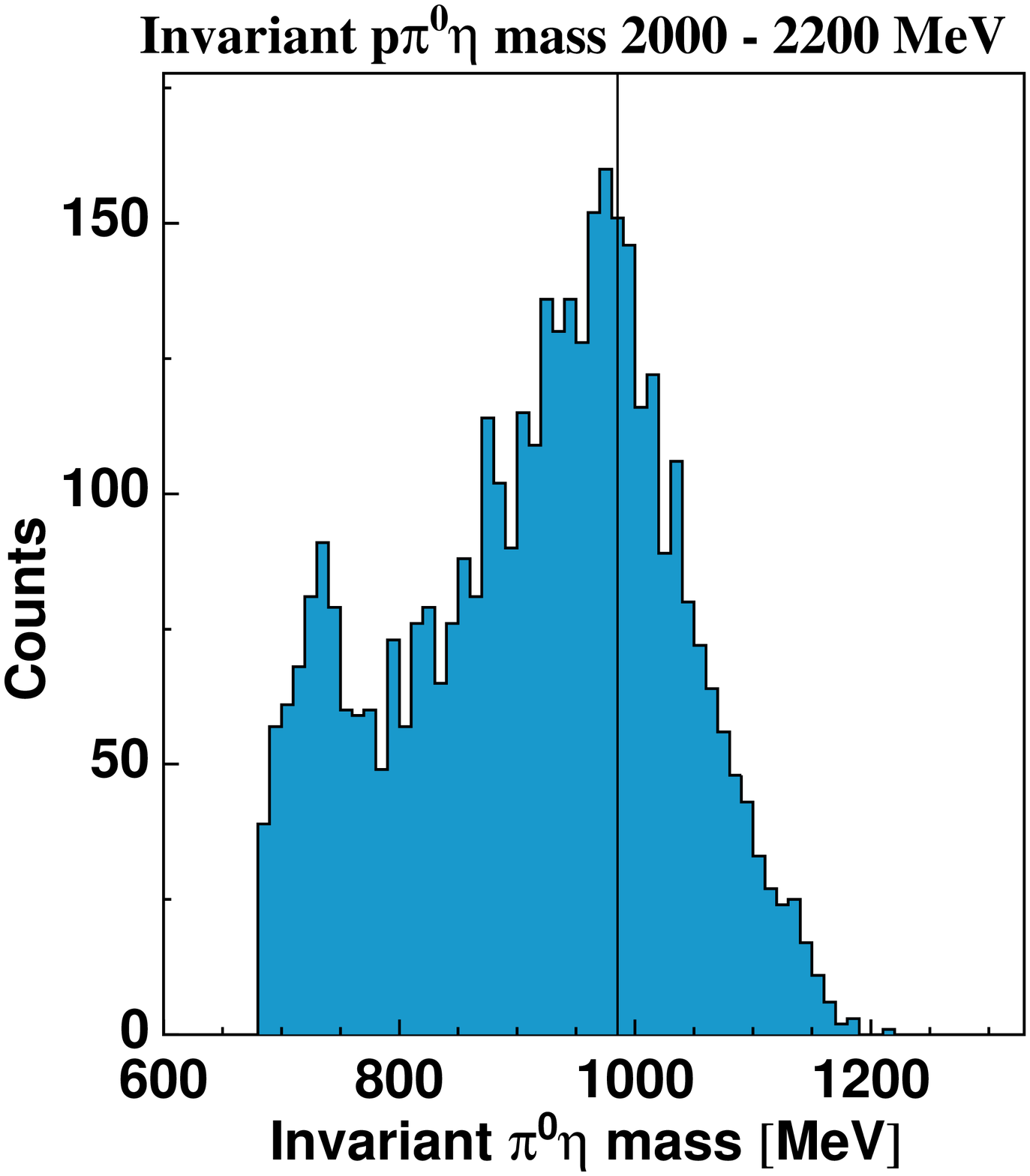}}
\caption{Invariant mass distributions of the $p\pi^o$ and $\eta\pi^o$ pairs 
observed in the reaction $\gamma p\rightarrow p\eta\pi^o$ for two different
ranges of initial states  \cite{Junkersfeld_02}. 
}
\label{fig_65}       
\end{figure}

\vspace*{-9.3cm}
\hspace*{6.0cm}
\begin{minipage}{11.8cm}   
As a further outlook, we present recent data for the decay of higher lying 
resonances via intermediate states resulting in multiple meson reactions. We 
have discussed such a case for the double $\pi^o$ decay of the \d ~resonance. 
Preliminary data from the CB-ELSA collaboration \cite{Junkersfeld_02} extend 
the double $\pi^o$ data to much higher energies. 
Figure~\ref{fig_64} shows the invariant mass spectrum for the 
proton - $\pi^o$ pairs at incident photon energies corresponding to excited 
nucleon states from 2.0 - 2.1 GeV. The $\Delta$ peak is clearly visible, 
and there is a peak corresponding to the second resonance region at
$\sqrt{s}\approx$ 1.5 GeV. Again, background terms are suppressed for neutral 
pions. This could be the first evidence for cascade decays of higher lying 
resonances via $\Delta$ and possibly \d  ~intermediate states. 

Furthermore, first results are available for the 
$\gamma p\rightarrow \pi^o\eta$ reaction.
Invariant mass distributions of the $p\pi^o$ and $\pi^o\eta$ pairs for two
ranges of incident photon energies are summarized in fig.~\ref{fig_65}.
At the low incident energies, only the $\Delta$ peak is seen in $m_{p\pi^o}$
and $m_{\eta\pi^o}$ shows a distribution similar to phase space.
But at higher energies $m_{p\pi^o}$ 
\end{minipage}

%
%
%
%
\noindent{develops} a peak in the second resonance region and 
$m_{\eta\pi^o}$ shows a clear signal from the $a_{o}(980)$ meson. The cascade 
decays involving the $I=0$ $\eta$ meson are particularly useful for the study 
of resonances since the isospin of the initial state is fixed when the isospin 
of the final state is known 
($\Delta\rightarrow \eta\Delta\rightarrow N\eta\pi^o$ or
$N^{\star}\rightarrow \eta N^{\star}\rightarrow N\eta\pi^o$). Predictions
for $\Delta\rightarrow\Delta\eta$ decays in the quark model are given 
by Capstick and Roberts \cite{Capstick_98b}.

In summary, the progress achieved in the field of meson photoproduction over 
the last decade has been substantial and contributes significantly to our
understanding of hadron structure. But we have also seen that the extraction
of resonance properties from the data is more complicated than the analog
problem in nuclear physics. In the future this problem will be attacked from two
directions: more complete experiments, in particular the measurement of
polarization observables, will deliver better constraints for the analyses, and
further improvements of the reaction models will decrease the model related
uncertainties. Meanwhile, questions concerning the in-medium properties of
hadrons, which have been intensely discussed in the context of heavy ion induced
reactions, move into focus. Photon induced reactions do not suffer from 
initial state interactions, so that they can induce reactions throughout
the entire nuclear volume (although final state interactions of the produced
hadrons must be considered). Several photon facilities are presently developing 
programs in this field.

\newpage
\Large
\noindent{{\bf Acknowledgments}}\\
\normalsize
~\\
We thank L. Tiator for the careful reading of part of the manuscript, many
inspiring discussions and instructive suggestions. We gratefully acknowledge
stimulating discussions with H. Arenh\"ovel and I. Strakovsky. 
Part of the figures were kindly provided by 
H.J. Arends, H. Arenh\"ovel, I. Aznauryan, R. Beck, C. Bennhold, J. Ernst,
M. Kotulla, T. Mart, V. Muccifora, L. Tiator, I. Strakovsky, and 
M. Vanderhaeghen.

\section{Appendix}

\large
{\bf 6.1 Properties of Light Unflavored Mesons}
\normalsize
%
%

%
%
%
\begin{table}[ht]
  \caption[Properties of meson]{
    \label{tab_09}
    Properties of non-strange light pseudo-scalar and 
    vector mesons \cite{PDG}. Only the most important decays are listed.
    $E_{thr}$ is the threshold energy for photoproduction from the proton.
    In case of the $\rho$ the threshold energy for the nominal mass is given,
    but due to the large width, production at much lower energies is possible. 
}
  \begin{center}
    \begin{tabular}{|c|c|c|c|c|c|c|lr|}
      \hline 
      & & 
      & {\bf mass}
      & {\bf E$_{thr}$}
      & {\bf width $\Gamma$}
      & {\bf life time}
      & \multicolumn{2}{c|}{\bf decays} \\
      & {\bf $I^G$} 
      & {\bf $J^{PC}$} 
      & {\bf [MeV]}   
      & {\bf [MeV]}
      & {\bf [MeV]}
      & {\bf (sec)}
      & \multicolumn{2}{c|}{\bf (\%)} \\
      \hline \hline 
      $\pi^{\pm}$ & 1$^-$ & 0$^-$ & 139.57 & 149.95 & 2.5$\times$10$^{-14}$
      & 2.6$\times$10$^{-8}$
      & $\mu^{\pm}\nu_{\mu}$ & 100.0 \\
      \hline
      $\pi^o$ & 1$^-$ & 0$^{-+}$ & 134.98 & 144.69 & 7.8$\times$10$^{-6}$
      & 8.4$\times$10$^{-17}$
      & $\gamma\gamma$ & 98.8 \\
      & & & & & & & $\gamma e^+ e^-$ & 1.2\\
      \hline
      $\eta$ & 0$^+$ & 0$^{-+}$ & 547.30 & 706.9 & 1.18$\times$10$^{-3}$
      & 5.6$\times$10$^{-19}$
      & $\gamma\gamma$ & 38.8 \\
      & & & & & & & $\pi^o\pi^o\pi^o$ & 31.9\\
      & & & & & & & $\pi^+\pi^-\pi^o$ & 23.6\\
      & & & & & & & $\pi^+\pi^-\gamma$ & 4.9\\
      \hline
      $\eta^{\prime}$ & 0$^+$ & 0$^{-+}$ & 957.78 & 1446.7 & 0.2
      & 3.3$\times$10$^{-21}$
      & $\pi^+\pi^-\eta$ & 43.7 \\
      & & & & & & & $\rho^o\gamma$ & 30.2\\
      & & & & & & & $\pi^o\pi^o\eta$ & 20.8\\
      \hline
      $\rho$ & 1$^+$ & 1$^{--}$ & 770.0  & (1086.) & 150.2
      & 4.4$\times$10$^{-24}$ 
      & $\pi\pi$ (not $\rho^o\rightarrow\pi^o\pi^o$) & $\approx$100\\
      \hline
      $\omega$ & 0$^-$ & 1$^{--}$ & 781.94  & 1108. & 8.44
      & 7.8$\times$10$^{-23}$ 
      & $\pi+\pi-\pi^o$ & 88.8\\
      & & & & & & & $\pi^o\gamma$ & 8.5\\
      & & & & & & & $\pi^+\pi-$ & 2.21\\
      \hline
      $\phi$ & 0$^-$ & 1$^{--}$ & 1019.413  & 1573. & 4.458
      & 1.5$\times$10$^{-22}$ 
      & $K^+K^-$ & 49.1\\
      & & & & & & & $K^o_L K^o_S$ & 34.1\\
      & & & & & & & $\rho\pi + \pi^+\pi^-\pi^o$ & 15.5\\
      \hline
    \end{tabular}
  \end{center}
\end{table}
%

\vspace*{0.3cm}
\large
\noindent{{\bf 6.2 Helicity Amplitudes}}
\normalsize
\vspace*{0.4cm}
%
%

\noindent{In} the reaction $\gamma N\rightarrow N m_{PS}$, where $m_{PS}$ is 
any pseudo-scalar meson, the helicities involved can have the values
$\lambda_{\gamma}=\pm1$ for the real photon and $\nu_i =\pm 1/2$, 
$\nu_f =\pm 1/2$ for the initial and final state nucleons.
Therefore, 8 matrix elements $H_{\nu_f,\mu =\nu_i-\lambda_{\gamma}}=$
$\langle\nu_f |T|\lambda_{\gamma}\nu_i\rangle$ are possible. 
They are reduced by parity conservation to the four independent 
helicity amplitudes $H_1$ - $H_4$:  
\begin{equation} 
\label{eq:A1}
H_1 =  H_{+1/2,+3/2} =  +H_{-1/2,-3/2}\;\;\;\;\;\;\;\;
H_2 =  H_{+1/2,+1/2} =  -H_{-1/2,-1/2}
\end{equation}
\begin{equation}
\nonumber
H_3 =  H_{-1/2,+3/2} =  -H_{+1/2,-3/2}\;\;\;\;\;\;\;\;
H_4 =  H_{+1/2,-1/2} =  +H_{-1/2,+1/2}.\nonumber
\end{equation}
The relation between CGLN- and helicity amplitudes is given by 
\cite{Walker_69}):
\begin{eqnarray}
\label{eq:A2}
H_1(\Theta ,\Phi) & = &
\frac{-1}{\sqrt{2}}e^{i\Phi}sin(\Theta^{\star})
cos\left(\frac{\Theta^{\star}}{2}\right)
(F_3+F_4)\\
H_2(\Theta ,\Phi) & = &
\sqrt{2}cos\left(\frac{\Theta^{\star}}{2}\right)
\left[(F_2-F_1)+\frac{1}{2}(1-cos(\Theta^{\star}))(F_3-F_4)\right]\nonumber\\
H_3(\Theta ,\Phi) & = &
\frac{1}{\sqrt{2}}e^{2i\Phi}sin(\Theta^{\star})
sin\left(\frac{\Theta^{\star}}{2}\right)
(F_3-F_4)\nonumber\\
H_4(\Theta ,\Phi) & = &
\sqrt{2}e^{i\Phi}sin\left(\frac{\Theta^{\star}}{2}\right)
\left[(F_1+F_2)+\frac{1}{2}(1+cos(\Theta^{\star}))(F_3+F_4)\right]\nonumber
\end{eqnarray}
The expression of the physical observables in terms of the helicity
amplitudes is particularly simple \cite{Walker_69}:
\begin{eqnarray}
\label{eq:A3}
\frac{d\sigma}{d\Omega} & = &
\frac{1}{2}\frac{q^{\star}}{k^{\star}}(H_1^2+H_2^2+H_3^2+H_4^2)\\
\Sigma & = &
\frac{q^{\star}}{k^{\star}}\;
Re(H_4^{\star}H_1-H_3^{\star}H_2)/\frac{d\sigma}{d\Omega}\\
R & = &
-\frac{q^{\star}}{k^{\star}}\;
Im((H_3^{\star}H_1+H_4^{\star}H_2)/\frac{d\sigma}{d\Omega}\\
T & = &
\frac{q^{\star}}{k^{\star}}\;
Im((H_2^{\star}H_1+H_4^{\star}H_3)/\frac{d\sigma}{d\Omega}
\end{eqnarray}   
where $q^{\star}$, $k^{\star}$ are the cm momenta of meson and photon, and
$\Sigma$, $R$ and $T$ are photon beam asymmetry, recoil polarization and
target asymmetry.

\noindent{Sometimes} `transversity' amplitudes are used instead of the 
helicity amplitudes (see e.g. \cite{Barker_75}). The only difference is, 
that in this case the axis of spin quantization is the transverse, rather 
than the particle's momentum direction. This means that the axis of quantization
is chosen perpendicular to the scattering plane. A certain advantage of this
representation is that the differential cross section and the single
polarization observables can be expressed as linear combinations of
the squares of the amplitudes rather than as bilinear functions of the 
amplitudes. 

\vspace*{0.6cm}
\large
\noindent{\bf{6.3 Multipole Expansions}}
\normalsize
\vspace*{0.3cm}
%
%

%
%
%
\begin{table}[hhh]
  \caption[Lowest order multipole amplitudes]{
    \label{tab_01}
    Lowest order multipole ampl. for the photoproduction
    of pseudo-scalar mesons (x=$cos(\Theta^{\star})$).
}
  \begin{center}
    \begin{tabular}{|c|c|c|c|c|c|}
      \hline 
      photon & 
      initial state & 
      interm. & 
      final state & 
      multi- & \\
      M-pole & 
      $(L_{\gamma}^P,J_N^P)$ &
      state $J_{N^{\star}}^P$ &
      $(J_N^P,L_{\eta}^P)$ &
      pole & 
      $(k^{\star}/q^{\star})d\sigma /d\Omega$ \\
      \hline\hline
      & & & & & \\
      E1 & (1$^-$,$\frac{1}{2}^+$) & $\frac{1}{2}^-$ & ($\frac{1}{2}^+$,0$^-$) &
      $E_{o+}$ & $|E_{o+}|^2$ \\ 
      & & & & & \\
      & & $\frac{3}{2}^-$ & ($\frac{1}{2}^+$,2$^-$) &
      $E_{2-}$ & $\frac{1}{2}|E_{2-}|^2(5-3x^2)$ \\ 
      & & & & & \\
      \hline
      & & & & & \\
      M1 & (1$^+$,$\frac{1}{2}^+$) & $\frac{1}{2}^+$ & ($\frac{1}{2}^+$,1$^+$) &
      $M_{1-}$ & $|M_{1-}|^2$ \\ 
      & & & & & \\
      & & $\frac{3}{2}^+$ & ($\frac{1}{2}^+$,1$^+$) &
      $M_{1+}$ & $\frac{1}{2}|M_{1+}|^2(5-3x^2)$ \\ 
      & & & & & \\
      \hline
      & & & & & \\
      E2 & (2$^+$,$\frac{1}{2}^+$) & $\frac{3}{2}^+$ & ($\frac{1}{2}^+$,1$^+$) &
      $E_{1+}$ & $\frac{9}{2}|E_{1+}|^2(1+x^2)$ \\ 
      & & & & & \\
      & & $\frac{5}{2}^+$ & ($\frac{1}{2}^+$,3$^+$) &
      $E_{3-}$ & $\frac{9}{2}|E_{3-}|^2(1+6x^2-5x^4)$ \\ 
      & & & & & \\
      \hline
      & & & & & \\
      M2 & (2$^-$,$\frac{1}{2}^+$) & $\frac{3}{2}^-$ & ($\frac{1}{2}^+$,2$^-$) &
      $M_{2-}$ & $\frac{9}{2}|M_{2-}|^2(1+x^2)$ \\ 
      & & & & & \\
      & & $\frac{5}{2}^-$ & ($\frac{1}{2}^+$,2$^-$) &
      $M_{2+}$ & $\frac{9}{2}|M_{2+}|^2(1+6x^2-5x^4)$ \\ 
      & & & & & \\
      \hline
    \end{tabular}
  \end{center}
\end{table}
%

\noindent{The} partial wave expansion of the helicity amplitudes is given by
\cite{Walker_69}:
\begin{eqnarray}
\label{eq:A4}
H_1(\Theta^{\star},\Phi) = &
\frac{1}{\sqrt{2}}e^{i\Phi} sin(\Theta^{\star}) cos(\frac{\Theta^{\star}}{2}) &
\sum_{l=0}^{\infty}[B_{l+}-B_{(l+1)-}] 
P_l^{\prime\prime}(cos(\Theta^{\star}))
-P_{l+1}^{\prime\prime}(cos(\Theta^{\star}))]\\
H_2(\Theta^{\star},\Phi) = &
\sqrt{2}cos(\frac{\Theta^{\star}}{2}) &
\sum_{l=0}^{\infty}[A_{l+}-A_{(l+1)-}] 
P_l^{\prime}(cos(\Theta^{\star}))
-P_{l+1}^{\prime}(cos(\Theta^{\star}))]\nonumber\\
H_3(\Theta^{\star},\Phi) = &
\frac{1}{\sqrt{2}}e^{2i\Phi}sin(\Theta^{\star})sin(\frac{\Theta^{\star}}{2}) &
\sum_{l=0}^{\infty}[B_{l+}-B_{(l+1)-}] 
P_l^{\prime\prime}(cos(\Theta^{\star}))
+P_{l+1}^{\prime\prime}(cos(\Theta^{\star}))]\nonumber\\
H_4(\Theta^{\star},\Phi) = &
\sqrt{2}e^{i\Phi}sin(\frac{\Theta^{\star}}{2}) &
\sum_{l=0}^{\infty}[A_{l+}+A_{(l+1)-}] 
P_l^{\prime}(cos(\Theta^{\star}))
+P_{l+1}^{\prime}(cos(\Theta^{\star}))]\nonumber
\end{eqnarray}
where the helicity elements $A_{l\pm}$, $B_{l\pm}$ correspond to transitions
with nucleon - meson relative orbital angular momentum $l$, final state total
angular momentum $J=l\pm 1/2$ and $\gamma N$ initial state helicity 1/2 for 
$A_{l\pm}$ and 3/2 for $B_{l\pm}$. 

\noindent{Examples} for the lowest order multipoles of the CGLN-amplitudes 
with the relevant quantum numbers and angular distributions are summarized 
in table \ref{tab_01}.
\newpage
\noindent{The} helicity elements are related to the multipole amplitudes via:
\begin{equation}
\label{eq:A5}
A_{l+}  =  \frac{1}{2}[(l+2)E_{l+}+lM_{l+}]\;\;\;\;\;\;\;\;\ 
A_{(l+1)-}  =  \frac{1}{2}[-lE_{(l+1)-}+(l+2)M_{(l+1)-}]
\end{equation}
\begin{equation}
B_{l+}  =  E_{l+}-M_{l+}\;\;\;\;\;\;\;\;\;
B_{(l+1)-}  =  E_{(l+1)-}+M_{(l+1)-}\;\;.\nonumber
\end{equation}

\noindent{An} advantage of this parameterization is the close connection between the
helicity elements and the electromagnetic resonance couplings:
\begin{eqnarray}
\label{eq:A6}
A_{1/2} & = &
\sqrt{2\pi\alpha /k^{\star}}\langle N^{\star},J_z=+\frac{1}{2}|J_{em}|
                                                    N,S_z=-\frac{1}{2}\rangle\\
A_{3/2} & = &
\sqrt{2\pi\alpha /k^{\star}}\langle N^{\star},J_z=+\frac{3}{2}|J_{em}|
                                             N,S_z=+\frac{1}{2}\rangle\nonumber
\end{eqnarray}
which for a Breit-Wigner form of the resonances is given by \cite{PDG}:
\begin{eqnarray}
\label{eq:A7}
A_{1/2} & = & \mp (1/C_{Nm})\sqrt{(2J+1)\pi 
\frac{q^{\star}}{k^{\star}}
\frac{M_R}{m_N}\frac{\Gamma_R^2}{\Gamma_m}}
\;Im[A_{l\pm}(W=M_R)]\\
A_{3/2} & = & \pm (1/C_{Nm})\sqrt{(2J+1)\pi 
\frac{q^{\star}}{k^{\star}}
\frac{M_R}{m_N}\frac{\Gamma_R^2}{\Gamma_m}}\;
\sqrt{(2J-1)(2J+3)/16}\;Im[B_{l\pm}(W=M_R)]\nonumber
\end{eqnarray}
where $m_N$ is the nucleon mass, $M_R$, $\Gamma_R$ are resonance position
and width, respectively, $\Gamma_m$ is the partial width for the used 
decay channel, and $J$ the momentum of the resonance. For pion photoproduction 
$C_{Nm}$ is the Clebsch-Gordan coefficient for the decay of the resonance
($I=1/2$ $N^{\star}$ or $I=3/2$ $\Delta$) into the relevant N${\pi}$ charge 
state (note that due to the phase convention eq.~\-(\ref{eq:A8}) an additional
minus sign appears for $\pi^+$ mesons). $C_{Nm}$ equals -1 for 
$\eta$-photoproduction.

\vspace*{0.6cm}
\large
\noindent{\bf{6.4 Isospin Amplitudes}}
\normalsize
\vspace*{0.3cm}
%
%
   
\noindent{In} the notation $|I,I_3\rangle$ the isospin part of the 
wave-functions for the nucleon and the pion (or any other isovector meson) is 
written as:
\begin{equation}
\label{eq:A8}
|p\rangle = |\frac{1}{2},+\frac{1}{2}\rangle
~~~~|n\rangle = |\frac{1}{2},-\frac{1}{2}\rangle 
\end{equation}
\begin{equation}
|\pi^+\rangle = -|1,+1\rangle 
~~~~|\pi^o\rangle = |1,0\rangle
~~~~|\pi^-\rangle = |1,-1\rangle\;.\nonumber 
\end{equation}
The nucleon - pion states are then given by:
\begin{eqnarray}
\label{eq:A9}
|\pi^+ p\rangle & = & -|\frac{3}{2},+\frac{3}{2}\rangle\\
|\pi^o p\rangle & = & \sqrt{\frac{2}{3}}|\frac{3}{2},+\frac{1}{2}\rangle
                  -\sqrt{\frac{1}{3}}|\frac{1}{2},+\frac{1}{2}\rangle\nonumber\\
|\pi^- p\rangle & = & \sqrt{\frac{1}{3}}|\frac{3}{2},-\frac{1}{2}\rangle
                  -\sqrt{\frac{2}{3}}|\frac{1}{2},-\frac{1}{2}\rangle\nonumber\\
|\pi^+ n\rangle & = & -\sqrt{\frac{1}{3}}|\frac{3}{2},+\frac{1}{2}\rangle
                  -\sqrt{\frac{2}{3}}|\frac{1}{2},+\frac{1}{2}\rangle\nonumber\\
|\pi^o n\rangle & = & \sqrt{\frac{2}{3}}|\frac{3}{2},-\frac{1}{2}\rangle
                  +\sqrt{\frac{1}{3}}|\frac{1}{2},-\frac{1}{2}\rangle\nonumber\\
|\pi^- n\rangle & = & |\frac{3}{2},-\frac{3}{2}\rangle\nonumber                   
\end{eqnarray}
where the first and last state cannot occur in photoproduction reactions.
The isospin amplitudes of eq.~\-(\ref{eq:iso_2}) follow from this wave-functions
and the definitions in eq.~\-(\ref{eq:iso_1}). 

\noindent{Different} sets of the isospin amplitudes are used in the literature.
Alternative sets of amplitudes are:
\begin{eqnarray}
\label{eq:A10}
A(\gamma p\rightarrow\pi^+ n) = &
\sqrt{2}\;(A^0+A^-)\;\;\;\;\;\;\;\;\;\;\; 
A(\gamma p\rightarrow\pi^o p) = &
(A^++A^0)\\
A(\gamma n\rightarrow\pi^- p) = &
\sqrt{2}\;(A^0-A^-)\;\;\;\;\;\;\;\;\;\;\; 
A(\gamma n\rightarrow\pi^o n) = &
(A^+-A^0)\;,\nonumber
\end{eqnarray}
with:
\begin{equation}
\label{eq:A11}
A^{IS} = -3\sqrt{3}A^0,\;\;\;\;\;\;
A^{IV} = \sqrt{\frac{1}{3}}(A^++2A^-),\;\;\;\;\;\;
A^{V3} = \sqrt{\frac{2}{3}}(A^+-A^-)\;,
\end{equation}
and:
\begin{eqnarray}
\label{eq:A12}
A(\gamma p\rightarrow\pi^+ n) & = &
\sqrt{2}\;(A^{(0)}+\frac{1}{3}A^{(\frac{1}{2})}-\frac{1}{3}A^{(\frac{3}{2})}) \\
A(\gamma p\rightarrow\pi^o p) & = &
~~~~~A^{(0)}+\frac{1}{3}A^{(\frac{1}{2})}+\frac{2}{3}A^{(\frac{3}{2})}\nonumber \\
A(\gamma n\rightarrow\pi^- p) & = &
\sqrt{2}\;(A^{(0)}-\frac{1}{3}A^{(\frac{1}{2})}+\frac{1}{3}A^{(\frac{3}{2})})\nonumber \\
A(\gamma n\rightarrow\pi^o n) & = &
~~~~~-A^{(0)}+\frac{1}{3}A^{(\frac{1}{2})}+\frac{2}{3}A^{(\frac{3}{2})}\;,\nonumber 
\end{eqnarray}
with:
\begin{equation}
\label{eq:A13}
A^{(0)} =  A^0,\;\;\;\;\;\;
A^{(\frac{1}{2})} =  A^++2A^-,\;\;\;\;\;\;
A^{(\frac{3}{2})} =  A^+-A^-\;,
\end{equation}
where sometimes the following abbreviations are used:
\begin{equation}
\label{eq:A14}
_pA^{1/2} = A^{(0)}+\frac{1}{3}A^{(\frac{1}{2})}\;\;\;\;\;\;\;
_nA^{1/2} = A^{(0)}-\frac{1}{3}A^{(\frac{1}{2})}\;.
\end{equation}

\end{document}